\documentclass[a4paper,11pt]{article}
\pdfoutput=1 % if your are submitting a pdflatex (i.e. if you have images in pdf, png or jpg format)

\usepackage{jheppub}
\usepackage[T1]{fontenc} % if needed
\usepackage{lmodern}
\usepackage{amsfonts}
\usepackage{graphicx}
\usepackage{amsthm}
\usepackage{amsmath}
\usepackage{amssymb}
\usepackage{mathrsfs}
\usepackage{bbold}
\usepackage{mathtools}
\usepackage{slashed}
\usepackage{float}
\usepackage{epstopdf}
\usepackage{etoolbox}
\makeatletter
\patchcmd{\maketitle}{\@fpheader}{$ $}{}{}
\makeatother
\usepackage{xspace}
\usepackage{tabularx}
\usepackage{amsbsy}
\usepackage{array}
\usepackage{bm}
\usepackage{cases}
\usepackage{lscape}
\usepackage{empheq}
\usepackage{longtable}
\usepackage{cancel}

\usepackage{subcaption}
\usepackage[labelformat=simple]{subcaption}

\newcolumntype{L}[1]{>{\raggedright\let\newline\\\arraybackslash\hspace{0pt}}m{#1}}
\newcolumntype{C}[1]{>{\centering\let\newline\\\arraybackslash\hspace{0pt}}m{#1}}
\newcolumntype{R}[1]{>{\raggedleft\let\newline\\\arraybackslash\hspace{0pt}}m{#1}}

\newcommand{\mathematica}{\textsc{Mathematica}\xspace}
\newcommand{\feyncalc}{\textsc{FeynCalc}\xspace}
\newcommand{\packageX}{\textsc{Package-X}\xspace}
\newcommand{\jaxodraw}{\textsc{JaxoDraw}\xspace}

\allowdisplaybreaks[2]

\title{OPE of Green Functions of Chiral Currents}

\author{Tom\'{a}\v{s} Kadav\'{y},}
\author{Karol Kampf}
\author{and Ji\v{r}\'{i} Novotn\'{y}}

\affiliation{Institute of Particle and Nuclear Physics, Charles University\\V Hole\v{s}ovi\v{c}k\'{a}ch 2, 180 00 Prague 8, Czech Republic}

\emailAdd{\{kadavy, kampf, novotny\}@ipnp.mff.cuni.cz}

\abstract{In this paper we investigate the high-energy behavior of two-point and three-point Green functions of the QCD chiral currents and densities using the framework of the operator product expansion in the chiral limit. In detail, we study the contributions of the quark, gluon, quark-gluon and four-quark condensates to all the relevant non-vanishing three-point correlators.}

\begin{document}

\maketitle

\flushbottom

%%%%%%%%%%%%%%%%%%%%%%%%%%%%%%%%%%%%%%%%%%%%%%%%%%%%%%%%%%%%%%%%%%%%%%%%%%%%%%%%%%%%%%%%%%%%%%%%%%%%%%%%%
%%%%%%%%%%%%%%%%%%%%%%%%%%%%%%%%%%%%%%%%%%%%%%%%%%%%%%%%%%%%%%%%%%%%%%%%%%%%%%%%%%%%%%%%%%%%%%%%%%%%%%%%%
%%% Section: Introduction
%%%%%%%%%%%%%%%%%%%%%%%%%%%%%%%%%%%%%%%%%%%%%%%%%%%%%%%%%%%%%%%%%%%%%%%%%%%%%%%%%%%%%%%%%%%%%%%%%%%%%%%%%
%%%%%%%%%%%%%%%%%%%%%%%%%%%%%%%%%%%%%%%%%%%%%%%%%%%%%%%%%%%%%%%%%%%%%%%%%%%%%%%%%%%%%%%%%%%%%%%%%%%%%%%%%

\section{Introduction}\label{sec:intro}
There is countless experimental evidence showing that Quantum chromodynamics (QCD) is the correct fundamental theory of strong interactions. However, the perturbative approach fails in the low-energy region of the hadronic spectrum, i.e. for energies less than 2 GeV, where QCD becomes non-perturbative.
A possible way is to use at low energies an effective field theory that would be built upon the relevant degrees of freedom, i.e. the mesons and the baryons. The situation here is not that simple, though, since such a theory in the full-energy region is not known from the first principles. Nevertheless, in the region of energies typically less than $M_{\rho}$, with $M_{\rho}$ being the mass of the $\rho(770)$ meson, we have an effective field theory of QCD, the Chiral perturbation theory ($\chi$PT) \cite{Coleman:1969sm,Weinberg:1978kz,Gasser:1983yg,Gasser:1984gg}.
Inspired by the large-$N_{c}$ limit, we can construct the effective theory for an intermediate energy region that also satisfies all symmetries of the underlying theory. This effective theory, Resonance chiral theory (R$\chi$T), is relevant for energies within the bounds of $M_{\rho}\leq E\leq 2\,\mathrm{GeV}$ \cite{Ecker:1988te,Cirigliano:2006hb}. For higher energies, R$\chi$T loses its convergence and can not be properly used because of the higher meson states that become significant in hadron dynamics.

The phenomenological Lagrangian approach based on large-$N_{c}$ and the chiral symmetry was first introduced in 1989 in \cite{Ecker:1988te}. It was further developed and enlarged both for the even-parity sector and the odd-parity sector \cite{Cirigliano:2006hb,Kampf:2006yf,Masjuan:2007ay,Geng:2008ag,Jiang:2009uf,Kampf:2011ty,Nieves:2011gb,Czyz:2017veo,Roig:2013baa}. Important questions connected with the renormalization within R$\chi$T were recently studied e.g. in \cite{Rosell:2009yb,SanzCillero:2009pt,Kampf:2009jh,Pich:2010sm,Terschlusen:2013iqa,Bruns:2013tja,Terschlusen:2016kje}.
R$\chi$T increases the number of degrees of freedom of Chiral perturbation theory by including massive $\mathrm{U}(3)$ multiplets of vector $V(1^{--})$, axial-vector $A(1^{++})$, scalar $S(0^{++})$ and pseudoscalar $P(0^{-+})$ resonances. Interactions within these types of channels can be studied with the help of the Green functions of the chiral currents that for such reason represent a powerful tool in order to obtain physical observables of the theory. Comparing the theoretical predictions with experimental measurements, we can determine the values of the parameters of the theory and obtain a more comprehensive understanding of the behaviour of the processes and the theory itself.

The motivation behind this paper is to be able to provide the matching of the QCD operator product expansion (OPE) with the R$\chi$T, which allows us to compare the effective field theory at the low energies with the description at high energies. However, the matching itself is not a subject of this work since it will be studied in detail in our future paper \cite{KadavyGF:2020a}. For a very recent application of the short distance constraints see e.g. \cite{Masjuan:2020jsf}.

Within the OPE framework, all the three-point Green functions are studied in terms of the QCD condensates with dimension $D\leq 6$, In this paper we provide a complete OPE description of all the three-point correlators to the first nonvanishing order in $\alpha_{s}$.
It is important to mention that to the best of our knowledge, the OPE has been studied extensively only for some of the Green functions and mainly only with the emphasis on the quark condensate (see e.g. \cite{Moussallam:1994xp} and \cite{Jamin:2008rm}). To this end, we have recalculated some of the known contributions independently, while the remaining contributions in this paper were calculated for the first time here.

This paper is organized as follows. After introducing our notation, we present a short review of the Green functions, operator product expansion framework and the QCD condensates in Section \ref{sec:greenfunctions}. In Section \ref{sec:ope-condensates}, we briefly remind the reader some basic concepts, such as the Fock-Schwinger gauge, which is crucial for the calculations of the contributions of the QCD condensates to the Green functions, and important propagation formulas that are needed to rewrite the nonlocal condensates into the local ones. Sections \ref{sec:perturbative}, \ref{sec:quark-condensate}, \ref{sec:gluon-condensate}, \ref{sec:quark-gluon-condensate} and \ref{sec:four-quark-condensate} are devoted to the calculations and the results of the perturbative contribution and the contributions of the quark, gluon, quark-gluon and four-quark condensates to the three-point Green functions. These sections are the main outcomes of this paper.

After the Conclusion, an extensive set of appendices follows. Appendix \ref{sec:fourier_transform} contains a review of the important Fourier transforms that are useful for our calculations. Appendix \ref{sec:derivation} is devoted to a detailed derivation of the propagation formulas that are necessary in order to obtain the effective contributions for some of the QCD condensates. Appendix \ref{sec:appendix-2pt} follows, in which we present the results for the QCD condensate contributions to all the two-point Green functions. Appendices \ref{sec:VVA_AAA_decompositions} and \ref{sec:AAV_VVV_decompositions} provide a detailed derivation of the decompositions of the $\langle VVA\rangle$, $\langle AAA\rangle$, $\langle AAV\rangle$ and $\langle VVV\rangle$ Green functions, since their tensor structures is not that trivial as in other cases, and thus deserve a special attention.

Symbolic computation have been performed with a use of \mathematica, \feyncalc \cite{Mertig:1990an,Shtabovenko:2016sxi} and \packageX \cite{Patel:2015tea,Patel:2016fam}. The Feynman diagrams have been drawn using \jaxodraw \cite{Binosi:2008ig}.

\subsection{Notation}
Here, we present the following notation that is used throughout the paper.
\begin{itemize}
\item The three-point Green functions are denoted generally as $\langle\mathcal{O}_{1}\mathcal{O}_{2}\mathcal{O}_{3}\rangle$, where $\mathcal{O}_{i}$ are composite local operators (see \eqref{eq:operator-definice} and the introductory paragraph in the next section for details). By this notation we assume that the operator $\mathcal{O}_{1}$ is evaluated at the space-time point $x$ and carries momentum $p$ and the flavor index $a$. If it also carries the Lorentz index, we assume it to be $\mu$. In other words, the first operator in the designation of the Green function, $\mathcal{O}_{1}$, is associated with a set of ($\mu$, $a$, $x$, $p$). Similarly, the operators $\mathcal{O}_{2}$ and $\mathcal{O}_{3}$ are associated with ($\nu$, $b$, $y$, $q$) and ($\rho$, $c$, $z$, $r$), respectively.
\item We consider all three momenta to be ingoing, which gives the momentum conservation
\begin{equation}
p+q+r=0\,,\label{eq:4-momentumConservation}
\end{equation}
that allows us to express the scalar products of $p$, $q$ and $r$ in terms of squares of momenta, such as
\begin{align}
p\hspace{-1pt}\cdot\hspace{-1pt}q&=\frac{1}{2}(-p^{2}-q^{2}+r^{2})\,,\label{eq:product1}\\
p\hspace{-1pt}\cdot\hspace{-1pt}r&=\frac{1}{2}(-p^{2}+q^{2}-r^{2})\,,\label{eq:product2}\\
q\hspace{-1pt}\cdot\hspace{-1pt}r&=\frac{1}{2}(p^{2}-q^{2}-r^{2})\,.\label{eq:product3}
\end{align}
\item Accordingly to the fact that all momenta are considered as ingoing, the Fourier transform is defined to be
\begin{align}
\widetilde{F}(p)&=\int\mathrm{d}^{4}x\,e^{-i p\cdot x}F(x)\,,\label{eq:FT-1}\\
F(x)&=\int\frac{\mathrm{d}^{4}p}{(2\pi)^{4}}e^{i p\cdot x}\widetilde{F}(p)\,.\label{eq:FT-2}
\end{align}
Useful Fourier transforms, relevant for this paper, are listed in the Appendix \ref{sec:fourier_transform}.
\item The powers of momenta are denoted as $p^{2n}\equiv(p^{2})^{n}=(p_{\mu}p^{\mu})^{n}$ with $n$ being the integer.
\item Spinor indices are denoted as small Latin letters $(i,j,k,l)$. Color indices are denoted as small Greek letters $(\alpha,\beta,\gamma,\delta)$. Flavor indices corresponding to the fundamental representation of the flavor group are denoted as capital Latin letters $(A,B,C,D)$. SU(3) indices of the adjoint representation are denoted as small Latin letters $(a,b,c,d)$.
\item We use $\varepsilon^{0123}=+1$.
\item A short-hand Veltman's Schoonschip notation for the contractions of Levi-Civita tensor with the components of momenta is used, for example $\varepsilon^{\mu\nu\alpha(p)}\equiv\varepsilon^{\mu\nu\alpha\beta}p_{\beta}$ or $\varepsilon^{\mu\nu(p)(q)}\equiv\varepsilon^{\mu\nu\alpha\beta}p_{\alpha}q_{\beta}$. Similarly for the sigma tensor $\sigma^{\alpha\beta}=\frac{i}{2}[\gamma^{\alpha},\gamma^{\beta}]$ we denote $\sigma^{(p)\mu}\equiv\sigma^{\alpha\mu}p_{\alpha}$ and $\sigma^{(p)(q)}\equiv\sigma^{\alpha\beta}p_{\alpha}q_{\beta}$.
\item SU(3) generators $T^{a}$ are defined as $T^{a}=\frac{1}{2}\lambda^{a}$, with $\lambda^{a}$ being the Gell-Mann matrices, $a=1,\ldots,8$. Such normalization implies that $\mathrm{Tr}(T^{a}T^{b})=\frac{1}{2}\delta^{ab}$.
\item Symbols $[\bullet,\bullet]$ and $\lbrace\bullet,\bullet\rbrace$ stand for the commutator and anticommutator, respectively.
\item The covariant derivative in the fundamental representation is taken to be
\begin{equation}
\nabla_{\mu}=\partial_{\mu}+i g_{s}\mathcal{A}_{\mu}\,,\label{eq:covariant_derivative}
\end{equation}
where $\mathcal{A}_{\mu}=\mathcal{A}_{\mu}^{a}T^{a}$ is the gluon field.
\item The commutator of these derivatives gives the gluon field strength tensor,
\begin{equation}
[\nabla_{\mu},\nabla_{\nu}]=i g_{s}G_{\mu\nu}\,,\label{eq:commutator_G}
\end{equation}
with $G_{\mu\nu}=G_{\mu\nu}^{a}T^{a}$ given as
\begin{align}
G_{\mu\nu}&=\partial_{\mu}\mathcal{A}_{\nu}-\partial_{\nu}\mathcal{A}_{\mu}+i g_{s}[\mathcal{A}_{\mu},\mathcal{A}_{\nu}]\,,\label{eq:gluon_field_strength_tensor}\\
G_{\mu\nu}^{a}&=\partial_{\mu}\mathcal{A}_{\nu}^{a}-\partial_{\nu}\mathcal{A}_{\mu}^{a}-g_{s}f^{abc}\mathcal{A}_{\mu}^{b}\mathcal{A}_{\nu}^{c}\,.\label{eq:gluon_field_strength_tensor_a}
\end{align}
\item The chiral limit is considered throughout the paper.
\end{itemize}

%%%%%%%%%%%%%%%%%%%%%%%%%%%%%%%%%%%%%%%%%%%%%%%%%%%%%%%%%%%%%%%%%%%%%%%%%%%%%%%%%%%%%%%%%%%%%%%%%%%%%%%%%
%%%%%%%%%%%%%%%%%%%%%%%%%%%%%%%%%%%%%%%%%%%%%%%%%%%%%%%%%%%%%%%%%%%%%%%%%%%%%%%%%%%%%%%%%%%%%%%%%%%%%%%%%
%%% Section: Green Functions of Chiral Currents
%%%%%%%%%%%%%%%%%%%%%%%%%%%%%%%%%%%%%%%%%%%%%%%%%%%%%%%%%%%%%%%%%%%%%%%%%%%%%%%%%%%%%%%%%%%%%%%%%%%%%%%%%
%%%%%%%%%%%%%%%%%%%%%%%%%%%%%%%%%%%%%%%%%%%%%%%%%%%%%%%%%%%%%%%%%%%%%%%%%%%%%%%%%%%%%%%%%%%%%%%%%%%%%%%%%

\section{Green Functions of Chiral Currents}\label{sec:greenfunctions}
The Green functions are defined as the vacuum expectation values of the time ordered products of the composite local operators. The standard definition of the $n$-point correlator reads\footnote{In our case, the symbol $\vert 0\rangle$ stands for the nonperturbative QCD vacuum. For clarity, however, we will omit showing the vacuum state from now on, i.e. we symbolically define $\langle 0\vert\bullet\vert 0\rangle\equiv\langle\bullet\rangle$.}
\begin{align}
\Pi&_{\mathcal{O}_{1}\ldots\mathcal{O}_{n}}(p_{1},\ldots,p_{n-1};p_{n})=\label{eq:gf-definition}\\
&=\int\mathrm{d}^{4}x_{1}\ldots\mathrm{d}^{4}x_{n-1}\,e^{-i(p_{1}\cdot x_{1}+\ldots+p_{n-1}\cdot x_{n-1})}\,\big\langle 0\big\vert\mathrm{T}\,\mathcal{O}_{1}(x_{1})\ldots\mathcal{O}_{n-1}(x_{n-1})\mathcal{O}_{n}(0)\big\vert 0\big\rangle\,,\nonumber
\end{align}
where the object on the left-hand side is the Green function in the momentum representation, which is obtained by performing Fourier transform on the Green function in the coordinate representation.
$\mathrm{T}$ stands for the time-ordering and all the indices are suppressed in \eqref{eq:gf-definition} for simplicity. The translation invariance was used to set the coordinate of the $n^{\mathrm{th}}$ operator into the origin.
As we will see in Subsection \ref{ssec:translation}, the translation invariance of the Green function may not always be apparent, so it is always useful to check whether such a requirement is satisfied. For this reason, we always take all the operators to be at nonzero space-time points in the intermediate stages of our calculations.

In our case, we consider the local composite operators in \eqref{eq:gf-definition} to be represented by the octets of the chiral scalar $S^{a}(x)=\overline{q}(x)T^{a}q(x)$ and pseudoscalar $P^{a}(x)=i\overline{q}(x)\gamma_{5}T^{a}q(x)$ densities, and the vector $V_{\mu}^{a}(x)=\overline{q}(x)\gamma_{\mu}T^{a}q(x)$ and axial-vector $A_{\mu}^{a}(x)=\overline{q}(x)\gamma_{\mu}\gamma_{5}T^{a}q(x)$ currents, where $q(x)$ stands for the triplet of the lightest quarks,
\begin{equation}
q=(u,d,s)^{\mathrm{T}}\,.\label{eq:quark_triplet}
\end{equation}
These operators can be rewritten generally as
\begin{equation}
\mathcal{O}_{1}^{a}(x)=\overline{q}_{i,\alpha}^{A}(x)(\Gamma_{1})_{ik}(T^{a})^{AB}q_{k,\alpha}^{B}(x)\,,\label{eq:operator-definice}
\end{equation}
where the spin matrix $\Gamma_{1}$ denotes the unit and the Dirac matrices, $\Gamma_{1}\in(1,\gamma_{\mu},i\gamma_{5},\gamma_{\mu}\gamma_{5})$. We have also explicitly written out the spinor ($i$, $k$), flavor ($A$, $B$) and color ($\alpha$) indices of the quark fields and the flavor matrix $T^{a}$. According to the notation presented above, we associate the momentum $p$ to the operator \eqref{eq:operator-definice}, since it is evaluated at space-time point $x$ and carries the flavor index $a$. Similarly, we associate momenta $q$ and $r=-(p+q)$ with the operators $\mathcal{O}_{2}^{b}(y)$ and $\mathcal{O}_{3}^{c}(z)$, respectively.

\subsection{Classification}
Formally, one can arrange 20 different combinations of three composite operators selected from the currents $V$ and $A$ and the densities $S$ and $P$. However, the Lorentz covariance and invariance of QCD with respect to parity and/or time reversal forbid an existence of some of the combinations. Specifically, the correlators $\langle SSP\rangle$, $\langle PPP\rangle$, $\langle VSP\rangle$, $\langle ASS\rangle$ and $\langle APP\rangle$ are not allowed to exist in QCD.

Considering the relevant 15 Green functions, we introduce a simple division of the correlators into two sets. Such division will be useful later when it comes to a study of the respective QCD condensates contributions, since all the correlators of the specific set have, in the chiral limit, nonvanishing contribution from the same QCD condensates.

The classification is as follows:
\begin{itemize}
\item Set 1: The correlators with the perturbative contribution in the chiral limit:
\begin{itemize}
\item $\langle ASP\rangle$, $\langle VSS\rangle$, $\langle VPP\rangle$, $\langle VVA\rangle$, $\langle AAA\rangle$, $\langle AAV\rangle$, $\langle VVV\rangle$.
\end{itemize}
\item Set 2: The correlators that are the order parameters of the chiral symmetry breaking in the chiral limit:
\begin{itemize}
\item $\langle SSS\rangle$, $\langle SPP\rangle$, $\langle VVP\rangle$, $\langle AAP\rangle$, $\langle VAS\rangle$, $\langle VVS\rangle$, $\langle AAS\rangle$, $\langle VAP\rangle$.
\end{itemize}
\end{itemize}

As indicated, the first set consists of the correlators that have the lowest possible contribution to the OPE. On the other hand, the latter set consists of the order parameters of the chiral symmetry breaking, and their OPE expansion thus starts with the nonperturbative contribution from the quark condensate.

To be precise, appropriate combinations of the correlators of the Set 1 also make up the order parameters. Specifically, these are as follows: $\langle VSS\rangle-\langle VPP\rangle$ and $\langle AAV\rangle-\langle VVV\rangle$. On the other hand, $\langle VVA\rangle-\langle AAA\rangle$ is not the order parameter. Neither are the combinations of $\langle ASP\rangle-\langle VSS\rangle$ and $\langle ASP\rangle-\langle VPP\rangle$, due to the different flavor structures.

Before we advance, let us briefly remind some basic properties of all the relevant Green functions. In the following subsections, we present a short review of the chiral Ward identities that allow us to establish the tensor decompositions of the correlators.

\subsection{Chiral Ward Identities}
The Green functions are connected through the chiral Ward identities that reflect the symmetry properties of a given theory on the quantum level. The knowledge of the identities allows us to determine the structure of the Green functions.

Let us take the definition \eqref{eq:gf-definition}, restrict ourselves only to the three-point Green functions, i.e. take $n=3$, and label the first operator $\mathcal{O}_{1}(x_{1})$ as $\mathcal{O}_{1}^{\mu}(x)$ for now.  Then, for the three-point Green functions in the chiral limit in the $x$-representation, the chiral Ward identity can be expressed as\footnote{Generally, there should also be another term with a divergence of the Noether current present in \eqref{eq:ward-identities-3pt}. However, in the chiral limit, the nonsinglet currents are conserved, i.e. $\partial^{\mu}V_{\mu}^{a}=0$ and $\partial^{\mu}A_{\mu}^{a}=0$.}
\begin{align}
\partial_{\mu}^{x}\big\langle\mathrm{T}\,\mathcal{O}_{1}^{\mu}(x)\mathcal{O}_{2}(y)\mathcal{O}_{3}(z)\big\rangle&=\delta(x^{0}-y^{0})\big\langle\mathrm{T}[\mathcal{O}_{1}^{0}(x),\mathcal{O}_{2}(y)]\mathcal{O}_{3}(z)\big\rangle\label{eq:ward-identities-3pt}\\
&+\delta(x^{0}-z^{0})\big\langle\mathrm{T}\,\mathcal{O}_{2}(y)[\mathcal{O}_{1}^{0}(x),\mathcal{O}_{3}(z)]\big\rangle\nonumber\\
&+\mathrm{anomaly}\,.\nonumber
\end{align}

To evaluate the chiral Ward identities it is necessary to know the equal-time commutation relations among the currents $V$, $A$ and the densities $S$, $P$. Since we are interested only in the nonsinglet currents and densities in this paper, we omit the contributions from the singlet ones. We also omit contributions of the Schwinger terms. Then, the commutators of the vector and axial-vector currents are as follows:
\begin{align}
[V_{0}^{a}(t,\bm{x}),V_{\mu}^{b}(t,\bm{y})]&=[A_{0}^{a}(t,\bm{x}),A_{\mu}^{b}(t,\bm{y})]=i\delta^{3}(\bm{x}-\bm{y})f^{abc}V_{\mu}^{c}(t,\bm{x})\,,\label{eq:commutation_relations_1}\\
[V_{0}^{a}(t,\bm{x}),A_{\mu}^{b}(t,\bm{y})]&=[A_{0}^{a}(t,\bm{x}),V_{\mu}^{b}(t,\bm{y})]=i\delta^{3}(\bm{x}-\bm{y})f^{abc}A_{\mu}^{c}(t,\bm{x})\,,\label{eq:commutation_relations_2}
\end{align}
with $\bm{x}$, $\bm{y}$ being the space coordinates. Similarly, the commutators of the vector or axial-vector currents and the scalar or pseudoscalar densities read
\begin{align}
[V_{0}^{a}(t,\bm{x}),S^{b}(t,\bm{y})]=&\quad\,i\delta^{3}(\bm{x}-\bm{y})f^{abc}S^{c}(t,\bm{x})\,,\label{eq:commutation_relations_3}\\
[V_{0}^{a}(t,\bm{x}),P^{b}(t,\bm{y})]=&\quad\,i\delta^{3}(\bm{x}-\bm{y})f^{abc}P^{c}(t,\bm{x})\,,\label{eq:commutation_relations_4}\\
[A_{0}^{a}(t,\bm{x}),S^{b}(t,\bm{y})]=&\quad\,i\delta^{3}(\bm{x}-\bm{y})d^{abc}P^{c}(t,\bm{x})\,,\label{eq:commutation_relations_5}\\
[A_{0}^{a}(t,\bm{x}),P^{b}(t,\bm{y})]=&-i\delta^{3}(\bm{x}-\bm{y})d^{abc}S^{c}(t,\bm{x})\,.\label{eq:commutation_relations_6}
\end{align}

In the previous expressions \eqref{eq:commutation_relations_1}-\eqref{eq:commutation_relations_6}, $f^{abc}$ is a totally antisymmetric $\mathrm{SU}(3)$ structure constant and $d^{abc}$ is totally symmetric $\mathrm{SU}(3)$ group invariant:
\begin{align}
f^{abc}=&-2i\,\mathrm{Tr}\big([T^{a},T^{b}]T^{c}\big)\,,\\
d^{abc}=&\quad\,2\,\mathrm{Tr}\big(\lbrace T^{a},T^{b}\rbrace T^{c}\big)\,.
\end{align}

Using the commutation relations above and performing the Fourier transform of \eqref{eq:ward-identities-3pt}, we are able to obtain the respective Ward identities in the momentum representation for all the Green functions. The results are shortly summarized below.

\subsubsection{Green Functions of Set 1}
As we have seen, the Ward identities of three-point Green functions give us the right-hand side of these identities in terms of two-point Green functions. For this reason, we refer the reader to the appendix \ref{sec:appendix-2pt}, where the definitions and decompositions \eqref{VV_definition_apendix}-\eqref{AP_definition_apendix} of two-point correlators are presented.

In what follows, we start with the Green functions that belong to the Set 1.

\subsubsection*{\texorpdfstring{\boldmath$\langle ASP\rangle$}{}, \texorpdfstring{$\langle VSS\rangle$}{} and \texorpdfstring{$\langle VPP\rangle$}{} Green Functions}
One immediately finds out that the right-hand side of the Ward identities of these correlators are proportional to two-point Green functions $\langle SS\rangle$ and $\langle PP\rangle$. Specifically, we have
\begin{align}
p^{\mu}\big[\Pi_{ASP}(p,q;r)\big]_{\mu}^{abc}&=\Big(\Pi_{PP}(r^{2})-\Pi_{SS}(q^{2})\Big)d^{abc}\,,\label{eq:ward_ASP_kopie}\\
p^{\mu}\big[\Pi_{VSS}(p,q;r)\big]_{\mu}^{abc}&=\Big(\Pi_{SS}(r^{2})-\Pi_{SS}(q^{2})\Big)f^{abc}\,,\label{eq:ward_VSS_kopie}\\
p^{\mu}\big[\Pi_{VPP}(p,q;r)\big]_{\mu}^{abc}&=\Big(\Pi_{PP}(r^{2})-\Pi_{PP}(q^{2})\Big)f^{abc}\,.\label{eq:ward_VPP_kopie}
\end{align}

\subsubsection*{\texorpdfstring{\boldmath$\langle VVA\rangle$}{} and \texorpdfstring{$\langle AAA\rangle$}{} Green Functions}
The correlators $\langle VVA\rangle$ and $\langle AAA\rangle$ are objects of utmost importance due to the presence of the anomaly, which is of perturbative nature.

The respective Ward identities for $\langle VVA\rangle$ read, due to the violation of conservation of the chiral axial-vector current on the quantum level,
\begin{align}
p^{\mu}\big[\Pi_{VVA}(p,q;r)\big]_{\mu\nu\rho}^{abc}=&\quad\,0\,,\label{eq:VVA-Ward_kopie_1}\\
q^{\nu}\big[\Pi_{VVA}(p,q;r)\big]_{\mu\nu\rho}^{abc}=&\quad\,0\,,\label{eq:VVA-Ward_kopie_2}\\
r^{\rho}\big[\Pi_{VVA}(p,q;r)\big]_{\mu\nu\rho}^{abc}=&-\frac{i N_{c}}{8\pi^{2}}\varepsilon^{\mu\nu(p)(q)}d^{abc}\,.\label{eq:VVA-Ward_kopie_3}
\end{align}
Similarly, for the $\langle AAA\rangle$ we have
\begin{align}
p^{\mu}\big[\Pi_{AAA}(p,q;r)\big]_{\mu\nu\rho}^{abc}=&-\frac{i N_{c}}{24\pi^{2}}\varepsilon^{\nu\rho(p)(q)}d^{abc}\,,\label{eq:AAA-Ward_kopie_1}\\
q^{\nu}\big[\Pi_{AAA}(p,q;r)\big]_{\mu\nu\rho}^{abc}=&\quad\,\frac{i N_{c}}{24\pi^{2}}\varepsilon^{\mu\rho(p)(q)}d^{abc}\,,\label{eq:AAA-Ward_kopie_2}\\
r^{\rho}\big[\Pi_{AAA}(p,q;r)\big]_{\mu\nu\rho}^{abc}=&-\frac{i N_{c}}{24\pi^{2}}\varepsilon^{\mu\nu(p)(q)}d^{abc}\,.\label{eq:AAA-Ward_kopie_3}
\end{align}

\subsubsection*{\texorpdfstring{\boldmath$\langle AAV\rangle$}{} and \texorpdfstring{$\langle VVV\rangle$}{} Green Functions}
The right-hand side of the Ward identities of the $\langle AAV\rangle$ and $\langle VVV\rangle$ Green functions are proportional to $\langle VV\rangle$ and $\langle AA\rangle$ correlators. The specific forms are as follows:
\begin{align}
p^{\mu}\big[\Pi_{AAV}(p,q;r)\big]_{\mu\nu\rho}^{abc}&=\Big(\big[\Pi_{VV}(r)\big]_{\nu\rho}-\big[\Pi_{AA}(q)\big]_{\nu\rho}\Big)f^{abc}\,,\label{eq:AAV-Ward_kopie_1}\\
q^{\nu}\big[\Pi_{AAV}(p,q;r)\big]_{\mu\nu\rho}^{abc}&=\Big(\big[\Pi_{AA}(p)\big]_{\mu\rho}-\big[\Pi_{VV}(r)\big]_{\mu\rho}\Big)f^{abc}\,,\label{eq:AAV-Ward_kopie_2}\\
r^{\rho}\big[\Pi_{AAV}(p,q;r)\big]_{\mu\nu\rho}^{abc}&=\Big(\big[\Pi_{AA}(q)\big]_{\mu\nu}-\big[\Pi_{AA}(p)\big]_{\mu\nu}\Big)f^{abc}\,,\label{eq:AAV-Ward_kopie_3}
\end{align}
and
\begin{align}
p^{\mu}\big[\Pi_{VVV}(p,q;r)\big]_{\mu\nu\rho}^{abc}&=\Big(\big[\Pi_{VV}(r)\big]_{\nu\rho}-\big[\Pi_{VV}(q)\big]_{\nu\rho}\Big)f^{abc}\,,\label{eq:VVV-Ward_kopie_1}\\
q^{\nu}\big[\Pi_{VVV}(p,q;r)\big]_{\mu\nu\rho}^{abc}&=\Big(\big[\Pi_{VV}(p)\big]_{\mu\rho}-\big[\Pi_{VV}(r)\big]_{\mu\rho}\Big)f^{abc}\,,\label{eq:VVV-Ward_kopie_2}\\
r^{\rho}\big[\Pi_{VVV}(p,q;r)\big]_{\mu\nu\rho}^{abc}&=\Big(\big[\Pi_{VV}(q)\big]_{\mu\nu}-\big[\Pi_{VV}(p)\big]_{\mu\nu}\Big)f^{abc}\,.\label{eq:VVV-Ward_kopie_3}
\end{align}

\subsubsection{Green Functions of Set 2}
Let us now focus on the correlators from the Set 2.

\subsubsection*{\boldmath\texorpdfstring{$\langle VVP\rangle$}{}, \texorpdfstring{$\langle AAP\rangle$}{} and \texorpdfstring{$\langle VAS\rangle$}{} Green Functions}
Right-hand sides of the Ward identities of the $\langle VVP\rangle$, $\langle AAP\rangle$ and $\langle VAS\rangle$ Green functions vanish due to the forbidden existence of the $\langle VP\rangle$ and $\langle AS\rangle$ correlators. Therefore,\footnote{Here and in \eqref{eq:VVS_ward}, the curly brackets do not represent the anticommutator, of course, but a shortened notation of writing down the Ward identities.}
\begin{align}
\big\lbrace p^{\mu},q^{\nu}\big\rbrace\big[\Pi_{VVP}(p,q;r)\big]_{\mu\nu}^{abc}&=\big\lbrace 0,0\big\rbrace\,,\\
\big\lbrace p^{\mu},q^{\nu}\big\rbrace\big[\Pi_{AAP}(p,q;r)\big]_{\mu\nu}^{abc}&=\big\lbrace 0,0\big\rbrace\,,\\
\big\lbrace p^{\mu},q^{\nu}\big\rbrace\big[\Pi_{VAS}(p,q;r)\big]_{\mu\nu}^{abc}&=\big\lbrace 0,0\big\rbrace\,.
\end{align}

\subsubsection*{\boldmath\texorpdfstring{$\langle VVS\rangle$}{}, \texorpdfstring{$\langle AAS\rangle$}{} and \texorpdfstring{$\langle VAP\rangle$}{} Green Functions}
For the $\langle VVS\rangle$ Green function, its Ward identities lead to the combinations of the $\langle VS\rangle$ correlators and vanish identically, too. On the other hand, $\langle AAS\rangle$ and $\langle VAP\rangle$ correlators are, within Ward identities, reduced to the combinations of the $\langle AP\rangle$ Green functions, which contribute nontrivially.

The respective Ward identities can be formally written down, according to \eqref{eq:commutation_relations_1}-\eqref{eq:commutation_relations_6}, as follows. The right-hand sides of the Ward identities for the $\langle VVS\rangle$  correlator vanish,
\begin{align}
\big\lbrace p^{\mu},q^{\nu}\big\rbrace\big[\Pi_{VVS}(p,q;r)\big]_{\mu\nu}^{abc}=\big\lbrace 0,0\big\rbrace\,,\label{eq:VVS_ward}
\end{align}
while for the $\langle AAS\rangle$ and $\langle VAP\rangle$ we have\footnote{The Ward identities \eqref{eq:AAS_ward_1}-\eqref{eq:VAP_ward_2} for the $\langle AAS\rangle$ and $\langle VAP\rangle$ Green functions differ with the ones presented in \cite{Jamin:2008rm} (eq.~4.12, page no.~12) or \cite{Cirigliano:2004ue} (eq.~3, page no.~3) either due to the different normalization of the quark condensate or due to the various normalization or conventional factors in the definitions of the correlators or the chiral currents/densities. See the end of Subsection \ref{ssec:quark-condensate_results} for details.}
\begin{alignat}{2}
p^{\mu}\big[\Pi_{AAS}(p,q;r)\big]_{\mu\nu}^{abc}=&\quad\,\big[\Pi_{AP}(q)\big]_{\nu}d^{abc}&&=-\frac{\langle\overline{q}q\rangle}{3q^{2}}q_{\nu}d^{abc}\,,\label{eq:AAS_ward_1}\\
q^{\nu}\big[\Pi_{AAS}(p,q;r)\big]_{\mu\nu}^{abc}=&\quad\,\big[\Pi_{AP}(p)\big]_{\mu}d^{abc}&&=-\frac{\langle\overline{q}q\rangle}{3p^{2}}p_{\mu}d^{abc}\,,\label{eq:AAS_ward_2}\\
p^{\mu}\big[\Pi_{VAP}(p,q;r)\big]_{\mu\nu}^{abc}=&-\Big(\big[\Pi_{AP}(r)\big]_{\nu}+\big[\Pi_{AP}(q)\big]_{\nu}\Big)f^{abc}&&=\,\,\,\,\,\frac{\langle\overline{q}q\rangle}{3}\bigg(\frac{q_{\nu}}{q^{2}}+\frac{r_{\nu}}{r^{2}}\bigg)f^{abc}\,,\label{eq:VAP_ward_1}\\
q^{\nu}\big[\Pi_{VAP}(p,q;r)\big]_{\mu\nu}^{abc}=&\quad\,\big[\Pi_{AP}(r)\big]_{\mu}f^{abc}&&=-\frac{\langle\overline{q}q\rangle}{3r^{2}}r_{\mu}f^{abc}\,.\label{eq:VAP_ward_2}
\end{alignat}
As mentioned above, unlike for any other Green functions above, the Ward identities reduce the correlators $\langle AAS\rangle$ and $\langle VAP\rangle$ to a combination of $\langle AP\rangle$. Since the $\langle AP\rangle$ correlator is saturated only by the Goldstone boson exchange, we have already substituted it in the exact form \eqref{eq:AP_contribution} (see also \eqref{AP_definition_apendix} for notation).

\subsection{Tensor Decomposition}
By knowing the Ward identities of the Green functions, one is able to construct the tensor structures of such correlators, which helps us to present the results of the OPE in a compact form. Generally, one may start writing down a complete basis of relevant tensors and then separate the tensor structure into a transversal and longitudinal part, which is fixed by the Ward identities. In our case, the results of the three-point Green functions are thus fixed by the respective results for the two-point Green functions, presented in Appendix \ref{sec:appendix-2pt}.

\subsubsection{Green Functions of Set 1}
In what follows we present the tensor decomposition of the correlators of the Set 1.

\subsubsection*{\boldmath\texorpdfstring{$\langle ASP\rangle$}{}, \texorpdfstring{$\langle VSS\rangle$}{} and \texorpdfstring{$\langle VPP\rangle$}{} Green Functions}
The Lorentz structure of these Green functions is easy to reproduce since the only available structure has to be made of momenta with one Lorentz index. Upon assuming the Ward identities \eqref{eq:ward_ASP_kopie}-\eqref{eq:ward_VSS_kopie}, it is straightforward to write down the decompositions in the following forms:
\begin{alignat}{2}
&\big[\Pi_{ASP}(p,q;r)\big]_{\mu}^{abc}&&=d^{abc}\big[\Pi_{ASP}(p,q;r)\big]_{\mu}\,,\label{eq:asp-definition}\\
&\big[\Pi_{VSS}(p,q;r)\big]_{\mu}^{abc}&&=f^{abc}\big[\Pi_{VSS}(p,q;r)\big]_{\mu}\,,\label{eq:vss-definition}\\
&\big[\Pi_{VPP}(p,q;r)\big]_{\mu}^{abc}&&=f^{abc}\big[\Pi_{VPP}(p,q;r)\big]_{\mu}\,,\label{eq:vpp-definition}
\end{alignat}
with the Lorentz parts given as
\begin{alignat}{3}
&\big[\Pi_{ASP}(p,q;r)\big]_{\mu}&&=\mathcal{F}_{ASP}(p^{2},q^{2};r^{2})\mathcal{T}_{\mu}(p,q;r)&&+\Big(\Pi_{PP}(r^{2})-\Pi_{SS}(q^{2})\Big)\frac{p_{\mu}}{p^{2}}\,,\label{eq:asp-definition_dekompozice}\\
&\big[\Pi_{VSS}(p,q;r)\big]_{\mu}&&=\mathcal{F}_{VSS}(p^{2},q^{2};r^{2})\mathcal{T}_{\mu}(p,q;r)&&+\Big(\Pi_{SS}(r^{2})-\Pi_{SS}(q^{2})\Big)\frac{p_{\mu}}{p^{2}}\,,\label{eq:vss-definition_dekompozice}\\
&\big[\Pi_{VPP}(p,q;r)\big]_{\mu}&&=\mathcal{F}_{VPP}(p^{2},q^{2};r^{2})\mathcal{T}_{\mu}(p,q;r)&&+\Big(\Pi_{PP}(r^{2})-\Pi_{PP}(q^{2})\Big)\frac{p_{\mu}}{p^{2}}\,,\label{eq:vpp-definition_dekompozice}
\end{alignat}
where we have introduced the tensor $\mathcal{T}_{\mu}$,
\begin{equation}
\mathcal{T}_{\mu}(p,q;r)=q_{\mu}+\frac{p^{2}+q^{2}-r^{2}}{2p^{2}}p_{\mu}\,,\label{eq:tensor_R}
\end{equation}
with the property
\begin{equation}
p^{\mu}\mathcal{T}_{\mu}(p,q;r)=0\,.
\end{equation}

The $\langle VSS\rangle$ and $\langle VPP\rangle$ Green functions must be symmetric under the interchange of $(q,b)\leftrightarrow(r,c)$ due to the Bose symmetry. Since the flavor factor of such correlators is antisymmetric, the Lorentz structures \eqref{eq:vss-definition_dekompozice} and \eqref{eq:vpp-definition_dekompozice} must be antisymmetric under such interchange, too. This, however, leads to the conclusion that the respective formfactors must be symmetric under exchanging the momenta $q$ and $r$,
\begin{align}
\mathcal{F}_{VSS}(p^{2},q^{2};r^{2})&=\mathcal{F}_{VSS}(p^{2},r^{2};q^{2})\,,\\
\mathcal{F}_{VPP}(p^{2},q^{2};r^{2})&=\mathcal{F}_{VPP}(p^{2},r^{2};q^{2})\,,
\end{align}
due to the antisymmetricity of the tensor \eqref{eq:tensor_R} in the last two arguments,
\begin{equation}
\mathcal{T}_{\mu}(p,q;r)=-\mathcal{T}_{\mu}(p,r;q)\,.
\end{equation}

\subsubsection*{\boldmath\texorpdfstring{$\langle VVA\rangle$}{} and \texorpdfstring{$\langle AAA\rangle$}{} Green Functions}
The $\langle VVA\rangle$ and $\langle AAA\rangle$ Green functions are remarkable due to the presence of the chiral anomaly, as mentioned in the previous subsection. The general form of these correlators can be written down, knowing \eqref{eq:VVA-Ward_kopie_1}-\eqref{eq:AAA-Ward_kopie_3}, as
\begin{alignat}{2}
&\big[\Pi_{VVA}(p,q;r)\big]_{\mu\nu\rho}^{abc}&&=d^{abc}\big[\Pi_{VVA}(p,q;r)\big]_{\mu\nu\rho}\,,\label{eq:vva-definition}\\
&\big[\Pi_{AAA}(p,q;r)\big]_{\mu\nu\rho}^{abc}&&=d^{abc}\big[\Pi_{AAA}(p,q;r)\big]_{\mu\nu\rho}\,,\label{eq:aaa-definition}
\end{alignat}
where the Lorentz structures can be further decomposed into the longitudinal and transversal parts,
\begin{alignat}{2}
&\big[\Pi_{VVA}(p,q;r)\big]_{\mu\nu\rho}&&=\big[\Pi_{VVA}^{(L)}(p,q;r)\big]_{\mu\nu\rho}+\big[\Pi_{VVA}^{(T)}(p,q;r)\big]_{\mu\nu\rho}\,,\label{eq:vva-definition_longitudinal_and_transversal}\\
&\big[\Pi_{AAA}(p,q;r)\big]_{\mu\nu\rho}&&=\big[\Pi_{AAA}^{(L)}(p,q;r)\big]_{\mu\nu\rho}+\big[\Pi_{AAA}^{(T)}(p,q;r)\big]_{\mu\nu\rho}\,,\label{eq:aaa-definition_longitudinal_and_transversal}
\end{alignat}
with the longitudinal components being fixed by the anomaly, i.e.
\begin{alignat}{2}
&\big[\Pi_{VVA}^{(L)}(p,q;r)\big]_{\mu\nu\rho}=&&-\frac{i N_{c}}{8\pi^{2}r^{2}}\varepsilon^{\mu\nu(p)(q)}r^{\rho}\,,\label{eq:vva-definition_longitudinal}\\
&\big[\Pi_{AAA}^{(L)}(p,q;r)\big]_{\mu\nu\rho}=&&-\frac{i N_{c}}{24\pi^{2}r^{2}}\varepsilon^{\mu\nu(p)(q)}r^{\rho}+\frac{i N_{c}}{24\pi^{2}q^{2}}\varepsilon^{\mu\rho(p)(q)}q^{\nu}-\frac{i N_{c}}{24\pi^{2}p^{2}}\varepsilon^{\nu\rho(p)(q)}p^{\mu}\,.\label{eq:aaa-definition_longitudinal}
\end{alignat}

The transversal parts, on the other hand, are a bit more complicated. We write them down as a sum of three terms,
\begin{alignat}{2}
&\big[\Pi_{VVA}^{(T)}(p,q;r)\big]_{\mu\nu\rho}=&&\quad\,\mathcal{F}_{VVA}(p^{2},q^{2};r^{2})\mathcal{T}_{\mu\nu\rho}^{(1)}(p,q;r)+\mathcal{G}_{VVA}(p^{2},q^{2};r^{2})\mathcal{T}_{\mu\nu\rho}^{(2)}(p,q;r)\label{eq:vva-definition_transversal}\\
& &&+\mathcal{H}_{VVA}(p^{2},q^{2};r^{2})\mathcal{T}_{\mu\nu\rho}^{(3)}(p,q;r)\,,\nonumber\\
&\big[\Pi_{AAA}^{(T)}(p,q;r)\big]_{\mu\nu\rho}=&&\quad\,\mathcal{F}_{AAA}(p^{2},q^{2};r^{2})\mathcal{T}_{\mu\nu\rho}^{(4)}(p,q;r)+\mathcal{G}_{AAA}(p^{2},q^{2};r^{2})\mathcal{T}_{\mu\nu\rho}^{(5)}(p,q;r)\label{eq:aaa-definition_transversal}\\
& &&+\mathcal{H}_{AAA}(p^{2},q^{2};r^{2})\mathcal{T}_{\mu\nu\rho}^{(6)}(p,q;r)\,,\nonumber
\end{alignat}
where the respective tensors are given as\footnote{To the best of our knowledge, the tensors \eqref{eq:VVA-transversal-tensor-1}-\eqref{eq:VVA-transversal-tensor-3} for the $\langle VVA\rangle$ correlator were introduced for the first time in \cite{Knecht:2003xy}. The reader, however, should be aware of a different normalization with respect to ours, see eq.~(2.1) and (2.14) in Ref.~\cite{Knecht:2003xy}.}
\begin{align}
\mathcal{T}_{\mu\nu\rho}^{(1)}(p,q;r)&=p^{\nu}\varepsilon^{\mu\rho(p)(q)}-q^{\mu}\varepsilon^{\nu\rho(p)(q)}-\frac{p^{2}+q^{2}-r^{2}}{r^{2}}\bigg(\varepsilon^{\mu\nu(p)(q)}r^{\rho}-\frac{1}{2}r^{2}\varepsilon^{\mu\nu\rho(p-q)}\bigg)\,,\label{eq:VVA-transversal-tensor-1}\\
\mathcal{T}_{\mu\nu\rho}^{(2)}(p,q;r)&=\varepsilon^{\mu\nu(p)(q)}(p-q)^{\rho}+\frac{p^{2}-q^{2}}{r^{2}}\varepsilon^{\mu\nu(p)(q)}r^{\rho}\,,\label{eq:VVA-transversal-tensor-2}\\
\mathcal{T}_{\mu\nu\rho}^{(3)}(p,q;r)&=p^{\nu}\varepsilon^{\mu\rho(p)(q)}+q^{\mu}\varepsilon^{\nu\rho(p)(q)}-\frac{p^{2}+q^{2}-r^{2}}{2}\varepsilon^{\mu\nu\rho(r)}\,,\label{eq:VVA-transversal-tensor-3}\\
\mathcal{T}_{\mu\nu\rho}^{(4)}(p,q;r)&=(p^{2}+q^{2}-r^{2})t_{\mu\nu\rho}^{(14)}(p,q;r)+t_{\mu\nu\rho}^{(15)}(p,q;r)\,,\label{eq:AAA_tensor_Q_main_text}\\
\mathcal{T}_{\mu\nu\rho}^{(5)}(p,q;r)&=(p^{2}-q^{2})t_{\mu\nu\rho}^{(13)}(p,q;r)-(p^{2}+q^{2}-2r^{2})t_{\mu\nu\rho}^{(14)}(p,q;r)\,,\label{eq:AAA_tensor_Q1_main_text}\\
\mathcal{T}_{\mu\nu\rho}^{(6)}(p,q;r)&=\frac{1}{3}(p^{2}+q^{2}-2r^{2})t_{\mu\nu\rho}^{(13)}(p,q;r)+(p^{2}-q^{2})t_{\mu\nu\rho}^{(14)}(p,q;r)\,,\label{eq:AAA_tensor_Q2_main_text}
\end{align}
where the tensors on the right hand sides of  \eqref{eq:AAA_tensor_Q_main_text}-\eqref{eq:AAA_tensor_Q2_main_text} defined as \eqref{eq:AAA_tensor_structure_part_1_v2}-\eqref{eq:AAA_tensor_structure_part_3_v2}.

The properties of the tensors above with respect to the Bose symmetries are given in \eqref{eq:VVA_tensor_1_symetrie}-\eqref{eq:VVA_tensor_3_symetrie} and \eqref{eq:AAA_tensor_Q_bose}-\eqref{eq:AAA_tensor_Q2_bose}. As a result,  the corresponding formfactors, satisfy the following relations:
\begin{align}
\mathcal{F}_{VVA}(p^{2},q^{2};r^{2})=\quad &\mathcal{F}_{VVA}(q^{2},p^{2};r^{2})\,,\label{eq:VVAformfactor1}\\
\mathcal{G}_{VVA}(p^{2},q^{2};r^{2})=-&\mathcal{G}_{VVA}(q^{2},p^{2};r^{2})\,,\label{eq:VVAformfactor2}\\
\mathcal{H}_{VVA}(p^{2},q^{2};r^{2})=-&\mathcal{H}_{VVA}(q^{2},p^{2};r^{2})\label{eq:VVAformfactor3}
\end{align}
and
\begin{alignat}{4}
&\mathcal{F}_{AAA}(p^{2},q^{2},r^{2})=-&&\mathcal{F}_{AAA}(q^{2},p^{2},r^{2})=-&&\mathcal{F}_{AAA}(r^{2},q^{2},p^{2})=-&&\mathcal{F}_{AAA}(p^{2},r^{2},q^{2})\,,\label{eq:AAAformfactor1}\\
&\mathcal{G}_{AAA}(p^{2},q^{2},r^{2})=-&&\mathcal{G}_{AAA}(q^{2},p^{2},r^{2})=-&&\mathcal{G}_{AAA}(r^{2},q^{2},p^{2})=-&&\mathcal{G}_{AAA}(p^{2},r^{2},q^{2})\,,\label{eq:AAAformfactor2}\\
&\mathcal{H}_{AAA}(p^{2},q^{2},r^{2})=&&\mathcal{H}_{AAA}(q^{2},p^{2},r^{2})=&&\mathcal{H}_{AAA}(r^{2},q^{2},p^{2})=&&\mathcal{H}_{AAA}(p^{2},r^{2},q^{2})\,.\label{eq:AAAformfactor3}
\end{alignat}

\subsubsection*{\boldmath\texorpdfstring{$\langle AAV\rangle$}{} and \texorpdfstring{$\langle VVV\rangle$}{} Green Functions}
The $\langle AAV\rangle$ and $\langle VVV\rangle$ Green functions have a complicated structure made of momenta and the metric tensor and their Ward identities include the $\langle VV\rangle$ and $\langle AA\rangle$ Green functions.
After separating the flavor and the Lorentz structures, the decomposition of these correlators can be written as
\begin{alignat}{2}
&\big[\Pi_{AAV}(p,q;r)\big]_{\mu\nu\rho}^{abc}&&=f^{abc}\big[\Pi_{AAV}(p,q;r)\big]_{\mu\nu\rho}\,,\label{eq:aav-definition}\\
&\big[\Pi_{VVV}(p,q;r)\big]_{\mu\nu\rho}^{abc}&&=f^{abc}\big[\Pi_{VVV}(p,q;r)\big]_{\mu\nu\rho}\,.\label{eq:vvv-definition}
\end{alignat}
We once again separate the Lorentz parts into the longitudinal and transversal components (see Appendix \ref{sec:AAV_VVV_decompositions} for a detailed derivation),
\begin{align}
\big[\Pi_{AAV}(p,q;r)\big]_{\mu\nu\rho}&=\big[\Pi_{AAV}^{(L)}(p,q;r)\big]_{\mu\nu\rho}+\big[\Pi_{AAV}^{(T)}(p,q;r)\big]_{\mu\nu\rho}\,,\label{eq:aav-definition_2}\\
\big[\Pi_{VVV}(p,q;r)\big]_{\mu\nu\rho}&=\big[\Pi_{VVV}^{(L)}(p,q;r)\big]_{\mu\nu\rho}+\big[\Pi_{VVV}^{(T)}(p,q;r)\big]_{\mu\nu\rho}\,,\label{eq:vvv-definition_2}
\end{align}
where the respective parts read
\begin{align}
\big[\Pi_{AAV}^{(L)}(p,q;r)\big]_{\mu\nu\rho}=&\quad\,\Pi_{AA}(p^{2})\mathcal{T}_{\mu\nu\rho}^{(7)}(p,q;r)-\Pi_{AA}(q^{2})\mathcal{T}_{\nu\mu\rho}^{(7)}(q,p;r)\label{eq:aav-definition_2_longitudinal}\\
&-\Pi_{VV}(r^{2})\mathcal{T}_{\rho\nu\mu}^{(7)}(r,q;p)\,,\nonumber\\
\big[\Pi_{VVV}^{(L)}(p,q;r)\big]_{\mu\nu\rho}=&\quad\,\Pi_{VV}(p^{2})\mathcal{T}_{\mu\nu\rho}^{(7)}(p,q;r)-\Pi_{VV}(q^{2})\mathcal{T}_{\nu\mu\rho}^{(7)}(q,p;r)\label{eq:vvv-definition_2_longitudinal}\\
&-\Pi_{VV}(r^{2})\mathcal{T}_{\rho\nu\mu}^{(7)}(r,q;p)\,,\nonumber
\end{align}
and
\begin{align}
\big[\Pi_{AAV}^{(T)}(p,q;r)\big]_{\mu\nu\rho}=&\quad\,\mathcal{F}_{AAV}(p^{2},q^{2};r^{2})\mathcal{T}_{\mu\nu\rho}^{(8)}(p,q;r)+\mathcal{G}_{AAV}(p^{2},q^{2};r^{2})\mathcal{T}_{\mu\nu\rho}^{(9)}(p,q;r)\label{eq:aav-definition_2_transversal}\\
&+\mathcal{H}_{AAV}(p^{2},q^{2};r^{2})\mathcal{T}_{\nu\rho\mu}^{(8)}(q,r;p)-\mathcal{H}_{AAV}(q^{2},p^{2};r^{2})\mathcal{T}_{\mu\rho\nu}^{(8)}(p,r;q)\,,\nonumber\\
\big[\Pi_{VVV}^{(T)}(p,q;r)\big]_{\mu\nu\rho}=&\quad\,\mathcal{F}_{VVV}(p^{2},q^{2};r^{2})\mathcal{T}_{\mu\nu\rho}^{(8)}(p,q;r)+\mathcal{F}_{VVV}(r^{2},p^{2};q^{2})\mathcal{T}_{\rho\mu\nu}^{(8)}(r,p;q)\label{eq:vvv-definition_2_transversal}\\
&+\mathcal{F}_{VVV}(q^{2},r^{2};p^{2})\mathcal{T}_{\nu\rho\mu}^{(8)}(q,r;p)+\mathcal{G}_{VVV}(p^{2},q^{2};r^{2})\mathcal{T}_{\mu\nu\rho}^{(9)}(p,q;r)\,,\nonumber
\end{align}
with the tensors defined as
\begin{align}
\mathcal{T}_{\mu\nu\rho}^{(7)}(p,q;r)=&-p_{\nu}g_{\mu\rho}+p_{\rho}g_{\mu\nu}+\frac{q_{\rho}\big[(p\hspace{-1pt}\cdot\hspace{-1pt}q)g_{\mu\nu}-q_{\mu}p_{\nu}\big]-r_{\nu}\big[(p\hspace{-1pt}\cdot\hspace{-1pt}r)g_{\mu\rho}-r_{\mu}p_{\rho}\big]}{q\hspace{-1pt}\cdot\hspace{-1pt}r}\,,\label{eq:AAV_VVV_tensor_1}\\
\mathcal{T}_{\mu\nu\rho}^{(8)}(p,q;r)=&-\frac{\big[(p\hspace{-1pt}\cdot\hspace{-1pt}q)g_{\mu\nu}-q_{\mu}p_{\nu}\big]\big[(q\hspace{-1pt}\cdot\hspace{-1pt}r)p_{\rho}-(p\hspace{-1pt}\cdot\hspace{-1pt}r)q_{\rho}\big]}{(p\hspace{-1pt}\cdot\hspace{-1pt}r)(q\hspace{-1pt}\cdot\hspace{-1pt}r)}\,,\label{eq:AAV_VVV_tensor_2}\\
\mathcal{T}_{\mu\nu\rho}^{(9)}(p,q;r)=&-q_{\mu}\big[(p\hspace{-1pt}\cdot\hspace{-1pt}r)g_{\nu\rho}-r_{\nu}p_{\rho}\big]+g_{\mu\nu}\big[(p\hspace{-1pt}\cdot\hspace{-1pt}r)q_{\rho}-(q\hspace{-1pt}\cdot\hspace{-1pt}r)p_{\rho}\big]\label{eq:AAV_VVV_tensor_3}\\
&+r_{\mu}\big[(p\hspace{-1pt}\cdot\hspace{-1pt}q)g_{\nu\rho}-p_{\nu}q_{\rho}\big]+g_{\mu\rho}\big[(q\hspace{-1pt}\cdot\hspace{-1pt}r)p_{\nu}-(p\hspace{-1pt}\cdot\hspace{-1pt}q)r_{\nu}\big]\,.\nonumber
\end{align}
We note that the formfactors have the following symmetry properties:
\begin{align}
\mathcal{F}_{AAV}(p^{2},q^{2};r^{2})&=\mathcal{F}_{AAV}(q^{2},p^{2};r^{2})\,,\\
\mathcal{G}_{AAV}(p^{2},q^{2};r^{2})&=\mathcal{G}_{AAV}(q^{2},p^{2};r^{2})\,,\\
\mathcal{F}_{VVV}(p^{2},q^{2};r^{2})&=\mathcal{F}_{VVV}(q^{2},p^{2};r^{2})\,,
\end{align}
while $\mathcal{G}_{VVV}$ is fully symmetric in all of its arguments.

Since the decompositions \eqref{eq:aav-definition_2}-\eqref{eq:vvv-definition_2} are given by the contributions \eqref{VV_definition_apendix}-\eqref{AA_definition_apendix} of the two-point correlators, the $\langle AAV\rangle$ Green function is entirely determined by three independent formfactors, while the $\langle VVV\rangle$ is given only by two.

\subsubsection{Green Functions of Set 2}
Now, we focus on the correlators that are the order parameters of the chiral symmetry breaking in the chiral limit, i.e. on those that belong to the Set 2. Before we advance, let us mention that some of these correlators have been studied in the past in various contexts, see for example \cite{Moussallam:1997xx,Knecht:2001xc,Cirigliano:2005xn,RuizFemenia:2003hm,Cirigliano:2004ue,Jamin:2008rm,Dai:2019lmj}.

\subsubsection*{\boldmath\texorpdfstring{$\langle SSS\rangle$}{} and \texorpdfstring{$\langle SPP\rangle$}{} Green Functions}
The first two Green functions, $\langle SSS\rangle$ and $\langle SPP\rangle$ are zero-rank tensors with trivial Lorentz structure that can be written down in a simple form
\begin{alignat}{2}
&\big[\Pi_{SSS}(p,q;r)\big]^{abc}&&=d^{abc}\mathcal{F}_{SSS}(p^{2},q^{2};r^{2})\,,\label{eq:sss-definition}\\
&\big[\Pi_{SPP}(p,q;r)\big]^{abc}&&=d^{abc}\mathcal{F}_{SPP}(p^{2},q^{2};r^{2})\,.\label{eq:spp-definition}
\end{alignat}

\subsubsection*{\boldmath\texorpdfstring{$\langle VVP\rangle$}{}, \texorpdfstring{$\langle AAP\rangle$}{} and \texorpdfstring{$\langle VAS\rangle$}{} Green Functions}
These three correlators belong to the odd-intrinsic parity sector of QCD. Their Lorentz structure is thus given by the Levi-Civita tensor with two indices contracted with external momenta. Similarly as above, these Green functions can be written down as
\begin{alignat}{2}
&\big[\Pi_{VVP}(p,q;r)\big]_{\mu\nu}^{abc}&&=d^{abc}\mathcal{F}_{VVP}(p^{2},q^{2};r^{2})\varepsilon^{\mu\nu(p)(q)}\,,\label{eq:vvp-definition}\\
&\big[\Pi_{AAP}(p,q;r)\big]_{\mu\nu}^{abc}&&=d^{abc}\mathcal{F}_{AAP}(p^{2},q^{2};r^{2})\varepsilon^{\mu\nu(p)(q)}\,,\label{eq:aap-definition}\\
&\big[\Pi_{VAS}(p,q;r)\big]_{\mu\nu}^{abc}&&=f^{abc}\mathcal{F}_{VAS}(p^{2},q^{2};r^{2})\varepsilon^{\mu\nu(p)(q)}\,.\label{eq:vas-definition}
\end{alignat}

\subsubsection*{\boldmath\texorpdfstring{$\langle VVS\rangle$}{}, \texorpdfstring{$\langle AAS\rangle$}{} and \texorpdfstring{$\langle VAP\rangle$}{} Green Functions}
Unlike in the previous case, these Green functions belong to the even sector of QCD. Their structure is a little bit more complicated and can be written down as
\begin{alignat}{2}
&\big[\Pi_{VVS}(p,q;r)\big]_{\mu\nu}^{abc}&&=d^{abc}\big[\Pi_{VVS}(p,q;r)\big]_{\mu\nu}\,,\label{eq:vvs-definition_0}\\
&\big[\Pi_{AAS}(p,q;r)\big]_{\mu\nu}^{abc}&&=d^{abc}\big[\Pi_{AAS}(p,q;r)\big]_{\mu\nu}\,,\label{eq:aas-definition_0}\\
&\big[\Pi_{VAP}(p,q;r)\big]_{\mu\nu}^{abc}&&=f^{abc}\big[\Pi_{VAP}(p,q;r)\big]_{\mu\nu}\,,\label{eq:vap-definition_0}
\end{alignat}
with
\begin{alignat}{2}
\big[\Pi_{VVS}(p,q;r)\big]_{\mu\nu}&=\mathcal{F}_{VVS}(p^{2},q^{2};r^{2})\mathcal{T}_{\mu\nu}^{(1)}(p,q;r)&&+\mathcal{G}_{VVS}(p^{2},q^{2};r^{2})\mathcal{T}_{\mu\nu}^{(2)}(p,q;r)\,,\label{eq:vvs-definition}\\
\big[\Pi_{AAS}(p,q;r)\big]_{\mu\nu}&=\mathcal{F}_{AAS}(p^{2},q^{2};r^{2})\mathcal{T}_{\mu\nu}^{(1)}(p,q;r)&&+\mathcal{G}_{AAS}(p^{2},q^{2};r^{2})\mathcal{T}_{\mu\nu}^{(2)}(p,q;r)\label{eq:aas-definition}\\
& &&-\langle\overline{q}q\rangle\frac{p_{\mu}q_{\nu}}{3p^{2}q^{2}}\,,\nonumber\\
\big[\Pi_{VAP}(p,q;r)\big]_{\mu\nu}&=\mathcal{F}_{VAP}(p^{2},q^{2};r^{2})\mathcal{T}_{\mu\nu}^{(1)}(p,q;r)&&+\mathcal{G}_{VAP}(p^{2},q^{2};r^{2})\mathcal{T}_{\mu\nu}^{(2)}(p,q;r)\label{eq:vap-definition}\\
& &&+\frac{\langle\overline{q}q\rangle}{3r^{2}}\bigg(\frac{(p_{\mu}+2q_{\mu})q_{\nu}}{q^{2}}-g_{\mu\nu}\bigg)\,,\nonumber
\end{alignat}
where the respective tensors $\mathcal{T}_{\mu\nu}^{(1)}$ and $\mathcal{T}_{\mu\nu}^{(2)}$,
\begin{align}
\mathcal{T}_{\mu\nu}^{(1)}(p,q;r)&=q_{\mu}p_{\nu}+\frac{1}{2}(p^{2}+q^{2}-r^{2})g_{\mu\nu}\,,\label{eq:tensor_P}\\
\mathcal{T}_{\mu\nu}^{(2)}(p,q;r)&=p^{2}q_{\mu}q_{\nu}+q^{2}p_{\mu}p_{\nu}+\frac{1}{2}(p^{2}+q^{2}-r^{2})p_{\mu}q_{\nu}-p^{2}q^{2}g_{\mu\nu}\,,\label{eq:tensor_Q}
\end{align}
are transversal,
\begin{alignat}{2}
p^{\mu}\mathcal{T}_{\mu\nu}^{(1)}(p,q;r)&=p^{\mu}\mathcal{T}_{\mu\nu}^{(2)}(p,q;r)&&=0\,,\\
q^{\nu}\mathcal{T}_{\mu\nu}^{(1)}(p,q;r)&=q^{\nu}\mathcal{T}_{\mu\nu}^{(2)}(p,q;r)&&=0\,,
\end{alignat}
and symmetrical upon exchanging $(p,\mu)\leftrightarrow(q,\nu)$,
\begin{align}
\mathcal{T}_{\mu\nu}^{(1)}(p,q;r)&=\mathcal{T}_{\mu\nu}^{(2)}(p,q;r)\,,\label{eq:tensor_P_symmetry}\\
\mathcal{T}_{\mu\nu}^{(1)}(p,q;r)&=\mathcal{T}_{\mu\nu}^{(2)}(p,q;r)\,.\label{eq:tensor_Q_symmetry}
\end{align}

The symmetricity \eqref{eq:tensor_P_symmetry}-\eqref{eq:tensor_Q_symmetry} of the tensors \eqref{eq:tensor_P}-\eqref{eq:tensor_Q} directly implies that the respective formfactors of the $\langle VVS\rangle$ and $\langle AAS\rangle$ correlators are symmetrical in the first two arguments,
\begin{align}
\mathcal{F}_{VVS}(p^{2},q^{2};r^{2})&=\mathcal{F}_{VVS}(q^{2},p^{2};r^{2})\,,\\
\mathcal{G}_{VVS}(p^{2},q^{2};r^{2})&=\mathcal{G}_{VVS}(q^{2},p^{2};r^{2})\,,
\end{align}
and
\begin{align}
\mathcal{F}_{AAS}(p^{2},q^{2};r^{2})&=\mathcal{F}_{AAS}(q^{2},p^{2};r^{2})\,,\\
\mathcal{G}_{AAS}(p^{2},q^{2};r^{2})&=\mathcal{G}_{AAS}(q^{2},p^{2};r^{2})\,.
\end{align}

As we have already mentioned, the terms proportional to the quark condensate in \eqref{eq:aas-definition}-\eqref{eq:vap-definition} are present there due to the requirement of satisfying the Ward identities \eqref{eq:AAS_ward_1}-\eqref{eq:VAP_ward_2} that are fixed by the contribution of the $\langle AP\rangle$ correlator. See Appendix \ref{ssec:quark_condensate_2pt} for details.

%%%%%%%%%%%%%%%%%%%%%%%%%%%%%%%%%%%%%%%%%%%%%%%%%%%%%%%%%%%%%%%%%%%%%%%%%%%%%%%%%%%%%%%%%%%%%%%%%%%%%%%%%
%%%%%%%%%%%%%%%%%%%%%%%%%%%%%%%%%%%%%%%%%%%%%%%%%%%%%%%%%%%%%%%%%%%%%%%%%%%%%%%%%%%%%%%%%%%%%%%%%%%%%%%%%
%%% Section: Operator Product Expansion and QCD Condensates
%%%%%%%%%%%%%%%%%%%%%%%%%%%%%%%%%%%%%%%%%%%%%%%%%%%%%%%%%%%%%%%%%%%%%%%%%%%%%%%%%%%%%%%%%%%%%%%%%%%%%%%%%
%%%%%%%%%%%%%%%%%%%%%%%%%%%%%%%%%%%%%%%%%%%%%%%%%%%%%%%%%%%%%%%%%%%%%%%%%%%%%%%%%%%%%%%%%%%%%%%%%%%%%%%%%

\section{Operator Product Expansion and QCD Condensates}\label{sec:ope-condensates}
Regarding the study of the correlators above, there are two regimes where the QCD dynamics of the current correlators is well understood. The first one is that of low external momenta where the dynamics is governed by means of $\chi$PT. The second one corresponds to the high energies where the asymptotic freedom allows us to use the perturbative approach in terms of the strong coupling constant $\alpha_{s}$ and the asymptotics of the correlators for large euclidean momenta is given by the operator product expansion \cite{Wilson:1970ag}.

\subsection{Operator Product Expansion}
Within the framework of the operator product expansion (OPE), short-distance behaviour of the Green functions can be studied. Since we are interested  in the three-point correlators, let us consider a product of three gauge-invariant composite operators, $\mathcal{O}_{1}^{a}(x)\mathcal{O}_{2}^{b}(y)\mathcal{O}_{3}^{c}(z)$. Then, if the coordinates of these operators are close to each other, the OPE allows us to rewrite such a product as a (possibly infinite) sum of gauge-invariant local operators, made of quark and gluon fields, with $c-$number Wilson coefficients. The vacuum averages of these local operators are purely nonperturbative parameters, called the QCD condensates. Their numerical values can not be calculated directly from the first principles. They need to be obtained by other means, such as using calculations in the lattice QCD or extracting them from experimental measurements.

\subsection{QCD Condensates}
If we consider an OPE of the Green functions in terms of the QCD condensates up to and including dimension 6 without derivative terms, an arbitrary three-point correlator in massless theory can be written down as follows:
\begin{align}
\Pi_{\mathcal{O}_{1}^{a}\mathcal{O}_{2}^{b}\mathcal{O}_{3}^{c}}(\lambda p,\lambda q;\lambda r)\xrightarrow{\lambda\rightarrow\infty}&\quad\,\,\,\lambda\,C_{\mathcal{O}_{1}^{a}\mathcal{O}_{2}^{b}\mathcal{O}_{3}^{c}}^{\mathbb{1}}(p,q;r)\label{eq:OPE-condensates2_1}\\
&+\frac{1}{\lambda^{2}}C_{\mathcal{O}_{1}^{a}\mathcal{O}_{2}^{b}\mathcal{O}_{3}^{c}}^{\langle\overline{q}q\rangle}(p,q;r)\langle\overline{q}q\rangle\label{eq:OPE-condensates2_2}\\
&+\frac{1}{\lambda^{3}}C_{\mathcal{O}_{1}^{a}\mathcal{O}_{2}^{b}\mathcal{O}_{3}^{c}}^{\langle G^{2}\rangle}(p,q;r)\langle G_{\mu\nu}^{a}G^{\mu\nu,a}\rangle\label{eq:OPE-condensates2_3}\\
&+\frac{1}{\lambda^{4}}C_{\mathcal{O}_{1}^{a}\mathcal{O}_{2}^{b}\mathcal{O}_{3}^{c}}^{\langle\overline{q}\sigma\cdot Gq\rangle}(p,q;r)\langle\overline{q}\sigma_{\mu\nu}G^{\mu\nu}q\rangle\label{eq:OPE-condensates2_4}\\
&+\frac{1}{\lambda^{5}}C_{\mathcal{O}_{1}^{a}\mathcal{O}_{2}^{b}\mathcal{O}_{3}^{c}}^{\langle\overline{q}q\rangle^{2}}(p,q;r)\langle(\overline{q}Xq)(\overline{q}Xq)\rangle+\ldots\,,\label{eq:OPE-condensates2_5}
\end{align}
where the first term corresponds to the perturbative contribution ($D=0$) and the subsequent ones stand for the contributions of the quark ($D=3$), gluon ($D=4$), quark-gluon ($D=5$) and four-quark ($D=6$) condensates, with the respective canonical dimension shown in the bracket. We do not consider contributions of higher-dimensional ($D>6$) condensates in this work.\footnote{Similarly to the four-quark condensate, there is also another QCD condensate with $D=6$, the three-gluon condensate $\langle G_{\mu}^{a,\nu}G_{\nu}^{b,\sigma}G_{\sigma}^{c,\mu}\rangle f^{abc}$. For simplicity, we intentionally omit this contribution in this paper because we will cover this topic in detail in the future work. Moreover, it is assumed that such contribution vanishes in the chiral limit and, therefore, is of no interest for us in this paper.}

The Wilson coefficients, denoted as $C_{\mathcal{O}_{1}^{a}\mathcal{O}_{2}^{b}\mathcal{O}_{3}^{c}}^{\cal{I}}(p,q;r)$ in \eqref{eq:OPE-condensates2_1}-\eqref{eq:OPE-condensates2_5}, contain all the information about short-distance physics, i.e. the dynamics above some scale $\mu$, and are calculable in perturbative QCD by means of the technique of Feynman diagrams. They  are labeled by the upper index ${\cal{I}}=\mathbb{1},\,\langle\overline{q}q\rangle,\dots$, symbolizing which QCD condensate contribution they belong to.

In \eqref{eq:OPE-condensates2_5}, $X$ stands for any combination of the basis of $4\times 4$ matrices $(1,\gamma_{\mu},\gamma_{5},\gamma_{\mu}\gamma_{5},\sigma_{\mu\nu})$ with the basis of $3\times 3$ matrices $(1,T^{a})$, that preserves the Lorentz invariance of the four-quark condensate.

The four-quark condensate is of particular interest for us, and not only based on the fact that evaluation of its contribution to the Green functions is the most complicated of all the cases. It is a vacuum averaged value of a dimension-six operator, constructed of four quark fields, which can be written down in the schematic form
\begin{equation}
\langle(\overline{q}Xq)(\overline{q}Xq)\rangle\,.\label{eq:4q-condensate}
\end{equation}

According to \eqref{eq:4q-condensate}, there are several types of four-quark condensates that differ in the present matrix structure. However, within the approximation scheme, based on the assumption of the dominance of an intermediate vacuum states in the large $N_{c}$ limit, the matrix elements \eqref{eq:4q-condensate} can be further simplified in terms of squares of the quark condensate. In detail, having explicitly written down the spinor, color and flavor indices of the quark fields, the factorization hypothesis says that at large-$N_{c}$ limit, a general four-quark condensate can be rewritten as \cite{Shifman:1978bx}
\begin{equation}
\big\langle\overline{q}_{i,\alpha}^{A}q_{j,\beta}^{B}\overline{q}_{k,\gamma}^{C}q_{l,\delta}^{D}\big\rangle=\frac{\langle\overline{q}q\rangle^{2}}{2^{4}\cdot 3^{4}}(\delta_{ij}\delta_{kl}\delta_{\alpha\beta}\delta_{\gamma\delta}\delta^{AB}\delta^{CD}-\delta_{il}\delta_{jk}\delta_{\alpha\delta}\delta_{\beta\gamma}\delta^{AD}\delta^{BC})\,,\label{eq:4quark-averaging}
\end{equation}
which follows from the normalization of the quark condensate (see \eqref{eq:quark-condensate-propagation} or \eqref{eq:quark-condensate-propagation-LO}).

Assuming the special case of $X$ in \eqref{eq:4q-condensate} being the combination of the $4\times 4$ matrices, $\Gamma\in(1,\gamma_{\mu},\gamma_{5},\gamma_{\mu}\gamma_{5},\sigma_{\mu\nu})$, and the Gell-Mann matrices $T^{a}$, the formula \eqref{eq:4quark-averaging} above allows us to write down the general four-quark condensate \eqref{eq:4q-condensate} simply as
\begin{equation}
\big\langle(\overline{q}\,\Gamma\,T^{a}q)(\overline{q}\,\Gamma\,T^{a}q)\big\rangle=-\frac{\langle\overline{q}q\rangle^{2}}{2^{2}\cdot 3^{3}}\mathrm{Tr}\big[\Gamma^{2}\big]\,,\label{eq:4quark-averaging_general}
\end{equation}
which we can illustratively evaluate for various matrices $\Gamma$ as
\begin{subnumcases}{\big\langle(\overline{q}\,\Gamma\,T^{a}q)(\overline{q}\,\Gamma\,T^{a}q)\big\rangle=-\frac{\langle\overline{q}q\rangle^{2}}{2^{2}\cdot 3^{3}}\times}
   \,\,\,2^{2} & for\,$\Gamma=1$\,,\label{eq:4quark-averaging_1}\\
   \,\,\,2^{4} & for\,$\Gamma=\gamma_{\mu}$\,,\label{eq:4quark-averaging_2}\\
   \,\,\,2^{2} & for\,$\Gamma=\gamma_{5}$\,,\label{eq:4quark-averaging_3}\\
   -2^{4} & for\,$\Gamma=\gamma_{\mu}\gamma_{5}$\,,\label{eq:4quark-averaging_4}\\
   2^{4}\cdot 3 & for\,$\Gamma=\sigma_{\mu\nu}$\,,\label{eq:4quark-averaging_5}
\end{subnumcases}
however, in our case, only the relation \eqref{eq:4quark-averaging_2} will be needed in further calculations, i.e.
\begin{equation}
\big\langle(\overline{q}\gamma_{\mu}T^{a}q)(\overline{q}\gamma^{\mu}T^{a}q)\big\rangle=-\frac{2^{2}}{3^{3}}\langle\overline{q}q\rangle^{2}\,.\label{eq:4q-faktorizace}
\end{equation}

We note that a similar formula to ours \eqref{eq:4quark-averaging_general} can be also found in Ref.~\cite{Shifman:1978bx} (see eq.~6.15, page no.~432) or in Ref.~\cite{Govaerts:1986ua} (see eq.~1.7, page no.~708). However, notice that the authors of \cite{Govaerts:1986ua} use a different normalization since they assume the $\mathrm{SU}(3)$ symmetry to be broken and, therefore, do not consider the flavor indices.

That said, any of our results \eqref{eq:4quark-averaging_1}-\eqref{eq:4quark-averaging_5} for the four-quark condensates are smaller by a factor of three with respect to the results found in any literature, which considers the quark field to be single flavor. For example, compare \eqref{eq:4quark-averaging_1}, \eqref{eq:4quark-averaging_2}, \eqref{eq:4quark-averaging_3} and \eqref{eq:4quark-averaging_5} with eq.~4.8, page no.~715; eq.~4.4a, page no.~712; eq.~5.2, page no.~719 and eq.~4.4b, page no.~712 in Ref.~\cite{Govaerts:1986ua}, respectively. See also \cite{Reinders:1984sr}.

\subsection{Fock-Schwinger Gauge}
In the paragraphs above, we have mentioned that QCD condensates are, among other properties, vacuum expectation values of local operators. However, while calculating the contributions of the condensates by the means of the corresponding Feynman diagrams, one does not necessarily obtain the local condensate immediately since the quark or gluon fields are, generally, in different space-time points.

One is thus obligated to expand the quark or gluon fields into the Taylor series around the origin, project out the Lorentz structure and form the local condensate, where all the fields forming the condensate are already in the same space-time point. To this end, a suitable framework for such manipulations with quark and gluon fields is the Fock-Schwinger gauge \cite{Fock:1937,Schwinger:1951nm}.

By denoting $\mathcal{A}_{\mu}^{a}(x)$ as the gluon field\footnote{We alert the reader not to mix up the gluon field $\mathcal{A}_{\mu}^{a}(x)$ with the axial-vector current $A_{\mu}^{a}(x)$.}, the Fock-Schwinger gauge corresponds to the choice
\begin{equation}
(x-x_{0})^{\mu}\mathcal{A}_{\mu}^{a}(x)=0\,,\label{eq:fock-schwinger}
\end{equation}
where $x_{0}$ is an arbitrary coordinate which plays the role of the gauge parameter. Throughout the paper we consider $x_{0}=0$. Using the Fock-Schwinger gauge, it is possible to obtain the following expressions for the expansion of the quark and gluon fields, given in terms evaluated at the origin:
\begin{align}
q(x)&=q(0)+x^{\mu}\nabla_{\mu}q(0)+\frac{1}{2}x^{\mu}x^{\nu}\nabla_{\mu}\nabla_{\nu}q(0)+\frac{1}{6}x^{\mu}x^{\nu}x^{\rho}\nabla_{\mu}\nabla_{\nu}\nabla_{\rho}q(0)+\ldots\,,\label{eq:fock-schwinger-quark-field}\\
\mathcal{A}_{\mu}(x)&=\frac{1}{2}x^{\nu}G_{\nu\mu}(0)+\frac{1}{3}x^{\rho}x^{\nu}D_{\rho}G_{\nu\mu}(0)+\frac{1}{8}x^{\rho}x^{\sigma}x^{\nu}D_{\rho}D_{\sigma}G_{\nu\mu}(0)+\ldots\,,\label{eq:fock-schwinger-gluon-field}
\end{align}
where $\nabla_{\mu}=\partial_{\mu}+i g_{s}\mathcal{A}_{\mu}$ and $(D_{\mu})^{ab}=\partial_{\mu}\delta^{ab}+g_{s}f^{abc}\mathcal{A}_{\mu}^{c}$ are covariant derivatives in fundamental and adjoint representations. In \eqref{eq:fock-schwinger-quark-field}-\eqref{eq:fock-schwinger-gluon-field}, we intentionally suppressed all the relevant spinor, color and flavor indices.

\subsection{Propagation of Nonlocal Condensates}
Using the Fock-Schwinger gauge, one can thus obtain expressions that convert the nonlocal condensates into local ones. In our case, we have derived the following propagation formulas for the nonlocal quark, gluon and quark-gluon condensates, that gives us contributions of the local QCD condensates:
\begin{align}
\big\langle\overline{q}_{i,\alpha}^{A}(x)q_{k,\beta}^{B}(y)\big\rangle&=\bigg(\frac{\langle\overline{q}q\rangle}{2^{2}\cdot 3^{2}}\delta_{ik}-\frac{g_{s}\langle\overline{q}\sigma\hspace{-1pt}\cdot\hspace{-1pt}G q\rangle}{2^{5}\cdot 3^{2}}\big[F^{\langle\overline{q}q\rangle}(x,y)\big]_{ki}\label{eq:quark-condensate-propagation}\\
&\hspace{67pt}+\frac{i\pi\alpha_{s}\langle\overline{q}q\rangle^{2}}{2^{3}\cdot 3^{7}}\big[G^{\langle\overline{q}q\rangle}(x,y)\big]_{ki}+\ldots\bigg)\delta_{\alpha\beta}\delta^{AB}\,,\nonumber\\
\alpha_{s}\big\langle\mathcal{A}_{\mu}^{a}(x)\mathcal{A}_{\nu}^{b}(y)\big\rangle&=\frac{\alpha_{s}\langle G^{2}\rangle}{2^{7}\cdot 3}H_{\mu\nu}^{\langle G^{2}\rangle}(x,y)\delta^{ab}+\ldots\,,\label{eq:propagation_gluon_condensate}\\
g_{s}\big\langle\overline{q}_{i,\alpha}^{A}(x)\mathcal{A}_{\mu}^{a}(u)q_{k,\beta}^{B}(y)\big\rangle&=\bigg(\frac{g_{s}\langle\overline{q}\sigma\hspace{-1pt}\cdot\hspace{-1pt}G q\rangle}{2^{7}\cdot 3^{2}}\big[F_{\mu}^{\langle\overline{q}\mathcal{A}q\rangle}(x,u,y)\big]_{ki}\label{eq:4q_propagace_qAq}\\
&\hspace{67pt}+\frac{\pi\alpha_{s}\langle\overline{q}q\rangle^{2}}{2^{3}\cdot 3^{5}}\big[G_{\mu}^{\langle\overline{q}\mathcal{A}q\rangle}(x,u,y)\big]_{ki}+\ldots\bigg)(T^{a})_{\beta\alpha}\delta^{AB}\,.\nonumber
\end{align}
Before we proceed, we emphasize that we have assumed the validity of the dominance of an intermediate vacuum states in the large $N_{c}$ limit \eqref{eq:4quark-averaging} when evaluating the propagation formulas proportional to the four-quark condensate.

As can be seen from above, the nonlocal quark condensates propagates itself as the local quark, quark-gluon and four-quark condensates and the respective functions in \eqref{eq:quark-condensate-propagation}, that collect the coordinate dependence, read
\begin{align}
F^{\langle\overline{q}q\rangle}(x,y)&=\frac{1}{2}(x-y)^{2}+\frac{i}{3}\sigma^{(x)(y)}\,,\label{eq:quark-condensate-propagation-function-quark-gluon}\\
G^{\langle\overline{q}q\rangle}(x,y)&=4(x\cdot y)(\slashed{x}-\slashed{y})-(x^{2}-y^{2})(\slashed{x}+\slashed{y})\,.\label{eq:quark-condensate-propagation-function}
\end{align}

The propagation of the nonlocal gluon condensate is straightforward since the only relevant term is the local gluon condensate, with the corresponding function $H_{\mu\nu}^{\langle G^{2}\rangle}(x,y)$ in \eqref{eq:propagation_gluon_condensate} in the form
\begin{equation}
H_{\mu\nu}^{\langle G^{2}\rangle}(x,y)=(x\hspace{-1pt}\cdot\hspace{-1pt}y)g_{\mu\nu}-y_{\mu}x_{\nu}\,.\label{eq:propagation_gluon_condensate_function}
\end{equation}

Finally, the nonlocal quark-gluon condensate propagates as the local quark-gluon and four-quark condensates, with the functions in \eqref{eq:4q_propagace_qAq} given as
\begin{align}
F_{\mu}^{\langle\overline{q}\mathcal{A}q\rangle}(x,u,y)&=\sigma^{(u)\mu}\,,\label{eq:quark-gluon-condensate-propagation-function-quark-gluon}\\
G_{\mu}^{\langle\overline{q}\mathcal{A}q\rangle}(x,u,y)&=\frac{1}{6}\gamma^{\mu}\big[3u\hspace{-1pt}\cdot\hspace{-1pt}(x+y)-4u^{2}\big]+\frac{1}{6}\slashed{u}\big[4u^{\mu}-3(x+y)^{\mu}\big]-\frac{i}{2}\varepsilon^{\mu(x-y)(u)\alpha}\gamma_{\alpha}\gamma_{5}\,.\label{eq:4q_propagace_qAq-function_contracted}
\end{align}

The propagation formulas \eqref{eq:quark-condensate-propagation}-\eqref{eq:4q_propagace_qAq} can be graphically illustrated as follows:
\begin{figure}[htb]
  \centering
    \includegraphics[scale=0.5]{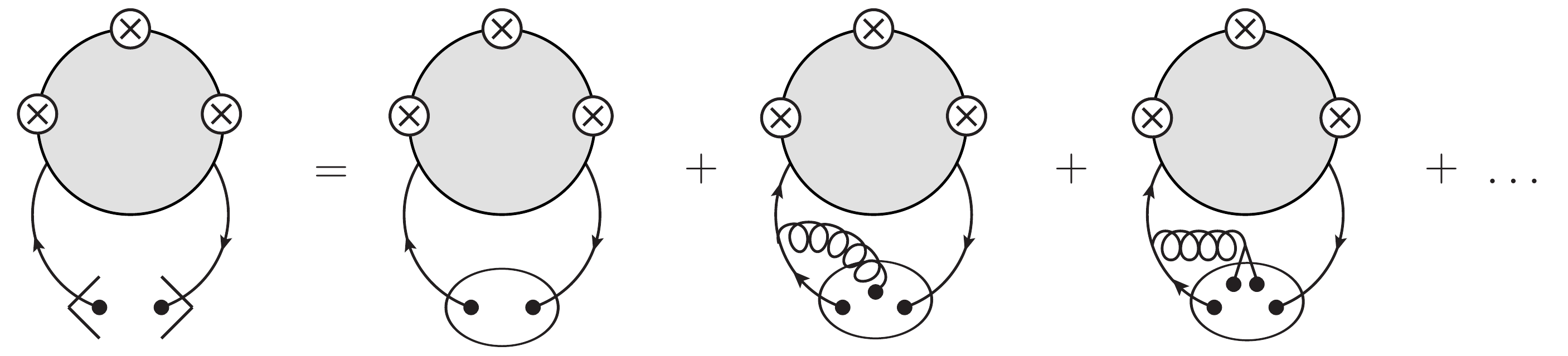}
    \caption{A graphical illustration of the propagation formula \eqref{eq:quark-condensate-propagation}. The black line stands for the quark line and the bold dot on its end represents the quark field that goes between the vacuum states. The curly line stands for the gluon field, if such line has a bold dot on its end, then it represents the gluon field that goes between the vacuum states, too. The symbol $\mathbf{\otimes}$ stands for an insertion of the current and the grey blob means any possible couplings to another currents or the quark-gluon vertices. On the left-hand side, the brackets around the quark fields means averaging over them according to the formula \eqref{eq:quark-condensate-propagation}, while on the right-hand side, the oval over the respective quark or gluon fields means a creation of the local condensates.}
    \label{fig:propagation_quark_condensate}
\end{figure}
\begin{figure}[htb]
  \centering
    \includegraphics[scale=0.5]{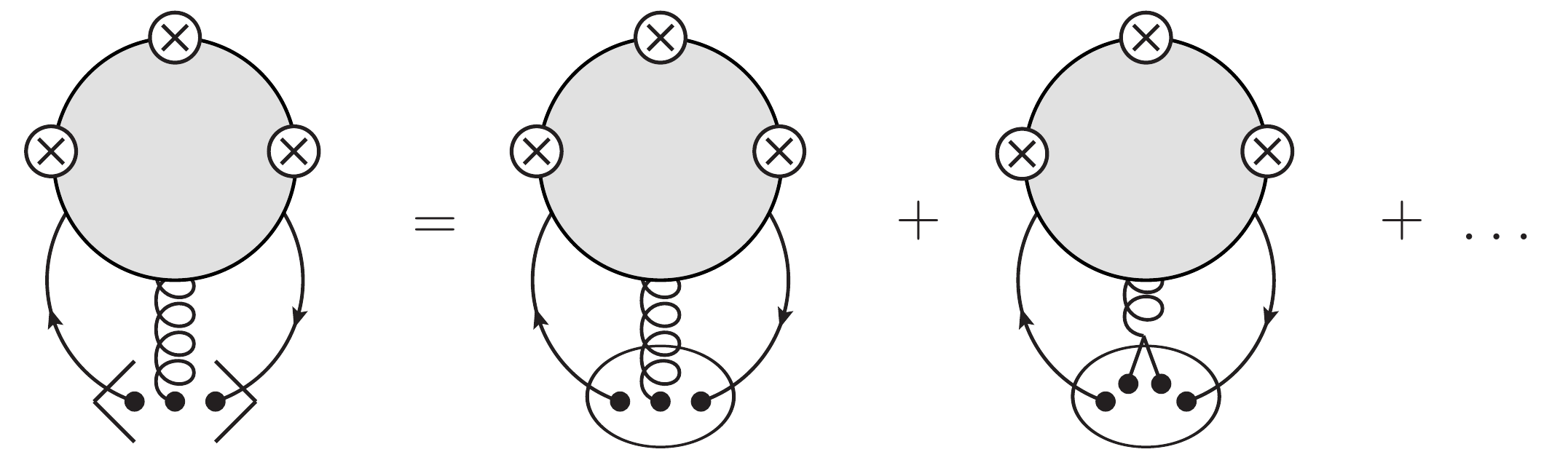}
    \caption{A graphical illustration of the propagation formula \eqref{eq:4q_propagace_qAq}. The notation corresponds to the one already explained in Fig.~\ref{fig:propagation_quark_condensate}.}
    \label{fig:propagation_quark-gluon_condensate}
\end{figure}

\subsection{Comparison with Literature}\label{ssec:comparison}
In the literature, the formulas \eqref{eq:quark-condensate-propagation}-\eqref{eq:4q_propagace_qAq} are often presented with one of the coordinates set into the origin. Recovering the general structure for all coordinates to be nonzero is not trivial, since the Fock-Schwinger gauge prevents us from simply shifting the coordinates. Also, some of the sources in the literature present formulas, equivalent to \eqref{eq:quark-condensate-propagation}-\eqref{eq:4q_propagace_qAq}, with several typos. Here we try to list such misprints for readers convenience.

Of course, due to the loss of information when setting one of the coordinates to zero, one is not able to sufficiently check the validity of our formulas, however, one may at least try to verify whether the numerical factors are in agreement.

First of all, we explicitly write out the flavor indices when expanding the quark fields. The propagation formulas \eqref{eq:quark-condensate-propagation}-\eqref{eq:4q_propagace_qAq} contain two quark fields between the vacuum states, so the flavor factor present in such expressions is $\frac{1}{3}\delta^{AB}$. On the other hand, most of the literature does not consider these indices explicitly. Thus, when comparing our formulas to the literature, one should be careful whether it is necessary to multiply the relevant propagation formulas with a factor of 3 and scratch the $\delta^{AB}$. Then, the comparison can be made quite easily.

Since the comparison with the known results is of utmost importance, due to the frequent typos or inconsistencies found in the literature, we pay a special care to it in the following paragraphs. However, let us emphasize that the comparison of our formulas with the literature is provided upon consideration that all the results of other authors have been rewritten into such form that the respective indices and coordinates are the same as ours. Only then the comparison is performed.

\subsection*{Propagation formula \eqref{eq:quark-condensate-propagation}}
Let us start with the propagation of the nonlocal quark condensate, specifically with the first term of such propagation,
\begin{align}
\big\langle\overline{q}_{i,\alpha}^{A}(x)q_{k,\beta}^{B}(y)\big\rangle\ni\frac{1}{2^{2}\cdot 3^{2}}\langle\overline{q}q\rangle\delta_{ik}\delta_{\alpha\beta}\delta^{AB}\,.\label{eq:quark-condensate-propagation_nula_1}
\end{align}

Since the derivation of the local quark condensate is straightforward, we only stress that omitting the flavor indices leads to the difference in the normalization, since the factor becomes greater by the factor of three. This fact further propagates itself, especially into the evaluation of the four-quark condensate within the factorization approximation. As an example of literature where the flavor indices are omitted, we mention \cite{Elias:1986vb} (eq.~2, page no.~3537), \cite{Elias:1987ac} (eq.~A5, page no.~1596) or \cite{Elias:1985qp} (eq.~10, page no.~186). Also, see \cite{Ioffe:2010zz} (eq.~6.227, page no.~255) and \cite{Narison:2007spa} (eq.~28.15, page no.~301).

The part of \eqref{eq:quark-condensate-propagation} proportional to the quark-gluon condensate reads
\begin{align}
\big\langle\overline{q}_{i,\alpha}^{A}(x)q_{k,\beta}^{B}(y)\big\rangle\ni -\frac{g_{s}\langle\overline{q}\sigma\hspace{-1pt}\cdot\hspace{-1pt}G q\rangle}{2^{5}\cdot 3^{2}}\bigg[\frac{1}{2}(x-y)^{2}+\frac{i}{3}\sigma^{(x)(y)}\bigg]_{ki}\delta_{\alpha\beta}\delta^{AB}\,,\label{eq:quark-condensate-propagation_nula_2}
\end{align}
which is in an agreement with Ref.~\cite{Elias:1986vb} (eq.~2, page no.~3537), \cite{Elias:1987ac} (eq.~A45, page no.~1600) and \cite{Elias:1985qp} (eq.~10, page no.~186), where the authors consider both coordinates of the quark fields to be nonzero.\footnote{Notice that eq.~2 at page no.~3537 in~\cite{Elias:1986vb} lacks a factor of $\frac{1}{3}$ in comparison with eq.~A45, page no.~1600 in~\cite{Elias:1987ac}, where the authors suppress the factor of $\frac{1}{3}\delta^{\alpha\beta}$ intentionally.} However, the authors use a different sign convention for the covariant derivative in the fundamental representation, which makes for a difference in an overall sign.

On the other hand, one-coordinate version of the formula \eqref{eq:quark-condensate-propagation_nula_2}, for $y=0$, coincides with the one given by Ioffe et al. in \cite{Ioffe:2010zz} (eq.~6.227, page no.~255) but differs by a factor of $i$ with the expression presented by Narison in \cite{Narison:2007spa} (eq.~28.15, page no.~301). 

The one-coordinate version of \eqref{eq:quark-condensate-propagation} for the propagation of the four-quark condensate,
\begin{align}
\big\langle\overline{q}_{i,\alpha}^{A}(x)q_{k,\beta}^{B}(0)\big\rangle\ni -i\frac{\pi\alpha_{s}\langle\overline{q}q\rangle^{2}}{2^{3}\cdot 3^{7}}x^{2}\big(\slashed{x}\big)_{ki}\delta_{\alpha\beta}\delta^{AB}\,,\label{eq:quark-condensate-propagation_nula_3}
\end{align}
differs from the expression presented by Ioffe et al. in Ref.~\cite{Ioffe:2010zz} (eq.~6.227, page no.~255) by a factor of four. However, we are in an agreement with the formula given by Narison in Ref.~\cite{Narison:2007spa} (eq.~28.15, page no.~301). Once again, we remind the reader that also the difference in the normalization of the four-quark condensate \eqref{eq:4q-faktorizace} has been taken into account while comparing this part.

\subsection*{Propagation formula \eqref{eq:propagation_gluon_condensate}}
The formula \eqref{eq:propagation_gluon_condensate} for converting the vacuum value of two gluon fields into the gluon condensate is trivial, can be easily derived and has been presented in the literature many times already. See for example \cite{Bagan:1992tg}, eq.~65 at page no.~165, or \cite{Narison:2007spa}, eq.~28.12, page no.~300.

\subsection*{Propagation formula \eqref{eq:4q_propagace_qAq}}
The propagation formula \eqref{eq:4q_propagace_qAq} for the nonlocal quark-gluon condensate reads
\begin{align}
g_{s}\big\langle\overline{q}_{i,\alpha}^{A}(x)\mathcal{A}_{\mu}^{a}(u)q_{k,\beta}^{B}(y)\big\rangle\ni\frac{g_{s}\langle\overline{q}\sigma\hspace{-1pt}\cdot\hspace{-1pt}G q\rangle}{2^{7}\cdot 3^{2}}\big(\sigma^{(u)\mu}\big)_{ki}(T^{a})_{\beta\alpha}\delta^{AB}\,,\label{eq:4q_propagace_qAq_nula_1}
\end{align}
which is in an agreement with \cite{Elias:1987ac}, eq.~A75 at page no.~1603, and also with Narison~\cite{Narison:2007spa} (eq.~28.16 at page no.~301), after contracting the formula with the color part.

Finally, the part of \eqref{eq:4q_propagace_qAq} relevant for the propagation of the four-quark condensate is
\begin{align}
g_{s}\big\langle\overline{q}_{i,\alpha}^{A}(x)&\mathcal{A}_{\mu}^{a}(u)q_{k,\beta}^{B}(0)\big\rangle\ni\frac{\pi\alpha_{s}\langle\overline{q}q\rangle^{2}}{2^{3}\cdot 3^{5}}\times\label{eq:4q_propagace_qAq_nula_2}\\
&\times\bigg(\frac{1}{6}\gamma^{\mu}(3u\hspace{-1pt}\cdot\hspace{-1pt}x-4u^{2})+\frac{1}{6}\slashed{u}(4u^{\mu}-3x^{\mu})-\frac{i}{2}\varepsilon^{\mu(x)(u)\alpha}\gamma_{\alpha}\gamma_{5}\bigg)_{ki}(T^{a})_{\beta\alpha}\delta^{AB}\,,\nonumber
\end{align}
which differs from Narison~\cite{Narison:2007spa} (eq.~28.16, page no.~301) by the factor of $i$, provided we take into account \eqref{eq:4q-faktorizace}.

\subsection{Translation Invariance}\label{ssec:translation}
Although the Fock-Schwinger gauge violates translation invariance, the final result of any calculation, obtained within this gauge, must be translation invariant. One thus has to make sure and verify that the whole contribution is indeed translation invariant after all the possible contributing diagrams (and their respective permutations) are accounted for. If this is the case, translation invariance allows us to make a shift by, say, the $z$-coordinate in all arguments\footnote{Using the computer brute force we have independently checked \eqref{eq:translation_invariance} for all the relevant Dirac matrices.}, which finally allows us to set $z=0$, i.e.
\begin{equation}
\Pi_{\mathcal{O}_{1}^{a}\mathcal{O}_{2}^{b}\mathcal{O}_{3}^{c}}(x,y,z)=\Pi_{\mathcal{O}_{1}^{a}\mathcal{O}_{2}^{b}\mathcal{O}_{3}^{c}}(x-z,y-z,0)\overset{(z=0)}{\equiv}\Pi_{\mathcal{O}_{1}^{a}\mathcal{O}_{2}^{b}\mathcal{O}_{3}^{c}}(x,y)\,.\label{eq:translation_invariance}
\end{equation}
The translation invariance, as outlined in the formula above, also simplifies the Fourier transform, because one thus integrates over only two coordinates, instead of three. This also allows us to overcome potentional difficulties related to derivatives of the delta function.

In our case, the Fock-Schwinger gauge is used to obtain the contributions of the quark, gluon, quark-gluon and four-quark condensates. In all cases, the translation invariance of the final result was carefully checked. For details, see respective Sections \ref{sec:gluon-condensate}, \ref{sec:quark-gluon-condensate} and \ref{sec:four-quark-condensate}.

%%%%%%%%%%%%%%%%%%%%%%%%%%%%%%%%%%%%%%%%%%%%%%%%%%%%%%%%%%%%%%%%%%%%%%%%%%%%%%%%%%%%%%%%%%%%%%%%%%%%%%%%%
%%%%%%%%%%%%%%%%%%%%%%%%%%%%%%%%%%%%%%%%%%%%%%%%%%%%%%%%%%%%%%%%%%%%%%%%%%%%%%%%%%%%%%%%%%%%%%%%%%%%%%%%%
%%% Section: Perturbative Contribution
%%%%%%%%%%%%%%%%%%%%%%%%%%%%%%%%%%%%%%%%%%%%%%%%%%%%%%%%%%%%%%%%%%%%%%%%%%%%%%%%%%%%%%%%%%%%%%%%%%%%%%%%%
%%%%%%%%%%%%%%%%%%%%%%%%%%%%%%%%%%%%%%%%%%%%%%%%%%%%%%%%%%%%%%%%%%%%%%%%%%%%%%%%%%%%%%%%%%%%%%%%%%%%%%%%%

\section{Perturbative Contribution}\label{sec:perturbative}

\subsection{General Remarks}
Perturbative contribution is the lowest possible contribution to the OPE, denoted as \eqref{eq:OPE-condensates2_1}. Formally, one could assign the unit operator as the respective QCD condensate to this contribution. Then, in the case of three-point Green functions, such contribution is given by a triangle diagram, with all the quark fields contracted.
\begin{figure}[htb]
  \centering
    \includegraphics[scale=0.5]{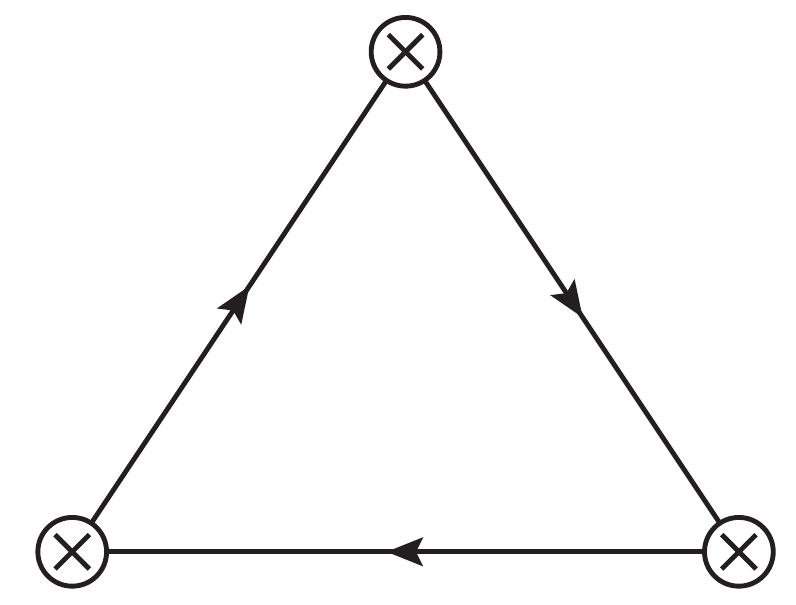}
    \caption{Feynman diagram of the perturbative contribution to the three-point Green functions at the leading order. Having to add the Bose-symmetrized diagram is tacitly assumed.}
    \label{fig:perturbative-contribution-leading-order}
\end{figure}

The relevant contribution to the three-point Green functions in the leading order $\mathcal{O}(1)$, depicted at Fig.~\ref{fig:perturbative-contribution-leading-order}, can be written down as\footnote{For clarity, we will omit the specific designation of the unit operator $\mathbb{1}$ of the perturbative contribution in \eqref{eq:perturbative-triangle} and in the subsequent results in Subsection \ref{ssec:perturbative_results}, however, we keep the symbol of this contribution in the upper indices of the respective formfactors, as shown in \eqref{eq:OPE-condensates2_1}.}
\begin{align}
\Pi_{\mathcal{O}_{1}^{a}\mathcal{O}_{2}^{b}\mathcal{O}_{3}^{c}}^{\mathbb{1}}(p,q;r)=&-N_{c}\,\mathrm{Tr}[T^{a}T^{b}T^{c}]\int\frac{\mathrm{d}^{4}\ell}{(2\pi)^{4}}\mathrm{Tr}\Big[\Gamma_{1}S_{0}(\ell)\Gamma_{2}S_{0}(\ell-q)\Gamma_{3}S_{0}(\ell+p)\Big]\label{eq:perturbative-triangle}\\
&+\big[(\Gamma_{1},a,p)\leftrightarrow(\Gamma_{2},b,q)\big]\,,\nonumber
\end{align}
where the minus sign in front of the integral sign is present there because of the closed fermion loop and
\begin{equation}
S_{0}(\ell)=\frac{i}{\slashed{\ell}}\label{eq:quark-propagator}
\end{equation}
denotes the free massless fermion propagator.\footnote{Including all the relevant indices, the massless quark propagator \eqref{eq:quark-propagator} reads
\begin{equation*}
\big[S_{0}(\ell)\big]_{ik,\alpha\beta}^{AB}=\bigg(\frac{i}{\slashed{\ell}}\bigg)_{ik}\delta_{\alpha\beta}\delta^{AB}\equiv\big[S_{0}(\ell)\big]_{ik}\delta_{\alpha\beta}\delta^{AB}\,,
\end{equation*}
since the free propagator does not change the flavor nor the color of the propagating quark.}

Before we present the results, let us briefly comment on an evaluation of the integral in \eqref{eq:perturbative-triangle}. As it is common, we work in dimensional regularisation and perform an integration over dimension $d=4-2\varepsilon$. As will be mentioned in the particular cases later, this procedure often leads to presence of divergent terms. Specifically in our case, which corresponds to the $\overline{\mathrm{MS}}$ scheme, it is useful to define the divergent term $1/\widehat{\varepsilon}$ in the form
\begin{equation}
\frac{1}{\widehat{\varepsilon}}=\frac{1}{\varepsilon}-\gamma_{\mathrm{E}}+\mathrm{ln}(4\pi)\,,
\end{equation}
where $\gamma_{E}=0.577\ldots$ is the Euler-Mascheroni constant.

We also define the following quantities. The Passarino-Veltman function for a scalar one-loop triangle integral with massless internal lines, $C_{0}(p^{2},q^{2},r^{2};0,0,0)$, is defined to be
\begin{equation}
\int\frac{\mathrm{d}^{4}\ell}{(2\pi)^{4}}\frac{1}{(\ell-q)^{2}\ell^{2}(\ell+p)^{2}}=\frac{i}{(4\pi)^{2}}C_{0}(p^{2},q^{2},r^{2};0,0,0)\,,
\end{equation}
where the explicit form of the $C_{0}$-function reads \cite{tHooft:1978jhc}
\begin{align}
C_{0}(p^{2},q^{2},r^{2};0,0,0)=\frac{1}{\lambda_{K}^{1/2}}\Bigg[&\mathrm{Li}_{2}\Bigg(\frac{p^{2}+q^{2}-r^{2}+\lambda_{K}^{1/2}}{p^{2}+q^{2}-r^{2}-\lambda_{K}^{1/2}}\Bigg)-\mathrm{Li}_{2}\Bigg(\frac{p^{2}+q^{2}-r^{2}-\lambda_{K}^{1/2}}{p^{2}+q^{2}-r^{2}+\lambda_{K}^{1/2}}\Bigg)\label{eq:C0function}\\
+&\mathrm{Li}_{2}\Bigg(\frac{p^{2}-q^{2}-r^{2}-\lambda_{K}^{1/2}}{p^{2}-q^{2}-r^{2}+\lambda_{K}^{1/2}}\Bigg)-\mathrm{Li}_{2}\Bigg(\frac{p^{2}-q^{2}-r^{2}+\lambda_{K}^{1/2}}{p^{2}-q^{2}-r^{2}-\lambda_{K}^{1/2}}\Bigg)\nonumber\\
+&\mathrm{Li}_{2}\Bigg(\frac{p^{2}-q^{2}+r^{2}+\lambda_{K}^{1/2}}{p^{2}-q^{2}+r^{2}-\lambda_{K}^{1/2}}\Bigg)-\mathrm{Li}_{2}\Bigg(\frac{p^{2}-q^{2}+r^{2}-\lambda_{K}^{1/2}}{p^{2}-q^{2}+r^{2}+\lambda_{K}^{1/2}}\Bigg)\Bigg]\,,\nonumber
\end{align}
with
\begin{align}
\lambda_{K}\equiv\lambda_{K}(p^{2},q^{2},r^{2})&=(p^{2}+q^{2}-r^{2})^{2}-4p^{2}q^{2}\,,\label{eq:kallen}\\
&=p^{4}+q^{4}+r^{4}-2p^{2}q^{2}-2p^{2}r^{2}-2q^{2}r^{2}\label{eq:kallen2}
\end{align}
being the fully symmetric K\"{a}ll\'{e}n (triangle) function and $\mathrm{Li}_{2}(x)$ the dilogarithm (also known as Spence's function), defined usually in the form
\begin{equation}
\mathrm{Li}_{2}(x)=-\int_{0}^{x}\frac{\ln(1-u)}{u}\,\mathrm{d}u\,.
\end{equation}

A treatment of the Dirac $\gamma_{5}$ matrix plays an important role in the calculations of loop integrals. Since we evaluated all the integrals shown below using the \mathematica tool \packageX, we refer the reader for a detailed description to the Ref.~\cite{Patel:2015tea,Patel:2016fam}. Here we only mention that we consider the naive implementation of the dimensional regularization, where $\gamma_{5}$ anticommutes with other Dirac matrices.

\subsection{Results}\label{ssec:perturbative_results}
In what follows we present all the obtained results divided into three parts, based on the similarity of the respective results.
We introduce the following notation especially for this subsection. We split the perturbative results, given in terms of the respective formfactors, into several parts,
\begin{align}
\mathcal{P}^{\mathbb{1}}(p^{2},q^{2};r^{2})=&\quad\,\mathcal{P}^{\mathbb{1},(c)}(p^{2},q^{2};r^{2})+\mathcal{P}^{\mathbb{1},(C_{0})}(p^{2},q^{2};r^{2})C_{0}(p^{2},q^{2},r^{2};0,0,0)\label{eq:perturbative_split}\\
&+\mathcal{P}^{\mathbb{1},(p)}(p^{2},q^{2};r^{2})\log\bigg(\frac{p^{2}}{r^{2}}\bigg)+\mathcal{P}^{\mathbb{1},(q)}(p^{2},q^{2};r^{2})\log\bigg(\frac{q^{2}}{r^{2}}\bigg)\nonumber\\
&+\mathcal{P}^{\mathbb{1},(\mu)}(p^{2},q^{2};r^{2})\log\bigg(\hspace{-4pt}-\frac{\mu^{2}}{r^{2}}\bigg)\,,\nonumber
\end{align}
with $\mathcal{P}$ being symbolically the general formfactors $\mathcal{F}$, $\mathcal{G}$ or $\mathcal{H}$ of all the correlators.

\subsubsection*{\boldmath\texorpdfstring{$\langle ASP\rangle$}{}, \texorpdfstring{$\langle VSS\rangle$}{} and \texorpdfstring{$\langle VPP\rangle$}{} Green Functions}
Although the correlators $\langle ASP\rangle$, $\langle VPP\rangle$ and $\langle VSS\rangle$ differ in the flavor structure, i.e. in the presence of the $d^{abc}$ and $f^{abc}$ symbols, one can notice that in the naive implementation of the dimensional regularization, due to the anticommuting $\gamma_{5}$, the Lorentz structure of \eqref{eq:perturbative-triangle} is the same for all of these correlators. In other words,
\begin{equation}
\big[\Pi_{ASP}^{\mathbb{1}}(p,q;r)\big]_{\mu}=\big[\Pi_{VSS}^{\mathbb{1}}(p,q;r)\big]_{\mu}=\big[\Pi_{VPP}^{\mathbb{1}}(p,q;r)\big]_{\mu}\,.\label{eq:perturbative-ASP-VPP-VSS_1}
\end{equation}

Since the two-point perturbative contributions of $\langle SS\rangle$ and $\langle PP\rangle$ are the same, see \eqref{eq:2pt_gluon_condensate_SS_PP}, the respective formfactors of the $\langle ASP\rangle$, $\langle VPP\rangle$ and $\langle VSS\rangle$ Green functions in \eqref{eq:asp-definition_dekompozice}-\eqref{eq:vpp-definition_dekompozice} are also the same, i.e.
\begin{align}
\mathcal{F}_{ASP}^{\mathbb{1}}(p^{2},q^{2};r^{2})=\mathcal{F}_{VPP}^{\mathbb{1}}(p^{2},q^{2};r^{2})=\mathcal{F}_{VSS}^{\mathbb{1}}(p^{2},q^{2};r^{2})\,,\label{eq:perturbative-ASP-VPP-VSS_2}
\end{align}
with the explicit result for the $\langle ASP\rangle$ correlator in the form
\begin{align}
\mathcal{F}_{ASP}^{\mathbb{1},(c)}(p^{2},q^{2};r^{2})=&-\frac{i N_{c}}{8\pi^{2}}\bigg(2+\frac{1}{\widehat{\varepsilon}}\bigg)\,,\\
\mathcal{F}_{ASP}^{\mathbb{1},(C_{0})}(p^{2},q^{2};r^{2})=&-\frac{i N_{c}}{4\pi^{2}\lambda_{K}}p^{2}q^{2}r^{2}\,,\\
\mathcal{F}_{ASP}^{\mathbb{1},(p)}(p^{2},q^{2};r^{2})=&\quad\,\frac{i N_{c}}{8\pi^{2}\lambda_{K}}p^{2}(p^{2}-q^{2}-r^{2})\,,\\
\mathcal{F}_{ASP}^{\mathbb{1},(q)}(p^{2},q^{2};r^{2})=&-\frac{i N_{c}}{8\pi^{2}\lambda_{K}}q^{2}(p^{2}-q^{2}+r^{2})\,,\\
\mathcal{F}_{ASP}^{\mathbb{1},(\mu)}(p^{2},q^{2};r^{2})=&-\frac{i N_{c}}{8\pi^{2}}\,.
\end{align}

\subsubsection*{\boldmath\texorpdfstring{$\langle VVA\rangle$}{} and \texorpdfstring{$\langle AAA\rangle$}{} Green Functions}
The $\langle VVA\rangle$ triangle has been an object of interest and utmost importance for decades. The reason is the presence of the axial anomaly that has been studied for the first time in 1969 by seminal papers of Adler \cite{Adler:1969gk} and Bell and Jackiw \cite{Bell:1969ts}.

Having the perturbative contribution to the $\langle VVA\rangle$ calculated, one can evaluate the $\langle AAA\rangle$ in the similar way. One possible approach \cite{Coriano:2007fw} consists of taking the cyclic permutation of the Dirac matrices in \eqref{eq:perturbative-triangle} into account, which leads to the contribution given as
\begin{equation}
\big[\Pi_{AAA}^{\mathbb{1}}(p,q;r)\big]_{\mu\nu\rho}^{abc}=\frac{1}{3}\Big(\big[\Pi_{VVA}^{\mathbb{1}}(p,q;r)\big]_{\mu\nu\rho}^{abc}+\big[\Pi_{VVA}^{\mathbb{1}}(r,q;p)\big]_{\rho\nu\mu}^{cba}+\big[\Pi_{VVA}^{\mathbb{1}}(p,r;q)\big]_{\mu\rho\nu}^{acb}\Big)\,.\label{eq:AAA_perturbative_permutations}
\end{equation}
This parametrization ensures satisfaction of the Bose symmetry of the $\langle AAA\rangle$ triangle in all the arguments and preserves the axial-vector Ward identities.

Then, the perturbative contributions for the $\langle VVA\rangle$ Green function can be written down as\footnote{In the Ref.~\cite{Armillis:2009sm}, the perturbative contribution to the $\langle VVA\rangle$ Green function in the massless limit is shown explicitly, and is in the form similar to ours. Although the authors do not consider a presence of the $\mathrm{SU}(3)$ generators and use a different normalization, we agree on the results for the respective formfactors up to one term, which is, however, most likely a typo. See the third line in eq.~50 at page no.~15 in Ref.~\cite{Armillis:2009sm}.}
\begin{align}
\mathcal{F}_{VVA}^{\mathbb{1},(c)}(p^{2},q^{2};r^{2})=&\quad\,\frac{i N_{c}}{8\pi^{2}\lambda_{K}}r^{2}\,,\label{eq:vva-perturbative_1_1}\\
\mathcal{F}_{VVA}^{\mathbb{1},(C_{0})}(p^{2},q^{2};r^{2})=&\quad\,\frac{i N_{c}}{8\pi^{2}\lambda_{K}^{2}}r^{2}\times\label{eq:vva-perturbative_1_2}\\
&\hspace{-50pt}\times\Big[(p^{2}+q^{2})\Big(r^{4}+(p^{2}-q^{2})^{2}\Big)-2r^{2}(p^{4}+q^{4}-p^{2}q^{2})\Big]\,,\nonumber\\
\mathcal{F}_{VVA}^{\mathbb{1},(p)}(p^{2},q^{2};r^{2})=&-\frac{i N_{c}}{16\pi^{2}\lambda_{K}^{2}}(p^{2}-q^{2}-r^{2})(\lambda_{K}+6p^{2}r^{2})\,,\label{eq:vva-perturbative_1_3}\\
\mathcal{F}_{VVA}^{\mathbb{1},(q)}(p^{2},q^{2};r^{2})=&\quad\,\frac{i N_{c}}{16\pi^{2}\lambda_{K}^{2}}(p^{2}-q^{2}+r^{2})(\lambda_{K}+6q^{2}r^{2})\,,\label{eq:vva-perturbative_1_4}\\
\mathcal{F}_{VVA}^{\mathbb{1},(\mu)}(p^{2},q^{2};r^{2})=&\quad\,0\,,\label{eq:vva-perturbative_1_5}\\
\mathcal{G}_{VVA}^{\mathbb{1},(c)}(p^{2},q^{2};r^{2})=&\quad\,\frac{i N_{c}}{8\pi^{2}\lambda_{K}}(p^{2}-q^{2})\,,\label{eq:vva-perturbative_2_1}\\
\mathcal{G}_{VVA}^{\mathbb{1},(C_{0})}(p^{2},q^{2};r^{2})=&\quad\,\frac{i N_{c}}{8\pi^{2}\lambda_{K}^{2}}(p^{2}-q^{2})r^{2}(\lambda_{K}+6p^{2}q^{2})\,,\label{eq:vva-perturbative_2_2}\\
\mathcal{G}_{VVA}^{\mathbb{1},(p)}(p^{2},q^{2};r^{2})=&-\frac{i N_{c}}{16\pi^{2}\lambda_{K}^{2}}\times\label{eq:vva-perturbative_2_3}\\
&\hspace{-50pt}\times\Big(r^{6}-r^{4}(7p^{2}+q^{2})+r^{2}(12p^{2}q^{2}+5p^{4}-q^{4})+(p^{2}-q^{2})^{2}(p^{2}+q^{2})\Big)\,,\nonumber\\
\mathcal{G}_{VVA}^{\mathbb{1},(q)}(p^{2},q^{2};r^{2})=&\quad\,\frac{i N_{c}}{16\pi^{2}\lambda_{K}^{2}}\times\label{eq:vva-perturbative_2_4}\\
&\hspace{-50pt}\times\Big(r^{6}-r^{4}(p^{2}+7q^{2})+r^{2}(12p^{2}q^{2}-p^{4}+5q^{4})+(p^{2}-q^{2})^{2}(p^{2}+q^{2})\Big)\,,\nonumber\\
\mathcal{G}_{VVA}^{\mathbb{1},(\mu)}(p^{2},q^{2};r^{2})=&\quad\,0\,,\label{eq:vva-perturbative_2_5}\\
\mathcal{H}_{VVA}^{\mathbb{1},(c)}(p^{2},q^{2};r^{2})=&-\frac{i N_{c}}{8\pi^{2}\lambda_{K}}(p^{2}-q^{2})\,,\label{eq:vva-perturbative_3_1}\\
\mathcal{H}_{VVA}^{\mathbb{1},(C_{0})}(p^{2},q^{2};r^{2})=&-\frac{i N_{c}}{8\pi^{2}\lambda_{K}^{2}}(p^{2}-q^{2})r^{2}(\lambda_{K}+6p^{2}q^{2})\,,\label{eq:vva-perturbative_3_2}\\
\mathcal{H}_{VVA}^{\mathbb{1},(p)}(p^{2},q^{2};r^{2})=&\quad\,\frac{i N_{c}}{16\pi^{2}\lambda_{K}^{2}}\times\label{eq:vva-perturbative_3_3}\\
&\hspace{-50pt}\times\Big(r^{6}-r^{4}(7p^{2}+q^{2})+r^{2}(12p^{2}q^{2}+5p^{4}-q^{4})+(p^{2}-q^{2})^{2}(p^{2}+q^{2})\Big)\,,\nonumber\\
\mathcal{H}_{VVA}^{\mathbb{1},(q)}(p^{2},q^{2};r^{2})=&-\frac{i N_{c}}{16\pi^{2}\lambda_{K}^{2}}\times\label{eq:vva-perturbative_3_4}\\
&\hspace{-50pt}\times\Big(r^{6}-r^{4}(p^{2}+7q^{2})+r^{2}(12p^{2}q^{2}-p^{4}+5q^{4})+(p^{2}-q^{2})^{2}(p^{2}+q^{2})\Big)\,,\nonumber\\
\mathcal{H}_{VVA}^{\mathbb{1},(\mu)}(p^{2},q^{2};r^{2})=&0\,,\label{eq:vva-perturbative_3_5}
\end{align}
and for the $\langle AAA\rangle$ correlator we have
\begin{align}
\mathcal{F}_{AAA}^{\mathbb{1},(c)}(p^{2},q^{2};r^{2})=&\quad\,0\,,\label{eq:aaa-perturbative_1_1}\\
\mathcal{F}_{AAA}^{\mathbb{1},(C_{0})}(p^{2},q^{2};r^{2})=&\quad\,0\,,\label{eq:aaa-perturbative_1_2}\\
\mathcal{F}_{AAA}^{\mathbb{1},(p)}(p^{2},q^{2};r^{2})=&\quad\,0\,,\label{eq:aaa-perturbative_1_3}\\
\mathcal{F}_{AAA}^{\mathbb{1},(q)}(p^{2},q^{2};r^{2})=&\quad\,0\,,\label{eq:aaa-perturbative_1_4}\\
\mathcal{F}_{AAA}^{\mathbb{1},(\mu)}(p^{2},q^{2};r^{2})=&\quad\,0\,,\label{eq:aaa-perturbative_1_5}\\
\mathcal{G}_{AAA}^{\mathbb{1},(c)}(p^{2},q^{2};r^{2})=&-\frac{i N_{c}}{16\pi^{2}\lambda_{K}}\frac{(p^{2}-q^{2})(p^{2}-r^{2})(q^{2}-r^{2})}{\lambda_{K}+p^{2}(q^{2}+r^{2})+q^{2}r^{2}}\,,\label{eq:aaa-perturbative_2_1}\\
\mathcal{G}_{AAA}^{\mathbb{1},(C_{0})}(p^{2},q^{2};r^{2})=&-\frac{3i N_{c}}{8\pi^{2}\lambda_{K}^{2}}\frac{p^{2}q^{2}r^{2}(p^{2}-q^{2})(p^{2}-r^{2})(q^{2}-r^{2})}{\lambda_{K}+p^{2}(q^{2}+r^{2})+q^{2}r^{2}}\,,\label{eq:aaa-perturbative_2_2}\\
\mathcal{G}_{AAA}^{\mathbb{1},(p)}(p^{2},q^{2};r^{2})=&\quad\,\frac{i N_{c}}{32\pi^{2}\lambda_{K}^{2}}\frac{p^{2}(q^{2}-r^{2})}{\lambda_{K}+p^{2}(q^{2}+r^{2})+q^{2}r^{2}}\times\label{eq:aaa-perturbative_2_3}\\
&\hspace{-50pt}\times\Big(r^{6}-r^{4}(p^{2}+7q^{2})-r^{2}(p^{4}-24p^{2}q^{2}+7q^{4})+(p^{2}-q^{2})^{2}(p^{2}+q^{2})\Big)\,,\nonumber\\
\mathcal{G}_{AAA}^{\mathbb{1},(q)}(p^{2},q^{2};r^{2})=&-\frac{i N_{c}}{32\pi^{2}\lambda_{K}^{2}}\frac{q^{2}(p^{2}-r^{2})}{\lambda_{K}+p^{2}(q^{2}+r^{2})+q^{2}r^{2}}\times\label{eq:aaa-perturbative_2_4}\\
&\hspace{-50pt}\times\Big(p^{6}-p^{4}(q^{2}+7r^{2})-p^{2}(q^{4}-24q^{2}r^{2}+7r^{4})+(q^{2}-r^{2})^{2}(q^{2}+r^{2})\Big)\,,\nonumber\\
\mathcal{G}_{AAA}^{\mathbb{1},(\mu)}(p^{2},q^{2};r^{2})=&\quad\,0\,,\label{eq:aaa-perturbative_2_5}\\
\mathcal{H}_{AAA}^{\mathbb{1},(c)}(p^{2},q^{2};r^{2})=&\quad\,\frac{i N_{c}}{16\pi^{2}\lambda_{K}}\frac{1}{\lambda_{K}+p^{2}(q^{2}+r^{2})+q^{2}r^{2}}\times\label{eq:aaa-perturbative_3_1}\\
&\hspace{-50pt}\times\Big[p^{6}-(q^{2}+r^{2})\Big(p^{4}-(q^{2}-r^{2})^{2}\Big)-p^{2}(-3q^{2}r^{2}+q^{4}+r^{4})\Big]\,,\nonumber\\
\mathcal{H}_{AAA}^{\mathbb{1},(C_{0})}(p^{2},q^{2};r^{2})=&\quad\,\frac{3i N_{c}}{8\pi^{2}\lambda_{K}^{2}}\frac{p^{2}q^{2}r^{2}}{\lambda_{K}+p^{2}(q^{2}+r^{2})+q^{2}r^{2}}\times\label{eq:aaa-perturbative_3_2}\\
&\hspace{-50pt}\times\Big[p^{6}-(q^{2}+r^{2})\Big(p^{4}-(q^{2}-r^{2})^{2}\Big)-p^{2}(-3q^{2}r^{2}+q^{4}+r^{4})\Big]\,,\nonumber\\
\mathcal{H}_{AAA}^{\mathbb{1},(p)}(p^{2},q^{2};r^{2})=&-\frac{3i N_{c}}{32\pi^{2}\lambda_{K}^{2}}\frac{p^{2}}{\lambda_{K}+p^{2}(q^{2}+r^{2})+q^{2}r^{2}}\times\label{eq:aaa-perturbative_3_3}\\
&\hspace{-50pt}\times\Big[-r^{8}+r^{6}(3p^{2}-4q^{2})+r^{4}(-5p^{2}q^{2}-3p^{4}+10q^{4})\nonumber\\
&\hspace{-32pt}+(p^{2}-q^{2})\Big(r^{2}(9p^{2}q^{2}+p^{4}+4q^{4})+q^{2}(p^{2}-q^{2})^{2}\Big)\Big]\,,\nonumber\\
\mathcal{H}_{AAA}^{\mathbb{1},(q)}(p^{2},q^{2};r^{2})=&-\frac{3i N_{c}}{32\pi^{2}\lambda_{K}^{2}}\frac{q^{2}}{\lambda_{K}+p^{2}(q^{2}+r^{2})+q^{2}r^{2}}\times\label{eq:aaa-perturbative_3_4}\\
&\hspace{-50pt}\times\Big[-p^{8}+p^{6}(3q^{2}-4r^{2})+p^{4}(-5q^{2}r^{2}-3q^{4}+10r^{4})\nonumber\\
&\hspace{-32pt}+(q^{2}-r^{2})\Big(p^{2}(9q^{2}r^{2}+q^{4}+4r^{4})+r^{2}(q^{2}-r^{2})^{2}\Big)\Big]\,,\nonumber\\
\mathcal{H}_{AAA}^{\mathbb{1},(\mu)}(p^{2},q^{2};r^{2})=&\quad\,0\,.\label{eq:aaa-perturbative_3_5}
\end{align}

\subsubsection*{\boldmath\texorpdfstring{$\langle AAV\rangle$}{} and \texorpdfstring{$\langle VVV\rangle$}{} Green Functions}
Since the $\langle VVV\rangle$ Green function does not contain any $\gamma_{5}$ matrices, the integration over the loop momenta goes along familiar lines. On the other hand, the $\langle AAV\rangle$ triangle contains two $\gamma_{5}$ matrices, however, these can be eliminated without a change of the overall sign in the naive implementation of the dimensional regularization. Therefore, this triangle is thus simply reduced to the one of the $\langle VVV\rangle$ correlator, i.e.
\begin{equation}
\big[\Pi_{VVV}^{\mathbb{1}}(p,q;r)\big]_{\mu\nu\rho}=\big[\Pi_{AAV}^{\mathbb{1}}(p,q;r)\big]_{\mu\nu\rho}\,.\label{eq:AAV_VVV_perturbative_equality}
\end{equation}

Since the perturbative contribution to the two-point Green functions $\langle VV\rangle$ and $\langle AA\rangle$ is the same, see \eqref{eq:2pt_perturbative_VV_AA}, the longitudinal parts \eqref{eq:aav-definition_2_longitudinal}-\eqref{eq:vvv-definition_2_longitudinal} of the $\langle AAV\rangle$ and $\langle VVV\rangle$ correlators are equal to each other, too. Then, obviously, \eqref{eq:AAV_VVV_perturbative_equality} implies the equality of the respective transversal parts \eqref{eq:aav-definition_2_transversal}-\eqref{eq:vvv-definition_2_transversal} of these three-point correlators. This fact, however, leads to a trivial property of the $\langle VVV\rangle$ formfactors for the perturbative contribution, since they can be expressed in the terms of the ones of the $\langle AAV\rangle$. In detail, we find the following relations:
\begin{align}
\mathcal{F}_{VVV}^{\mathbb{1}}(p^{2},q^{2};r^{2})&=\mathcal{F}_{AAV}^{\mathbb{1}}(p^{2},q^{2};r^{2})\,,\label{eq:AAV_VVV_perturbative_equality_formfactor_1}\\
\mathcal{G}_{VVV}^{\mathbb{1}}(p^{2},q^{2};r^{2})&=\mathcal{G}_{AAV}^{\mathbb{1}}(p^{2},q^{2};r^{2})\,.\label{eq:AAV_VVV_perturbative_equality_formfactor_2}
\end{align}

To present the perturbative results in a compact form, we once again employ here the notation \eqref{eq:perturbative_split} and write down only the results for the $\langle AAV\rangle$ correlator, due to \eqref{eq:AAV_VVV_perturbative_equality_formfactor_1}-\eqref{eq:AAV_VVV_perturbative_equality_formfactor_2}. The results are as follows:
\begin{align}
\mathcal{F}_{AAV}^{\mathbb{1},(c)}(p^{2},q^{2};r^{2})=&-\frac{i N_{c}}{24\pi^{2}}\frac{1}{\widehat{\varepsilon}}-\frac{i N_{c}}{72\pi^{2}\lambda_{K}^{2}}\times\\
&\hspace{-50pt}\times\Big(-23r^{6}(p^{2}+q^{2})-17r^{2}(p^{2}+q^{2})(p^{2}-q^{2})^{2}\nonumber\\
&\hspace{-32pt}+5r^{4}(3p^{2}+q^{2})(p^{2}+3q^{2})+11(p^{2}-q^{2})^{4}+14r^{8}\Big)\,,\nonumber\\
\mathcal{F}_{AAV}^{\mathbb{1},(C_{0})}(p^{2},q^{2};r^{2})=&-\frac{i N_{c}}{8\pi^{2}\lambda_{K}^{3}}r^{2}(p^{2}-q^{2}-r^{2})(p^{2}-q^{2}+r^{2})\times\\
&\hspace{-50pt}\times\Big(-p^{4}(6q^{4}-3r^{4}-q^{2}r^{2})+p^{6}(2q^{2}-3r^{2})\nonumber\\
&\hspace{-32pt}+p^{2}(q^{2}-r^{2})(q^{2}+r^{2})(2q^{2}+r^{2})+p^{8}+q^{2}(q^{2}-r^{2})^{3}\Big)\,,\nonumber\\
\mathcal{F}_{AAV}^{\mathbb{1},(p)}(p^{2},q^{2};r^{2})=&\quad\,\frac{i N_{c}}{48\pi^{2}\lambda_{K}^{3}}\times\\
&\hspace{-50pt}\times\Big(-r^{12}+6r^{2}(p^{2}-q^{2})^{3}(7p^{2}q^{2}+2p^{4}+q^{4})+(p^{2}-q^{2})^{5}(3p^{2}+q^{2})\nonumber\\
&\hspace{-32pt}-2r^{10}(8p^{2}+3q^{2})+3r^{8}(-2p^{2}q^{2}+11p^{4}+11q^{4})\nonumber\\
&\hspace{-32pt}-r^{4}(p^{2}-q^{2})(-41p^{4}q^{2}+37p^{2}q^{4}+35p^{6}+33q^{6})\nonumber\\
&\hspace{-32pt}+4r^{6}(-14p^{4}q^{2}+10p^{2}q^{4}+p^{6}-13q^{6})\Big)\,,\nonumber\\
\mathcal{F}_{AAV}^{\mathbb{1},(q)}(p^{2},q^{2};r^{2})=&-\frac{i N_{c}}{48\pi^{2}\lambda_{K}^{3}}\times\\
&\hspace{-50pt}\times\Big(-r^{4}(p^{2}-q^{2})(37p^{4}q^{2}-41p^{2}q^{4}+33p^{6}+35q^{6})\nonumber\\
&\hspace{-32pt}+6r^{2}(p^{2}-q^{2})^{3}(7p^{2}q^{2}+p^{4}+2q^{4})+(p^{2}-q^{2})^{5}(p^{2}+3q^{2})+r^{12}\nonumber\\
&\hspace{-32pt}+4r^{6}(-10p^{4}q^{2}+14p^{2}q^{4}+13p^{6}-q^{6})\nonumber\\
&\hspace{-32pt}-3r^{8}(-2p^{2}q^{2}+11p^{4}+11q^{4})+2r^{10}(3p^{2}+8q^{2})\Big)\,,\nonumber\\
\mathcal{F}_{AAV}^{\mathbb{1},(\mu)}(p^{2},q^{2};r^{2})=&-\frac{i N_{c}}{24\pi^{2}}\,,\\
\mathcal{G}_{AAV}^{\mathbb{1},(c)}(p^{2},q^{2};r^{2})=&\quad\,\frac{5i N_{c}}{24\pi^{2}\lambda_{K}^{2}}(p^{2}-q^{2}-r^{2})(p^{2}+q^{2}-r^{2})(p^{2}-q^{2}+r^{2})\,,\\
\mathcal{G}_{AAV}^{\mathbb{1},(C_{0})}(p^{2},q^{2};r^{2})=&\quad\,\frac{i N_{c}}{4\pi^{2}\lambda_{K}^{3}}\times\\
&\hspace{-50pt}\times\Big(-4p^{8}(q^{4}+r^{4})+p^{10}(q^{2}+r^{2})+p^{6}(q^{2}+r^{2})(-7q^{2}r^{2}+6q^{4}+6r^{4})\nonumber\\
&\hspace{-32pt}-p^{4}(q^{2}-r^{2})^{2}(9q^{2}r^{2}+4q^{4}+4r^{4})+p^{2}(-q^{6}r^{4}-q^{4}r^{6}+q^{10}+r^{10})\nonumber\\
&\hspace{-32pt}+q^{2}r^{2}(q^{2}-r^{2})^{4}\Big)\,,\nonumber\\
\mathcal{G}_{AAV}^{\mathbb{1},(p)}(p^{2},q^{2};r^{2})=&-\frac{i N_{c}}{24\pi^{2}\lambda_{K}^{3}}\times\\
&\hspace{-50pt}\times\Big(-5p^{8}(q^{2}+r^{2})-2p^{6}(-20q^{2}r^{2}+9q^{4}+9r^{4})+24p^{4}(q^{2}-r^{2})^{2}(q^{2}+r^{2})\nonumber\\
&\hspace{-32pt}-4p^{2}(q^{2}-r^{2})^{2}(7q^{2}r^{2}+q^{4}+r^{4})+6p^{10}-3(q^{2}-r^{2})^{4}(q^{2}+r^{2})\Big)\,,\nonumber\\
\mathcal{G}_{AAV}^{\mathbb{1},(q)}(p^{2},q^{2};r^{2})=&\quad\,\frac{i N_{c}}{24\pi^{2}\lambda_{K}^{3}}\times\\
&\hspace{-50pt}\times\Big(-(q^{2}-r^{2})^{3}(13q^{2}r^{2}+6q^{4}+3r^{4})+6p^{4}(4q^{4}r^{2}-8q^{2}r^{4}+3q^{6}+r^{6})\nonumber\\
&\hspace{-32pt}+p^{2}(q^{2}-r^{2})(-35q^{4}r^{2}-11q^{2}r^{4}+5q^{6}+9r^{6})+3p^{10}\nonumber\\
&\hspace{-32pt}+2p^{6}(-12q^{4}+3r^{4}+10q^{2}r^{2})+p^{8}(4q^{2}-9r^{2})\Big)\,,\nonumber\\
\mathcal{G}_{AAV}^{\mathbb{1},(\mu)}(p^{2},q^{2};r^{2})=&\quad\,0\,,\\
\mathcal{H}_{AAV}^{\mathbb{1},(c)}(p^{2},q^{2};r^{2})=&-\frac{i N_{c}}{24\pi^{2}}\frac{1}{\widehat{\varepsilon}}-\frac{i N_{c}}{72\pi^{2}\lambda_{K}^{2}}\times\\
&\hspace{-50pt}\times\Big(-23p^{6}(q^{2}+r^{2})+5p^{4}(3q^{2}+r^{2})(q^{2}+3r^{2})+14p^{8}\nonumber\\
&\hspace{-32pt}-17p^{2}(q^{2}-r^{2})^{2}(q^{2}+r^{2})+11(q^{2}-r^{2})^{4}\Big)\,,\nonumber\\
\mathcal{H}_{AAV}^{\mathbb{1},(C_{0})}(p^{2},q^{2};r^{2})=&-\frac{i N_{c}}{8\pi^{2}\lambda_{K}^{3}}p^{2}(p^{2}+q^{2}-r^{2})(p^{2}-q^{2}+r^{2})\times\\
&\hspace{-50pt}\times\Big(-r^{4}(p^{2}q^{2}+3p^{4}-6q^{4})+r^{2}(p^{2}-q^{2})(p^{2}+q^{2})(p^{2}+2q^{2})\nonumber\\
&\hspace{-32pt}+r^{6}(3p^{2}-2q^{2})+q^{2}(p^{2}-q^{2})^{3}-r^{8}\Big)\,,\nonumber\\
\mathcal{H}_{AAV}^{\mathbb{1},(p)}(p^{2},q^{2};r^{2})=&\quad\,\frac{i N_{c}}{24\pi^{2}\lambda_{K}^{3}}p^{2}(p^{2}+q^{2}-r^{2})(p^{2}-q^{2}+r^{2})\times\\
&\hspace{-50pt}\times\Big(5p^{4}(q^{2}+r^{2})-4p^{2}(-5q^{2}r^{2}+4q^{4}+4r^{4})+2p^{6}+9(q^{2}-r^{2})^{2}(q^{2}+r^{2})\Big)\,,\nonumber\\
\mathcal{H}_{AAV}^{\mathbb{1},(q)}(p^{2},q^{2};r^{2})=&-\frac{i N_{c}}{48\pi^{2}\lambda_{K}^{3}}(p^{2}+q^{2}-r^{2})\times\\
&\hspace{-50pt}\times\Big(-p^{2}(q^{2}-r^{2})^{2}(44q^{2}r^{2}+9q^{4}+7r^{4})-(q^{2}-r^{2})^{4}(3q^{2}+r^{2})+p^{10}\nonumber\\
&\hspace{-32pt}-2p^{6}(-7q^{2}r^{2}+24q^{4}+13r^{4})+p^{8}(15q^{2}+7r^{2})\nonumber\\
&\hspace{-32pt}+p^{4}(-6q^{4}r^{2}+44q^{6}+26r^{6})\Big)\,,\nonumber\\
\mathcal{H}_{AAV}^{\mathbb{1},(\mu)}(p^{2},q^{2};r^{2})=&-\frac{i N_{c}}{24\pi^{2}}\,.
\end{align}

\section{Quark Condensate}\label{sec:quark-condensate}

\subsection{General Remarks}
The quark condensate, $\langle\overline{q}q\rangle$, plays a crucial role since its presence is responsible for the breakdown of the chiral symmetry in QCD. Moreover, within the context of the Fock-Schwinger gauge, the nonlocal quark condensate  propagates not only as a local quark condensate, but also generates higher QCD condensates.

In this section, however, only the first term of \eqref{eq:quark-condensate-propagation} is relevant. Therefore, the formula for a conversion of the nonlocal quark condensate into the local one reads
\begin{equation}
\big\langle\overline{q}_{i,\alpha}^{A}(x)q_{k,\beta}^{B}(y)\big\rangle\ni\frac{\langle\overline{q}q\rangle}{2^{2}\cdot 3^{2}}\delta_{ik}\delta_{\alpha\beta}\delta^{AB}\,.\label{eq:quark-condensate-propagation-LO}
\end{equation}
Apparently, the expansion of the nonlocal quark condensate to the local one is independent of coordinates in the chiral limit, which simplifies the calculations significantly, because one can evaluate the contributing diagrams directly in the momentum representation.

\begin{figure}[htb]
  \centering
    \includegraphics[scale=0.5]{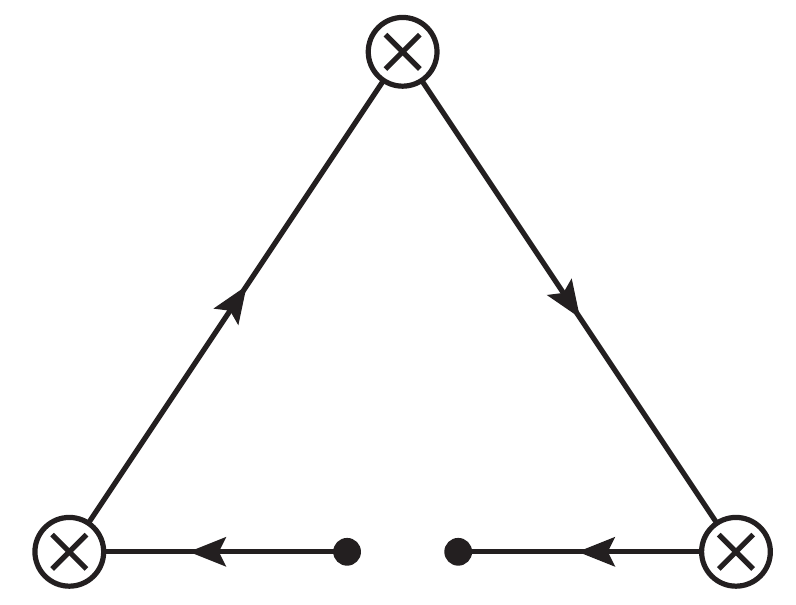}
    \caption{Feynman diagram of the quark condensate contribution to the three-point Green functions at the leading order $\mathcal{O}(1)$.}
    \label{fig:quark-condensate-leading-order}
\end{figure}

The leading order contribution of the quark condensate is given by the set of tree-level Feynman graphs depicted symbolically at Fig.~\ref{fig:quark-condensate-leading-order}, formed by two contractions between the three corresponding currents or densities. Altogether, there are six contributing diagrams and the calculations can be performed in the momentum representation, according to \eqref{eq:quark-condensate-propagation-LO}. Then, the contribution of the quark condensate at the leading order is given as
\begin{align}
\Pi_{\mathcal{O}_{1}^{a}\mathcal{O}_{2}^{b}\mathcal{O}_{3}^{c}}^{\langle\overline{q}q\rangle}(p,q;r)=\frac{1}{2^{2}\cdot3}\langle\overline{q}q\rangle\mathrm{Tr}\big[T^{a}T^{b}T^{c}\big]\mathrm{Tr}\Big[\Gamma_{1}S_{0}(r+q)\Gamma_{2}S_{0}(r)\Gamma_{3}\Big]+(5\,\mathrm{terms})\,.\label{eq:quark-condensate-LO-formula}
\end{align}

As the next-to-leading order $\mathcal{O}(\alpha_{s})$ correction to the quark condensate contribution, one is obligated to consider the corrections to the diagrams at Fig.~\ref{fig:quark-condensate-leading-order} either by the means of the gluon or the quark loops. Although the latter do not contribute at $\mathcal{O}(\alpha_{s})$ due to the conservation of color in QCD, the former represent a very important contribution.

A motivation behind studying the contribution of the gluonic corrections to the quark condensate at $\mathcal{O}(\alpha_{s})$ is an opportunity to explore the renormalisation dependence of such condensate in full QCD, i.e. the dependence on the renormalisation scale $\mu$.

Since we have not studied the gluonic corrections and the renormalisation of the quark condensate explicitly, we refer the reader to the original work \cite{Jamin:2008rm} for a comprehensive description. Therein, the authors studied both the two-point and the three-point Green functions.

\subsection{Results}\label{ssec:quark-condensate_results}
Here we present the results for the leading order contribution of the quark condensate to the Green functions, obtained from the formula \eqref{eq:quark-condensate-LO-formula}. We have
\begin{alignat}{3}
&\mathcal{F}_{SSS}^{\langle\overline{q}q\rangle}(p^{2},q^{2};r^{2})&&=&&\quad\,\frac{\langle\overline{q}q\rangle}{12p^{2}q^{2}r^{2}}\lambda_{K}\,,\label{eq:sss-quark}\\
&\mathcal{F}_{SPP}^{\langle\overline{q}q\rangle}(p^{2},q^{2};r^{2})&&=&&-\frac{\langle\overline{q}q\rangle}{12p^{2}q^{2}r^{2}}\big[p^{4}-(q^{2}-r^{2})^{2}\big]\,,\label{eq:spp-quark}\\
&\mathcal{F}_{VVP}^{\langle\overline{q}q\rangle}(p^{2},q^{2};r^{2})&&=&&\quad\,\frac{\langle\overline{q}q\rangle}{6p^{2}q^{2}r^{2}}(p^{2}+q^{2}+r^{2})\,,\label{eq:vvp-quark}\\
&\mathcal{F}_{AAP}^{\langle\overline{q}q\rangle}(p^{2},q^{2};r^{2})&&=&&\quad\,\frac{\langle\overline{q}q\rangle}{6p^{2}q^{2}r^{2}}(p^{2}+q^{2}-r^{2})\,,\label{eq:aap-quark}\\
&\mathcal{F}_{VAS}^{\langle\overline{q}q\rangle}(p^{2},q^{2};r^{2})&&=&&\quad\,\frac{\langle\overline{q}q\rangle}{6p^{2}q^{2}r^{2}}(p^{2}-q^{2}-r^{2})\,,\label{eq:vas-quark}\\
&\mathcal{F}_{VVS}^{\langle\overline{q}q\rangle}(p^{2},q^{2};r^{2})&&=&&-\frac{\langle\overline{q}q\rangle}{6p^{2}q^{2}r^{2}}(p^{2}+q^{2}+r^{2})\,,\label{eq:vvs-quark-1}\\
&\mathcal{G}_{VVS}^{\langle\overline{q}q\rangle}(p^{2},q^{2};r^{2})&&=&&-\frac{\langle\overline{q}q\rangle}{3p^{2}q^{2}r^{2}}\,,\label{eq:vvs-quark-2}\\
&\mathcal{F}_{AAS}^{\langle\overline{q}q\rangle}(p^{2},q^{2};r^{2})&&=&&-\frac{\langle\overline{q}q\rangle}{6p^{2}q^{2}r^{2}}(p^{2}+q^{2}-r^{2})\,,\label{eq:aas-quark-1}\\
&\mathcal{G}_{AAS}^{\langle\overline{q}q\rangle}(p^{2},q^{2};r^{2})&&=&&-\frac{\langle\overline{q}q\rangle}{3p^{2}q^{2}r^{2}}\,,\label{eq:aas-quark-2}\\
&\mathcal{F}_{VAP}^{\langle\overline{q}q\rangle}(p^{2},q^{2};r^{2})&&=&&\quad\,\frac{\langle\overline{q}q\rangle}{6p^{2}q^{2}r^{2}}(p^{2}-q^{2}-r^{2})\,,\label{eq:vap-quark-1}\\
&\mathcal{G}_{VAP}^{\langle\overline{q}q\rangle}(p^{2},q^{2};r^{2})&&=&&-\frac{\langle\overline{q}q\rangle}{3p^{2}q^{2}r^{2}}\,.\label{eq:vap-quark-2}
\end{alignat}

The results above deserve a detailed discussion. We remind the reader that we understand the quark field $q$ to be the flavor triplet, as defined by \eqref{eq:quark_triplet}. Therefore, the quark condensate $\langle\overline{q}q\rangle$ is thus a sum over all flavor-diagonal quark condensates,
\begin{equation}
\langle\overline{q}q\rangle=\sum_{\psi=u,d,s}\langle\overline{\psi}\psi\rangle=\langle\overline{u}u\rangle+\langle\overline{d}d\rangle+\langle\overline{s}s\rangle\,,\label{eq:quark_condensate_sum_1}
\end{equation}
where, in the chiral limit, all the values of the condensates on the right-hand side of \eqref{eq:quark_condensate_sum_1} are equaled to each other, i.e.
\begin{equation}
\langle\overline{q}q\rangle=3\langle\overline{\psi}\psi\rangle\,.\label{eq:quark_condensate_sum_2}
\end{equation}

As we have mentioned, several authors have studied the role of the quark condensate contribution to the OPE of Green functions already. However, as it seems, most of them have taken the quark condensate to be a condensate of quarks of the single flavor. In such case, in spite of differences in notation, their quark condensates would be denoted as $\langle\overline{\psi}\psi\rangle$ in our designation.
Therefore, our results differ by the factor of three with respect to the results of such authors. These include the authors of Ref.~\cite{Jamin:2008rm} or \cite{Cirigliano:2004ue}. These cases are discussed below in detail. However, in what follows, we already take into account the difference in the factor of three and discuss only whether an agreement between our results and other authors has been achieved.

Let us start with Ref.~\cite{Jamin:2008rm}. Apart from the normalization of the quark condensate, the authors use a different normalization of the scalar and pseudoscalar density. For every occurrence of any of these densities, a factor of two must be take into account. Also, the authors use a conventional factor of $i^{2}$ in the definition of the three-point Green function.
Therefore, for the $\langle SSS\rangle$ and $\langle SPP\rangle$ correlators, a factor of $-24$ must be considered. Then, we are in an agreement with their results, see eq.~4.7 at page no.~11. For the $\langle VVP\rangle$, $\langle AAP\rangle$ and $\langle VAS\rangle$ Green functions, a factor of $-6$ must be taken into account. However, their result is different by an overall sign, see eq.~4.11 at page no.~12.

The results for the $\langle VVS\rangle$, $\langle AAS\rangle$ and $\langle VAP\rangle$ Green functions are not given in terms of the respective formfactors in \cite{Jamin:2008rm} but in the form of suitable contractions so that rational scalar functions are built. For this reason, we present our results of such contractions below:
\begin{align}
g^{\mu\nu}\big[\Pi_{VVS}^{\langle\overline{q}q\rangle}(p,q;r)\big]_{\mu\nu}=&-\frac{\langle\overline{q}q\rangle}{6p^{2}q^{2}r^{2}}\big[(p^{2}-q^{2})^{2}+(p^{2}+q^{2})r^{2}-2r^{4}\big]\,,\label{eq:vvs-quark-1_contraction}\\
q^{\mu}p^{\nu}\big[\Pi_{VVS}^{\langle\overline{q}q\rangle}(p,q;r)\big]_{\mu\nu}=&\quad\,\frac{\langle\overline{q}q\rangle}{12p^{2}q^{2}}\lambda_{K}\,\label{eq:vvs-quark-2_contraction},\\
g^{\mu\nu}\big[\Pi_{AAS}^{\langle\overline{q}q\rangle}(p,q;r)\big]_{\mu\nu}=&-\frac{\langle\overline{q}q\rangle}{6p^{2}q^{2}r^{2}}\big[(p^{2}-q^{2})^{2}-3(p^{2}+q^{2})r^{2}+2r^{4}\big]\,,\label{eq:aas-quark-1_contraction}\\
q^{\mu}p^{\nu}\big[\Pi_{AAS}^{\langle\overline{q}q\rangle}(p,q;r)\big]_{\mu\nu}=&-\frac{\langle\overline{q}q\rangle}{12p^{2}q^{2}}(\lambda_{K}+4p^{2}q^{2})\,,\label{eq:aas-quark-2_contraction}\\
g^{\mu\nu}\big[\Pi_{VAP}^{\langle\overline{q}q\rangle}(p,q;r)\big]_{\mu\nu}=&\quad\,\frac{\langle\overline{q}q\rangle}{6p^{2}q^{2}r^{2}}(p^{2}-q^{2}-2r^{2})(p^{2}+q^{2}-r^{2})\,,\label{eq:vap-quark-1_contraction}\\
q^{\mu}p^{\nu}\big[\Pi_{VAP}^{\langle\overline{q}q\rangle}(p,q;r)\big]_{\mu\nu}=&\quad\,\frac{\langle\overline{q}q\rangle}{12p^{2}q^{2}r^{2}}\big[r^{2}\lambda_{K}+2p^{2}q^{2}(p^{2}-q^{2}+r^{2})\big]\,.\label{eq:vap-quark-2_contraction}
\end{align}
All of these are in an agreement with the results in \cite{Jamin:2008rm}, see eq.~4.18 at page no.~14, upon the conversion factor $-6$ is taken into account.

Our results \eqref{eq:vap-quark-1}-\eqref{eq:vap-quark-2} for the $\langle VAP\rangle$ are in an agreement with the ones in \cite{Cirigliano:2004ue}, see eq.~8 at page no.~4 therein.

In calculations within the framework of Chiral perturbation theory and Resonance chiral theory, the quark condensate is often rewritten in terms of the chiral parameter $B_{0}$ and the pion decay constant in the chiral limit $F_{0}$. Specifically, due to \eqref{eq:quark_condensate_sum_2}, we have
\begin{equation}
\langle\overline{q}q\rangle=-3B_{0}F_{0}^{2}\,.\label{eq:quark_condensate_chpt_rcht}
\end{equation}

Assuming the validity of \eqref{eq:quark_condensate_chpt_rcht}, our result \eqref{eq:vvp-quark} for the contribution of the quark condensate to the $\langle VVP\rangle$ Green function coincides with the one in Ref.~\cite{Moussallam:1994xp}, see eq.~7 at page no.~4 therein.

On the other hand, the result \eqref{eq:vvp-quark} for the $\langle VVP\rangle$ correlator differs by an overall sign with the results in \cite{Kampf:2011ty}, see eq.~31 at page no.~12. The difference in an overall sign is also present in \cite{Mateu:2007tr}, see eq.~8 at page no.~3, upon taking into account that the authors introduce a conventional factor of $i^{2}$ in the definition of the $\langle VVP\rangle$ Green function, and they use a different normalization of the pseudoscalar density and the quark condensate (see eq.~3 and eq.~4 at page no.~2 and eq.~6 at page no.~3 therein, respectively). This may be because of the different convention for the Levi-Civita tensor.

Similarly , the result \eqref{eq:vas-quark} for the $\langle VAS\rangle$ Green function also differs by an overall sign with the result in \cite{Kampf:2011ty}, see eq.~82 at page no.~21, due to the same reason as above.

%%%%%%%%%%%%%%%%%%%%%%%%%%%%%%%%%%%%%%%%%%%%%%%%%%%%%%%%%%%%%%%%%%%%%%%%%%%%%%%%%%%%%%%%%%%%%%%%%%%%%%%%%
%%%%%%%%%%%%%%%%%%%%%%%%%%%%%%%%%%%%%%%%%%%%%%%%%%%%%%%%%%%%%%%%%%%%%%%%%%%%%%%%%%%%%%%%%%%%%%%%%%%%%%%%%
%%% Section: Gluon Condensate
%%%%%%%%%%%%%%%%%%%%%%%%%%%%%%%%%%%%%%%%%%%%%%%%%%%%%%%%%%%%%%%%%%%%%%%%%%%%%%%%%%%%%%%%%%%%%%%%%%%%%%%%%
%%%%%%%%%%%%%%%%%%%%%%%%%%%%%%%%%%%%%%%%%%%%%%%%%%%%%%%%%%%%%%%%%%%%%%%%%%%%%%%%%%%%%%%%%%%%%%%%%%%%%%%%%

\section{Gluon Condensate}\label{sec:gluon-condensate}

\subsection{General Remarks}
Next nonperturbative contribution to the OPE of Green functions stems from the operator with dimension 4, i.e. the gluon condensate $\langle G_{\mu\nu}^{a}G^{\mu\nu,a}\rangle$, which we will denote simply as $\langle G^{2}\rangle$ from now on.

Having three chiral currents or densities, it is obvious that the gluon condensate can be formed from the triangle graphs by coupling two gluon fields to the quark lines. To evaluate such diagrams, one would write down the contribution in the coordinate representation and then use the propagation formula \eqref{eq:propagation_gluon_condensate}. After integrating over the coordinates of the two gluons, it would be possible to perform the Fourier transform, which would give us the final result in the momentum representation. However, this procedure is quite lengthy and time-consuming. Instead, it is much efficient to use the strategy described below.

To calculate the contribution of the gluon condensate to the three-point Green functions, it is easier to use an approach based on the Fock-Schwinger gauge in the coordinate representation, in which the gluon condensate arises in an obvious way.

Since we work in the chiral limit, the basic ingredient for our calculation represents the massless quark propagator in external gluon field \cite{Novikov:1983gd}, which is formulated already in terms of the gluon field strength tensors and, for this reason, is much more suitable in calculating the contribution of the gluon condensate. This propagator, within the Fock-Schwinger gauge, reads
\begin{align}
S(x,y)&=S_{0}(x,y)+S_{1}^{\alpha\beta}(x,y)G_{\alpha\beta}(0)+S_{2}(x,y)G_{\alpha\beta}(0)G^{\alpha\beta}(0)+\ldots\,,\label{eq:quark-propagator-x-representation-short}
\end{align}
with the coordinate dependence summarized in the following factors:\footnote{The first contribution \eqref{eq:quark-propagator-x-representation-short-factor-1} stands for the free massless quark propagator in the coordinate representation. Its connection to the propagator in the momentum representation \eqref{eq:quark-propagator} reads $S_{0}(x,y)=\int\frac{\mathrm{d}^{4}\ell}{(2\pi)^{4}}e^{-i\ell(x-y)}S_{0}(\ell)$.}
\begin{alignat}{3}
& S_{0}(x,y)&&=&&\frac{i}{2\pi^{2}}\frac{\slashed x-\slashed y}{(x-y)^{4}}\,,\label{eq:quark-propagator-x-representation-short-factor-1}\\
& S_{1}^{\alpha\beta}(x,y)&&=-&&\frac{g_{s}}{4\pi^{2}}\bigg(\frac{i(x-y)_{\mu}}{4(x-y)^{2}}\gamma_{\nu}\gamma_{5}\varepsilon^{\mu\nu\alpha\beta}-\frac{\slashed x-\slashed y}{(x-y)^{4}}x^{\alpha}y^{\beta}\bigg)\,,\label{eq:quark-propagator-x-representation-short-factor-2}\\
& S_{2}(x,y)&&=-&&\frac{ig_{s}^{2}}{192\pi^{2}}\frac{\slashed x-\slashed y}{(x-y)^{4}}\big[x^{2}y^{2}-(x\hspace{-1pt}\cdot\hspace{-1pt}y)^{2}\big]\,.\label{eq:quark-propagator-x-representation-short-factor-3}
\end{alignat}
Since the Fock-Schwinger gauge violates translation invariance, the propagator \eqref{eq:quark-propagator-x-representation-short} is not invariant with respect to translation either, see last term in \eqref{eq:quark-propagator-x-representation-short-factor-2}. As an illustration, this propagator can be graphically denoted as follows:
\begin{figure}[htb]
  \centering
    \includegraphics[scale=0.5]{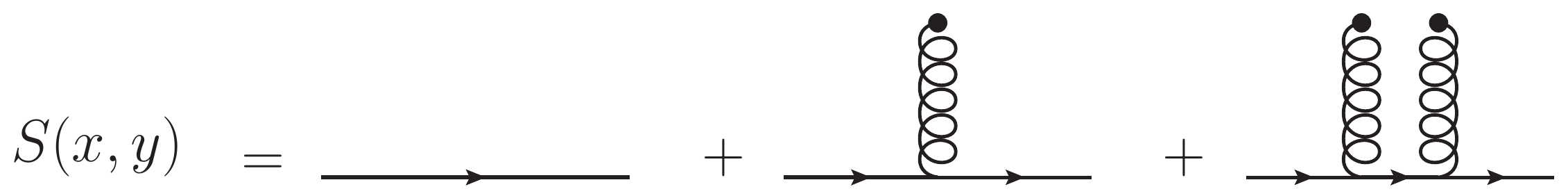}
    \caption{An illustration of the quark propagator in the external gluon field \eqref{eq:quark-propagator-x-representation-short}. On the right-hand side, the figures stand for \eqref{eq:quark-propagator-x-representation-short-factor-1}, \eqref{eq:quark-propagator-x-representation-short-factor-2} and \eqref{eq:quark-propagator-x-representation-short-factor-3}, respectively.}
    \label{fig:fig_quark_propagator}
\end{figure}

\begin{figure}[htb]
\centering
    \begin{subfigure}[t]{0.27\linewidth}
        \hspace{4pt}\includegraphics[width=37mm]{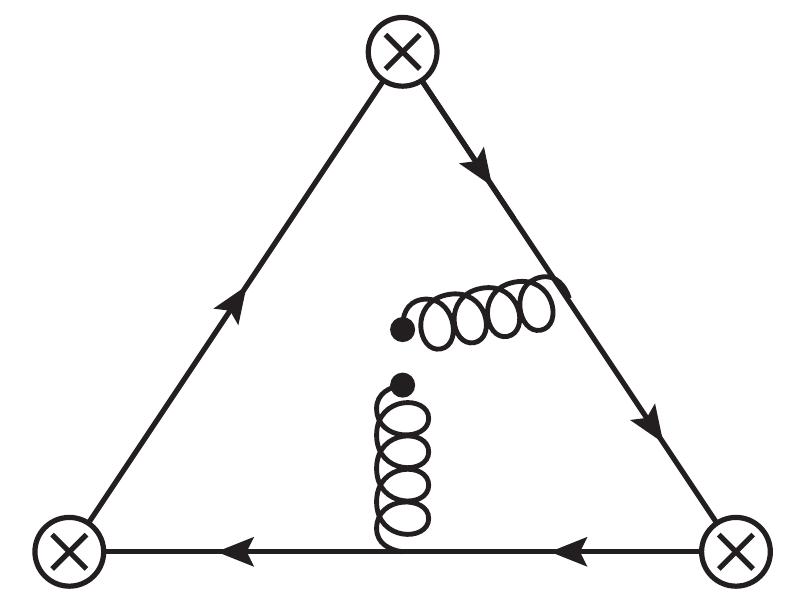}
        \caption{}
        \label{fig:fig_gluon_condensate_v2_a}
    \end{subfigure}
    \begin{subfigure}[t]{0.27\linewidth}
        \hspace{4pt}\includegraphics[width=37mm]{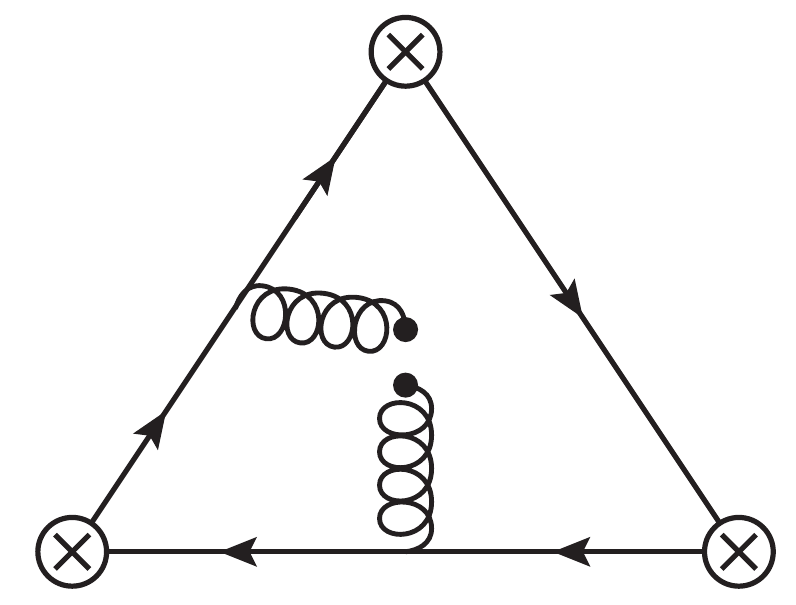}
        \caption{}
        \label{fig:fig_gluon_condensate_v2_b}
    \end{subfigure}
    \begin{subfigure}[t]{0.27\linewidth}
        \hspace{4pt}\includegraphics[width=37mm]{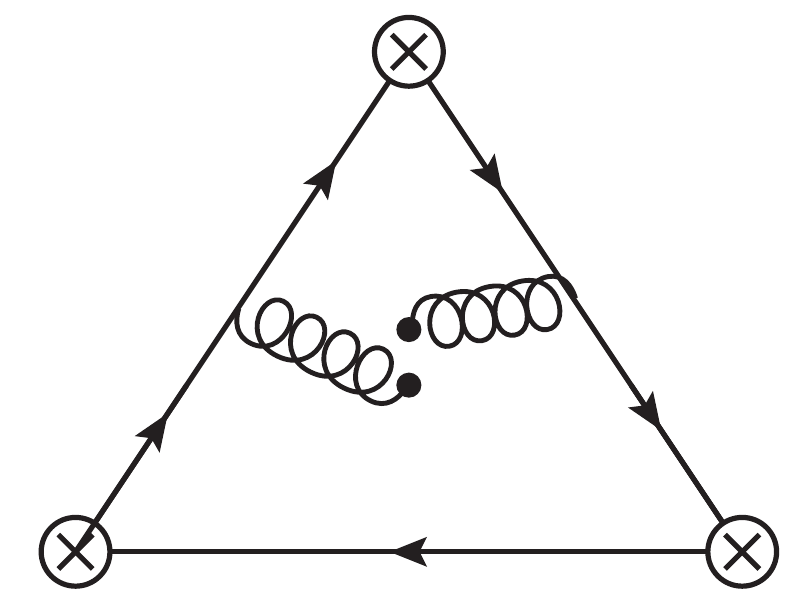}
        \caption{}
        \label{fig:fig_gluon_condensate_v2_c}
    \end{subfigure}\vspace{10pt}
    
    \begin{subfigure}[t]{0.27\textwidth}
        \hspace{4pt}\includegraphics[width=37mm]{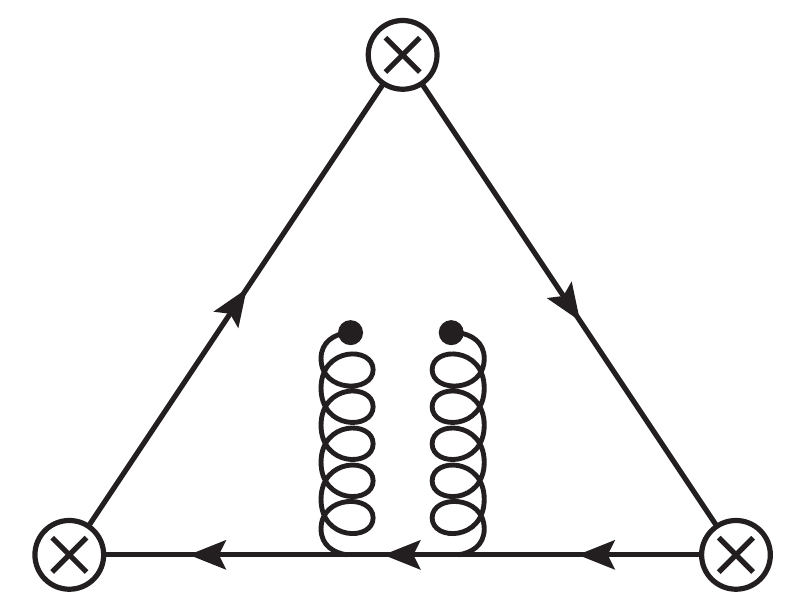}
        \caption{}
        \label{fig:fig_gluon_condensate_v2_d}
    \end{subfigure}
    \begin{subfigure}[t]{0.27\textwidth}
        \hspace{4pt}\includegraphics[width=37mm]{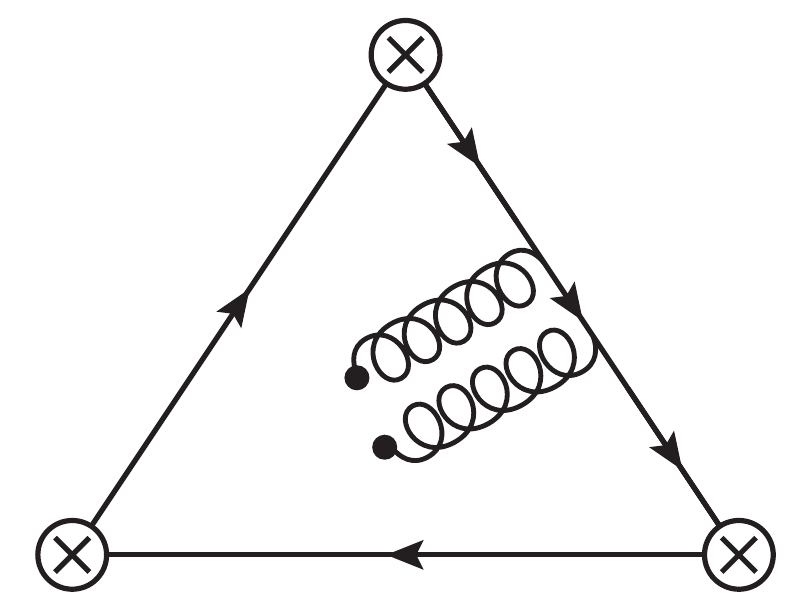}
        \caption{}
        \label{fig:fig_gluon_condensate_v2_e}
    \end{subfigure}
    \begin{subfigure}[t]{0.27\textwidth}
        \hspace{4pt}\includegraphics[width=37mm]{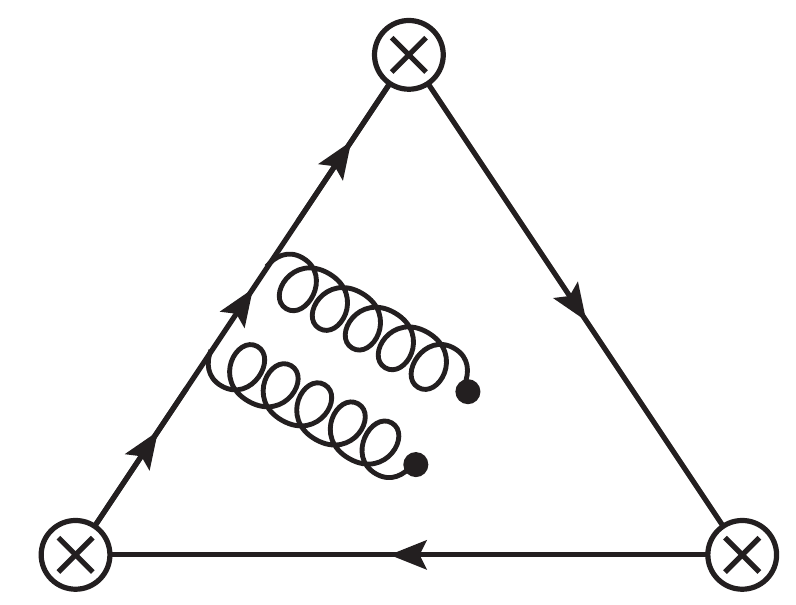}
        \caption{}
        \label{fig:fig_gluon_condensate_v2_f}
    \end{subfigure}
    \caption{Feynman diagrams of the contributions of the gluon condensate to the three-point Green functions. Having to add the Bose-symmetrized diagrams is tacitly assumed. The diagrams \ref{fig:fig_gluon_condensate_v2_a}-\ref{fig:fig_gluon_condensate_v2_c} contain two propagators \eqref{eq:quark-propagator-x-representation-short-factor-2} and the diagrams \ref{fig:fig_gluon_condensate_v2_d}-\ref{fig:fig_gluon_condensate_v2_f} one propagator \eqref{eq:quark-propagator-x-representation-short-factor-3}.}
    \label{fig:fig_gluon_condensate_v2}
\end{figure}

Then, within the Fock-Schwinger gauge, the whole contribution of the gluon condensate to the three-point Green functions is thus given only by the six diagrams (and the Bose-symmetrized ones) denoted at Fig.~\ref{fig:fig_gluon_condensate_v2}.

Easily, one can then write down the total contribution of the diagrams at Fig. \ref{fig:fig_gluon_condensate_v2} as
\begin{align}
\Pi_{\mathcal{O}_{1}^{a}\mathcal{O}_{2}^{b}\mathcal{O}_{3}^{c}}^{\langle G^{2}\rangle}(x,y,z)=&-\mathrm{Tr}\big[T^{a}T^{b}T^{c}\big]\mathrm{Tr}\Big[\Gamma_{1}S(x,y)\Gamma_{2}S(y,z)\Gamma_{3}S(z,x)\Big]_{\langle G^{2}\rangle}\label{eq:gluon-condensate-full}\\
&+\big[(\Gamma_{1},a,x)\leftrightarrow(\Gamma_{2},b,y)\big]\,,\nonumber
\end{align}
where the subscript at the end of the first line refers to the fact that, after inserting \eqref{eq:quark-propagator-x-representation-short} for the individual propagators, we keep only the terms with two gluon tensors, and the second line stands for the Bose-symmetrized contributions. To form the gluon condensate, the following formula is needed:
\begin{align}
\big\langle\mathrm{Tr}[G_{\alpha\beta}G_{\gamma\delta}]\big\rangle=\frac{1}{24}(g_{\alpha\gamma}g_{\beta\delta}-g_{\alpha\delta}g_{\beta\gamma})\langle G^{2}\rangle\,,\label{eq:gluon-condensate-projection}
\end{align}
which can be easily obtained from applying color matrices to the projection formula for the composite operator \cite{Ioffe:1982qb},
\begin{equation}
\big\langle G_{\alpha\beta}^{a}G_{\gamma\delta}^{b}\rangle=\frac{1}{96}\delta^{ab}(g_{\alpha\gamma}g_{\beta\delta}-g_{\alpha\delta}g_{\beta\gamma})\langle G^{2}\rangle\,,\label{eq:gluon-condensate-projection_v2}
\end{equation}
and then taking the trace.

The total contribution of \eqref{eq:gluon-condensate-full} thus represents six individual contributions plus the Bose-symmetrized ones. However, upon making sure that \eqref{eq:gluon-condensate-full} is indeed invariant under translation, one can set the third coordinate $z$ to the origin. This is greatly convenient because such choice simplifies the calculations significantly by eliminating the second term in \eqref{eq:quark-propagator-x-representation-short-factor-2} and \eqref{eq:quark-propagator-x-representation-short-factor-3} completely. Then we are left with only four contributing topologies and four Bose-symmetrized ones. From now on, we are thus allowed to use a notation of \eqref{eq:translation_invariance} with only two coordinates.

The next part consists of converting the previous results from the coordinate representation into the momentum one by performing the Fourier transform. This represents a quite lengthy process to go through. However, after some algebra, one can write down the total contribution in the form
\begin{align}
\Pi_{\mathcal{O}_{1}^{a}\mathcal{O}_{2}^{b}\mathcal{O}_{3}^{c}}^{\langle G^{2}\rangle}(p,q;r)&=\int\mathrm{d}^{4}x\,\mathrm{d}^{4}y\,e^{-i(p\cdot x+q\cdot y)}\Pi_{\mathcal{O}_{1}^{a}\mathcal{O}_{2}^{b}\mathcal{O}_{3}^{c}}^{\langle G^{2}\rangle}(x,y)\,,\\
&=\frac{i\alpha_{s}\langle G^{2}\rangle}{768\pi^{5}}\mathrm{Tr}\big[T^{a}T^{b}T^{c}\big]\sum_{i=1}^{4}\Pi_{i}^{\langle G^{2}\rangle}(p,q;r)+\big[(\Gamma_{1},a,p)\leftrightarrow(\Gamma_{2},b,q)\big]\,,
\end{align}
where
\begin{align}
\Pi_{1}^{\langle G^{2}\rangle}(p,q;r)&=\mathrm{Tr}\big[\Gamma_{1}\gamma^{\alpha}\gamma_{5}\Gamma_{2}\gamma^{\beta}\gamma_{5}\Gamma_{3}\gamma^{\gamma}\big]\big[\widetilde{F}_{1}^{\langle G^{2}\rangle}(p,q;r)\big]_{\alpha\beta\gamma}\,,\\
\Pi_{2}^{\langle G^{2}\rangle}(p,q;r)&=\mathrm{Tr}\big[\Gamma_{1}\gamma^{\alpha}\gamma_{5}\Gamma_{2}\gamma^{\beta}\Gamma_{3}\gamma^{\gamma}\gamma_{5}\big]\big[\widetilde{F}_{2}^{\langle G^{2}\rangle}(p,q;r)\big]_{\alpha\beta\gamma}\,,\\
\Pi_{3}^{\langle G^{2}\rangle}(p,q;r)&=\mathrm{Tr}\big[\Gamma_{1}\gamma^{\alpha}\Gamma_{2}\gamma^{\beta}\gamma_{5}\Gamma_{3}\gamma^{\gamma}\gamma_{5}\big]\big[\widetilde{F}_{3}^{\langle G^{2}\rangle}(p,q;r)\big]_{\alpha\beta\gamma}\,,\\
\Pi_{4}^{\langle G^{2}\rangle}(p,q;r)&=2\,\mathrm{Tr}\big[\Gamma_{1}\gamma^{\alpha}\Gamma_{2}\gamma^{\beta}\Gamma_{3}\gamma^{\gamma}\big]\big[\widetilde{F}_{4}^{\langle G^{2}\rangle}(p,q;r)\big]_{\alpha\beta\gamma}\,.
\end{align}

The functions $\widetilde{F}_{i}^{\langle G^{2}\rangle}$ are remnants of the Fourier transform, and can be obtained in a closed form as
\begin{align}
\big[\widetilde{F}_{1}^{\langle G^{2}\rangle}(p,q;r)\big]_{\alpha\beta\gamma}=&-g_{\alpha\beta}\int\frac{\mathrm{d}^{4}\ell}{(2\pi)^{4}}\widetilde{F}_{1}(\ell-p)\big[\widetilde{F}_{2}(\ell+r)\big]_{\mu}\big[\widetilde{F}_{6}(\ell)\big]_{\mu\gamma}\label{eq:gluon_condensate_integral_1}\\
&-\int\frac{\mathrm{d}^{4}\ell}{(2\pi)^{4}}\big[\widetilde{F}_{2}(\ell+r)\big]_{\alpha}\big[\widetilde{F}_{2}(\ell-p)\big]_{\beta}\big[\widetilde{F}_{5}(\ell)\big]_{\gamma}+g_{\alpha\beta}\widetilde{F}_{1}(q)\big[\widetilde{F}_{5}(r)\big]_{\gamma}\,,\nonumber\\
\big[\widetilde{F}_{2}^{\langle G^{2}\rangle}(p,q;r)\big]_{\alpha\beta\gamma}=&-g_{\alpha\gamma}\int\frac{\mathrm{d}^{4}\ell}{(2\pi)^{4}}\widetilde{F}_{1}(\ell-p)\big[\widetilde{F}_{2}(\ell)\big]_{\mu}\big[\widetilde{F}_{6}(\ell+r)\big]_{\mu\beta}\label{eq:gluon_condensate_integral_2}\\
&-g_{\alpha\gamma}\widetilde{F}_{1}(p)\big[\widetilde{F}_{5}(r)\big]_{\beta}-\int\frac{\mathrm{d}^{4}\ell}{(2\pi)^{4}}\big[\widetilde{F}_{2}(\ell)\big]_{\alpha}\big[\widetilde{F}_{5}(\ell+r)\big]_{\beta}\big[\widetilde{F}_{2}(\ell-p)\big]_{\gamma}\,,\nonumber\\
\big[\widetilde{F}_{3}^{\langle G^{2}\rangle}(p,q;r)\big]_{\alpha\beta\gamma}=&-\int\frac{\mathrm{d}^{4}\ell}{(2\pi)^{4}}\big[\widetilde{F}_{5}(\ell-p)\big]_{\alpha}\big[\widetilde{F}_{2}(\ell)\big]_{\beta}\big[\widetilde{F}_{2}(\ell+r)\big]_{\gamma}\label{eq:gluon_condensate_integral_3}\\
&+g_{\beta\gamma}\int\frac{\mathrm{d}^{4}\ell}{(2\pi)^{4}}\big[\widetilde{F}_{2}(\ell)\big]_{\mu}\big[\widetilde{F}_{2}(\ell+r)\big]_{\mu}\big[\widetilde{F}_{5}(\ell-p)\big]_{\alpha}\,,\nonumber\\
\big[\widetilde{F}_{4}^{\langle G^{2}\rangle}(p,q;r)\big]_{\alpha\beta\gamma}=&-\int\frac{\mathrm{d}^{4}\ell}{(2\pi)^{4}}\big[\widetilde{F}_{5}(\ell-p)\big]_{\alpha}\big[\widetilde{F}_{7}(\ell+r)\big]_{\mu\nu\beta}\big[\widetilde{F}_{7}(\ell)\big]_{\mu\nu\gamma}\label{eq:gluon_condensate_integral_4}\\
&+\int\frac{\mathrm{d}^{4}\ell}{(2\pi)^{4}}\big[\widetilde{F}_{5}(\ell-p)\big]_{\alpha}\big[\widetilde{F}_{2}(\ell+r)\big]_{\beta}\big[\widetilde{F}_{2}(\ell)\big]_{\gamma}\,,\nonumber
\end{align}
where the functions $\widetilde{F}_{i}$ are defined in the Appendix \ref{sec:fourier_transform}.

After performing the contractions and some algebraic manipulations, we arrive at
\begin{align}
\big[\widetilde{F}_{1}^{\langle G^{2}\rangle}(p,q;r)\big]_{\alpha\beta\gamma}&=\int\frac{\mathrm{d}^{4}\ell}{(2\pi)^{4}}\big[I_{1}^{\langle G^{2}\rangle}(\ell,p,q;r)\big]_{\alpha\beta\gamma}+\frac{8i\pi^{4}}{q^{2}r^{2}}g^{\alpha\beta}r^{\gamma}\,,\label{eq:gluon_condensate_integral_1_v2}\\
\big[\widetilde{F}_{2}^{\langle G^{2}\rangle}(p,q;r)\big]_{\alpha\beta\gamma}&=\int\frac{\mathrm{d}^{4}\ell}{(2\pi)^{4}}\big[I_{2}^{\langle G^{2}\rangle}(\ell,p,q;r)\big]_{\alpha\beta\gamma}-\frac{8i\pi^{4}}{p^{2}r^{2}}g^{\alpha\gamma}r^{\beta}\,,\label{eq:gluon_condensate_integral_2_v2}\\
\big[\widetilde{F}_{3}^{\langle G^{2}\rangle}(p,q;r)\big]_{\alpha\beta\gamma}&=\int\frac{\mathrm{d}^{4}\ell}{(2\pi)^{4}}\big[I_{3}^{\langle G^{2}\rangle}(\ell,p,q;r)\big]_{\alpha\beta\gamma}\,,\label{eq:gluon_condensate_integral_3_v2}\\
\big[\widetilde{F}_{4}^{\langle G^{2}\rangle}(p,q;r)\big]_{\alpha\beta\gamma}&=\int\frac{\mathrm{d}^{4}\ell}{(2\pi)^{4}}\big[I_{4}^{\langle G^{2}\rangle}(\ell,p,q;r)\big]_{\alpha\beta\gamma}\,,\label{eq:gluon_condensate_integral_4_v2}
\end{align}
with the integrands in the form
\begin{align}
\big[I_{1}^{\langle G^{2}\rangle}(\ell,p,q;r)\big]_{\alpha\beta\gamma}&=64\pi^{6}\bigg(\frac{2(\ell+r)^{\alpha}(\ell-p)^{\beta}\ell^{\gamma}}{\ell^{2}(\ell-p)^{4}(\ell+r)^{4}}-g^{\alpha\beta}\frac{\ell^{\gamma}\big[r^{2}-(\ell+r)^{2}\big]+\ell^{2}r^{\gamma}}{\ell^{4}(\ell-p)^{2}(\ell+r)^{4}}\bigg)\,,\\
\big[I_{2}^{\langle G^{2}\rangle}(\ell,p,q;r)\big]_{\alpha\beta\gamma}&=64\pi^{6}\bigg(g^{\alpha\gamma}\frac{r^{\beta}(\ell+r)^{2}+(\ell^{2}-r^{2})(\ell+r)^{\beta}}{\ell^{4}(\ell-p)^{2}(\ell+r)^{4}}+\frac{2\ell^{\alpha}(\ell+r)^{\beta}(\ell-p)^{\gamma}}{\ell^{4}(\ell-p)^{4}(\ell+r)^{2}}\bigg)\,,\\
\big[I_{3}^{\langle G^{2}\rangle}(\ell,p,q;r)\big]_{\alpha\beta\gamma}&=64\pi^{6}\bigg(\frac{2(\ell-p)^{\alpha}\ell^{\beta}(\ell+r)^{\gamma}}{\ell^{4}(\ell-p)^{2}(\ell+r)^{4}}-g^{\beta\gamma}\frac{(\ell-p)^{\alpha}\big[(\ell+r)^{2}+\ell^{2}-r^{2}\big]}{\ell^{4}(\ell-p)^{2}(\ell+r)^{4}}\bigg)\,,\\
\big[I_{4}^{\langle G^{2}\rangle}(\ell,p,q;r)\big]_{\alpha\beta\gamma}&=-128\pi^{6}\frac{(\ell-p)^{\alpha}(\ell+r)^{\beta}\ell^{\gamma}}{\ell^{4}(\ell-p)^{2}(\ell+r)^{4}}+\frac{32\pi^{6}(\ell-p)^{\alpha}}{\ell^{6}(\ell-p)^{2}(\ell+r)^{6}}\times\\
&\hspace{-30pt}\times\bigg(\big[(\ell+r)^{2}+\ell^{2}-r^{2}\big]\Big[\ell^{2}g^{\beta\gamma}(\ell+r)^{2}-4r^{\beta}\Big(\ell^{\gamma}\big[r^{2}-(\ell+r)^{2}\big]+\ell^{2}r^{\gamma}\Big)\Big]\nonumber\\
&\hspace{-30pt}+2\ell^{\beta}\Big(\big[\ell^{\gamma}(\ell^{2}-2r^{2})-\ell^{2}r^{\gamma}\big](\ell+r)^{2}-2(\ell^{2}-r^{2})(r^{2}l^{\gamma}+\ell^{2}r^{\gamma})\Big)\bigg)\,.\nonumber
\end{align}

\subsection{Results}
Here we present the results, obtained by the procedure described above. Since the propagator \eqref{eq:quark-propagator-x-representation-short} is odd in the number of gamma matrices, one can easily move the $\gamma_{5}$ matrices contained in the currents and densities of the correlators. After easy manipulations one can find an apparent relations between these correlators such as
\begin{alignat}{3}
\big[\Pi_{ASP}^{\langle G^{2}\rangle}(p,q;r)\big]_{\mu}\quad&=\big[\Pi_{VSS}^{\langle G^{2}\rangle}(p,q;r)\big]_{\mu}&&=\big[\Pi_{VPP}^{\langle G^{2}\rangle}(p,q;r)\big]_{\mu}\,,\label{eq:gluon-condensate-ASP-VPP-VSS-p-repr_KOPIE}\\
\big[\Pi_{VVA}^{\langle G^{2}\rangle}(p,q;r)\big]_{\mu\nu\rho}&=\big[\Pi_{AAA}^{\langle G^{2}\rangle}(p,q;r)\big]_{\mu\nu\rho}&&\,,\label{eq:VVA_AAA_gluon_condensate_equality}\\
\big[\Pi_{AAV}^{\langle G^{2}\rangle}(p,q;r)\big]_{\mu\nu\rho}&=\big[\Pi_{VVV}^{\langle G^{2}\rangle}(p,q;r)\big]_{\mu\nu\rho}&&\,,\label{eq:AAV_VVV_gluon_condensate_equality}
\end{alignat}
that are somewhat equivalent to the ones in the case of the perturbative contribution, apart from \eqref{eq:VVA_AAA_gluon_condensate_equality}, which appears here for the first time.

In the context of the relations \eqref{eq:VVA_AAA_gluon_condensate_equality} and \eqref{eq:AAV_VVV_gluon_condensate_equality} we note that the gluon condensate contributions to the $\langle VVA\rangle$ and $\langle AAV\rangle$ thus accidently have even higher symmetry then it is necessary, since they satisfy also the additional Bose symmetries of the $\langle AAA\rangle$ and $\langle VVV\rangle$ correlators, respectively, which they are equal to.

We also remind the reader that the longitudinal parts \eqref{eq:vva-definition_longitudinal}-\eqref{eq:aaa-definition_longitudinal} of the $\langle VVA\rangle$ and $\langle AAA\rangle$ Green functions are not affacted by the gluon condensate contribution and, therefore, the relation \eqref{eq:VVA_AAA_gluon_condensate_equality} thus automatically represents the transversal parts \eqref{eq:vva-definition_transversal}-\eqref{eq:aaa-definition_transversal} of these correlators.

The contribution of the gluon condensate to the $\langle SS\rangle$ and $\langle PP\rangle$ Green functions is the same, see \eqref{eq:2pt_gluon_condensate_SS_PP}. Then, once again we find out that
\begin{equation}
\mathcal{F}_{ASP}^{\langle G^{2}\rangle}(p^{2},q^{2};r^{2})=\mathcal{F}_{VSS}^{\langle G^{2}\rangle}(p^{2},q^{2};r^{2})=\mathcal{F}_{VPP}^{\langle G^{2}\rangle}(p^{2},q^{2};r^{2})\,,
\end{equation}
which is similar to \eqref{eq:perturbative-ASP-VPP-VSS_2} for the case of the perturbative contribution.

However, the contribution of the gluon condensate to the $\langle VV\rangle$ and $\langle AA\rangle$ correlators is the same, too. This allows us to express the formfactors of the $\langle AAA\rangle$ and $\langle VVV\rangle$ correlators in the terms of the ones of the $\langle VVA\rangle$ and $\langle AAV\rangle$, respectively. Specifically, we find
\begin{align}
\mathcal{F}_{AAA}^{\langle G^{2}\rangle}(p^{2},q^{2};r^{2})=&-\mathcal{G}_{VVA}^{\langle G^{2}\rangle}(p^{2},q^{2};r^{2})-\mathcal{H}_{VVA}^{\langle G^{2}\rangle}(p^{2},q^{2};r^{2})\,,\\
\mathcal{G}_{AAA}^{\langle G^{2}\rangle}(p^{2},q^{2};r^{2})=&-\frac{1}{2}\mathcal{H}_{VVA}^{\langle G^{2}\rangle}(p^{2},q^{2};r^{2})-\frac{1}{4}\frac{1}{\lambda_{K}+p^{2}(q^{2}+r^{2})+q^{2}r^{2}}\times\\
&\times\Big((p^{2}-q^{2})(p^{2}+q^{2}+r^{2})\mathcal{F}_{VVA}^{\langle G^{2}\rangle}(p^{2},q^{2};r^{2})\nonumber\\
&\hspace{6pt}+(p^{2}+q^{2}-2r^{2})(p^{2}+q^{2}-r^{2})\mathcal{G}_{VVA}^{\langle G^{2}\rangle}(p^{2},q^{2};r^{2})\Big)\,,\nonumber\\
\mathcal{H}_{AAA}^{\langle G^{2}\rangle}(p^{2},q^{2};r^{2})=&-\frac{3}{4}\frac{1}{\lambda_{K}+p^{2}(q^{2}+r^{2})+q^{2}r^{2}}\times\\
&\times\bigg[\Big(\mathcal{F}_{VVA}^{\langle G^{2}\rangle}(p^{2},q^{2};r^{2})-\mathcal{G}_{VVA}^{\langle G^{2}\rangle}(p^{2},q^{2};r^{2})\Big)p^{2}(p^{2}-r^{2})\nonumber\\
&\hspace{6pt}+\Big(\mathcal{F}_{VVA}^{\langle G^{2}\rangle}(p^{2},q^{2};r^{2})+\mathcal{G}_{VVA}^{\langle G^{2}\rangle}(p^{2},q^{2};r^{2})\Big)q^{2}(q^{2}-r^{2})\bigg]\,,\nonumber
\end{align}
and
\begin{align}
\mathcal{F}_{VVV}^{\langle G^{2}\rangle}(p^{2},q^{2};r^{2})&=\mathcal{F}_{AAV}^{\langle G^{2}\rangle}(p^{2},q^{2};r^{2})\,,\label{eq:AAV_VVV_gluon_condensate_equality_formfactor_1}\\
\mathcal{G}_{VVV}^{\langle G^{2}\rangle}(p^{2},q^{2};r^{2})&=\mathcal{G}_{AAV}^{\langle G^{2}\rangle}(p^{2},q^{2};r^{2})\,.\label{eq:AAV_VVV_gluon_condensate_equality_formfactor_2}
\end{align}

In terms of the formfactors, the results of the contributions of the gluon condensate to the three-point Green functions can be written down as follows:
\begin{alignat}{3}
&\mathcal{F}_{ASP}^{\langle G^{2}\rangle}(p^{2},q^{2};r^{2})&&=&&-\frac{i\alpha_{s}\langle G^{2}\rangle}{48\pi}\frac{(3p^{2}-q^{2}-r^{2})}{p^{2}q^{2}r^{2}}\,,\\
&\mathcal{F}_{VVA}^{\langle G^{2}\rangle}(p^{2},q^{2};r^{2})&&=&&-\frac{i\alpha_{s}\langle G^{2}\rangle}{96\pi}\frac{p^{2}(r^{2}-4q^{2})+p^{4}+q^{2}(q^{2}+r^{2})}{p^{4}q^{4}r^{2}}\,,\label{eq:vva-gluon_1}\\
&\mathcal{G}_{VVA}^{\langle G^{2}\rangle}(p^{2},q^{2};r^{2})&&=&&-\frac{i\alpha_{s}\langle G^{2}\rangle}{96\pi}\frac{(p^{2}-q^{2})(p^{2}+q^{2}+r^{2})}{p^{4}q^{4}r^{2}}\,,\label{eq:vva-gluon_2}\\
&\mathcal{H}_{VVA}^{\langle G^{2}\rangle}(p^{2},q^{2};r^{2})&&=&&\quad\,\frac{i\alpha_{s}\langle G^{2}\rangle}{96\pi}\frac{(p^{2}-q^{2})(p^{2}+q^{2}+r^{2})}{p^{4}q^{4}r^{2}}\,,\label{eq:vva-gluon_3}\\
&\mathcal{F}_{AAA}^{\langle G^{2}\rangle}(p^{2},q^{2};r^{2})&&=&&\quad\,0\,,\label{eq:aaa-gluon_1}\\
&\mathcal{G}_{AAA}^{\langle G^{2}\rangle}(p^{2},q^{2};r^{2})&&=&&\quad\,0\,,\label{eq:aaa-gluon_2}\\
&\mathcal{H}_{AAA}^{\langle G^{2}\rangle}(p^{2},q^{2};r^{2})&&=&&-\frac{i\alpha_{s}\langle G^{2}\rangle}{32\pi}\frac{1}{p^{2}q^{2}r^{2}}\,,\label{eq:aaa-gluon_3}\\
&\mathcal{F}_{AAV}^{\langle G^{2}\rangle}(p^{2},q^{2};r^{2})&&=&&\quad\,\frac{i\alpha_{s}\langle G^{2}\rangle}{96\pi}\frac{p^{2}+q^{2}-r^{2}}{p^{4}q^{4}r^{2}}\big[\lambda_{K}+r^{2}(p^{2}+q^{2}-r^{2})\big]\,,\label{eq:aav-gluon_1}\\
&\mathcal{G}_{AAV}^{\langle G^{2}\rangle}(p^{2},q^{2};r^{2})&&=&&-\frac{i\alpha_{s}\langle G^{2}\rangle}{48\pi}\frac{(q^{2}+r^{2})(p^{4}+q^{2}r^{2})+p^{2}(q^{2}-r^{2})^{2}}{p^{4}q^{4}r^{4}}\,,\label{eq:aav-gluon_2}\\
&\mathcal{H}_{AAV}^{\langle G^{2}\rangle}(p^{2},q^{2};r^{2})&&=&&-\frac{i\alpha_{s}\langle G^{2}\rangle}{96\pi}\frac{p^{2}-q^{2}-r^{2}}{p^{2}q^{4}r^{4}}\big[\lambda_{K}-p^{2}(p^{2}-q^{2}-r^{2})\big]\,,\label{eq:aav-gluon_3}\\
&\mathcal{F}_{VVV}^{\langle G^{2}\rangle}(p^{2},q^{2};r^{2})&&=&&\quad\,\frac{i\alpha_{s}\langle G^{2}\rangle}{96\pi}\frac{p^{2}+q^{2}-r^{2}}{p^{4}q^{4}r^{2}}\big[\lambda_{K}+r^{2}(p^{2}+q^{2}-r^{2})\big]\,,\label{eq:vvv-gluon_1}\\
&\mathcal{G}_{VVV}^{\langle G^{2}\rangle}(p^{2},q^{2};r^{2})&&=&&-\frac{i\alpha_{s}\langle G^{2}\rangle}{48\pi}\frac{(q^{2}+r^{2})(p^{4}+q^{2}r^{2})+p^{2}(q^{2}-r^{2})^{2}}{p^{4}q^{4}r^{4}}\,.\label{eq:vvv-gluon_2}
\end{alignat}

A very important note is in order here. As one can see, in contrast to the perturbative contribution and $\mathcal{O}(\alpha_{s})$ contribution to the quark condensate, the results for the gluon condensate do not contain any logarithmic terms although it is also a one-loop calculation. Although the Fourier transform of individual terms gives arise to such terms, all logarithms cancel each other completely and we are left with a simple rational result.\footnote{Similar fact was observed also in \cite{Ioffe:1982qb}, where the authors studied Borel-transformed invariant amplitudes of mesons.}

%%%%%%%%%%%%%%%%%%%%%%%%%%%%%%%%%%%%%%%%%%%%%%%%%%%%%%%%%%%%%%%%%%%%%%%%%%%%%%%%%%%%%%%%%%%%%%%%%%%%%%%%%
%%%%%%%%%%%%%%%%%%%%%%%%%%%%%%%%%%%%%%%%%%%%%%%%%%%%%%%%%%%%%%%%%%%%%%%%%%%%%%%%%%%%%%%%%%%%%%%%%%%%%%%%%
%%% Section: Quark-gluon Condensate
%%%%%%%%%%%%%%%%%%%%%%%%%%%%%%%%%%%%%%%%%%%%%%%%%%%%%%%%%%%%%%%%%%%%%%%%%%%%%%%%%%%%%%%%%%%%%%%%%%%%%%%%%
%%%%%%%%%%%%%%%%%%%%%%%%%%%%%%%%%%%%%%%%%%%%%%%%%%%%%%%%%%%%%%%%%%%%%%%%%%%%%%%%%%%%%%%%%%%%%%%%%%%%%%%%%

\section{Quark-gluon Condensate}\label{sec:quark-gluon-condensate}

\subsection{General Remarks}
The only QCD condensate with canonical dimension 5 in the chiral limit is the quark-gluon condensate, $\langle\overline{q}\sigma_{\mu\nu}G^{\mu\nu}q\rangle$, which we will denote by suppressing the Lorentz indices simply as $\langle\overline{q}\sigma\hspace{-1pt}\cdot\hspace{-1pt}Gq\rangle$ from now on.

There are two classes of contributions. The first one consists of propagation of the nonlocal quark condensate and the second one of propagation of the nonlocal quark-gluon condensate.

\subsection*{Propagation of Nonlocal Quark Condensate}
The first class of contributions of the quark-gluon condensate to the three-point Green functions is given by the diagrams where the soft gluon is attached to the quark line that carries zero momentum and as such these graphs can not be evaluated with standard perturbative methods. However, within the Fock-Schwinger gauge, this contribution stems from the propagation of the nonlocal quark condensate, as given by the formula \eqref{eq:quark-condensate-propagation} with the relevant part in the form
\begin{equation}
\langle\overline{q}_{i,\alpha}^{A}(x)q_{k,\beta}^{B}(y)\rangle\ni-\frac{g_{s}\langle\overline{q}\sigma\hspace{-1pt}\cdot\hspace{-1pt}Gq\rangle}{2^{5}\cdot 3^{2}}\big[F^{\langle\overline{q}q\rangle}(x,y)\big]_{ki}\delta_{\alpha\beta}\delta^{AB}\,,\label{eq:quark-condensate-propagation-kap6}
\end{equation}
with the function $F^{\langle\overline{q}q\rangle}(x,y)$ defined in \eqref{eq:quark-condensate-propagation-function-quark-gluon}.

\begin{figure}[htb]
    \centering
    \begin{subfigure}[t]{0.27\textwidth}
        \hspace{-1pt}\includegraphics[width=1\textwidth]{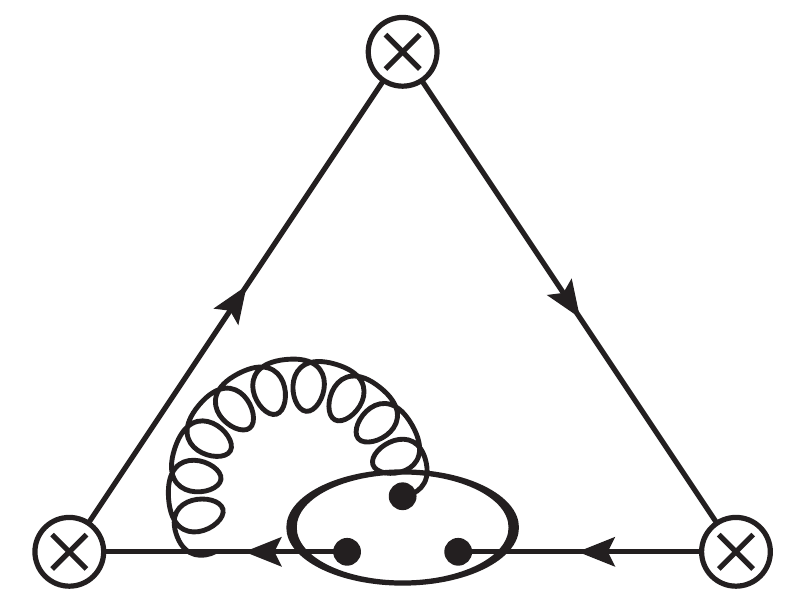}
        \caption{}
        \label{fig:fig_quark-gluon_condensate_propagation_1_a}
    \end{subfigure}
    \begin{subfigure}[t]{0.27\textwidth}
        \hspace{-1pt}\includegraphics[width=1\textwidth]{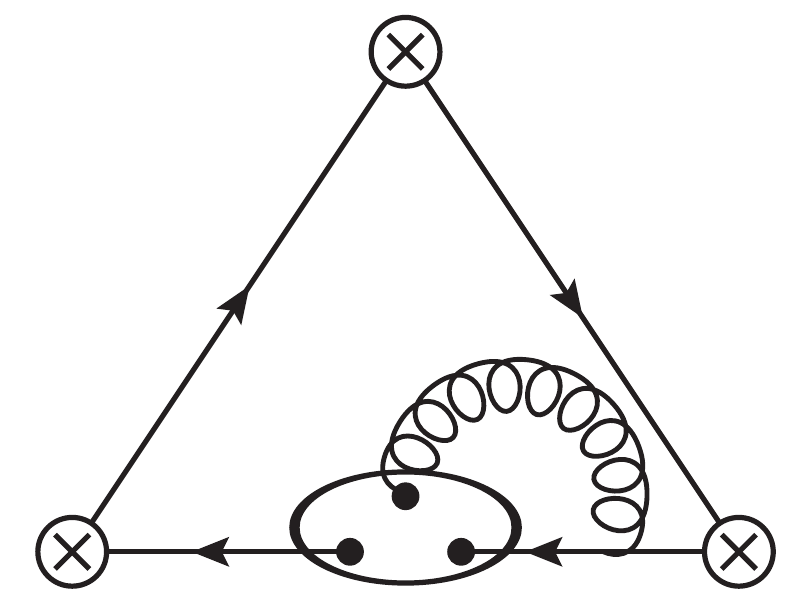}
        \caption{}
        \label{fig:fig_quark-gluon_condensate_propagation_1_b}
    \end{subfigure}
    \caption{Feynman diagrams of the contributions of the quark-gluon condensate to the three-point Green functions due to the effective propagation of the non-local quark condensate \eqref{eq:quark-condensate-propagation-kap6}.}
    \label{fig:fig_quark-gluon_condensate_propagation_1}
\end{figure}

There are six relevant diagrams, see Fig.~\ref{fig:fig_quark-gluon_condensate_propagation_1} and the appendix \ref{ssec:appendix_quark_gluon_condensate} for details. The contribution of the diagrams can be written down in the form
\begin{align}
\Pi_{\mathcal{O}_{1}^{a}\mathcal{O}_{2}^{b}\mathcal{O}_{3}^{c}}^{\langle\overline{q}q\rangle\rightarrow\langle\overline{q}\sigma\cdot Gq\rangle}&(x,y,z)=-\frac{g_{s}\langle\overline{q}\sigma\hspace{-1pt}\cdot\hspace{-1pt}Gq\rangle}{2^{5}\cdot 3}\mathrm{Tr}\big[T^{c}T^{b}T^{a}\big]\times\label{eq:quark-gluon-class-2}\\
&\times\mathrm{Tr}\Big[F^{\langle\overline{q}q\rangle}(z,x)\Gamma_{3}S_{0}(z,y)\Gamma_{2}S_{0}(y,x)\Gamma_{1}\Big]+\,(5\,\mathrm{permutations})\,.\nonumber
\end{align}

\subsection*{Propagation of Nonlocal Quark-gluon Condensate}
The second class of contributions of the quark-gluon condensate to the three-point Green functions consists of the diagrams with the soft gluon line attached to the quark line that carries nonzero momentum. These are graphs calculable by standard Feynman diagram techniques with the only difference in the fact that the local quark-gluon condensate is obtained, using the Fock-Schwinger gauge, from the propagation of the nonlocal quark-gluon condensate. The contribution is given by the formula \eqref{eq:4q_propagace_qAq} with the relevant part
\begin{equation}
g_{s}\langle\overline{q}_{i,\alpha}^{A}(x)\mathcal{A}_{\mu}^{a}(u)q_{k,\beta}^{B}(y)\rangle\ni\frac{g_{s}\langle\overline{q}\sigma\hspace{-1pt}\cdot\hspace{-1pt}G q\rangle}{2^{7}\cdot 3^{2}}\big[F_{\mu}^{\langle\overline{q}\mathcal{A}q\rangle}(x,u,y)\big]_{ki}(T^{a})_{\beta\alpha}\delta^{AB}\,,\label{eq:4q_propagace_qAq-kap6}
\end{equation}
with the function $F_{\mu}^{\langle\overline{q}\mathcal{A}q\rangle}(x,u,y)$ given by \eqref{eq:quark-gluon-condensate-propagation-function-quark-gluon}.

\begin{figure}[htb]
    \centering
    \begin{subfigure}[t]{0.27\textwidth}
        \hspace{-1pt}\includegraphics[width=1\textwidth]{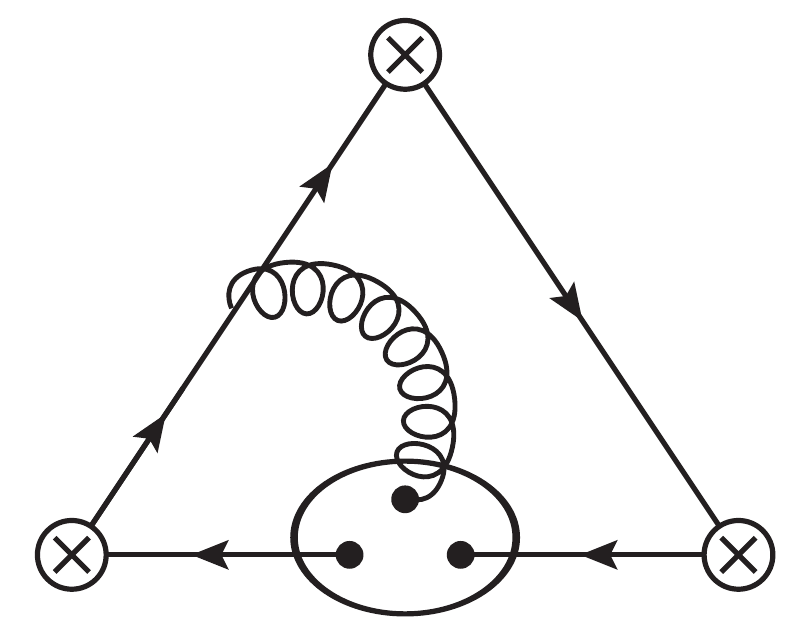}
        \caption{}
        \label{fig:fig_quark-gluon_condensate_propagation_2_a}
    \end{subfigure}
    \begin{subfigure}[t]{0.27\textwidth}
        \hspace{-2pt}\includegraphics[width=1\textwidth]{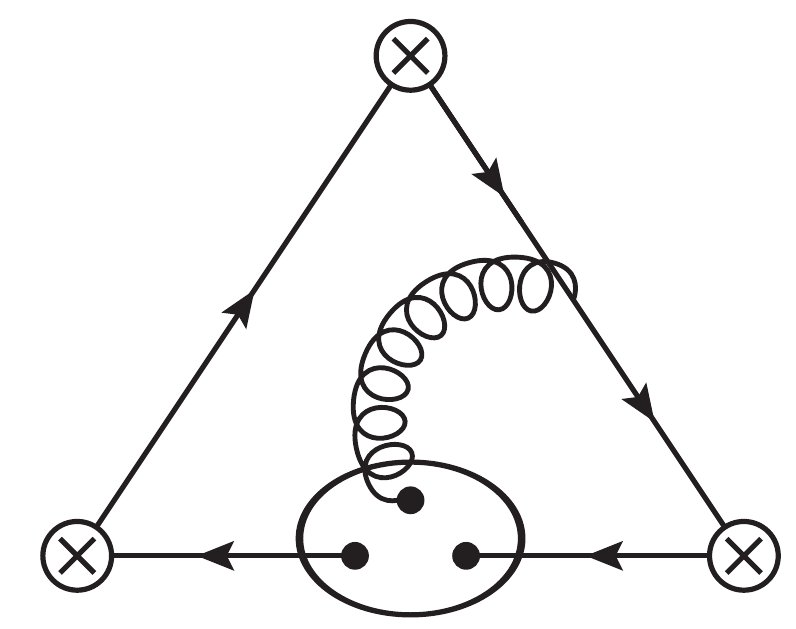}
        \caption{}
        \label{fig:fig_quark-gluon_condensate_propagation_2_b}
    \end{subfigure}
    \caption{Feynman diagrams of the contributions of the quark-gluon condensate to the three-point Green functions due to the effective propagation of the non-local quark-gluon condensate \eqref{eq:4q_propagace_qAq-kap6}.}
    \label{fig:fig_quark-gluon_condensate_propagation_2}
\end{figure}

In the case of the three-point Green functions, there are two possibilities of attaching the soft gluon to the fermion line between two chiral currents, which can be seen at Fig.~\ref{fig:fig_quark-gluon_condensate_propagation_2}. Due to six permutations of ordering these currents, there are 12 different contributing diagrams to this class.

The contributions of the diagrams can be written down, with a use of \eqref{eq:4q_propagace_qAq-kap6}, as
\begin{align}
\big[\Pi_{\mathcal{O}_{1}^{a}\mathcal{O}_{2}^{b}\mathcal{O}_{3}^{c}}^{\langle\overline{q}\mathcal{A}q\rangle\rightarrow\langle\overline{q}\sigma\cdot Gq\rangle}&(x,y,z)\big]_{(a)}=-\frac{i g_{s}\langle\overline{q}\sigma\hspace{-1pt}\cdot\hspace{-1pt}G q\rangle}{2^{5}\cdot 3^{2}}\mathrm{Tr}\big[T^{c}T^{b}T^{a}\big]\times\label{eq:quark-gluon-class-1_diag1}\\
&\hspace{-40pt}\times\int\mathrm{d}^{4}u\,\mathrm{Tr}\Big[F_{\alpha}^{\langle\overline{q}\mathcal{A}q\rangle}(z,u,x)\Gamma_{3}S_{0}(z,y)\Gamma_{2}S_{0}(y,u)\gamma^{\alpha}S_{0}(u,x)\Gamma_{1}\Big]+\,(5\,\mathrm{permutations})\,,\nonumber\\
\big[\Pi_{\mathcal{O}_{1}^{a}\mathcal{O}_{2}^{b}\mathcal{O}_{3}^{c}}^{\langle\overline{q}\mathcal{A}q\rangle\rightarrow\langle\overline{q}\sigma\cdot Gq\rangle}&(x,y,z)\big]_{(b)}=-\frac{i g_{s}\langle\overline{q}\sigma\hspace{-1pt}\cdot\hspace{-1pt}G q\rangle}{2^{5}\cdot 3^{2}}\mathrm{Tr}\big[T^{c}T^{b}T^{a}\big]\times\label{eq:quark-gluon-class-1_diag2}\\
&\hspace{-40pt}\times\int\mathrm{d}^{4}u\,\mathrm{Tr}\Big[F_{\alpha}^{\langle\overline{q}\mathcal{A}q\rangle}(z,u,x)\Gamma_{3}S_{0}(z,u)\gamma^{\alpha}S_{0}(u,y)\Gamma_{2}S_{0}(y,x)\Gamma_{1}\Big]+\,(5\,\mathrm{permutations})\,.\nonumber
\end{align}

To perform an integration over the coordinate of the gluon field, we use the following trick. Although the gluon is soft, such integration can be carried out simply as a standard Fourier transform also, if a nonzero momentum $k$ is temporarily assigned to the gluon field. This simplifies the manipulations significantly since it allows us to appropriately differentiate with respect to $k$, when needed, provided that we set $k=0$ back at the end of the calculations.\footnote{Similar argument has been used already in the literature. See, for example, pages 386 - 387 in Ref.~\cite{Ioffe:1982qb}.} Then, integrating out over the coordinate $u$, we obtain the following results:
\begin{align}
\big[\Pi_{\mathcal{O}_{1}^{a}\mathcal{O}_{2}^{b}\mathcal{O}_{3}^{c}}^{\langle\overline{q}\mathcal{A}q\rangle\rightarrow\langle\overline{q}\sigma\cdot Gq\rangle}(x,y,z)\big]_{(a)}&=-\frac{g_{s}\langle\overline{q}\sigma\hspace{-1pt}\cdot\hspace{-1pt}G q\rangle}{2^{5}\cdot 3^{2}}\mathrm{Tr}\big[T^{c}T^{b}T^{a}\big]\times\label{eq:quark-gluon-class-1_diag1_integrated_over}\\
&\times\mathrm{Tr}\big[\sigma_{\alpha\nu}\Gamma_{3}S_{0}(z,y)\Gamma_{2}\gamma^{\mu}\gamma^{\nu}\gamma^{\rho}\Gamma_{1}\big]\times\nonumber\\
&\times\bigg(\frac{1}{8\pi^{2}}g_{\alpha\rho}\big[F_{2}(x-y)\big]_{\mu}-i x_{\alpha}\big[F_{8}(x-y)\big]_{\mu\rho}+\big[F_{9}(x-y)\big]_{\alpha\mu\rho}\bigg)\nonumber\\
&+\,(5\,\mathrm{permutations})\,,\nonumber\\
\big[\Pi_{\mathcal{O}_{1}^{a}\mathcal{O}_{2}^{b}\mathcal{O}_{3}^{c}}^{\langle\overline{q}\mathcal{A}q\rangle\rightarrow\langle\overline{q}\sigma\cdot Gq\rangle}(x,y,z)\big]_{(b)}&=-\frac{g_{s}\langle\overline{q}\sigma\hspace{-1pt}\cdot\hspace{-1pt}G q\rangle}{2^{5}\cdot 3^{2}}\mathrm{Tr}\big[T^{c}T^{b}T^{a}\big]\times\label{eq:quark-gluon-class-1_diag2_integrated_over}\\
&\times\mathrm{Tr}\big[\sigma_{\alpha\nu}\Gamma_{3}\gamma^{\mu}\gamma^{\nu}\gamma^{\rho}\Gamma_{2}S_{0}(y,x)\Gamma_{1}\big]\times\nonumber\\
&\times\bigg(\frac{1}{8\pi^{2}}g_{\alpha\rho}\big[F_{2}(y-z)\big]_{\mu}-i y_{\alpha}\big[F_{8}(y-z)\big]_{\mu\rho}+\big[F_{9}(y-z)\big]_{\alpha\mu\rho}\bigg)\nonumber\\
&+\,(5\,\mathrm{permutations})\,.\nonumber
\end{align}

Now one can make sure that it is safe to set $z=0$. Then, it is a trivial task to perform the Fourier transform and convert the results to the momentum representation, such as
\begin{align}
\big[\Pi_{\mathcal{O}_{1}^{a}\mathcal{O}_{2}^{b}\mathcal{O}_{3}^{c}}^{\langle\overline{q}\mathcal{A}q\rangle\rightarrow\langle\overline{q}\sigma\cdot Gq\rangle}(p,q;r)\big]_{(a)}&=-\frac{g_{s}\langle\overline{q}\sigma\hspace{-1pt}\cdot\hspace{-1pt}G q\rangle}{2^{5}\cdot 3^{2}}\mathrm{Tr}\big[T^{c}T^{b}T^{a}\big]\mathrm{Tr}\big[\sigma_{\alpha\nu}\Gamma_{3}S_{0}(p+q)\Gamma_{2}\gamma^{\mu}\gamma^{\nu}\gamma^{\rho}\Gamma_{1}\big]\times\nonumber\\
&\times\bigg(\hspace{-4pt}-\frac{1}{8\pi^{2}}g_{\alpha\mu}\big[\widetilde{F}_{2}(p)\big]_{\rho}-\big[\widetilde{F}_{9}(p)\big]_{\alpha\mu\rho}\bigg)+\,(5\,\mathrm{permutations})\,,\label{eq:quark-gluon-class-1_diag1_p-repr}\\
\big[\Pi_{\mathcal{O}_{1}^{a}\mathcal{O}_{2}^{b}\mathcal{O}_{3}^{c}}^{\langle\overline{q}\mathcal{A}q\rangle\rightarrow\langle\overline{q}\sigma\cdot Gq\rangle}(p,q;r)\big]_{(b)}&=-\frac{g_{s}\langle\overline{q}\sigma\hspace{-1pt}\cdot\hspace{-1pt}G q\rangle}{2^{5}\cdot 3^{2}}\mathrm{Tr}\big[T^{c}T^{b}T^{a}\big]\mathrm{Tr}\big[\sigma_{\alpha\nu}\Gamma_{3}\gamma^{\mu}\gamma^{\nu}\gamma^{\rho}\Gamma_{2}S_{0}(p)\Gamma_{1}\big]\times\nonumber\\
&\times\bigg(\frac{1}{8\pi^{2}}g_{\alpha\mu}\big[\widetilde{F}_{2}(r)\big]_{\rho}+\big[F_{9}(r)\big]_{\alpha\mu\rho}\bigg)+\,(5\,\mathrm{permutations})\,.\label{eq:quark-gluon-class-1_diag2_p-repr}
\end{align}

\subsection{Results}
Here we present all the results obtained for the contribution of the quark-gluon condensate to the relevant three-point Green functions.
\begin{align}
\mathcal{F}_{SSS}^{\langle\overline{q}\sigma\cdot G q\rangle}(p^{2},q^{2};r^{2})=&\quad\,\frac{g_{s}\langle\overline{q}\sigma\hspace{-1pt}\cdot\hspace{-1pt}Gq\rangle}{144p^{4}q^{4}r^{4}}\times\label{eq:sss-quark-gluon}\\
&\hspace{-50pt}\times\Big(-r^{4}(4p^{2}q^{2}+7p^{4}+7q^{4})+8r^{6}(p^{2}+q^{2})+6r^{2}(p^{2}-q^{2})^{2}(p^{2}+q^{2})\nonumber\\
&\hspace{-32pt}-3(p^{2}-q^{2})^{2}(p^{4}+q^{4})-4r^{8}\Big)\,,\nonumber\\
\mathcal{F}_{SPP}^{\langle\overline{q}\sigma\cdot G q\rangle}(p^{2},q^{2};r^{2})=&\quad\,\frac{g_{s}\langle\overline{q}\sigma\hspace{-1pt}\cdot\hspace{-1pt}Gq\rangle}{144p^{4}q^{4}r^{4}}\times\label{eq:spp-quark-gluon}\\
&\hspace{-50pt}\times\Big(-r^{4}(4p^{2}q^{2}+p^{4}+7q^{4})+8r^{6}(p^{2}+q^{2})-6r^{2}(p^{2}-q^{2})(p^{4}+q^{4})\nonumber\\
&\hspace{-32pt}+3(p^{2}-q^{2})^{3}(p^{2}+q^{2})-4r^{8}\Big)\,,\nonumber\\
\mathcal{F}_{VVP}^{\langle\overline{q}\sigma\cdot G q\rangle}(p^{2},q^{2};r^{2})=&-\frac{g_{s}\langle\overline{q}\sigma\hspace{-1pt}\cdot\hspace{-1pt}Gq\rangle}{72p^{4}q^{4}r^{4}}\Big(r^{2}(p^{4}+q^{4})+3(p^{2}-q^{2})^{2}(p^{2}+q^{2})+4r^{6}\Big)\,,\label{eq:vvp-quark-gluon}\\
\mathcal{F}_{AAP}^{\langle\overline{q}\sigma\cdot G q\rangle}(p^{2},q^{2};r^{2})=&-\frac{g_{s}\langle\overline{q}\sigma\hspace{-1pt}\cdot\hspace{-1pt}Gq\rangle}{72p^{4}q^{4}r^{4}}\Big(r^{2}(p^{4}+q^{4})+3(p^{2}-q^{2})^{2}(p^{2}+q^{2})-4r^{6}\Big)\,,\label{eq:aap-quark-gluon}\\
\mathcal{F}_{VAS}^{\langle\overline{q}\sigma\cdot G q\rangle}(p^{2},q^{2};r^{2})=&-\frac{g_{s}\langle\overline{q}\sigma\hspace{-1pt}\cdot\hspace{-1pt}Gq\rangle}{72p^{4}q^{4}r^{4}}\Big(r^{2}(p^{4}-q^{4})+3(p^{2}-q^{2})(p^{4}+q^{4})-4r^{6}\Big)\,,\label{eq:vas-quark-gluon}\\
\mathcal{F}_{VVS}^{\langle\overline{q}\sigma\cdot G q\rangle}(p^{2},q^{2};r^{2})=&\quad\,\frac{g_{s}\langle\overline{q}\sigma\hspace{-1pt}\cdot\hspace{-1pt}Gq\rangle}{72p^{4}q^{4}r^{4}\lambda_{K}^{2}}\times\label{eq:vvs-quark-gluon_F}\\
&\hspace{-50pt}\times\Big(-(11p^{12}+7p^{10}q^{2}+17p^{8}q^{4}+74p^{6}q^{6}+17p^{4}q^{8}+7p^{2}q^{10}+11q^{12})r^{2}\nonumber\\
&\hspace{-32pt}-(p^{2}+q^{2})(17p^{4}-35p^{2}q^{2}+17q^{4})r^{8}+(25p^{4}+16p^{2}q^{2}+25q^{4})r^{10}\nonumber\\
&\hspace{-32pt}-16(p^{2}+q^{2})r^{12}-(2p^{8}+45p^{6}q^{2}+6p^{4}q^{4}+45p^{2}q^{6}+2q^{8})r^{6}\nonumber\\
&\hspace{-32pt}+(p^{2}+q^{2})(14p^{8}+29p^{6}q^{2}+70p^{4}q^{4}+29p^{2}q^{6}+14q^{8})r^{4}\nonumber\\
&\hspace{-32pt}+3(p^{2}-q^{2})^{4}(p^{2}+q^{2})(p^{4}+q^{4})+4r^{14}\Big)\,,\nonumber\\
\mathcal{G}_{VVS}^{\langle\overline{q}\sigma\cdot G q\rangle}(p^{2},q^{2};r^{2})=&\quad\,\frac{g_{s}\langle\overline{q}\sigma\hspace{-1pt}\cdot\hspace{-1pt}Gq\rangle}{12p^{4}q^{4}r^{4}\lambda_{K}^{2}}\times\label{eq:vvs-quark-gluon_G}\\
&\hspace{-50pt}\times\Big(-(p^{2}+q^{2})(4p^{4}+3p^{2}q^{2}+4q^{4})r^{6}+(p^{4}+q^{4})(p^{2}-q^{2})^{4}+(p^{4}+q^{4})r^{8}\nonumber\\
&\hspace{-32pt}-(p^{2}+q^{2})(4p^{8}-5p^{6}q^{2}+14p^{4}q^{4}-5p^{2}q^{6}+4q^{8})r^{2}\nonumber\\
&\hspace{-32pt}+2(3p^{8}+5p^{6}q^{2}+12p^{4}q^{4}+5p^{2}q^{6}+3q^{8})r^{4}\Big)\,,\nonumber\\
\mathcal{F}_{AAS}^{\langle\overline{q}\sigma\cdot G q\rangle}(p^{2},q^{2};r^{2})=&\quad\,\frac{g_{s}\langle\overline{q}\sigma\hspace{-1pt}\cdot\hspace{-1pt}Gq\rangle}{72p^{4}q^{4}r^{4}\lambda_{K}^{2}}\times\label{eq:aas-quark-gluon_F}\\
&\hspace{-50pt}\times\Big(-(9q^{2}+11r^{2})p^{12}+(q^{2}-r^{2})(9q^{8}-8q^{6}r^{2}+91q^{4}r^{4}+37q^{2}r^{6}+23r^{8})p^{4}\nonumber\\
&\hspace{-32pt}-(3q^{6}+17q^{4}r^{2}-43q^{2}r^{4}+10r^{6})p^{8}+(9q^{4}-7q^{2}r^{2}+14r^{4})p^{10}\nonumber\\
&\hspace{-32pt}+(-3q^{8}-74q^{6}r^{2}+99q^{4}r^{4}-13q^{2}r^{6}+15r^{8})p^{6}\nonumber\\
&\hspace{-32pt}-(q^{2}-r^{2})^{3}(9q^{6}+34q^{4}r^{2}+32q^{2}r^{4}+16r^{6})p^{2}\nonumber\\
&\hspace{-32pt}+(q^{2}-r^{2})^{5}(3q^{4}+4q^{2}r^{2}+4r^{4})+3p^{14}\Big)\,,\nonumber\\
\mathcal{G}_{AAS}^{\langle\overline{q}\sigma\cdot G q\rangle}(p^{2},q^{2};r^{2})=&\quad\,\frac{g_{s}\langle\overline{q}\sigma\hspace{-1pt}\cdot\hspace{-1pt}Gq\rangle}{12p^{4}q^{4}r^{4}\lambda_{K}^{2}}\times\label{eq:aas-quark-gluon_G}\\
&\hspace{-50pt}\times\Big(-(p^{2}+q^{2})(4p^{4}+3p^{2}q^{2}+4q^{4})r^{6}+(p^{4}+q^{4})(p^{2}-q^{2})^{4}\nonumber\\
&\hspace{-32pt}+(p^{4}+q^{4})r^{8}+2(3p^{8}+5p^{6}q^{2}+12p^{4}q^{4}+5p^{2}q^{6}+3q^{8})r^{4}\nonumber\\
&\hspace{-32pt}-(p^{2}+q^{2})(4p^{8}-5p^{6}q^{2}+14p^{4}q^{4}-5p^{2}q^{6}+4q^{8})r^{2}\Big)\,,\nonumber\\
\mathcal{F}_{VAP}^{\langle\overline{q}\sigma\cdot G q\rangle}(p^{2},q^{2};r^{2})=&\quad\,\frac{g_{s}\langle\overline{q}\sigma\hspace{-1pt}\cdot\hspace{-1pt}Gq\rangle}{72p^{4}q^{4}r^{4}\lambda_{K}^{2}}\times\label{eq:vap-quark-gluon_F}\\
&\hspace{-50pt}\times\Big(-(p^{2}-q^{2})(14p^{8}+57p^{6}q^{2}+110p^{4}q^{4}+57p^{2}q^{6}+14q^{8})r^{4}+4r^{14}\nonumber\\
&\hspace{-32pt}+(23p^{4}+16p^{2}q^{2}+25q^{4})r^{10}-(p^{2}-q^{2})(15p^{4}+p^{2}q^{2}-17q^{4})r^{8}\nonumber\\
&\hspace{-32pt}+(p^{2}-q^{2})(10p^{6}+23p^{4}q^{2}+47p^{2}q^{4}+2q^{6})r^{6}-16(p^{2}+q^{2})r^{12}\nonumber\\
&\hspace{-32pt}+(p^{2}-q^{2})(p^{2}+q^{2})(11p^{8}+7p^{6}q^{2}+7p^{2}q^{6}+11q^{8})r^{2}\nonumber\\
&\hspace{-32pt}-3(p^{2}-q^{2})^{5}(p^{2}+q^{2})^{2}\Big)\,,\nonumber\\
\mathcal{G}_{VAP}^{\langle\overline{q}\sigma\cdot G q\rangle}(p^{2},q^{2};r^{2})=&-\frac{g_{s}\langle\overline{q}\sigma\hspace{-1pt}\cdot\hspace{-1pt}Gq\rangle}{12p^{4}q^{4}r^{4}\lambda_{K}^{2}}(p^{2}-q^{2})\times\label{eq:vap-quark-gluon_G}\\
&\hspace{-50pt}\times\Big(-(p^{2}+q^{2})^{2}(4p^{4}-5p^{2}q^{2}+4q^{4})r^{2}+2r^{4}(p^{2}+q^{2})(3p^{4}+5p^{2}q^{2}+3q^{4})\nonumber\\
&\hspace{-32pt}-(4p^{4}+11p^{2}q^{2}+4q^{4})r^{6}+(p^{2}+q^{2})r^{8}+(p^{2}+q^{2})(p^{2}-q^{2})^{4}\Big)\,.\nonumber
\end{align}

Since these contributions are given by tree diagrams, all logarithmic terms, that emerge from the Fourier transform, have to cancel each other out, which serves as a nontrivial check of consistency of our calculations, besides the fulfillment of the Ward identities. As there are no contributions of the quark-gluon condensate to the two-point Green functions, right-hand side of all the relevant Ward identities of the three-point contributions \eqref{eq:vvp-quark-gluon}-\eqref{eq:vap-quark-gluon_G} must vanish. It is easy to verify that this fact is indeed satisfied, which also serves as a consistency check of the results above.

%%%%%%%%%%%%%%%%%%%%%%%%%%%%%%%%%%%%%%%%%%%%%%%%%%%%%%%%%%%%%%%%%%%%%%%%%%%%%%%%%%%%%%%%%%%%%%%%%%%%%%%%%
%%%%%%%%%%%%%%%%%%%%%%%%%%%%%%%%%%%%%%%%%%%%%%%%%%%%%%%%%%%%%%%%%%%%%%%%%%%%%%%%%%%%%%%%%%%%%%%%%%%%%%%%%
%%% Section: Four-quark Condensate
%%%%%%%%%%%%%%%%%%%%%%%%%%%%%%%%%%%%%%%%%%%%%%%%%%%%%%%%%%%%%%%%%%%%%%%%%%%%%%%%%%%%%%%%%%%%%%%%%%%%%%%%%
%%%%%%%%%%%%%%%%%%%%%%%%%%%%%%%%%%%%%%%%%%%%%%%%%%%%%%%%%%%%%%%%%%%%%%%%%%%%%%%%%%%%%%%%%%%%%%%%%%%%%%%%%

\section{Four-quark Condensate}\label{sec:four-quark-condensate}

\subsection{General Remarks}
There are three classes of contributing diagrams to the three-point Green functions that generate the four-quark condensate contribution. The first class is given by the diagrams that are calculable by standard technique of Feynman diagrams in a perturbative regime. The other two classes contain diagrams with soft gluons, that are attached to the quark lines that carry zero momentum, and as such can not be evaluated within the pertubative approach. These graphs are thus obtained effectively, given by the propagations of the nonlocal quark and quark-gluon condensates.

\subsection*{Direct Contribution}
The first class of contributions of the four-quark condensate to the three-point Green functions consists of diagrams with two separate fermion lines connected via a hard gluon line, i.e. via the gluon propagator with nonzero momentum. As such, these diagrams are evaluated within standard techniques of Feynman diagrams, the only difference with respect to a conventional perturbative theory is that the averaging over the uncontracted quark fields is performed according to \eqref{eq:4quark-averaging}.

\begin{figure}[htb]
    \centering
    \begin{subfigure}[t]{0.27\textwidth}
        \hspace{9pt}\includegraphics[width=0.85\textwidth]{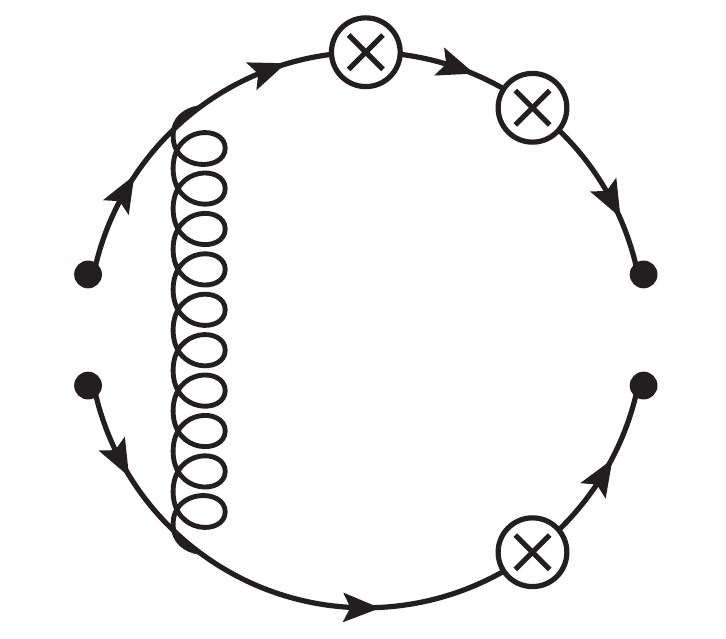}
        \caption{}
        \label{fig:fig_four-quark_condensate_propagation_1_a}
    \end{subfigure}
    \begin{subfigure}[t]{0.27\textwidth}
        \hspace{9pt}\includegraphics[width=0.85\textwidth]{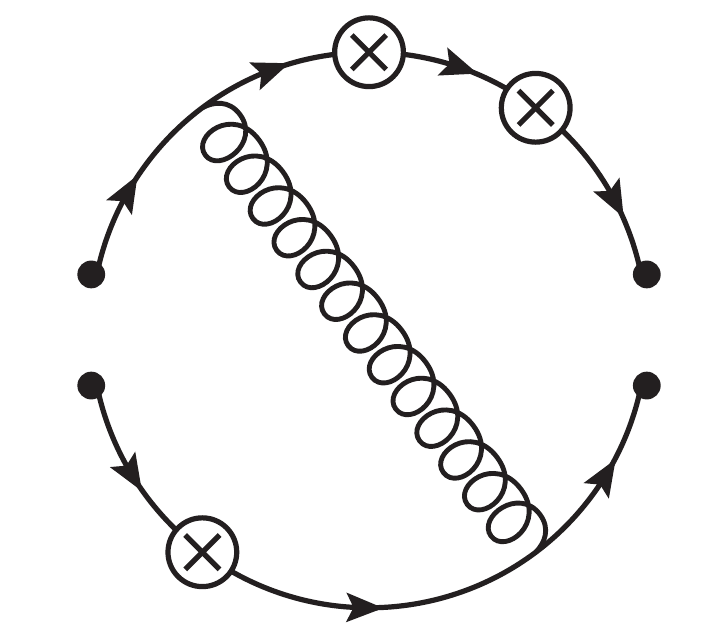}
        \caption{}
        \label{fig:fig_four-quark_condensate_propagation_1_b}
    \end{subfigure}
    \begin{subfigure}[t]{0.27\textwidth}
        \hspace{8pt}\includegraphics[width=0.85\textwidth]{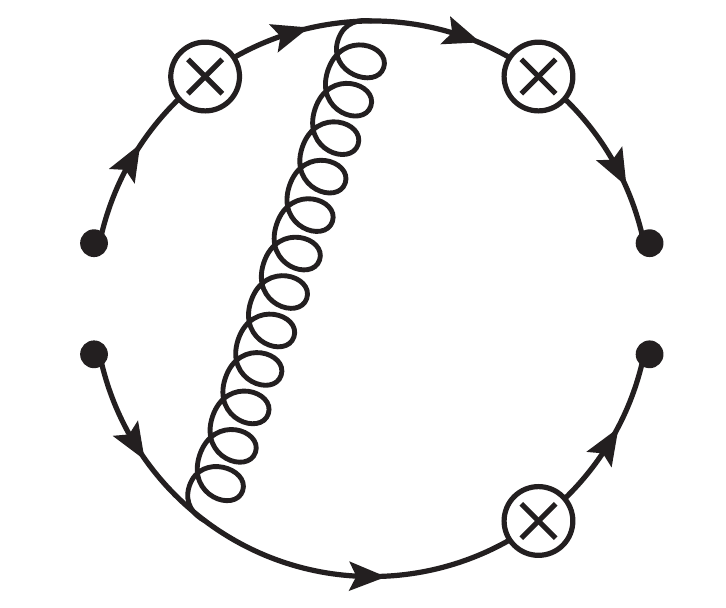}
        \caption{}
        \label{fig:fig_four-quark_condensate_propagation_1_c}
    \end{subfigure}\vspace{10pt}
    
    \begin{subfigure}[t]{0.27\textwidth}
        \hspace{9pt}\includegraphics[width=0.85\textwidth]{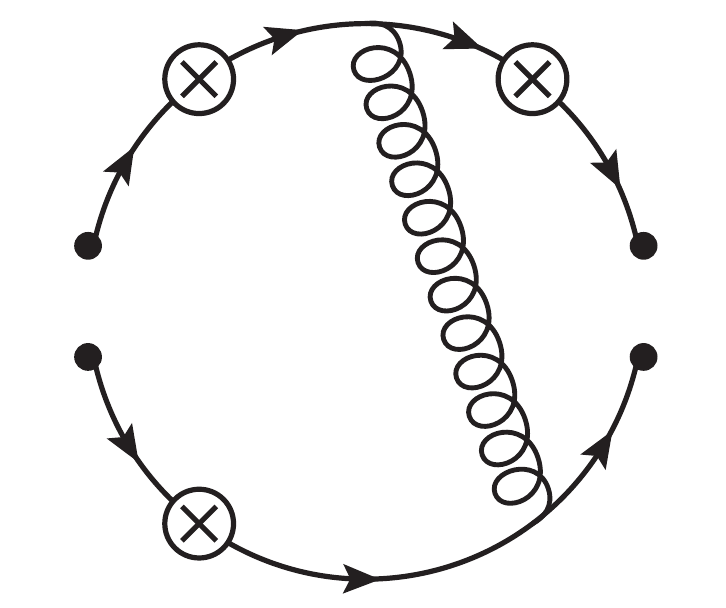}
        \caption{}
        \label{fig:fig_four-quark_condensate_propagation_1_d}
    \end{subfigure}
    \begin{subfigure}[t]{0.27\textwidth}
        \hspace{9pt}\includegraphics[width=0.85\textwidth]{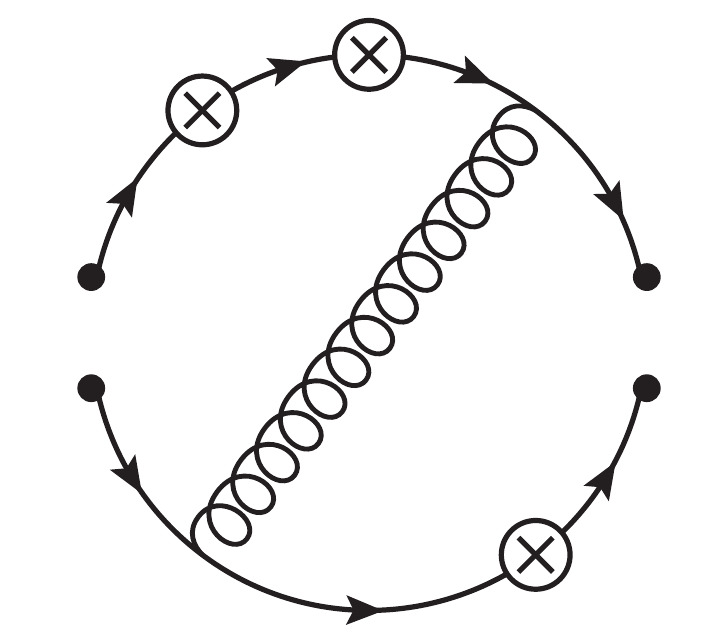}
        \caption{}
        \label{fig:fig_four-quark_condensate_propagation_1_e}
    \end{subfigure}
    \begin{subfigure}[t]{0.27\textwidth}
        \hspace{8pt}\includegraphics[width=0.85\textwidth]{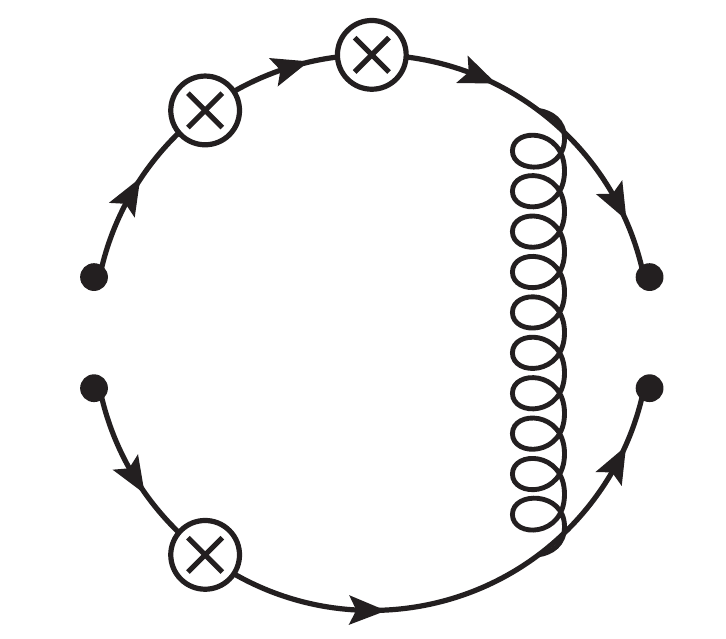}
        \caption{}
        \label{fig:fig_four-quark_condensate_propagation_1_f}
    \end{subfigure}
    \caption{Feynman diagrams of the direct contribution of the four-quark condensate to the three-point Green functions.}
    \label{fig:fig_four-quark_condensate_propagation_1}
\end{figure}

There are six unique graph topologies, depicted at Fig.~\ref{fig:fig_four-quark_condensate_propagation_1}, with six individual diagrams for each topology due to the number of permutations. Therefore, there are 36 contributing diagrams in total.

Once we take the gluon propagator in the form
\begin{equation}
\big[S_{g}(p)\big]_{\mu\nu}^{ab}=-\frac{i}{p^{2}}g_{\mu\nu}\delta^{ab}\equiv S_{g}(p)g_{\mu\nu}\delta^{ab}\,,
\end{equation}
by standard well-known approach, and a use of \eqref{eq:4quark-averaging}, one finds the contributions of the diagrams \ref{fig:fig_four-quark_condensate_propagation_1_a}-\ref{fig:fig_four-quark_condensate_propagation_1_f} at Fig.~\ref{fig:fig_four-quark_condensate_propagation_1} to be of the forms
\begin{align}
\big[\Pi_{\mathcal{O}_{1}^{a}\mathcal{O}_{2}^{b}\mathcal{O}_{3}^{c}}^{\mathrm{pert.}\rightarrow\langle\overline{q}q\rangle^{2}}(p,q&;r)\big]_{(a)}=\frac{\pi\alpha_{s}\langle\overline{q}q\rangle^{2}}{81}\mathrm{Tr}\big[T^{b}T^{a}T^{c}\big]S_{g}(r)\times\label{eq:4q-class-1_diag_a}\\
&\times\mathrm{Tr}\Big[\Gamma_{2}S_{0}(p+r)\Gamma_{1}S_{0}(r)\gamma^{\alpha}\Gamma_{3}S_{0}(-r)\gamma_{\alpha}\Big]+(5\,\mathrm{terms})\,,\nonumber\\
\big[\Pi_{\mathcal{O}_{1}^{a}\mathcal{O}_{2}^{b}\mathcal{O}_{3}^{c}}^{\mathrm{pert.}\rightarrow\langle\overline{q}q\rangle^{2}}(p,q&;r)\big]_{(b)}=\frac{\pi\alpha_{s}\langle\overline{q}q\rangle^{2}}{81}\mathrm{Tr}\big[T^{b}T^{a}T^{c}\big]S_{g}(r)\times\label{eq:4q-class-1_diag_b}\\
&\times\mathrm{Tr}\Big[\Gamma_{2}S_{0}(p+r)\Gamma_{1}S_{0}(r)\gamma^{\alpha}\gamma_{\alpha}S_{0}(r)\Gamma_{3}\Big]+(5\,\mathrm{terms})\,,\nonumber\\
\big[\Pi_{\mathcal{O}_{1}^{a}\mathcal{O}_{2}^{b}\mathcal{O}_{3}^{c}}^{\mathrm{pert.}\rightarrow\langle\overline{q}q\rangle^{2}}(p,q&;r)\big]_{(c)}=\frac{\pi\alpha_{s}\langle\overline{q}q\rangle^{2}}{81}\mathrm{Tr}\big[T^{b}T^{a}T^{c}\big]S_{g}(r)\times\label{eq:4q-class-1_diag_c}\\
&\times\mathrm{Tr}\Big[\Gamma_{2}S_{0}(p+r)\gamma^{\alpha}S_{0}(p)\Gamma_{1}\Gamma_{3}S_{0}(-r)\gamma_{\alpha}\Big]+(5\,\mathrm{terms})\,,\nonumber\\
\big[\Pi_{\mathcal{O}_{1}^{a}\mathcal{O}_{2}^{b}\mathcal{O}_{3}^{c}}^{\mathrm{pert.}\rightarrow\langle\overline{q}q\rangle^{2}}(p,q&;r)\big]_{(d)}=\frac{\pi\alpha_{s}\langle\overline{q}q\rangle^{2}}{81}\mathrm{Tr}\big[T^{b}T^{a}T^{c}\big]S_{g}(r)\times\label{eq:4q-class-1_diag_d}\\
&\times\mathrm{Tr}\Big[\Gamma_{2}S_{0}(p+r)\gamma^{\alpha}S_{0}(p)\Gamma_{1}\gamma_{\alpha}S_{0}(r)\Gamma_{3}\Big]+(5\,\mathrm{terms})\,,\nonumber\\
\big[\Pi_{\mathcal{O}_{1}^{a}\mathcal{O}_{2}^{b}\mathcal{O}_{3}^{c}}^{\mathrm{pert.}\rightarrow\langle\overline{q}q\rangle^{2}}(p,q&;r)\big]_{(e)}=\frac{\pi\alpha_{s}\langle\overline{q}q\rangle^{2}}{81}\mathrm{Tr}\big[T^{b}T^{a}T^{c}\big]S_{g}(r)\times\label{eq:4q-class-1_diag_e}\\
&\times\mathrm{Tr}\Big[\gamma^{\alpha}S_{0}(p+q)\Gamma_{2}S_{0}(p)\Gamma_{1}\Gamma_{3}S_{0}(-r)\gamma_{\alpha}\Big]+(5\,\mathrm{terms})\,,\nonumber\\
\big[\Pi_{\mathcal{O}_{1}^{a}\mathcal{O}_{2}^{b}\mathcal{O}_{3}^{c}}^{\mathrm{pert.}\rightarrow\langle\overline{q}q\rangle^{2}}(p,q&;r)\big]_{(f)}=\frac{\pi\alpha_{s}\langle\overline{q}q\rangle^{2}}{81}\mathrm{Tr}\big[T^{b}T^{a}T^{c}\big]S_{g}(r)\times\label{eq:4q-class-1_diag_f}\\
&\times\mathrm{Tr}\Big[\gamma_{\alpha}S_{0}(p+q)\Gamma_{2}S_{0}(p)\Gamma_{1}\gamma^{\alpha}S_{0}(r)\Gamma_{3}\Big]+(5\,\mathrm{terms})\,.\nonumber
\end{align}
One can notice the fact that the first term in \eqref{eq:4quark-averaging} does not contribute at all since it leads to terms with traces of single Gell-Mann matrices that vanishes identically.

\subsection*{Propagation of Nonlocal Quark Condensate}
\begin{figure}[htb]
    \centering
    \begin{subfigure}[t]{0.27\textwidth}
        \hspace{-1pt}\includegraphics[width=1\textwidth]{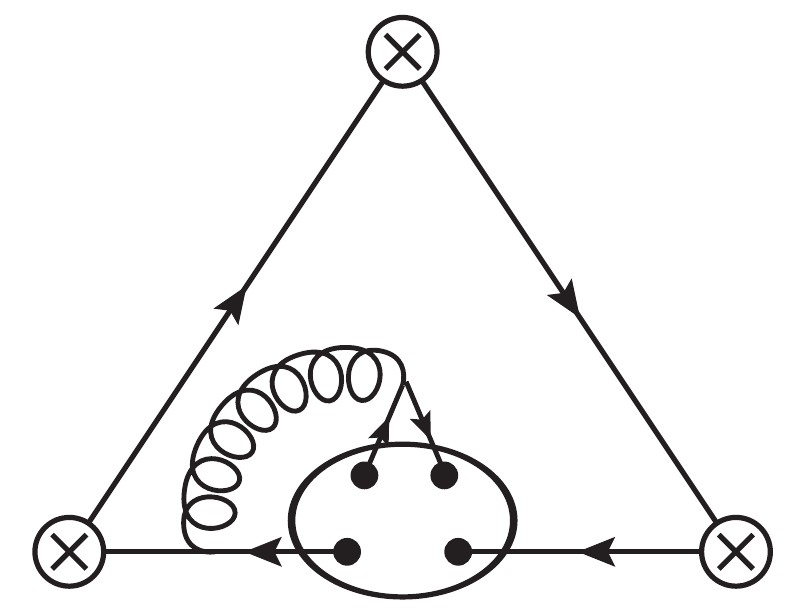}
        \caption{}
        \label{fig:fig_four-quark_condensate_propagation_2_a}
    \end{subfigure}
    \begin{subfigure}[t]{0.27\textwidth}
        \hspace{-1pt}\includegraphics[width=1\textwidth]{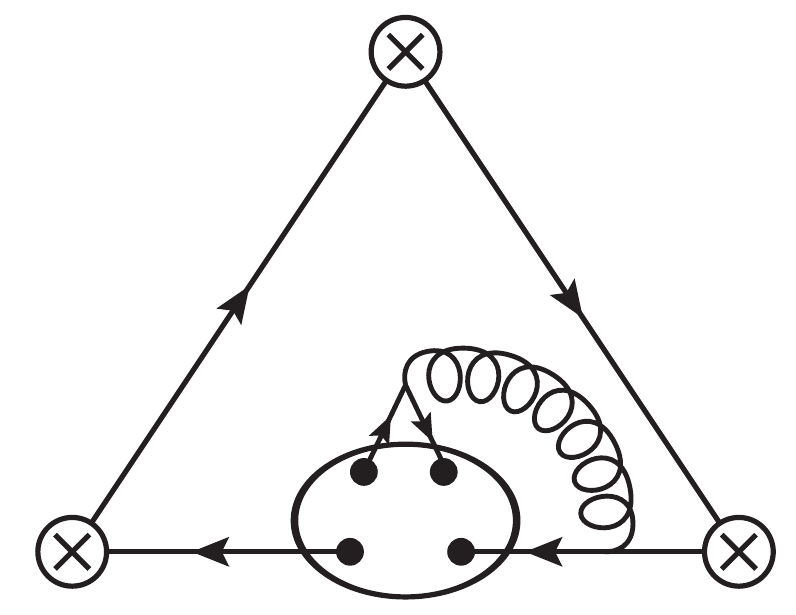}
        \caption{}
        \label{fig:fig_four-quark_condensate_propagation_2_b}
    \end{subfigure}
    \caption{Feynman diagrams of the contributions of the four-quark condensate to the three-point Green functions due to the effective propagation of the nonlocal quark condensate \eqref{eq:quark-condensate-propagation-kap7}.}
    \label{fig:fig_four-quark_condensate_propagation_2}
\end{figure}

The second class of contributions of the four-quark condensate to the three-point Green functions is given by the diagrams similar to the ones of the quark condensate with the difference in that the soft gluon is connected to the quark line with zero momentum. The relevant graphs are depicted at Fig.~\ref{fig:fig_four-quark_condensate_propagation_2}. As such, this contribution stems from the expansion of the quark condensate \eqref{eq:quark-condensate-propagation}. The contributing part is given as
\begin{equation}
\big\langle\overline{q}_{i,\alpha}^{A}(x)q_{k,\beta}^{B}(y)\big\rangle\ni\frac{i\pi\alpha_{s}\langle\overline{q}q\rangle^{2}}{2^{3}\cdot 3^{7}}\delta^{AB}\delta_{\alpha\beta}\big[G^{\langle\overline{q}q\rangle}(x,y)\big]_{ki}\,,\label{eq:quark-condensate-propagation-kap7}
\end{equation}
where the function $G^{\langle\overline{q}q\rangle}(x,y)$ is defined in \eqref{eq:quark-condensate-propagation-function}.

The contribution can be written down as
\begin{align}
\Pi_{\mathcal{O}_{1}^{a}\mathcal{O}_{2}^{b}\mathcal{O}_{3}^{c}}^{\langle\overline{q}q\rangle\rightarrow\langle\overline{q}q\rangle^{2}}(x,y,z)=&\quad\,\frac{i\pi\alpha_{s}\langle\overline{q}q\rangle^{2}}{2^{3}\cdot 3^{6}}\mathrm{Tr}\big[T^{c}T^{b}T^{a}\big]\mathrm{Tr}\Big[G^{\langle\overline{q}q\rangle}(z,x)\Gamma_{3}S_{0}(z,y)\Gamma_{2}S_{0}(y,x)\Gamma_{1}\Big]\label{eq:4q-class-2}\\
&+(5\,\mathrm{permutations})\,.\nonumber
\end{align}

\subsection*{Propagation of Nonlocal Quark-gluon Condensate}
\begin{figure}[htb]
    \centering
    \begin{subfigure}[t]{0.27\textwidth}
        \hspace{-1pt}\includegraphics[width=1\textwidth]{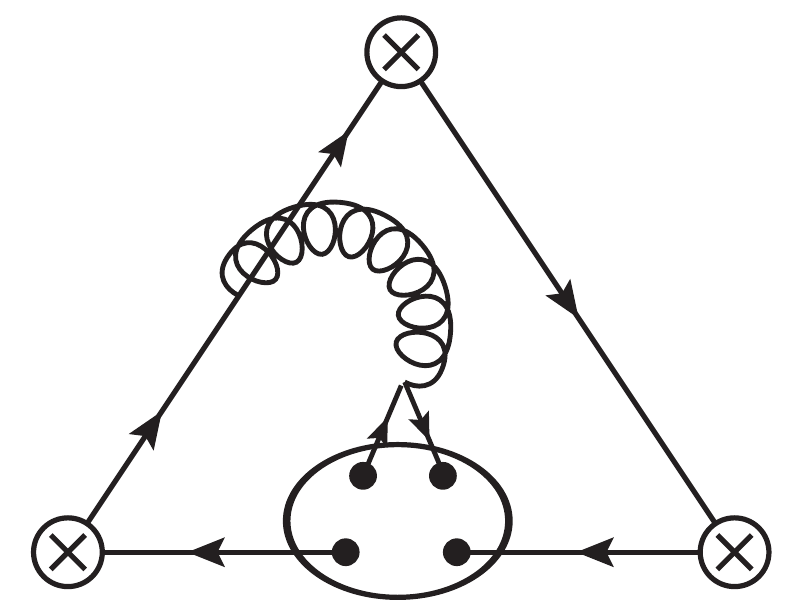}
        \caption{}
        \label{fig:fig_four-quark_condensate_propagation_3_a}
    \end{subfigure}
    \begin{subfigure}[t]{0.27\textwidth}
        \hspace{-1pt}\includegraphics[width=1\textwidth]{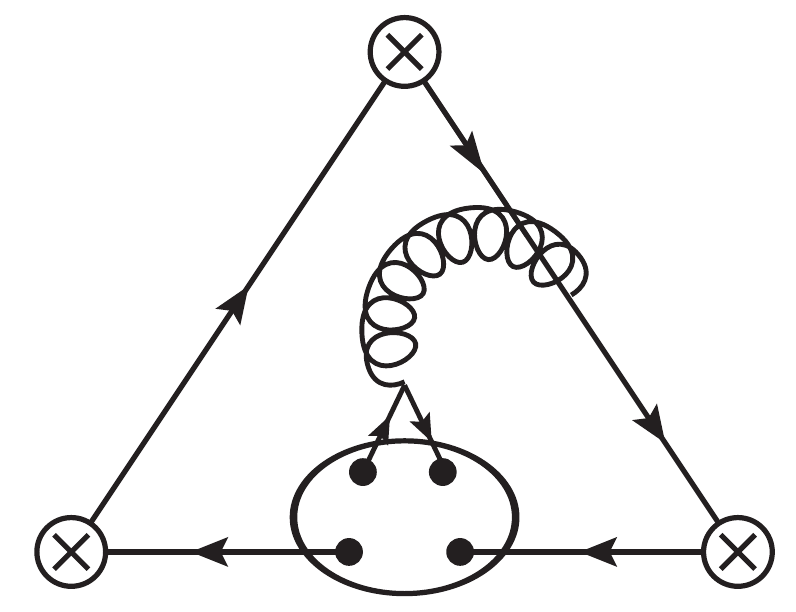}
        \caption{}
        \label{fig:fig_four-quark_condensate_propagation_3_b}
    \end{subfigure}
    \caption{Feynman diagrams of the contributions of the four-quark condensate to the three-point Green functions due to the effective propagation of the nonlocal quark condensate \eqref{eq:4q_propagace_qAq-kap7}.}
    \label{fig:fig_four-quark_condensate_propagation_3}
\end{figure}

The third class of contributions of the four-quark condensate to the three-point Green functions is given by the diagrams also similar to the ones of the quark condensate, however, here the soft gluon is inserted between the currents, as shown at Fig.~\ref{fig:fig_four-quark_condensate_propagation_3}. The contribution is then obtained effectively, using the expansion of the nonlocal quark-gluon condensate \eqref{eq:4q_propagace_qAq}.

Then, the contribution of the relevant diagrams is given as
\begin{equation}
g_{s}\langle\overline{q}_{i,\alpha}^{A}(x)\mathcal{A}_{\mu}^{a}(y)q_{k,\beta}^{B}(z)\rangle\ni\frac{\pi\alpha_{s}\langle\overline{q}q\rangle^{2}}{2^{3}\cdot 3^{5}}\delta^{AB}(T^{a})_{\beta\alpha}\big[G_{\mu}^{\langle\overline{q}\mathcal{A}q\rangle}(x,u,y)\big]_{ki}\,,\label{eq:4q_propagace_qAq-kap7}
\end{equation}
with $G_{\mu}^{\langle\overline{q}\mathcal{A}q\rangle}(x,u,y)$ given as \eqref{eq:4q_propagace_qAq-function_contracted}.

After performing familiar manipulations, the graphs at Fig.~\ref{fig:fig_four-quark_condensate_propagation_3} contribute as
\begin{align}
\big[\Pi_{\mathcal{O}_{1}^{a}\mathcal{O}_{2}^{b}\mathcal{O}_{3}^{c}}^{\langle\overline{q}\mathcal{A}q\rangle\rightarrow\langle\overline{q}q\rangle^{2}}&(x,y,z)\big]_{(a)}=-\frac{i\pi\alpha_{s}\langle\overline{q}q\rangle^{2}}{2\cdot 3^{5}}\mathrm{Tr}\big[T^{c}T^{b}T^{a}\big]\times\label{eq:four-quark_condensate_propagation_2_diag_1}\\
&\hspace{-40pt}\times\int\mathrm{d}^{4}u\,\mathrm{Tr}\Big[G_{\alpha}^{\langle\overline{q}\mathcal{A}q\rangle}(z,u,x)\Gamma_{3}S_{0}(z,y)\Gamma_{2}S_{0}(y,u)\gamma_{\alpha}S_{0}(u,x)\Gamma_{1}\Big]+\,(5\,\mathrm{permutations})\,,\nonumber\\
\big[\Pi_{\mathcal{O}_{1}^{a}\mathcal{O}_{2}^{b}\mathcal{O}_{3}^{c}}^{\langle\overline{q}\mathcal{A}q\rangle\rightarrow\langle\overline{q}q\rangle^{2}}&(x,y,z)\big]_{(b)}=-\frac{i\pi\alpha_{s}\langle\overline{q}q\rangle^{2}}{2\cdot 3^{5}}\mathrm{Tr}\big[T^{c}T^{b}T^{a}\big]\times\label{eq:four-quark_condensate_propagation_2_diag_2}\\
&\hspace{-40pt}\times\int\mathrm{d}^{4}u\,\mathrm{Tr}\Big[G_{\alpha}^{\langle\overline{q}\mathcal{A}q\rangle}(z,u,x)\Gamma_{3}S_{0}(z,u)\gamma_{\alpha}S_{0}(u,y)\Gamma_{2}S_{0}(y,x)\Gamma_{1}\Big]+\,(5\,\mathrm{permutations})\,,\nonumber
\end{align}
where we are required to integrate over the coordinate of the soft gluon.

Integrating over the soft gluon coordinate is by far the most complex task of all the evaluations presented in this thesis. For a detailed discussion of the calculation itself, see Appendix \ref{ssec:four-quark_condensate}, where the evaluation is described thoroughly for the case of the two-point correlators.

Similarly to the previous section, after performing the necessary integration in \eqref{eq:four-quark_condensate_propagation_2_diag_1}-\eqref{eq:four-quark_condensate_propagation_2_diag_2}, one can make sure that it is indeed safe to set $z=0$, which simplifies the subsequent Fourier transform.

\subsection{Results}
Finally, the results for the formfactors of the contribution of the four-quark condensate to the three-point Green functions are as follows:
\begin{align}
\mathcal{F}_{ASP}^{\langle\overline{q}q\rangle^{2}}(p^{2},q^{2};r^{2})=&\quad\,\frac{8i\pi\alpha_{s}\langle\overline{q}q\rangle^{2}}{729p^{4}q^{4}r^{4}}\times\label{eq:asp-four-quark}\\
&\hspace{-50pt}\times\Big[4p^{6}+12p^{4}(q^{2}-2r^{2})-9p^{2}q^{2}r^{2}-(q^{2}+r^{2})(7q^{2}r^{2}+2q^{4}+2r^{4}\Big]\,,\nonumber\\
\mathcal{F}_{VSS}^{\langle\overline{q}q\rangle^{2}}(p^{2},q^{2};r^{2})=&\quad\,\frac{8i\pi\alpha_{s}\langle\overline{q}q\rangle^{2}}{729p^{4}q^{4}r^{4}}\times\label{eq:vss-four-quark}\\
&\hspace{-50pt}\times\Big[4p^{6}-(q^{2}+r^{2})(24p^{4}-11q^{2}r^{2}+2q^{4}+2r^{4})+9p^{2}q^{2}r^{2}\Big]\,,\nonumber\\
\mathcal{F}_{VPP}^{\langle\overline{q}q\rangle^{2}}(p^{2},q^{2};r^{2})=&\quad\,\frac{8i\pi\alpha_{s}\langle\overline{q}q\rangle^{2}}{729p^{4}q^{4}r^{4}}\times\label{eq:vpp-four-quark}\\
&\hspace{-50pt}\times\Big[4p^{6}+(q^{2}+r^{2})(12p^{4}+11q^{2}r^{2}-2q^{4}-2r^{4})+9p^{2}q^{2}r^{2}\Big]\,,\nonumber\\
\mathcal{F}_{VVA}^{\langle\overline{q}q\rangle^{2}}(p^{2},q^{2};r^{2})=&\quad\,\frac{4i\pi\alpha_{s}\langle\overline{q}q\rangle^{2}}{729p^{6}q^{6}r^{4}}\times\label{eq:vva-four-quark_1}\\
&\hspace{-50pt}\times\Big[-9r^{4}(p^{4}+q^{4})+(p^{2}+q^{2})r^{2}\Big(2r^{4}-9(p^{2}-q^{2})^{2}\Big)\nonumber\\
&\hspace{-32pt}-2(2p^{6}q^{2}+9p^{4}q^{4}+2p^{2}q^{6}-p^{8}-q^{8})\Big]\,,\nonumber\\
\mathcal{G}_{VVA}^{\langle\overline{q}q\rangle^{2}}(p^{2},q^{2};r^{2})=&-\frac{4i\pi\alpha_{s}\langle\overline{q}q\rangle^{2}}{729p^{6}q^{6}r^{4}}\times\label{eq:vva-four-quark_2}\\
&\hspace{-50pt}\times(p^{2}-q^{2})(p^{2}+q^{2}+r^{2})\Big[(p^{2}+q^{2})(11r^{2}-2p^{2}-2q^{2})-2r^{4}\Big]\,,\nonumber\\
\mathcal{H}_{VVA}^{\langle\overline{q}q\rangle^{2}}(p^{2},q^{2};r^{2})=&\quad\,\frac{4i\pi\alpha_{s}\langle\overline{q}q\rangle^{2}}{729p^{6}q^{6}r^{4}}\times\label{eq:vva-four-quark_3}\\
&\hspace{-50pt}\times(p^{2}-q^{2})(p^{2}+q^{2}+r^{2})\Big[(p^{2}+q^{2})(11r^{2}-2p^{2}-2q^{2})-2r^{4}\Big]\,,\nonumber\\
\mathcal{F}_{AAA}^{\langle\overline{q}q\rangle^{2}}(p^{2},q^{2};r^{2})=&\quad\,0\,,\label{eq:aaa-four-quark_1}\\
\mathcal{G}_{AAA}^{\langle\overline{q}q\rangle^{2}}(p^{2},q^{2};r^{2})=&\quad\,\frac{2i\pi\alpha_{s}\langle\overline{q}q\rangle^{2}}{243p^{4}q^{4}r^{4}}\frac{(p^{2}-q^{2})(p^{2}-r^{2})(q^{2}-r^{2})(p^{2}+q^{2}+r^{2})}{\lambda_{K}+p^{2}(q^{2}+r^{2})+q^{2}r^{2}}\,,\label{eq:aaa-four-quark_2}\\
\mathcal{H}_{AAA}^{\langle\overline{q}q\rangle^{2}}(p^{2},q^{2};r^{2})=&\quad\,\frac{2i\pi\alpha_{s}\langle\overline{q}q\rangle^{2}}{243p^{4}q^{4}r^{4}}\frac{1}{\lambda_{K}+p^{2}(q^{2}+r^{2})+q^{2}r^{2}}\times\label{eq:aaa-four-quark_3}\\
&\hspace{-50pt}\times\Big[-18p^{4}q^{2}r^{2}+p^{2}(q^{2}+r^{2})\Big(7(p^{4}+q^{4}+r^{4})-25q^{2}r^{2}\Big)\nonumber\\
&\hspace{-32pt}+4(p^{8}+q^{8}+r^{8})+7q^{2}r^{2}(q^{4}+r^{4})\Big]\,,\nonumber\\
\mathcal{F}_{AAV}^{\langle\overline{q}q\rangle^{2}}(p^{2},q^{2};r^{2})=&-\frac{4i\pi\alpha_{s}\langle\overline{q}q\rangle^{2}}{729p^{6}q^{6}r^{4}}\times\label{eq:aav-four-quark_1}\\
&\hspace{-50pt}\times\Big[-2r^{4}(p^{2}+q^{2})\Big(13(p^{4}+q^{4})-r^{4}-14p^{2}q^{2}\Big)+r^{6}(2p^{2}q^{2}+11p^{4}+11q^{4})\nonumber\\
&\hspace{-32pt}+(p^{2}-q^{2})^{2}\Big(r^{2}(20p^{2}q^{2}+11p^{4}+11q^{4})+2(p^{6}+q^{6})\Big)\Big]\,,\nonumber\\
\mathcal{G}_{AAV}^{\langle\overline{q}q\rangle^{2}}(p^{2},q^{2};r^{2})=&\quad\,\frac{8i\pi\alpha_{s}\langle\overline{q}q\rangle^{2}}{729p^{6}q^{6}r^{6}}\times\label{eq:aav-four-quark_2}\\
&\hspace{-50pt}\times\Big[-r^{6}(2p^{2}q^{2}-9p^{4}-9q^{4})+2r^{8}(p^{2}+q^{2})+9r^{4}(p^{6}+q^{6})\nonumber\\
&\hspace{-32pt}+2r^{2}(p^{2}-q^{2})^{2}(p^{2}q^{2}+p^{4}+q^{4})+p^{2}q^{2}(p^{2}+q^{2})(-11p^{2}q^{2}+2p^{4}+2q^{4})\Big]\,,\nonumber\\
\mathcal{H}_{AAV}^{\langle\overline{q}q\rangle^{2}}(p^{2},q^{2};r^{2})=&-\frac{4i\pi\alpha_{s}\langle\overline{q}q\rangle^{2}}{729p^{4}q^{6}r^{6}}\times\label{eq:aav-four-quark_3}\\
&\hspace{-50pt}\times\Big[-p^{2}(2q^{6}r^{2}+2q^{2}r^{6}+7q^{8}-11r^{8})+2p^{8}(q^{2}+r^{2})+p^{6}(2q^{2}r^{2}-7q^{4}+11r^{4})\nonumber\\
&\hspace{-32pt}+2p^{4}(q^{4}r^{2}+q^{2}r^{4}+5q^{6}-13r^{6})+2(q^{2}-r^{2})^{2}(q^{6}+r^{6})\Big]\,,\nonumber\\
\mathcal{F}_{VVV}^{\langle\overline{q}q\rangle^{2}}(p^{2},q^{2};r^{2})=&-\frac{4i\pi\alpha_{s}\langle\overline{q}q\rangle^{2}}{729p^{6}q^{6}r^{4}}\times\label{eq:vvv-four-quark_1}\\
&\hspace{-50pt}\times\Big[-r^{2}(p^{2}-q^{2})^{2}(16p^{2}q^{2}+7p^{4}+7q^{4})+r^{6}(2p^{2}q^{2}-7p^{4}-7q^{4})\nonumber\\
&\hspace{-32pt}+2r^{4}(p^{2}+q^{2})(r^{4}-4p^{2}q^{2}+5p^{4}+5q^{4})+2(p^{2}-q^{2})^{2}(p^{6}+q^{6})\Big]\,,\nonumber\\
\mathcal{G}_{VVV}^{\langle\overline{q}q\rangle^{2}}(p^{2},q^{2};r^{2})=&\quad\,\frac{8i\pi\alpha_{s}\langle\overline{q}q\rangle^{2}}{729p^{6}q^{6}r^{6}}\times\label{eq:vvv-four-quark_2}\\
&\hspace{-50pt}\times\Big[-9p^{4}(q^{6}+r^{6})+2p^{8}(q^{2}+r^{2})+2p^{2}(q^{2}-r^{2})^{2}(q^{2}r^{2}+q^{4}+r^{4})\nonumber\\
&\hspace{-32pt}-p^{6}(2q^{2}r^{2}+9q^{4}+9r^{4})+q^{2}r^{2}(q^{2}+r^{2})(-11q^{2}r^{2}+2q^{4}+2r^{4})\Big]\,.\nonumber
\end{align}

%%%%%%%%%%%%%%%%%%%%%%%%%%%%%%%%%%%%%%%%%%%%%%%%%%%%%%%%%%%%%%%%%%%%%%%%%%%%%%%%%%%%%%%%%%%%%%%%%%%%%%%%%
%%%%%%%%%%%%%%%%%%%%%%%%%%%%%%%%%%%%%%%%%%%%%%%%%%%%%%%%%%%%%%%%%%%%%%%%%%%%%%%%%%%%%%%%%%%%%%%%%%%%%%%%%
%%% Section: Summary
%%%%%%%%%%%%%%%%%%%%%%%%%%%%%%%%%%%%%%%%%%%%%%%%%%%%%%%%%%%%%%%%%%%%%%%%%%%%%%%%%%%%%%%%%%%%%%%%%%%%%%%%%
%%%%%%%%%%%%%%%%%%%%%%%%%%%%%%%%%%%%%%%%%%%%%%%%%%%%%%%%%%%%%%%%%%%%%%%%%%%%%%%%%%%%%%%%%%%%%%%%%%%%%%%%%

\section{Summary}\label{sec:conclusion}
In this paper we have  presented the complete survey of leading order  contributions of the  QCD condensates up to dimension six to all relevant three-point Green functions of the chiral currents and densities within the framework of the operator product expansion. 

 We give a detailed derivation of the formulas for an expansion of the nonlocal quark and quark-gluon condensates in terms of local QCD condensates in a general case, where all the coordinates of the quark and gluon fields are nonzero (see eq. \eqref{eq:quark-condensate-propagation} and \eqref{eq:4q_propagace_qAq} and Appendix \ref{sec:derivation}). We hope that our complete list of corresponding formulas done in unified parameterization and convention will bring clarification and unification of equivalent expressions scattered in the literature. 

After a short review of well-known perturbative  contributions, which we recalculated independently, we studied higher-order QCD condensates. We present the results for the gluon and four-quark condensates contributions to the $\langle VVA\rangle$, $\langle AAA\rangle$, $\langle VVV\rangle$, $\langle ASP\rangle$, $\langle AAV\rangle$, $\langle VSS\rangle$ and $\langle VPP\rangle$ correlators and the results for the contribution of the quark and quark-gluon condensates to the $\langle SSS\rangle$, $\langle SPP\rangle$, $\langle VVP\rangle$, $\langle AAP\rangle$, $\langle VAS\rangle$, $\langle VVS\rangle$, $\langle AAS\rangle$ and $\langle VAP\rangle$ Green functions.

To our knowledge, the complete results for the gluon, quark-gluon and four-quark condensates have not yet been presented in the literature. We believe that our work can be useful for further theoretical and phenomenological studies of the Green functions. 

%%%%%%%%%%%%%%%%%%%%%%%%%%%%%%%%%%%%%%%%%%%%%%%%%%%%%%%%%%%%%%%%%%%%%%%%%%%%%%%%%%%%%%%%%%%%%%%%%%%%%%%%%
%%%%%%%%%%%%%%%%%%%%%%%%%%%%%%%%%%%%%%%%%%%%%%%%%%%%%%%%%%%%%%%%%%%%%%%%%%%%%%%%%%%%%%%%%%%%%%%%%%%%%%%%%
%%% Appendix
%%%%%%%%%%%%%%%%%%%%%%%%%%%%%%%%%%%%%%%%%%%%%%%%%%%%%%%%%%%%%%%%%%%%%%%%%%%%%%%%%%%%%%%%%%%%%%%%%%%%%%%%%
%%%%%%%%%%%%%%%%%%%%%%%%%%%%%%%%%%%%%%%%%%%%%%%%%%%%%%%%%%%%%%%%%%%%%%%%%%%%%%%%%%%%%%%%%%%%%%%%%%%%%%%%%

\appendix

%%%%%%%%%%%%%%%%%%%%%%%%%%%%%%%%%%%%%%%%%%%%%%%%%%%%%%%%%%%%%%%%%%%%%%%%%%%%%%%%%%%%%%%%%%%%%%%%%%%%%%%%%
%%%%%%%%%%%%%%%%%%%%%%%%%%%%%%%%%%%%%%%%%%%%%%%%%%%%%%%%%%%%%%%%%%%%%%%%%%%%%%%%%%%%%%%%%%%%%%%%%%%%%%%%%
%%% Section: Fourier Transform
%%%%%%%%%%%%%%%%%%%%%%%%%%%%%%%%%%%%%%%%%%%%%%%%%%%%%%%%%%%%%%%%%%%%%%%%%%%%%%%%%%%%%%%%%%%%%%%%%%%%%%%%%
%%%%%%%%%%%%%%%%%%%%%%%%%%%%%%%%%%%%%%%%%%%%%%%%%%%%%%%%%%%%%%%%%%%%%%%%%%%%%%%%%%%%%%%%%%%%%%%%%%%%%%%%%

\section{Fourier Transform}\label{sec:fourier_transform}
In this appendix we present useful formulas for the Fourier transform, that has been used in our calculations. Let us only remind the reader that we use the convention \eqref{eq:FT-1}-\eqref{eq:FT-2}. We have
\begin{alignat}{4}
&F_{1}(x)&&=\frac{1}{x^{2}}\,&&\longleftrightarrow\,\widetilde{F}_{1}(p)&&=-\frac{4i\pi^{2}}{p^{2}}\,,\label{eq:fourier-2-1}\\
&\big[F_{2}(x)\big]_{\mu}&&=\frac{x_{\mu}}{x^{2}}\,&&\longleftrightarrow\,\big[\widetilde{F}_{2}(p)\big]_{\mu}&&=-\frac{8\pi^{2}}{p^{4}}p_{\mu}\,,\label{eq:fourier-2-2}\\
&\big[F_{3}(x)\big]_{\mu\nu}&&=\frac{x_{\mu}x_{\nu}}{x^{2}}\,&&\longleftrightarrow\,\big[\widetilde{F}_{3}(p)\big]_{\mu\nu}&&=-\frac{8i\pi^{2}}{p^{4}}\bigg(g_{\mu\nu}-\frac{4p_{\mu}p_{\nu}}{p^{2}}\bigg)\,,\label{eq:fourier-2-3}\\
&F_{4}(x)&&=\frac{1}{x^{4}}\,&&\longleftrightarrow\,\widetilde{F}_{4}(p)&&=\quad i\pi^{2}\log(-p^{2})+\mathrm{const.}\,,\label{eq:fourier-4-1}\\
&\big[F_{5}(x)\big]_{\mu}&&=\frac{x_{\mu}}{x^{4}}\,&&\longleftrightarrow\,\big[\widetilde{F}_{5}(p)\big]_{\mu}&&=-\frac{2\pi^{2}}{p^{2}}p_{\mu}\,,\label{eq:fourier-4-2}\\
&\big[F_{6}(x)\big]_{\mu\nu}&&=\frac{x_{\mu}x_{\nu}}{x^{4}}\,&&\longleftrightarrow\,\big[\widetilde{F}_{6}(p)\big]_{\mu\nu}&&=-\frac{2i\pi^{2}}{p^{2}}\bigg(g_{\mu\nu}-\frac{2p_{\mu}p_{\nu}}{p^{2}}\bigg)\,,\label{eq:fourier-4-3}\\
&\big[F_{7}(x)\big]_{\mu\nu\rho}&&=\frac{x_{\mu}x_{\nu}x_{\rho}}{x^{4}}\,&&\longleftrightarrow\,\big[\widetilde{F}_{7}(p)\big]_{\mu\nu\rho}&&=-\frac{4\pi^{2}}{p^{4}}\bigg(g_{\mu\nu}p_{\rho}+g_{\mu\rho}p_{\nu}+g_{\nu\rho}p_{\mu}-\frac{4p_{\mu}p_{\nu}p_{\rho}}{p^{2}}\bigg)\,.\label{eq:fourier-4-4}
\end{alignat}

For evaluation of the contribution of the quark-gluon condensate the following formulas, apart from \eqref{eq:fourier-2-1}-\eqref{eq:fourier-2-2}, are useful  in order to make the manipulations as compact as possible:
\begin{alignat}{5}
&\big[F_{8}(x)\big]_{\mu\nu}&&=-&&\frac{i}{4\pi^{2}}\bigg(\frac{x_{\mu}x_{\nu}}{x^{4}}-\frac{g_{\mu\nu}}{2x^{2}}\bigg)\,&&\longleftrightarrow\,\big[\widetilde{F}_{8}(p)\big]_{\mu\nu}&&=\frac{p_{\mu}p_{\nu}}{p^{4}}\,,\label{eq:fourier-8}\\
&\big[F_{9}(x)\big]_{\mu\nu\rho}&&=&&\frac{1}{8\pi^{2}}\bigg(\frac{x_{\mu}x_{\nu}x_{\rho}}{x^{4}}-\frac{g_{\mu\nu}x_{\rho}+g_{\mu\rho}x_{\nu}+g_{\nu\rho}x_{\mu}}{2x^{2}}\bigg)\,&&\longleftrightarrow\,\big[\widetilde{F}_{9}(p)\big]_{\mu\nu\rho}&&=\frac{2p_{\mu}p_{\nu}p_{\rho}}{p^{6}}\,.\label{eq:fourier-9}
\end{alignat}

In order to evaluate the contribution of the four-quark condensate, it is necessary to introduce another Fourier transforms. In detail, the relations below arise due to the integration over the coordinate of the soft gluon in the case of the effective propagation of the non-local quark-gluon condensate. We have:
\begin{alignat}{5}
&F_{10}(x)&&=&&\log(-x^{2})\,&&\longleftrightarrow\,\widetilde{F}_{10}(p)&&=\frac{16i\pi^{2}}{p^{4}}\,,\label{eq:fourier-10}\\
&\big[F_{11}(x)\big]_{\mu\nu}&&=-&&\frac{i}{16\pi^{2}}\bigg(\frac{x_{\mu}x_{\nu}}{x^{2}}+\frac{1}{2}g_{\mu\nu}\log(-x^{2})\bigg)\,&&\longleftrightarrow\,\big[\widetilde{F}_{11}(p)\big]_{\mu\nu}&&=\frac{2p_{\mu}p_{\nu}}{p^{6}}\,,\label{eq:fourier-11}\\
&\big[F_{12}(x)\big]_{\mu\nu\rho\sigma}&& &&\,&&\longleftrightarrow\,\big[\widetilde{F}_{12}(p)\big]_{\mu\nu\rho\sigma}&&=\frac{8p_{\mu}p_{\nu}p_{\rho}p_{\sigma}}{p^{8}}\,,\label{eq:fourier-12}
\end{alignat}
where
\begin{align}
\big[F_{12}(x)\big]_{\mu\nu\rho\sigma}=\frac{i}{12\pi^{2}}\bigg[&-8i\pi^{2}\Big([F_{11}(x)]_{\mu\nu}g_{\rho\sigma}+[F_{11}(x)]_{\mu\rho}g_{\nu\sigma}+[F_{11}(x)]_{\nu\rho}g_{\mu\sigma}\\
&+[F_{11}(x)]_{\mu\sigma}g_{\nu\rho}+[F_{11}(x)]_{\nu\sigma}g_{\mu\rho}+[F_{11}(x)]_{\rho\sigma}g_{\mu\nu}\Big)\nonumber\\
&+\frac{1}{4}F_{10}(x)(g_{\mu\sigma}g_{\nu\rho}+g_{\mu\rho}g_{\nu\sigma}+g_{\mu\nu}g_{\rho\sigma})+\frac{x_{\mu}x_{\nu}x_{\rho}x_{\sigma}}{x^{4}}\bigg]\,.\nonumber
\end{align}

%%%%%%%%%%%%%%%%%%%%%%%%%%%%%%%%%%%%%%%%%%%%%%%%%%%%%%%%%%%%%%%%%%%%%%%%%%%%%%%%%%%%%%%%%%%%%%%%%%%%%%%%%
%%%%%%%%%%%%%%%%%%%%%%%%%%%%%%%%%%%%%%%%%%%%%%%%%%%%%%%%%%%%%%%%%%%%%%%%%%%%%%%%%%%%%%%%%%%%%%%%%%%%%%%%%
%%% Section: Derivation of Propagation Formulas
%%%%%%%%%%%%%%%%%%%%%%%%%%%%%%%%%%%%%%%%%%%%%%%%%%%%%%%%%%%%%%%%%%%%%%%%%%%%%%%%%%%%%%%%%%%%%%%%%%%%%%%%%
%%%%%%%%%%%%%%%%%%%%%%%%%%%%%%%%%%%%%%%%%%%%%%%%%%%%%%%%%%%%%%%%%%%%%%%%%%%%%%%%%%%%%%%%%%%%%%%%%%%%%%%%%

\section{Derivation of Propagation Formulas}\label{sec:derivation}
 In this appendix we present a complete derivation of the propagation formulas \eqref{eq:quark-condensate-propagation}, \eqref{eq:propagation_gluon_condensate} and \eqref{eq:4q_propagace_qAq}. It is important to note that derivation of some parts of these formulas can be found in the literature already, however, we present this derivation here with the intent of having all the procedures explained here in detail at one place, with all formal necessities and with every aspect belonging to the fact that we take into account the flavor indices.

A detailed comparison of our propagation formulas with the known results presented already in the literature has been discussed in Subsection \ref{ssec:comparison}. Therefore, in what follows, we will not point out differences between this work and the work of other authors. However, we only shortly mention some references, where the derivation of propagation formulas can be also found.

A comprehensive derivation of the propagations giving arise to the quark and the quark-gluon condensates is presented in \cite{Elias:1987ac}, where the authors work beyond the chiral limit. Further, some fragments of the derivation of the propagations of the four-quark condensate can be found in \cite{Ioffe:1982qb}. Both  these issues are also  discussed in detail in \cite{Pascual:1984zb}.

Although the authors of \cite{Ioffe:1982qb} do not present the direct derivation of the propagation formulas, they show the way of how to obtain the contributions of the four-quark condensate from the respective vacuum expectation values of the relevant operators. In detail, our derivation of \eqref{eq:four-quark_condensate_derivation_19}, \eqref{eq:4quark_qAq_derivation_9} and \eqref{eq:4quark_qAq_derivation_20} can be compared with the one of these authors, see eq.~21 at page no.~386, eq.~24 at page no.~387 and eq.~23 at page no.~386 therein, respectively.

The same derivation as in \cite{Ioffe:1982qb} is presented also in the book \cite{Ioffe:2010zz}.

\subsection{Preliminaries}
In order to provide as complete derivation of the propagational formulas as possible, we first recapitulate some basic facts that we will build the procedure upon. Among these preliminary facts we include the equations of motion and minimal Lorentz basis that nonlocal QCD condensates are made of.

\subsubsection{Equations of Motion}
In the chiral limit, the QCD Lagrangian reads
\begin{equation}
\mathcal{L}_{\mathrm{QCD}}=i\overline{q}\slashed{\nabla}q-\frac{1}{4}G_{\mu\nu}^{a}G^{\mu\nu,a}\,,\label{eq:lagrangian_QCD}
\end{equation}
with the covariant derivative and the gluon field strength tensor given as \eqref{eq:covariant_derivative} and \eqref{eq:gluon_field_strength_tensor_a}, respectively. Having the Lagrangian \eqref{eq:lagrangian_QCD}, the equations of motion can be obtained.

For the gluon field, we have the equation of motion in the form
\begin{equation}
(D^{\mu}G_{\mu\nu})^{a}=g_{s}\overline{q}\gamma_{\nu}T^{a}q\,,\label{eq:qcd_eom_3}
\end{equation}
where the covariant derivative in the adjoint representation is
\begin{equation}
(D_{\mu})^{ab}=\partial_{\mu}\delta^{ab}+g_{s}f^{abc}\mathcal{A}_{\mu}^{c}\,.
\end{equation}

On the other hand, the equations of motion for the quark fields are in the form of the Dirac equations:
\begin{align}
\slashed{\nabla}q&=0\,,\label{eq:qcd_eom_1}\\
\overline{q}\overleftarrow{\slashed{\nabla}}&=0\,,\label{eq:qcd_eom_2}
\end{align}
with $\slashed{\nabla}$ acting to the right, as usual, and $\overleftarrow{\slashed{\nabla}}$ acting to the left, as indicated.

To avoid confusion, we clarify how the covariant derivative in \eqref{eq:qcd_eom_2} should be understood. Apparently, making $\overline{q}$ out of $q$ necessarily leads to the Hermitian conjugation of the covariant derivative, which then acts to the left. To highlight this, we have denoted
\begin{equation}
\overleftarrow{\nabla}_{\mu}\equiv\nabla_{\mu}^{\dagger}\,.\label{eq:covariant_derivative_left}
\end{equation}
Then, the covariant derivative in \eqref{eq:qcd_eom_2} is taken to be
\begin{equation}
\overleftarrow{\slashed{\nabla}}\equiv\nabla_{\mu}^{\dagger}\gamma^{\mu}\,.\label{eq:covariant_derivative_left_slashed}
\end{equation}

\subsubsection{Minimal Lorentz Structure of Nonlocal QCD Condensates}
In the forthcoming sections, we will need to construct structures that carry either two or three Lorentz indices. These can be made of appropriate combinations of the metric, Levi-Civita tensor and Dirac matrices, such that the parity is conserved.

The requirement of parity conservation forces us to accompany the Levi-Civita tensor always with the $\gamma_{5}$ matrix, and vice versa. Then, one can write down the following two- and three-index structures (for now, we do not explicitly write down the respective permutations of the indices):
\begin{itemize}
\item $g^{\mu\nu}$, $\sigma^{\mu\nu}$, $\gamma^{\mu}\gamma^{\nu}$, $\varepsilon^{\mu\nu\alpha\beta}\gamma_{\alpha}\gamma_{\beta}\gamma^{5}$, $\varepsilon^{\mu\nu\alpha\beta}\sigma_{\alpha\beta}\gamma^{5}$, \ldots
\item $g^{\mu\nu}\gamma^{\rho}$, $\sigma^{\mu\nu}\gamma^{\rho}$, $\gamma^{\mu}\gamma^{\nu}\gamma^{\rho}$, $\varepsilon^{\mu\nu\rho\alpha}\gamma_{\alpha}\gamma^{5}$, \ldots,
\end{itemize}
where the ellipsis stand for any other terms such that the already shown structures would be accompanied by additional tensors or matrices so that the indices would be contracted properly.

However, as it turns out, not every term of those above is necessary for constructing the minimal Lorentz structure of the nonlocal QCD condensates. To this end, let us remind the following property. The basis of all $4\times 4$ matrices is determined by 16 matrices: $1$, $\gamma^{\mu}$, $\gamma^{5}$, $\gamma^{\mu}\gamma^{5}$ and $\sigma^{\mu\nu}$. Therefore, one can decompose any $4\times 4$ matrix $X$ easily as
\begin{equation}
X=\frac{1}{4}\mathrm{Tr}\big[X\big]+\frac{1}{4}\mathrm{Tr}\big[X\gamma^{\mu}\big]\gamma_{\mu}+\frac{1}{4}\mathrm{Tr}\big[X\gamma^{5}\big]\gamma^{5}-\frac{1}{4}\mathrm{Tr}\big[X\gamma^{\mu}\gamma^{5}\big]\gamma_{\mu}\gamma^{5}+\frac{1}{8}\mathrm{Tr}\big[X\sigma^{\mu\nu}\big]\sigma_{\mu\nu}\,.\label{eq:decomposition_4x4}
\end{equation}

Taking the decomposition formula \eqref{eq:decomposition_4x4} into account, we are allowed to eliminate some of the structures shown above,\footnote{To be thorough, we show the results that some of the structures above can be rewritten to. We have $\gamma^{\mu}\gamma^{\nu}=g^{\mu\nu}-i\sigma^{\mu\nu}$, $\varepsilon^{\mu\nu\alpha\beta}\gamma_{\alpha}\gamma_{\beta}\gamma^{5}=-2\sigma^{\mu\nu}$, $\varepsilon^{\mu\nu\alpha\beta}g_{\alpha\beta}\gamma^{5}=0$, $\varepsilon^{\mu\nu\alpha\beta}\sigma_{\alpha\beta}\gamma^{5}=-2i\sigma^{\mu\nu}$, $\gamma^{\mu}\gamma^{\nu}\gamma^{\rho}=g^{\nu\rho}\gamma^{\mu}-g^{\mu\rho}\gamma^{\nu}+g^{\mu\nu}\gamma^{\rho}+i\varepsilon^{\mu\nu\rho\alpha}\gamma_{\alpha}\gamma^{5}$ and $\sigma^{\mu\nu}\gamma^{\rho}=i(g^{\nu\rho}\gamma^{\mu}-g^{\mu\rho}\gamma^{\nu})-\varepsilon^{\mu\nu\rho\alpha}\gamma_{\alpha}\gamma^{5}$. In obtaining these relations we have made a use of a formula $\varepsilon^{\mu\nu\alpha\beta}\varepsilon_{\rho\sigma\alpha\beta}=-2(g_{\rho}^{\mu}g_{\sigma}^{\nu}-g_{\sigma}^{\mu}g_{\rho}^{\nu})$.\label{footnote_decompositions}} which leads us to the conclusion, that only the tensors
\begin{equation}
(g^{\mu\nu}\,,\,\sigma^{\mu\nu})\label{eq:2_index_basis}
\end{equation}
and
\begin{equation}
(g^{\mu\nu}\gamma^{\rho}\,,\,g^{\mu\rho}\gamma^{\nu}\,,\,g^{\nu\rho}\gamma^{\mu}\,,\,\varepsilon^{\mu\nu\rho\alpha}\gamma_{\alpha}\gamma^{5})\label{eq:3_index_basis}
\end{equation}
are needed to build the Lorentz structure with two or three indices, respectively.

\subsubsection{Expansion of Quark and Gluon Fields}
To be able to provide the propagation of the local QCD condensates through the nonlocal ones, we need the expansions \eqref{eq:fock-schwinger-quark-field}-\eqref{eq:fock-schwinger-gluon-field} of the quark and gluon fields within the Fock-Schwinger gauge. Let us rewrite these formulas here, including all the necessary indices, such as the spinor ($i=1\ldots 4$), color ($\alpha=1\ldots 3$) and the flavor ($A=1\ldots 3$) ones.

The expansion of the quark field \eqref{eq:fock-schwinger-quark-field} reads
\begin{align}
q_{i,\alpha}^{A}(x)&=q_{i,\alpha}^{A}+x^{\mu}\big(\nabla_{\mu}q_{i,\alpha}^{A}\big)+\frac{1}{2}x^{\mu}x^{\nu}\big(\nabla_{\mu}\nabla_{\nu}q_{i,\alpha}^{A}\big)+\frac{1}{6}x^{\mu}x^{\nu}x^{\rho}\big(\nabla_{\mu}\nabla_{\nu}\nabla_{\rho}q_{i,\alpha}^{A}\big)+\ldots\,,\label{eq:fock-schinger_expansion_1}
\end{align}
where we have intentionally omitted the explicit indication of the fact that the expanded fields on the right-hand side are evaluated at the origin, i.e. $q_{i,\alpha}^{A}\equiv q_{i,\alpha}^{A}(0)$. Moreover, any field without the designation of the space-time coordinate is considered to be evaluated at zero from now on.

Similarly, the expansion of the Dirac-conjugated quark field is
\begin{align}
\overline{q}_{i,\alpha}^{A}(x)&=\overline{q}_{i,\alpha}^{A}+\big(x^{\mu}\overline{q}_{i,\alpha}^{A}\overleftarrow{\nabla}_{\mu}\big)+\frac{1}{2}x^{\mu}x^{\nu}\big(\overline{q}_{i,\alpha}^{A}\overleftarrow{\nabla}_{\mu}\overleftarrow{\nabla}_{\nu}\big)+\frac{1}{6}x^{\mu}x^{\nu}x^{\rho}\big(\overline{q}_{i,\alpha}^{A}\overleftarrow{\nabla}_{\mu}\overleftarrow{\nabla}_{\nu}\overleftarrow{\nabla}_{\rho}\big)\,,\label{eq:fock-schinger_expansion_2}
\end{align}
where we have used the definition \eqref{eq:covariant_derivative_left} of the derivative acting to the left and where we have changed the order of the indices of such derivatives in the same order as in \eqref{eq:fock-schinger_expansion_1}, since the individual terms are symmetrical due to the presence of the coordinates.

Finally, we reintroduce the expansion \eqref{eq:fock-schwinger-gluon-field} of the gluon field as
\begin{align}
\mathcal{A}_{\mu}^{a}(x)=\frac{1}{2}x^{\nu}G_{\nu\mu}^{a}+\frac{1}{3}x^{\rho}x^{\nu}\big(D_{\rho}G_{\nu\mu}\big)^{a}+\ldots\,,\label{eq:fock-schinger_expansion_3}
\end{align}
where only these first two terms will be needed.

Having the relations \eqref{eq:fock-schinger_expansion_1}, \eqref{eq:fock-schinger_expansion_2} and \eqref{eq:fock-schinger_expansion_3} at our disposal, we can start with forming the nonlocal QCD condensates and then with the derivation of the individual propagation formulas.

\subsection{Derivation of Propagation Formula \eqref{eq:quark-condensate-propagation}}
We start with the propagation formula \eqref{eq:quark-condensate-propagation}, i.e. with the propagation of the local QCD condensates through the nonlocal quark condensate. Taking the formulas \eqref{eq:fock-schinger_expansion_1} and \eqref{eq:fock-schinger_expansion_2} and performing trivial manipulations, we obtain a series of terms out of which we consider futher only those terms that contain vacuum expectation values of the canonical dimensions 3, 5 and 6. In detail, we obtain the following relevant terms:
\begin{align}
    \big\langle\overline{q}_{i,\alpha}^{A}(x)q_{k,\beta}^{B}(y)\big\rangle &\begin{array}{l}\hspace{3pt} \ni \big\langle\overline{q}_{i,\alpha}^{A}q_{k,\beta}^{B}\big\rangle\vspace{2pt}\label{eq:quark_condensate_expansion_1}\\
          \end{array} \\
        & \left.\begin{array}{l}
           +\,\displaystyle\frac{1}{2}x^{\mu}x^{\nu}\big\langle\big(\overline{q}_{i,\alpha}^{A}\overleftarrow{\nabla}_{\mu}\overleftarrow{\nabla}_{\nu}\big)q_{k,\beta}^{B}\big\rangle\vspace{5pt}\\
           +\,\displaystyle x^{\mu}y^{\nu}\big\langle\big(\overline{q}_{i,\alpha}^{A}\overleftarrow{\nabla}_{\mu}\big)\big(\nabla_{\nu}q_{k,\beta}^{B}\big)\big\rangle\vspace{5pt}\\
           +\,\displaystyle\frac{1}{2}y^{\mu}y^{\nu}\big\langle\overline{q}_{i,\alpha}^{A}\big(\nabla_{\mu}\nabla_{\nu}q_{k,\beta}^{B}\big)\big\rangle\vspace{5pt}
         \end{array}\right\}\label{eq:quark_condensate_expansion_2}\\
        & \left.\begin{array}{l}
           +\,\displaystyle\frac{1}{6}x^{\mu}x^{\nu}x^{\rho}\big\langle\big(\overline{q}_{i,\alpha}^{A}\overleftarrow{\nabla}_{\mu}\overleftarrow{\nabla}_{\nu}\overleftarrow{\nabla}_{\rho}\big)q_{k,\beta}^{B}\big\rangle\vspace{5pt}\\
           +\,\displaystyle\frac{1}{2}x^{\mu}x^{\nu}y^{\rho}\big\langle\big(\overline{q}_{i,\alpha}^{A}\overleftarrow{\nabla}_{\mu}\overleftarrow{\nabla}_{\nu}\big)\big(\nabla_{\rho}q_{k,\beta}^{B}\big)\big\rangle\vspace{5pt}\\
           +\,\displaystyle\frac{1}{2}x^{\mu}y^{\nu}y^{\rho}\big\langle\big(\overline{q}_{i,\alpha}^{A}\overleftarrow{\nabla}_{\mu}\big)\big(\nabla_{\nu}\nabla_{\rho}q_{k,\beta}^{B}\big)\big\rangle\vspace{5pt}\\
           +\,\displaystyle\frac{1}{6}y^{\mu}y^{\nu}y^{\rho}\big\langle\overline{q}_{i,\alpha}^{A}\big(\nabla_{\mu}\nabla_{\nu}\nabla_{\rho}q_{k,\beta}^{B}\big)\big\rangle\,,
         \end{array}\right\}\label{eq:quark_condensate_expansion_3}
\end{align}
with \eqref{eq:quark_condensate_expansion_1}, \eqref{eq:quark_condensate_expansion_2} and \eqref{eq:quark_condensate_expansion_3} giving arise of the contributions of the local quark, quark-gluon and four-quark condensates, respectively.

\subsubsection{Propagation of Quark Condensate}
The lowest contribution of the nonlocal quark condensate is the local quark condensate. Expanding the left-hand sice of \eqref{eq:quark-condensate-propagation}, according to \eqref{eq:fock-schinger_expansion_1}-\eqref{eq:fock-schinger_expansion_2}, gives the relevant part in the form of \eqref{eq:quark_condensate_expansion_1}. Projecting out the normalization of the spinor, color and flavor structure gives us
\begin{align}
\big\langle\overline{q}_{i,\alpha}^{A}(x)q_{k,\beta}^{B}(y)\big\rangle\ni\big\langle\overline{q}_{i,\alpha}^{A}q_{k,\beta}^{B}\big\rangle=\bigg(\frac{1}{4}\delta_{ik}\bigg)\bigg(\frac{1}{3}\delta_{\alpha\beta}\bigg)\bigg(\frac{1}{3}\delta^{AB}\bigg)\langle\overline{q}q\rangle\,.
\end{align}
Therefore, the propagation formula for the quark condensate is simply
\begin{align}
\big\langle\overline{q}_{i,\alpha}^{A}(x)q_{k,\beta}^{B}(y)\big\rangle\ni\frac{1}{2^{2}\cdot 3^{2}}\langle\overline{q}q\rangle\delta_{ik}\delta_{\alpha\beta}\delta^{AB}\,.
\end{align}

\subsubsection{Propagation of Quark-gluon Condensate}
The propagation of the quark-gluon condensate through the nonlocal quark condensate requires to expand the quark field up to terms with two derivatives, which compensates for the difference in dimensions of the nonlocal quark condensate and the local quark-gluon condensates. Specifically, the contributing part \eqref{eq:quark_condensate_expansion_2} can be rewritten to
\begin{align}
\big\langle\overline{q}_{i,\alpha}^{A}(x)q_{k,\beta}^{B}(y)\big\rangle\ni\bigg(\frac{1}{2}x^{\mu}x^{\nu}-x^{\mu}y^{\nu}+\frac{1}{2}y^{\mu}y^{\nu}\bigg)\big\langle\overline{q}_{i,\alpha}^{A}\big(\nabla_{\mu}\nabla_{\nu}q_{k,\beta}^{B}\big)\big\rangle\,.\label{eq:propagace_qq_derivation_2}
\end{align}
The task is thus simplified into extracting the quark-gluon condensate out of the expectation value on the right-hand side of \eqref{eq:propagace_qq_derivation_2}. To do so, let us perform the following steps.

\begin{enumerate}
\item According to \eqref{eq:2_index_basis}, we write down the general structure of the vacuum expectation value in the form
\begin{equation}
\big\langle\overline{q}_{i,\alpha}^{A}\big(\nabla_{\mu}\nabla_{\nu}q_{k,\beta}^{B}\big)\big\rangle=\delta^{AB}\delta_{\alpha\beta}\big(C g_{\mu\nu}+D\sigma_{\mu\nu}\big)_{ki}\,,\label{eq:propagace_qq_derivation_3}
\end{equation}
with $C$ and $D$ being the unknown functions that need to be identified.
\item Using an obvious identity $\gamma^{\mu}\gamma^{\nu}=g^{\mu\nu}-i\sigma^{\mu\nu}$ (see the footnote \ref{footnote_decompositions}), we have
\begin{equation}
\slashed{\nabla}\slashed{\nabla}=\nabla^{2}-i\sigma^{\mu\nu}\nabla_{\mu}\nabla_{\nu}\,.\label{eq:propagace_qq_derivation_4}
\end{equation}
On the other hand, rewriting $\nabla_{\mu}\nabla_{\nu}$ into the symmetric and the antisymmetric part and using \eqref{eq:commutator_G} gives
\begin{equation}
\nabla_{\mu}\nabla_{\nu}=\frac{1}{2}\lbrace\nabla_{\mu},\nabla_{\nu}\rbrace+\frac{1}{2}ig_{s}G_{\mu\nu}\,,\label{eq:propagace_qq_derivation_4yy}
\end{equation}
which, after substituting back into \eqref{eq:propagace_qq_derivation_4}, gives
\begin{equation}
\slashed{\nabla}\slashed{\nabla}=\nabla^{2}+\frac{1}{2}g_{s}\sigma^{\mu\nu}G_{\mu\nu}\equiv\nabla^{2}+\frac{1}{2}g_{s}\sigma\hspace{-1pt}\cdot\hspace{-1pt}G\,,\label{eq:propagace_qq_derivation_4xx}
\end{equation}
where we have introduced a simplified notation of the contraction of the sigma tensor with the gluon field strength tensor. The relation \eqref{eq:propagace_qq_derivation_4yy} gives us also
\begin{equation}
\sigma^{\mu\nu}\nabla_{\mu}\nabla_{\nu}=\frac{1}{2} i g_{s}\sigma\hspace{-1pt}\cdot\hspace{-1pt}G\,.\label{eq:propagace_qq_derivation_4xxx}
\end{equation}
\item Now we multiply both sides of \eqref{eq:propagace_qq_derivation_3} with $(g^{\mu\nu})_{ik}\delta^{AB}\delta_{\alpha\beta}$. This leads easily to
\begin{equation}
\big\langle\overline{q}\big(\nabla^{2}q\big)\big\rangle=2^{4}\hspace{-1pt}\cdot\hspace{-1pt}3^{2}\hspace{-1pt}\cdot\hspace{-1pt}C\,,
\end{equation}
and after using \eqref{eq:propagace_qq_derivation_4xx} and the Dirac equation \eqref{eq:qcd_eom_1}, we get
\begin{equation}
C=-\frac{g_{s}}{2^{5}\hspace{-1pt}\cdot\hspace{-1pt}3^{2}}\langle\overline{q}\sigma\hspace{-1pt}\cdot\hspace{-1pt}Gq\rangle\,.\label{eq:propagace_qq_derivation_C}
\end{equation}
\item To obtain a constraint for the remaining function $D$, we now multiply both sides of \eqref{eq:propagace_qq_derivation_3} with $(\sigma^{\mu\nu})_{ik}\delta^{AB}\delta_{\alpha\beta}$. Similarly to the previous step, we obtain
\begin{equation}
\big\langle\overline{q}\big(\sigma^{\mu\nu}\nabla_{\mu}\nabla_{\nu}q\big)\big\rangle=2^{4}\hspace{-1pt}\cdot\hspace{-1pt}3^{3}\hspace{-1pt}\cdot\hspace{-1pt}D\,,
\end{equation}
which, using \eqref{eq:propagace_qq_derivation_4xxx}, leads to
\begin{equation}
D=\frac{i g_{s}}{2^{5}\hspace{-1pt}\cdot\hspace{-1pt}3^{3}}\langle\overline{q}\sigma\hspace{-1pt}\cdot\hspace{-1pt}Gq\rangle\,.\label{eq:propagace_qq_derivation_D}
\end{equation}
\item Knowing the relations \eqref{eq:propagace_qq_derivation_C} and \eqref{eq:propagace_qq_derivation_D} for the functions $C$ and $D$, we put them back into \eqref{eq:propagace_qq_derivation_2}, which gives us the vacuum expectation value in the form
\begin{equation}
\big\langle\overline{q}_{i,\alpha}^{A}\big(\nabla_{\mu}\nabla_{\nu}q_{k,\beta}^{B}\big)\big\rangle=-\frac{g_{s}\langle\overline{q}\sigma\hspace{-1pt}\cdot\hspace{-1pt}Gq\rangle}{2^{5}\cdot 3^{2}}\delta^{AB}\delta_{\alpha\beta}\bigg(g_{\mu\nu}-\frac{i}{3}\sigma_{\mu\nu}\bigg)_{ki}\,.\label{eq:propagace_qq_derivation_6}
\end{equation}
\end{enumerate}

Substituting \eqref{eq:propagace_qq_derivation_6} back into \eqref{eq:propagace_qq_derivation_2} gives us the final form for the propagation formula of the local quark-gluon condensate through the nonlocal quark condensate:
\begin{align}
\big\langle\overline{q}_{i,\alpha}^{A}(x)q_{k,\beta}^{B}(y)\big\rangle&\ni-\frac{g_{s}\langle\overline{q}\sigma\hspace{-1pt}\cdot\hspace{-1pt}Gq\rangle}{2^{5}\cdot 3^{2}}\delta^{AB}\delta_{\alpha\beta}\bigg[\frac{1}{2}(x-y)^{2}+\frac{i}{3}\sigma^{(x)(y)}\bigg]_{ki}\,.
\end{align}

\subsubsection{Propagation of Four-quark Condensate}
To obtain the propagation formula for the local four-quark condensate through the nonlocal quark condensate, one is required to expand the quark field up to three derivatives. Such procedure leads to \eqref{eq:quark_condensate_expansion_3}, which can be rewritten to
\begin{align}
\big\langle\overline{q}_{i,\alpha}^{A}(x)q_{k,\beta}^{B}(y)\big\rangle\ni-\frac{1}{2}\bigg[x_{\mu}x_{\nu}\bigg(\frac{1}{3}x_{\rho}-y_{\rho}\bigg)-\bigg(\frac{1}{3}y_{\mu}-x_{\mu}\bigg)y_{\nu}y_{\rho}\bigg]\big\langle\overline{q}_{i,\alpha}^{A}\big(\nabla_{\mu}\nabla_{\nu}\nabla_{\rho}q_{k,\beta}^{B}\big)\big\rangle\,.\label{eq:four-quark_condensate_derivation_1}
\end{align}
The following steps are needed to be performed to obtain the desired propagation formula.

\begin{enumerate}
\item In what follows, we take the general structure of the vacuum expectation value in \eqref{eq:four-quark_condensate_derivation_1} to be of the form
\begin{equation}
\big\langle\overline{q}_{i,\alpha}^{A}\big(\nabla^{\mu}\nabla^{\nu}\nabla^{\rho}q_{k,\beta}^{B}\big)\big\rangle=\delta^{AB}\delta_{\alpha\beta}\big(E g^{\mu\nu}\gamma^{\rho}+F g^{\mu\rho}\gamma^{\nu}+G g^{\nu\rho}\gamma^{\mu}+H\varepsilon^{\mu\nu\rho\alpha}\gamma_{\alpha}\gamma_{5}\big)_{ki}\,.\label{eq:four-quark_condensate_derivation_2}
\end{equation}
\item The term proportional to $\varepsilon^{\mu\nu\rho\alpha}\gamma_{\alpha}\gamma_{5}$ would not contribute to the propagation formula \eqref{eq:four-quark_condensate_derivation_1} due to the contractions with the coordinate part of the formula.
\item Having the Lorentz structure \eqref{eq:four-quark_condensate_derivation_2}, we now multiply both sides of this equation with $(\gamma_{\rho})_{ik}$ and compare the term proportional to the metic tensor $g^{\mu\nu}$. Then, the left-hand side vanishes due to the Dirac equation \eqref{eq:qcd_eom_1} and we obtain the condition
\begin{equation}
4E+F+G=0\,.\label{eq:four-quark_condensate_derivation_4}
\end{equation}
\item We now rewrite the left-hand side of \eqref{eq:four-quark_condensate_derivation_2} to make the derivative act to the left:
\begin{equation}
\big\langle\overline{q}_{i,\alpha}^{A}\big(\nabla^{\mu}\nabla^{\nu}\nabla^{\rho}q_{k,\beta}^{B}\big)\big\rangle=-\big\langle\big(\overline{q}_{i,\alpha}^{A}\overleftarrow{\nabla}^{\mu}\big)\big(\nabla^{\nu}\nabla^{\rho}q_{k,\beta}^{B}\big)\big\rangle\,,\label{eq:four-quark_condensate_derivation_5}
\end{equation}
Then, we multiply right-hand sides of \eqref{eq:four-quark_condensate_derivation_2} and \eqref{eq:four-quark_condensate_derivation_5} with $(\gamma_{\mu})_{ik}$. Similarly to the previous step, due to the Dirac equation \eqref{eq:qcd_eom_2}, we obtain the condition
\begin{equation}
E+F+4G=0\,.\label{eq:four-quark_condensate_derivation_6}
\end{equation}
\item The solution of the system of equations \eqref{eq:four-quark_condensate_derivation_4} and \eqref{eq:four-quark_condensate_derivation_6} is simply
\begin{equation}
(E,F)=(G,-5G)\,,\label{eq:four-quark_condensate_derivation_7}
\end{equation}
which leaves us only with one parameter, $G$, i.e
\begin{equation}
\big\langle\overline{q}_{i,\alpha}^{A}\big(\nabla^{\mu}\nabla^{\nu}\nabla^{\rho}q_{k,\beta}^{B}\big)\big\rangle=\delta^{AB}\delta_{\alpha\beta}G\big(g^{\mu\nu}\gamma^{\rho}-5g^{\mu\rho}\gamma^{\nu}+g^{\nu\rho}\gamma^{\mu}\big)_{ki}\,.\label{eq:four-quark_condensate_derivation_8}
\end{equation}
\item As a next step, we multiply both sides of \eqref{eq:four-quark_condensate_derivation_8} with $(\gamma_{\nu})_{ik}g_{\mu\rho}\delta^{AB}\delta_{\alpha\beta}$, which gives us the solution for $G$ in the form
\begin{align}
G=-\frac{1}{2^{5}\cdot 3^{4}}\big\langle\overline{q}\big(\nabla^{\mu}\slashed{\nabla}\nabla_{\mu}q\big)\big\rangle\,.\label{eq:four-quark_condensate_derivation_9}
\end{align}
\item Let us now rewrite the derivatives in the previous expression a bit. A useful formula
\begin{equation}
\slashed{\nabla}\nabla_{\mu}=\nabla_{\mu}\slashed{\nabla}-[\nabla_{\mu},\slashed{\nabla}]\label{eq:four-quark_condensate_derivation_10}
\end{equation}
allows us to appropriately modify \eqref{eq:four-quark_condensate_derivation_9}, since the first term on the right-hand side of \eqref{eq:four-quark_condensate_derivation_10} does not contribute due to the Dirac equation, and the commutator eventually leads to the presence of the gluon-field strength tensor, according to \eqref{eq:commutator_G}. In detail, we have
\begin{equation}
G=\frac{1}{2^{5}\cdot 3^{4}}\big\langle\overline{q}\big(\nabla^{\mu}[\nabla_{\mu},\slashed{\nabla}]q\big)\big\rangle=\frac{i g_{s}}{2^{5}\cdot 3^{4}}\big\langle\overline{q}\gamma^{\nu}T^{a}\nabla^{\mu}\big(G_{\mu\nu}^{a}q\big)\big\rangle\,,\label{eq:four-quark_condensate_derivation_11}
\end{equation}
which can be further rewritten to
\begin{equation}
G=\frac{i g_{s}}{2^{5}\cdot 3^{4}}\big\langle\big(\overline{q}\gamma^{\nu}T^{a}q\big)\big(D^{\mu}G_{\mu\nu}\big)^{a}\big\rangle+\frac{i g_{s}}{2^{5}\cdot 3^{4}}\big\langle\overline{q}\gamma^{\nu}T^{a}G_{\mu\nu}^{a}\big(\nabla^{\mu}q\big)\big\rangle\,.\label{eq:four-quark_condensate_derivation_12}
\end{equation}
\item We now get back to \eqref{eq:four-quark_condensate_derivation_9}, however, we rewrite it with the derivatives acting to the left,
\begin{equation}
G=-\frac{1}{2^{5}\cdot 3^{4}}\big\langle\big(\overline{q}\overleftarrow{\nabla\vphantom{\slashed{\nabla}}}^{\mu}\overleftarrow{\slashed{\nabla}}\big)\big(\nabla_{\mu}q\big)\big\rangle\,,\label{eq:four-quark_condensate_derivation_13}
\end{equation}
and use a formula equivalent to \eqref{eq:four-quark_condensate_derivation_10}, i.e.
\begin{equation}
\overleftarrow{\nabla\vphantom{\slashed{\nabla}}}_{\mu}\overleftarrow{\slashed{\nabla}}=\overleftarrow{\slashed{\nabla}}\overleftarrow{\nabla\vphantom{\slashed{\nabla}}}_{\mu}+[\overleftarrow{\nabla\vphantom{\slashed{\nabla}}}_{\mu},\overleftarrow{\slashed{\nabla}}]\,,\label{eq:four-quark_condensate_derivation_14}
\end{equation}
where the first term on the right-hand side once again vanishes due to the Dirac equation. Applying \eqref{eq:four-quark_condensate_derivation_14} to \eqref{eq:four-quark_condensate_derivation_13}, we obtain
\begin{equation}
G=-\frac{i g_{s}}{2^{5}\cdot 3^{4}}\big\langle\overline{q}\gamma^{\nu}T^{a}G_{\mu\nu}^{a}\big(\nabla^{\mu}q\big)\big\rangle\,.\label{eq:four-quark_condensate_derivation_15}
\end{equation}
\item Comparing \eqref{eq:four-quark_condensate_derivation_12} with \eqref{eq:four-quark_condensate_derivation_15}, we get
\begin{equation}
\big\langle\overline{q}\gamma^{\nu}T^{a}G_{\mu\nu}^{a}\big(\nabla^{\mu}q\big)\big\rangle=-\frac{1}{2}\big\langle\big(\overline{q}\gamma^{\nu}T^{a}q\big)\big(D^{\mu}G_{\mu\nu}\big)^{a}\big\rangle\,,\label{eq:four-quark_condensate_derivation_16}
\end{equation}
which can be simplified using the equation of motion \eqref{eq:qcd_eom_3} and \eqref{eq:4q-faktorizace} as
\begin{equation}
\big\langle\overline{q}\gamma^{\nu}T^{a}G_{\mu\nu}^{a}\big(\nabla^{\mu}q\big)\big\rangle=-\frac{1}{2}g_{s}\big\langle\big(\overline{q}\gamma^{\nu}T^{a}q\big)\big(\overline{q}\gamma_{\nu}T^{a}q\big)\big\rangle=\frac{2}{3^{3}}g_{s}\langle\overline{q}q\rangle^{2}\,.\label{eq:four-quark_condensate_derivation_17}
\end{equation}
\item Inserting \eqref{eq:four-quark_condensate_derivation_17} into \eqref{eq:four-quark_condensate_derivation_15}, we finally get
\begin{equation}
G=-i\frac{g_{s}^{2}}{2^{4}\cdot 3^{7}}\langle\overline{q}q\rangle^{2}\,,\label{eq:four-quark_condensate_derivation_18}
\end{equation}
i.e. the relevant part of the vacuum expectation value \eqref{eq:four-quark_condensate_derivation_2} can be written down as
\begin{equation}
\big\langle\overline{q}_{i,\alpha}^{A}\big(\nabla^{\mu}\nabla^{\nu}\nabla^{\rho}q_{k,\beta}^{B}\big)\big\rangle\ni -i\frac{g_{s}^{2}}{2^{4}\cdot 3^{7}}\langle\overline{q}q\rangle^{2}\delta^{AB}\delta_{\alpha\beta}\big(g^{\mu\nu}\gamma^{\rho}-5g^{\mu\rho}\gamma^{\nu}+g^{\nu\rho}\gamma^{\mu}\big)_{ki}\,.\label{eq:four-quark_condensate_derivation_19}
\end{equation}
\end{enumerate}

Finally, after inserting \eqref{eq:four-quark_condensate_derivation_19} back into \eqref{eq:four-quark_condensate_derivation_1}, the propagation formula is thus derived to be in the form
\begin{align}
\big\langle\overline{q}_{i,\alpha}^{A}(x)q_{k,\beta}^{B}(y)\big\rangle\ni\frac{i g_{s}^{2}}{2^{5}\cdot 3^{7}}\langle\overline{q}q\rangle^{2}\delta^{AB}\delta_{\alpha\beta}\Big[4(x\hspace{-1pt}\cdot\hspace{-1pt}y)(\slashed{x}-\slashed{y})-(x^{2}-y^{2})(\slashed{x}+\slashed{y})\Big]_{ki}\,.\label{eq:four-quark_condensate_derivation_20}
\end{align}

\subsection{Derivation of Propagation Formula \eqref{eq:propagation_gluon_condensate}}
A derivation of the propagation formula \eqref{eq:propagation_gluon_condensate} is trivial. However, we include the derivation here to be as thorough in our explanation as possible.

To evaluate \eqref{eq:propagation_gluon_condensate}, it is sufficient to take the first term of the expansion of the gluon field and then use the projection formula \eqref{eq:gluon-condensate-projection_v2}. Specifically, we have
\begin{equation}
\big\langle\mathcal{A}_{\mu}^{a}(x)\mathcal{A}_{\nu}^{b}(y)\big\rangle\ni\frac{1}{4}x^{\rho}y^{\sigma}\big\langle G_{\mu\rho}^{a}G_{\nu\sigma}^{b}\big\rangle=\bigg(\frac{1}{4}x^{\rho}y^{\sigma}\bigg)\bigg(\frac{1}{96}(g_{\mu\nu}g_{\rho\sigma}-g_{\mu\sigma}g_{\nu\rho})\langle G^{2}\rangle\delta^{ab}\bigg)\,,
\end{equation}
i.e.
\begin{equation}
\alpha_{s}\big\langle\mathcal{A}_{\mu}^{a}(x)\mathcal{A}_{\nu}^{b}(y)\big\rangle\ni\frac{\alpha_{s}\langle G^{2}\rangle}{2^{7}\cdot 3}\big[(x\hspace{-1pt}\cdot\hspace{-1pt}y)g_{\mu\nu}-y_{\mu}x_{\nu}\big]\delta^{ab}\,.
\end{equation}

\subsection{Derivation of Propagation Formula \eqref{eq:4q_propagace_qAq}}
Finally, we evaluate the propagation formula \eqref{eq:4q_propagace_qAq}, i.e. the propagation of the local quark-gluon and the four-quark condensates through the nonlocal quark-gluon condensate. We start with the former.
\begin{align}
    \big\langle\overline{q}_{i,\alpha}^{A}(x)\mathcal{A}_{\mu}^{a}(u)q_{k,\beta}^{B}(y)\big\rangle &\begin{array}{l}\hspace{3pt}\ni-\displaystyle\frac{1}{2}u_{\nu}\big\langle\overline{q}_{i,\alpha}^{A}G_{\mu\nu}^{a}q_{k,\beta}^{B}\big\rangle\vspace{5pt}\label{eq:quark-gluon_condensate_expansion_1}\\
          \end{array}\\
        & \left.\begin{array}{l}
          \hspace{10pt}-\,\displaystyle\frac{1}{2}u^{\nu}x^{\rho}\big\langle\big(\overline{q}_{i,\alpha}^{A}\overleftarrow{\nabla}_{\rho}\big)G_{\mu\nu}^{a}q_{k,\beta}^{B}\big\rangle\vspace{5pt}\\
          \hspace{10pt}-\,\displaystyle\frac{1}{3}u^{\nu}u^{\rho}\big\langle\overline{q}_{i,\alpha}^{A}\big(D_{\rho}G_{\mu\nu}\big)^{a}q_{k,\beta}^{B}\big\rangle\vspace{5pt}\\
          \hspace{10pt}-\,\displaystyle\frac{1}{2}u^{\nu}y^{\rho}\big\langle\overline{q}_{i,\alpha}^{A}G_{\mu\nu}^{a}\big(\nabla_{\rho}q_{k,\beta}^{B}\big)\big\rangle\,.       
         \end{array}\right\}\label{eq:quark-gluon_condensate_expansion_2}
\end{align}

\subsubsection{Propagation of Quark-gluon Condensate}
According to \eqref{eq:quark-gluon_condensate_expansion_1}, the relevant term for propagation of the local quark-gluon condensate from the nonlocal one is
\begin{equation}
\big\langle\overline{q}_{i,\alpha}^{A}(x)\mathcal{A}_{\mu}^{a}(u)q_{k,\beta}^{B}(y)\big\rangle\ni-\frac{1}{2}u_{\nu}\big\langle\overline{q}_{i,\alpha}^{A}G_{\mu\nu}^{a}q_{k,\beta}^{B}\big\rangle\,.\label{eq:propagace_qAq_derivation_1}
\end{equation}

Because the vacuum expectation value on the righ-hand side of \eqref{eq:propagace_qAq_derivation_1} is antisymmetric in the Lorentz indices due to the presence of the gluon-field strength tensor, we take its general form as
\begin{equation}
\big\langle\overline{q}_{i,\alpha}^{A}G_{\mu\nu}^{a}q_{k,\beta}^{B}\big\rangle=I\delta^{ab}(T^{a})_{\beta\alpha}(\sigma_{\mu\nu})_{ki}\,.\label{eq:propagace_qAq_derivation_2}
\end{equation}

In this case, it is only needed to contract both sides of \eqref{eq:propagace_qAq_derivation_2} with $(\sigma_{\mu\nu})_{ik}\delta^{AB}(T^{a})_{\alpha\beta}$, which gives gives
\begin{equation}
I=\frac{1}{2^{6}\hspace{-1pt}\cdot\hspace{-1pt}3^{2}}\langle\overline{q}\sigma\hspace{-1pt}\cdot\hspace{-1pt}Gq\rangle\,,\label{eq:propagace_qAq_derivation_4}
\end{equation}
i.e.
\begin{equation}
\big\langle\overline{q}_{i,\alpha}^{A}G_{\mu\nu}^{a}q_{k,\beta}^{B}\big\rangle=\frac{1}{2^{6}\hspace{-1pt}\cdot\hspace{-1pt}3^{2}}\langle\overline{q}\sigma\hspace{-1pt}\cdot\hspace{-1pt}Gq\rangle\delta^{ab}(T^{a})_{\beta\alpha}\big(\sigma_{\mu\nu}\big)_{ki}\,.\label{eq:propagace_qAq_derivation_5}
\end{equation}

Substituting \eqref{eq:propagace_qAq_derivation_5} back into \eqref{eq:propagace_qAq_derivation_1} gives us the final form of the propagation formula for the local quark-gluon condensate in the form
\begin{align}
g_{s}\big\langle\overline{q}_{i,\alpha}^{A}(x)\mathcal{A}_{\mu}^{a}(u)q_{k,\beta}^{B}(y)\big\rangle\ni\frac{g_{s}}{2^{7}\cdot 3^{2}}\langle\overline{q}\sigma\hspace{-1pt}\cdot\hspace{-1pt}Gq\rangle\delta^{AB}(T^{a})_{\beta\alpha}\big(\sigma_{(u)\mu}\big)_{ki}\,.\label{eq:propagace_qAq_derivation_6}
\end{align}

\subsubsection{Propagation of Four-quark Condensate}
The three contributing terms in \eqref{eq:quark-gluon_condensate_expansion_2} can be rewritten to the following two:
\begin{align}
\big\langle\overline{q}_{i,\alpha}^{A}(x)&\mathcal{A}_{\mu}^{a}(u)q_{k,\beta}^{B}(y)\big\rangle\ni\label{eq:4quark_qAq_derivation_1}\\
&\ni\frac{1}{2}u_{\nu}(x_{\rho}-y_{\rho})\big\langle\overline{q}_{i,\alpha}^{A}G_{\mu\nu}^{a}\big(\nabla_{\rho}q_{k,\beta}^{B}\big)\big\rangle+u_{\nu}\bigg(\frac{1}{2}x_{\rho}-\frac{1}{3}u_{\rho}\bigg)\big\langle\overline{q}_{i,\alpha}^{A}\big(D_{\rho}G_{\mu\nu}\big)^{a}q_{k,\beta}^{B}\big\rangle\,.\nonumber
\end{align}
To obtain the propagation formula, we will now carry on in the procedure below. We start with the first vacuum expectation value on the right-hand side of \eqref{eq:4quark_qAq_derivation_1}.

\begin{enumerate}
\item 
\begin{enumerate}
\item First of all, the Lorentz structure of such vacuum value must be antisymmetric in the Lorentz indices $\mu,\nu$. Also, unlike in the case of \eqref{eq:four-quark_condensate_derivation_2}, we are required to keep the term proportional to the Levi-Civita tensor. Therefore, we have
\begin{equation}
\big\langle\overline{q}_{i,\alpha}^{A}G_{\mu\nu}^{a}\big(\nabla_{\rho}q_{k,\beta}^{B}\big)\big\rangle=\delta^{AB}(T^{a})_{\beta\alpha}\big[J\big(g^{\nu\rho}\gamma^{\mu}-g^{\mu\rho}\gamma^{\nu}\big)+K\varepsilon^{\mu\nu\rho\alpha}\gamma_{\alpha}\gamma_{5}\big]_{ki}\,.\label{eq:4quark_qAq_derivation_5}
\end{equation}
\item Now we multiply both sides of \eqref{eq:4quark_qAq_derivation_5} with $(\gamma_{\rho}\gamma_{\mu}\gamma_{\nu})_{ik}$. Once again, left-hand side vanishes due to the Dirac equation \eqref{eq:qcd_eom_1}, and we find out that
\begin{equation}
K=-iJ\,,\label{eq:4quark_qAq_derivation_6}
\end{equation}
i.e. we now have only one unknown function $J$ to obtain, and
\begin{equation}
\big\langle\overline{q}_{i,\alpha}^{A}G_{\mu\nu}^{a}\big(\nabla_{\rho}q_{k,\beta}^{B}\big)\big\rangle=J\delta^{AB}(T^{a})_{\beta\alpha}\big[g^{\nu\rho}\gamma^{\mu}-g^{\mu\rho}\gamma^{\nu}-i\varepsilon^{\mu\nu\rho\alpha}\gamma_{\alpha}\gamma_{5}\big]_{ki}\,.\label{eq:4quark_qAq_derivation_7}
\end{equation}
\item In this step, we multiply both sides of \eqref{eq:4quark_qAq_derivation_7} with $(\gamma_{\nu})_{ik}g_{\mu\rho}\delta^{AB}(T^{a})_{\alpha\beta}$, which gives us
\begin{equation}
J=-\frac{1}{2^{6}\cdot 3^{2}}\big\langle\overline{q}\gamma^{\nu}T^{a}G_{\mu\nu}^{a}\big(\nabla^{\mu}q\big)\big\rangle=-\frac{g_{s}}{2^{5}\cdot 3^{5}}\langle\overline{q}q\rangle^{2}\,,\label{eq:4quark_qAq_derivation_8}
\end{equation}
where we have used the already known result \eqref{eq:four-quark_condensate_derivation_17}.
\item Substituting the solution for $J$ above back into \eqref{eq:4quark_qAq_derivation_7} thus leaves us with the result for the first vacuum expectation value in \eqref{eq:4quark_qAq_derivation_1} in the form
\begin{equation}
\big\langle\overline{q}_{i,\alpha}^{A}G_{\mu\nu}^{a}\big(\nabla_{\rho}q_{k,\beta}^{B}\big)\big\rangle=\frac{g_{s}\langle\overline{q}q\rangle^{2}}{2^{5}\cdot 3^{5}}\delta^{AB}(T^{a})_{\beta\alpha}\big[g^{\mu\rho}\gamma^{\nu}-g^{\nu\rho}\gamma^{\mu}+i\varepsilon^{\mu\nu\rho\alpha}\gamma_{\alpha}\gamma_{5}\big]_{ki}\,.\label{eq:4quark_qAq_derivation_9}
\end{equation}
\end{enumerate}
\item 
\begin{enumerate}
\item Similarly as in the previous case, the general Lorentz structure of the second vacuum expectation value in \eqref{eq:4quark_qAq_derivation_1} is
\begin{equation}
\big\langle\overline{q}_{i,\alpha}^{A}\big(D_{\rho}G_{\mu\nu}\big)^{a}q_{k,\beta}^{B}\big\rangle=\delta^{AB}(T^{a})_{\beta\alpha}\big[K(g^{\mu\rho}\gamma^{\nu}-g^{\nu\rho}\gamma^{\mu})+L\varepsilon^{\mu\nu\rho\alpha}\gamma_{\alpha}\gamma_{5}\big]_{ki}\,,\label{eq:4quark_qAq_derivation_10}
\end{equation}
where we have once again omitted the term proportional to $g^{\mu\nu}\gamma^{\rho}$ since the structure must be antisymmetric in the Lorentz indices $\mu,\nu$. Also, we formally keep the term proportional to $L$, although it vanishes, as we will see in the next step.
\item Having the derivative acting on the gluon field strength tensor, we exploit the Bianchi identity
\begin{equation}
D_{\rho}G_{\mu\nu}+D_{\nu}G_{\rho\mu}+D_{\mu}G_{\nu\rho}=0\,,\label{eq:4quark_qAq_derivation_11}
\end{equation}
which leads to the constraint on the right-hand side of \eqref{eq:4quark_qAq_derivation_10} in the form
\begin{equation}
3L\varepsilon^{\mu\nu\rho\alpha}\gamma_{\alpha}\gamma_{5}=0\,,\label{eq:4quark_qAq_derivation_12}
\end{equation}
i.e.
\begin{equation}
L=0\,,\label{eq:4quark_qAq_derivation_13}
\end{equation}
which simplifies \eqref{eq:4quark_qAq_derivation_10} into the form
\begin{equation}
\big\langle\overline{q}_{i,\alpha}^{A}\big(D_{\rho}G_{\mu\nu}\big)^{a}q_{k,\beta}^{B}\big\rangle=K\delta^{AB}(T^{a})_{\beta\alpha}\big(g^{\mu\rho}\gamma^{\nu}-g^{\nu\rho}\gamma^{\mu}\big)_{ki}\,,\label{eq:4quark_qAq_derivation_14}
\end{equation}
which is already antisymmetric in the indices $\mu,\nu$.
\item Now we freely rearrange the left-hand side of \eqref{eq:4quark_qAq_derivation_14} by moving the second quark field in front of the derivative term,
\begin{equation}
\big\langle\overline{q}_{i,\alpha}^{A}\big(D_{\rho}G_{\mu\nu}\big)^{a}q_{k,\beta}^{B}\big\rangle=\big\langle\overline{q}_{i,\alpha}^{A}q_{k,\beta}^{B}\big(D_{\rho}G_{\mu\nu}\big)^{a}\big\rangle\,,\label{eq:4quark_qAq_derivation_17}
\end{equation}
which does not qualitatively change anything, and then we multiply right-hand sides of \eqref{eq:4quark_qAq_derivation_14} and \eqref{eq:4quark_qAq_derivation_17} with $(\gamma_{\nu})_{ik}g_{\mu\rho}\delta^{AB}(T^{a})_{\alpha\beta}$, which leads to
\begin{equation}
\big\langle\big(\overline{q}\gamma^{\nu}T^{a}q\big)\big(D^{\mu}G_{\mu\nu}\big)^{a}\big\rangle=2^{6}\hspace{-1pt}\cdot\hspace{-1pt}3^{2}\hspace{-1pt}\cdot\hspace{-1pt}K\,,\label{eq:4quark_qAq_derivation_18}
\end{equation}
which, after using the equation of motion \eqref{eq:qcd_eom_3} and \eqref{eq:4q-faktorizace}, gives
\begin{equation}
K=-\frac{g_{s}}{2^{4}\cdot 3^{5}}\langle\overline{q}q\rangle^{2}\,.\label{eq:4quark_qAq_derivation_19}
\end{equation}
\item Finally, we insert \eqref{eq:4quark_qAq_derivation_19} back into \eqref{eq:4quark_qAq_derivation_14} and we get
\begin{equation}
\big\langle\overline{q}_{i,\alpha}^{A}\big(D_{\rho}G_{\mu\nu}\big)^{a}q_{k,\beta}^{B}\big\rangle=\frac{g_{s}\langle\overline{q}q\rangle^{2}}{2^{4}\cdot 3^{5}}\delta^{AB}(T^{a})_{\beta\alpha}\big(g^{\nu\rho}\gamma^{\mu}-g^{\mu\rho}\gamma^{\nu}\big)_{ki}\,.\label{eq:4quark_qAq_derivation_20}
\end{equation}
\end{enumerate}
\end{enumerate}

Finally, after inserting \eqref{eq:4quark_qAq_derivation_9} and \eqref{eq:4quark_qAq_derivation_20} back into \eqref{eq:4quark_qAq_derivation_1}, we can get the final form of the propagation formula for the local quark condensate through the nonlocal quark-gluon condsensate:
\begin{align}
\big\langle\overline{q}_{i,\alpha}^{A}(x)&\mathcal{A}_{\mu}^{a}(u)q_{k,\beta}^{B}(y)\big\rangle\ni\frac{g_{s}\langle\overline{q}q\rangle^{2}}{2^{5}\cdot 3^{5}}\delta^{AB}(T^{a})_{\beta\alpha}\times\label{eq:4quark_qAq_derivation_24}\\
&\times\bigg(\frac{1}{6}\gamma^{\mu}\big[3u\hspace{-1pt}\cdot\hspace{-1pt}(x+y)-4u^{2}\big]+\frac{1}{6}\slashed{u}\big[4u^{\mu}-3(x+y)^{\mu}\big]-\frac{i}{2}\varepsilon^{\mu(x-y)(u)\alpha}\gamma_{\alpha}\gamma_{5}\bigg)_{ki}\,.\nonumber
\end{align}

%%%%%%%%%%%%%%%%%%%%%%%%%%%%%%%%%%%%%%%%%%%%%%%%%%%%%%%%%%%%%%%%%%%%%%%%%%%%%%%%%%%%%%%%%%%%%%%%%%%%%%%%%
%%%%%%%%%%%%%%%%%%%%%%%%%%%%%%%%%%%%%%%%%%%%%%%%%%%%%%%%%%%%%%%%%%%%%%%%%%%%%%%%%%%%%%%%%%%%%%%%%%%%%%%%%
%%% Section: OPE for Two-point Green Functions
%%%%%%%%%%%%%%%%%%%%%%%%%%%%%%%%%%%%%%%%%%%%%%%%%%%%%%%%%%%%%%%%%%%%%%%%%%%%%%%%%%%%%%%%%%%%%%%%%%%%%%%%%
%%%%%%%%%%%%%%%%%%%%%%%%%%%%%%%%%%%%%%%%%%%%%%%%%%%%%%%%%%%%%%%%%%%%%%%%%%%%%%%%%%%%%%%%%%%%%%%%%%%%%%%%%

\section{OPE for Two-point Green Functions}\label{sec:appendix-2pt}
We include this appendix to show a clear connection between the two-point and the three-point Green functions through the Ward identities. Therefore, the results presented here are useful to check whether the right-hand side of the Ward identities of the three-point Green functions are indeed expressed as linear combinations of the two-point correlators of the respective QCD condensate contributions.

Similarly to Section \ref{sec:greenfunctions}, we present the following classification of all the two-point Green functions, relevant in the chiral limit:
\begin{itemize}
\item Set 1: The correlators with the perturbative contribution in the chiral limit:
\begin{itemize}
\item $\langle VV\rangle$, $\langle AA\rangle$, $\langle SS\rangle$, $\langle PP\rangle$.
\end{itemize}
\item Set 2: The correlator that is the order parameter of the chiral symmetry breaking in the chiral limit:
\begin{itemize}
\item $\langle AP\rangle$.
\end{itemize}
\end{itemize}
This classification tells us that the $\langle VV\rangle$, $\langle AA\rangle$, $\langle SS\rangle$ and $\langle PP\rangle$ Green functions have, apart from the perturbative contribution, also contributions from the gluon and four-quark condensates in the chiral limit. On the other hand, the $\langle AP\rangle$ is the order parameter of the chiral symmetry breaking and has, in the chiral limit, contribution from the quark condensate only.

We also note that the combinations of the correlators $\langle VV\rangle-\langle AA\rangle$ and $\langle SS\rangle-\langle PP\rangle$ are order parameters of the chiral symmetry breaking, too.

The decomposition of all the nonvanishing two-point correlators is given as
\begin{alignat}{3}
\big[&\Pi_{VV}(p)\big]_{\mu\nu}^{ab}&&=\big[\Pi_{VV}(p)\big]_{\mu\nu}\delta^{ab}&&=\Pi_{VV}(p^{2})(p^{2}g_{\mu\nu}-p_{\mu}p_{\nu})\delta^{ab}\,,\label{VV_definition_apendix}\\
\big[&\Pi_{AA}(p)\big]_{\mu\nu}^{ab}&&=\big[\Pi_{AA}(p)\big]_{\mu\nu}\delta^{ab}&&=\Pi_{AA}(p^{2})(p^{2}g_{\mu\nu}-p_{\mu}p_{\nu})\delta^{ab}\,,\label{AA_definition_apendix}\\
\big[&\Pi_{SS}(p)\big]^{ab}&&=\Pi_{SS}(p^{2})\delta^{ab}\,,&&\label{SS_definition_apendix}\\
\big[&\Pi_{PP}(p)\big]^{ab}&&=\Pi_{PP}(p^{2})\delta^{ab}&&\label{PP_definition_apendix}
\end{alignat}
and
\begin{equation}
\big[\Pi_{AP}(p)\big]_{\mu}^{ab}=\big[\Pi_{AP}(p)\big]_{\mu}\delta^{ab}=\Pi_{AP}(p^{2})p_{\mu}\delta^{ab}\,.\label{AP_definition_apendix}
\end{equation}
%Notice that all the correlators of Set 1 are even functions of momentum, while the $\langle AP\rangle$ Green function is odd.

\subsection{Perturbative Contribution}
Perturbative contribution is of remarkably simple form in the case of two-point correlators, for which it can be written down as
\begin{equation}
\Pi_{\mathcal{O}_{1}^{a}\mathcal{O}_{2}^{b}}^{\mathbb{1}}(p)=-\frac{N_{c}}{2}\delta^{ab}\int\frac{\mathrm{d}^{4}\ell}{(2\pi)^{4}}\mathrm{Tr}\Big[\Gamma_{1}S_{0}(\ell)\Gamma_{2}S_{0}(\ell+p)\Big]\,,
\end{equation}
and, after inserting for the Dirac matrices, one finds the individual contributions to be
\begin{alignat}{3}
\Pi_{VV}^{\mathbb{1}}(p^{2})&=\Pi_{AA}^{\mathbb{1}}(p^{2})&&=&&\frac{i N_{c}}{72\pi^{2}}\bigg[\frac{3}{\widehat{\varepsilon}}+3\log\bigg(\hspace{-4pt}-\frac{\mu^{2}}{p^{2}}\bigg)+5\bigg]\,,\label{eq:2pt_perturbative_VV_AA}\\
\Pi_{SS}^{\mathbb{1}}(p^{2})&=\Pi_{PP}^{\mathbb{1}}(p^{2})&&=-&&\frac{i N_{c}p^{2}}{16\pi^{2}}\bigg[\frac{1}{\widehat{\varepsilon}}+\log\bigg(\hspace{-4pt}-\frac{\mu^{2}}{p^{2}}\bigg)+2\bigg]\,.\label{eq:2pt_perturbative_SS_PP}
\end{alignat}

Obviously, the fact that perturbative contributions of $\langle VV\rangle$ and $\langle AA\rangle$, or $\langle SS\rangle$ and $\langle PP\rangle$, are the same is not surprising since both $\langle VV\rangle-\langle AA\rangle$ and $\langle SS\rangle-\langle PP\rangle$ are order parameters of the chiral symmetry breaking.
\begin{figure}[htb]
    \centering
    \begin{subfigure}[t]{0.27\textwidth}
        \hspace{20pt}\includegraphics[width=0.65\textwidth]{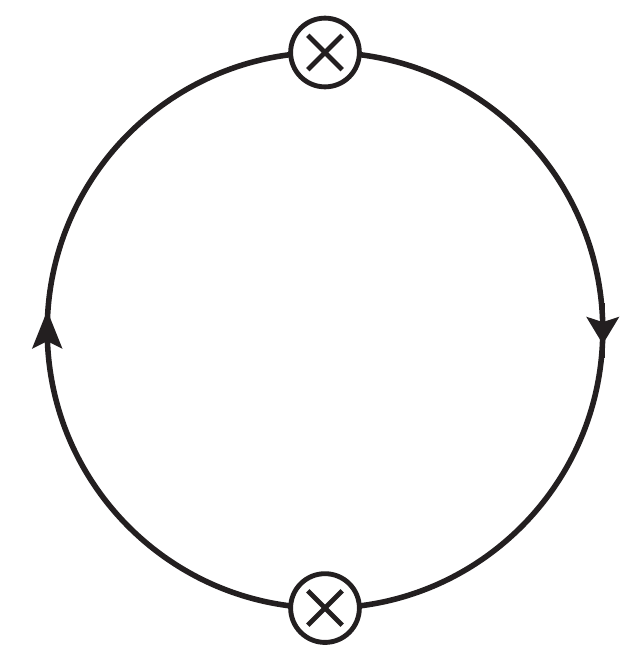}
        \caption{}
        \label{fig:fig_2pt_perturbative_contribution}
    \end{subfigure}
    \begin{subfigure}[t]{0.27\textwidth}
        \hspace{15pt}\includegraphics[width=0.71\textwidth]{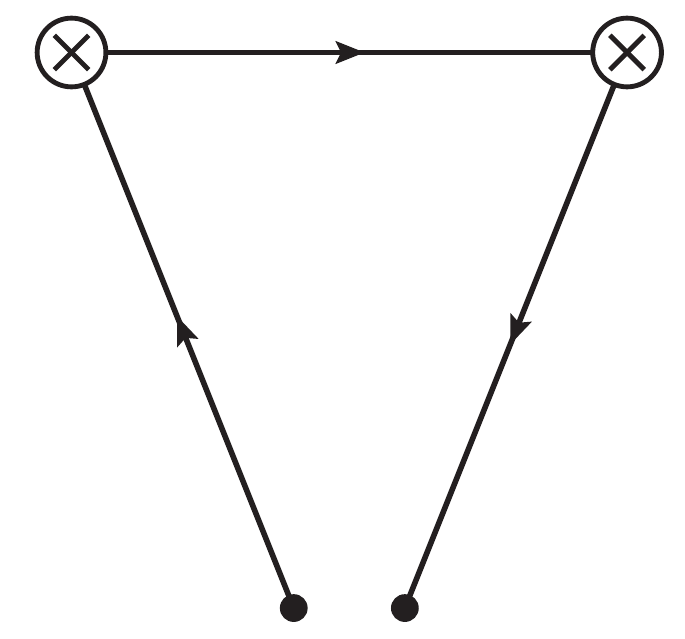}
        \caption{}
        \label{fig:fig_2pt_quark_condensate}
    \end{subfigure}
    \caption{Feynman diagrams of the perturbative contribution (left) and the contribution of the quark condensate (right) to the two-point Green functions.}
    \label{}
\end{figure}

\subsection{Quark Condensate}\label{ssec:quark_condensate_2pt}
In the chiral limit, Ward identities fix the $\langle AP\rangle$ Green function, to be fully saturated by the single Goldstone boson exchange. As a consequence, the OPE expansion of the $\langle AP\rangle$ correlator is given exactly:
\begin{align}
\Pi_{AP}^{\langle\overline{q}q\rangle}(p^{2})=-\frac{\langle\overline{q}q\rangle}{3p^{2}}\,.\label{eq:AP_contribution}
\end{align}

It is useful to point out that the authors of \cite{Jamin:2008rm} use a different normalizations of the quark condensate and the pseudoscalar density in their paper. Also, they use a conventional factor of $i$ in the deinition of the two-point correlators with the opposite of the Fourier transform. This leads to the result that differs in an overall factor of $-6i$ with respect to our result. For details, see eq.~1.2 at page no.~2 and eq.~3.2 at page no.~5 therein.

\subsection{Gluon Condensate}\label{ssec:appendix_gluon_condensate}
The contribution of the gluon condensate to the two-point Green functions is due to the diagrams shown below. Such contribution can be obtained by the means described already in the Section \ref{sec:gluon-condensate} or due to the effective propagation through the nonlocal gluon condensate. In what follows we shortly comment on these calculations.

\begin{figure}[htb]
    \centering
    \begin{subfigure}[t]{0.27\textwidth}
        \hspace{17pt}\includegraphics[width=0.7\textwidth]{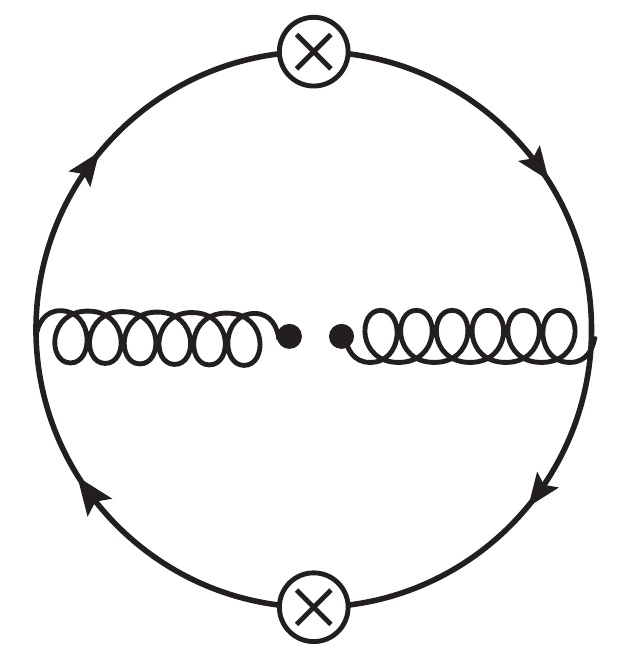}
        \caption{}
        \label{fig:fig_2pt_gluon_condensate_v2_a}
    \end{subfigure}
    \begin{subfigure}[t]{0.27\textwidth}
        \hspace{16pt}\includegraphics[width=0.7\textwidth]{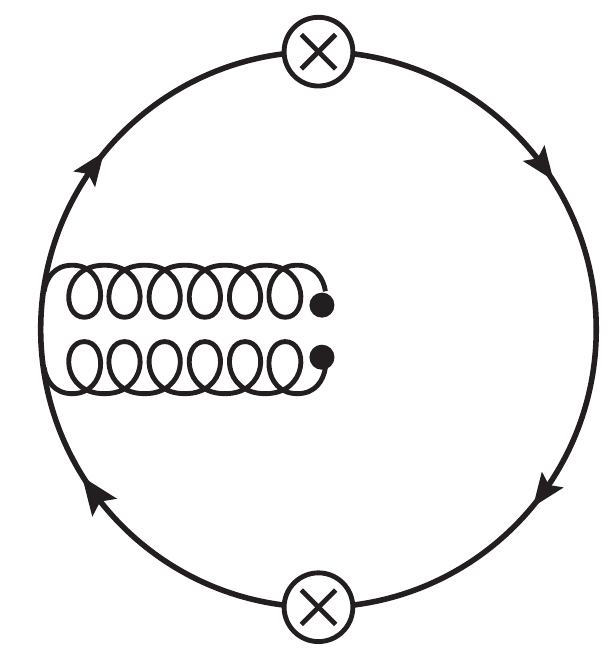}
        \caption{}
        \label{fig:fig_2pt_gluon_condensate_v2_b}
    \end{subfigure}
    \begin{subfigure}[t]{0.27\textwidth}
        \hspace{15pt}\includegraphics[width=0.7\textwidth]{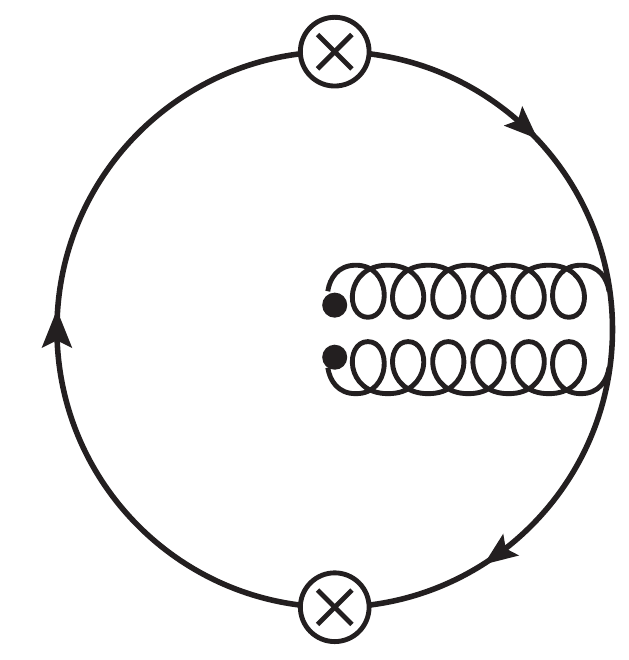}
        \caption{}
        \label{fig:fig_2pt_gluon_condensate_v2_c}
    \end{subfigure}
    \caption{Feynman diagrams of the contributions of the gluon condensate to the two-point Green functions. The diagram \ref{fig:fig_2pt_gluon_condensate_v2_a} contains two propagators \eqref{eq:quark-propagator-x-representation-short-factor-2} and the diagrams \ref{fig:fig_2pt_gluon_condensate_v2_b}-\ref{fig:fig_2pt_gluon_condensate_v2_c} one propagator \eqref{eq:quark-propagator-x-representation-short-factor-3}.}
    \label{fig:fig_2pt_gluon_condensate_v2}
\end{figure}

The contribution of the gluon condensate to the two-point correlators can be computed in the coordinate representation using the propagator \eqref{eq:quark-propagator-x-representation-short}. The contribution can be written down in the form
\begin{equation}
\Pi_{\mathcal{O}_{1}^{a}\mathcal{O}_{2}^{b}}^{\langle G^{2}\rangle}(x,y)=-\frac{1}{2}\delta^{ab}\mathrm{Tr}\Big[\Gamma_{1}S(x,y)\Gamma_{2}S(y,x)\Big]_{\langle G^{2}\rangle}\,,\label{eq:gluon_condensate_2pt}
\end{equation}
which gives us three possible diagrams, as shown at Fig.~\ref{fig:fig_2pt_gluon_condensate_v2}.

The gluon condensate, that arises from the diagram \ref{fig:fig_2pt_gluon_condensate_v2_a} at Fig.~\ref{fig:fig_2pt_gluon_condensate_v2}, is given by propagation of two individual gluon tensors due to \eqref{eq:quark-propagator-x-representation-short-factor-2}. Its contribution reads
\begin{align}
\big[\Pi_{\mathcal{O}_{1}^{a}\mathcal{O}_{2}^{b}}^{\langle G^{2}\rangle}(x,y)\big]_{(a)}=&-\frac{\alpha_{s}\langle G^{2}\rangle\delta^{ab}}{192\pi^{3}(x-y)^{4}}\bigg(\frac{x^{2}y^{2}-(x\hspace{-1pt}\cdot\hspace{-1pt}y)^{2}}{(x-y)^{4}}\mathrm{Tr}\big[\Gamma_{1}(\slashed{x}-\slashed{y})\Gamma_{2}(\slashed{x}-\slashed{y})\big]\label{eq:2pt_gluon_condensate_1}\\
&-\frac{1}{4}\big[(x-y)^{2}g_{\rho\sigma}-(x-y)_{\rho}(x-y)_{\sigma}\big]\mathrm{Tr}\big[\Gamma_{1}\gamma^{\rho}\gamma_{5}\Gamma_{2}\gamma^{\sigma}\gamma_{5}\big]\bigg)\,.\nonumber
\end{align}
Although the bottom line of the formula above is invariant with respect to translation, the upper line is not. This part gets cancelled by the contributions of the diagrams \ref{fig:fig_2pt_gluon_condensate_v2_b} and \ref{fig:fig_2pt_gluon_condensate_v2_c}. In fact, these graphs contribute equally and their sum reads
\begin{align}
\big[\Pi_{\mathcal{O}_{1}^{a}\mathcal{O}_{2}^{b}}^{\langle G^{2}\rangle}(x,y)\big]_{(b)}+\big[\Pi_{\mathcal{O}_{1}^{a}\mathcal{O}_{2}^{b}}^{\langle G^{2}\rangle}(x,y)\big]_{(c)}=\frac{\alpha_{s}\langle G^{2}\rangle\delta^{ab}}{192\pi^{3}}\frac{x^{2}y^{2}-(x\hspace{-1pt}\cdot\hspace{-1pt}y)^{2}}{(x-y)^{8}}\mathrm{Tr}\big[\Gamma_{1}(\slashed{x}-\slashed{y})\Gamma_{2}(\slashed{x}-\slashed{y})\big]\,.\label{eq:2pt_gluon_condensate_2}
\end{align}
Adding both contributions \eqref{eq:2pt_gluon_condensate_1}-\eqref{eq:2pt_gluon_condensate_2} together thus leaves us with the total contribution in the translation-invariant form
\begin{align}
\Pi_{\mathcal{O}_{1}^{a}\mathcal{O}_{2}^{b}}^{\langle G^{2}\rangle}(x,y)&=\frac{\alpha_{s}\langle G^{2}\rangle\delta^{ab}}{768\pi^{3}(x-y)^{4}}\big[(x-y)^{2}g_{\rho\sigma}-(x-y)_{\rho}(x-y)_{\sigma}\big]\mathrm{Tr}\big[\Gamma_{1}\gamma^{\rho}\gamma_{5}\Gamma_{2}\gamma^{\sigma}\gamma_{5}\big]\,.\label{eq:2pt_gluon_condensate_total}
\end{align}

Now we are allowed to perform a shift of the coordinations, symbolically enoted as $(x,y)\rightarrow(x-y,0)\equiv(x,0)$. In other words, we can set $y=0$ in \eqref{eq:2pt_gluon_condensate_total}, which effectively means that contribution of diagrams \ref{fig:fig_2pt_gluon_condensate_v2_b} and \ref{fig:fig_2pt_gluon_condensate_v2_c} vanishes identically. Then, after performing the Fourier transform, one finds the result for all the relevant correlators in the momentum representation as follows:
\begin{equation}
\Pi_{\mathcal{O}_{1}^{a}\mathcal{O}_{2}^{b}}^{\langle G^{2}\rangle}(p)=-\frac{i\alpha_{s}\langle G^{2}\rangle\delta^{ab}}{384\pi p^{2}}\bigg(\mathrm{Tr}[\Gamma_{1}\gamma^{\alpha}\gamma_{5}\Gamma_{2}\gamma_{\alpha}\gamma_{5}]+\frac{2}{p^{2}}\mathrm{Tr}[\Gamma_{1}\slashed{p}\gamma_{5}\Gamma_{2}\slashed{p}\gamma_{5}]\bigg)\,.\label{eq:2pt_gluon_condensate_total_x-repr_to_repr}
\end{equation}

\subsection*{Propagation of Nonlocal Gluon Condensate}
For a curious reader, let us only emphasize the complexity of the straightforward evaluation of the gluon condensate contribution to the two-point Green functions, when the vacuum gluons are treated as an external fields, which require us to apply the propagation formula \eqref{eq:propagation_gluon_condensate} and perform the integration over the coordinates of the two gluon fields.

However, since the diagrams \ref{fig:fig_2pt_gluon_condensate_v2_b}-\ref{fig:fig_2pt_gluon_condensate_v2_c} do not contribute, we present here the calculation only for the diagram \ref{fig:fig_2pt_gluon_condensate_v2_a}. We obtain the contribution of the corresponding diagram in the coordinate representation in the following form:\footnote{The evaluation based on this approach is of course trivial and has been discussed in the literature many times. For example, see Ref.~\cite{Nikolaev:1981ff}.}
\begin{align}
\big[\Pi_{\mathcal{O}_{1}^{a}\mathcal{O}_{2}^{b}}^{\langle G^{2}\rangle}&(x,y)\big]_{(a)}=\frac{\pi\alpha_{s}\langle G^{2}\rangle}{48}\delta^{ab}\times\label{eq:2pt_gluon_integration}\\
&\times\int\mathrm{d}^{4}u\,\mathrm{d}^{4}v\,H_{\alpha\beta}^{\langle G^{2}\rangle}(u,v)\,\mathrm{Tr}\Big[\Gamma_{1}S_{0}(x,u)\gamma^{\alpha}S_{0}(u,y)\Gamma_{2}S_{0}(y,v)\gamma^{\beta}S_{0}(v,x)\Big]\,,\nonumber
\end{align}
with the function $H_{\alpha\beta}^{\langle G^{2}\rangle}(u,v)$ given in \eqref{eq:propagation_gluon_condensate}.

As it turns out, it is much efficient to only perform the integration over the coordinates $u$, $v$ of the two gluon fields, leave the expression in the coordinate representation and make the Fourier transform of the results only after inserting the relevant Dirac matrices.

To integrate over the gluon coordinates, we employed once again the trick of assigning arbitrary momenta to the gluon fields, interchanging the integration for the Fourier transform and setting those momenta to zero at the end of the calculation. After making sure that the contribution \eqref{eq:2pt_gluon_integration} above is translation-invariant, it is possible to set $y=0$, which simplifies the expression a bit. After some algebraic manipulations, the contribution reads
\begin{align}
\big[\Pi_{\mathcal{O}_{1}^{a}\mathcal{O}_{2}^{b}}^{\langle G^{2}\rangle}&(x)\big]_{(a)}=-\frac{\pi\alpha_{s}\langle G^{2}\rangle}{48}\delta^{ab}\mathrm{Tr}\big[\Gamma_{1}\gamma^{\alpha}\gamma^{\beta}\gamma^{\gamma}\Gamma_{2}\gamma^{\mu}\gamma^{\nu}\gamma^{\rho}\big]\times\\
\times\bigg[&-\frac{1}{8\pi^{2}}[F_{2}(x)]_{\alpha}\Big(g_{\beta\nu}[F_{9}(x)]_{\gamma\mu\rho}-g_{\gamma\nu}[F_{9}(x)]_{\beta\mu\rho}\Big)-g_{\beta\nu}[F_{9}(x)]_{\mu\rho\sigma}[F_{9}(x)]_{\alpha\gamma\sigma}\nonumber\\
&+\frac{1}{8\pi^{2}}[F_{2}(x)]_{\mu}\Big(g_{\beta\rho}[F_{9}(x)]_{\alpha\gamma\nu}-g_{\beta\nu}[F_{9}(x)]_{\alpha\gamma\rho}\Big)+[F_{9}(x)]_{\beta\mu\rho}[F_{9}(x)]_{\alpha\gamma\nu}\nonumber\\
&-i[F_{8}(x)]_{\mu\rho}\Big(x_{\beta}[F_{9}(x)]_{\alpha\gamma\nu}-g_{\beta\nu}x_{\sigma}[F_{9}(x)]_{\alpha\gamma\sigma}\Big)-\frac{1}{64\pi^{4}}[F_{2}(x)]_{\mu}[F_{2}(x)]_{\alpha}\times\nonumber\\
&\times(g_{\beta\nu}g_{\gamma\rho}-g_{\beta\rho}g_{\gamma\nu})+\frac{i}{8\pi^{2}}[F_{2}(x)]_{\alpha}[F_{8}(x)]_{\mu\rho}(x_{\gamma}g_{\beta\nu}-x_{\beta}g_{\gamma\nu})\bigg]\,,\nonumber
\end{align}
which, after inserting for the respective Dirac matrices and performing necessary algebraical manipulations, gives
\begin{alignat}{4}
\big[\Pi_{VV}^{\langle G^{2}\rangle}(x)\big]_{\mu\nu}^{ab}&=\big[\Pi_{AA}^{\langle G^{2}\rangle}(x)\big]_{\mu\nu}^{ab}&&=-&&\frac{\alpha_{s}\langle G^{2}\rangle}{192\pi^{3}}\delta^{ab}\bigg(\frac{g_{\mu\nu}}{x^{2}}+\frac{2x_{\mu}x_{\nu}}{x^{4}}\bigg)\,,\label{eq:2pt_gluon_condensate_VV_AA_propagation}\\
\big[\Pi_{SS}^{\langle G^{2}\rangle}(x)\big]^{ab}&=\big[\Pi_{PP}^{\langle G^{2}\rangle}(x)\big]^{ab}&&=-&&\frac{\alpha_{s}\langle G^{2}\rangle}{64\pi^{3}}\delta^{ab}\frac{1}{x^{2}}\,,\label{eq:2pt_gluon_condensate_SS_PP_propagation}
\end{alignat}
which is now trivial to convert to the momentum representation.

\subsection*{Results}
Regardless of approach used, the formulas \eqref{eq:2pt_gluon_condensate_total_x-repr_to_repr} and \eqref{eq:2pt_gluon_condensate_VV_AA_propagation}-\eqref{eq:2pt_gluon_condensate_SS_PP_propagation} lead to the same final results in the momentum representation for the individual Green functions as follows:
\begin{align}
\Pi_{VV}^{\langle G^{2}\rangle}(p^{2})&=\Pi_{AA}^{\langle G^{2}\rangle}(p^{2})=\frac{i}{24\pi}\frac{\alpha_{s}\langle G^{2}\rangle}{p^{4}}\,,\label{eq:2pt_gluon_condensate_VV_AA}\\
\Pi_{SS}^{\langle G^{2}\rangle}(p^{2})&=\Pi_{PP}^{\langle G^{2}\rangle}(p^{2})=\frac{i}{16\pi}\frac{\alpha_{s}\langle G^{2}\rangle}{p^{2}}\,.\label{eq:2pt_gluon_condensate_SS_PP}
\end{align}

\subsection{Quark-gluon Condensate}\label{ssec:appendix_quark_gluon_condensate}
As stated above, one expects the contribution of the quark-gluon condensate to the $\langle AP\rangle$ correlator to vanish since the only nonperturbative contribution is given by the quark condensate. Verifying that such contribution vanishes serves as a reliable confirmation that our propagation formulas are correct.

As we have seen in the Section \ref{sec:quark-gluon-condensate}, the contribution of the local quark-gluon condensate is given by effective propagations of the nonlocal quark and quark-gluon condensates. In what follows we will show that both contributions to the $\langle AP\rangle$ correlator indeed cancel each other, as expected. On top of that, to verify such cancellation, it will suffice to stay in the coordinate representation.

\subsection*{Propagation of Nonlocal Quark Condensate}
\begin{figure}[htb]
    \centering
    \begin{subfigure}[t]{0.27\textwidth}
        \hspace{7pt}\includegraphics[width=0.85\textwidth]{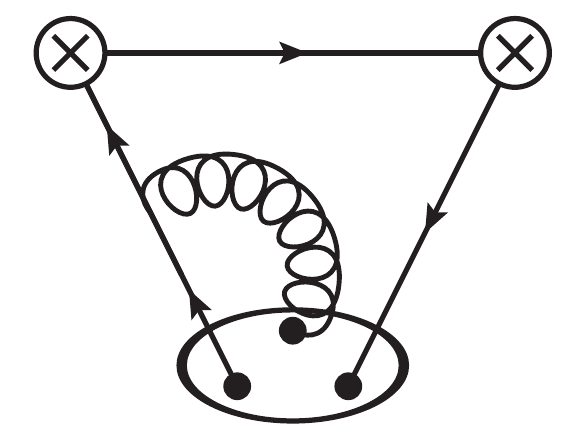}
        \caption{}
        \label{fig:fig_2pt_quark-gluon_condensate_1_a}
    \end{subfigure}
    \begin{subfigure}[t]{0.27\textwidth}
        \hspace{7pt}\includegraphics[width=0.85\textwidth]{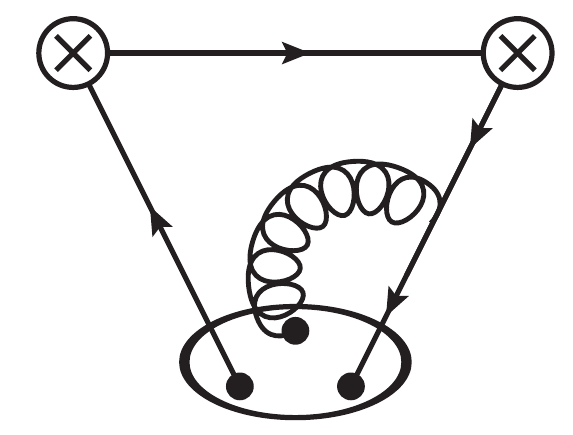}
        \caption{}
        \label{fig:fig_2pt_quark-gluon_condensate_1_b}
    \end{subfigure}
    \caption{Feynman diagrams of the contribution of the quark-gluon condensate to the two-point Green functions due to the effective propagation of the non-local quark condensate.}
    \label{fig:fig_2pt_quark-gluon_condensate_1}
\end{figure}

Let us start with the contribution to the two-point Green functions given by the propagation of the nonlocal quark condensate. The relevant topologies of contributing diagrams are depicted at Fig.~\ref{fig:fig_2pt_quark-gluon_condensate_1}. Since the contribution is effective due to the fact that the soft gluon is attached to the quark lines wit zero momentum, one should not think of the diagrams at Fig.~\ref{fig:fig_2pt_quark-gluon_condensate_1} as of standard Feynman graphs.

Therefore, when evaluating the contribution, one should ignore the gluon line and simply average over the uncontracted quark fields according to the part of the formula \eqref{eq:quark-condensate-propagation} proportional to the function $F^{\langle\overline{q}q\rangle}(x,y)$, given by \eqref{eq:quark-condensate-propagation-function-quark-gluon}.

Using the strategy explained above, one obtains the contribution of the diagrams at Fig.~\ref{fig:fig_2pt_quark-gluon_condensate_1} to be of the form
\begin{align}
\Pi_{\mathcal{O}_{1}^{a}\mathcal{O}_{2}^{b}}^{\langle\overline{q}q\rangle\rightarrow\langle\overline{q}\sigma\cdot Gq\rangle}(x,y)=&-\frac{g_{s}\langle\overline{q}\sigma\hspace{-1pt}\cdot\hspace{-1pt}Gq\rangle}{2^{6}\cdot 3}\delta^{ab}\mathrm{Tr}\Big[F^{\langle\overline{q}q\rangle}(y,x)\Gamma_{2}S_{0}(y,x)\Gamma_{1}\Big]\label{eq:2pt_quark-gluon_condensate_propagation_1}\\
&+\big[(\Gamma_{1},a,x)\leftrightarrow(\Gamma_{2},b,y)\big]\,.\nonumber
\end{align}

An inserting all the relevant combinations of the Dirac matrices and performing the trace leads us to the conclusion, that the only nonvanishing contribution is for the case of the $\langle AP\rangle$ correlator:
\begin{align}
\big[\Pi_{AP}^{\langle\overline{q}q\rangle\rightarrow\langle\overline{q}\sigma\cdot Gq\rangle}(x,y)\big]_{\mu}^{ab}&=-\frac{g_{s}\langle\overline{q}\sigma\hspace{-1pt}\cdot\hspace{-1pt}Gq\rangle\delta^{ab}}{288\pi^{2}(x-y)^{4}}\times\label{eq:AP_quark-gluon_result_1}\\
&\times\Big[x_{\mu}\Big(2(x-y)^{2}+x^{2}-y^{2}\Big)-y_{\mu}\Big(4(x-y)^{2}+x^{2}-y^{2}\Big)\Big]\,.\nonumber
\end{align}

\subsection*{Propagation of Nonlocal Quark-gluon Condensate}
Similarly, using the formula \eqref{eq:4q_propagace_qAq} with the part proportional to $F_{\mu}^{\langle\overline{q}\mathcal{A}q\rangle}(x,y,z)$, given with \eqref{eq:quark-gluon-condensate-propagation-function-quark-gluon}, one finds the contribution stemming from the propagation of the nonlocal quark-gluon condensate to be
\begin{align}
\Pi_{\mathcal{O}_{1}^{a}\mathcal{O}_{2}^{b}}^{\langle\overline{q}\mathcal{A}q\rangle\rightarrow\langle\overline{q}\sigma\cdot Gq\rangle}(x,y)=&-\frac{i g_{s}\langle\overline{q}\sigma\hspace{-1pt}\cdot\hspace{-1pt}Gq\rangle}{2^{6}\cdot 3^{2}}\delta^{ab}\int\mathrm{d}^{4}u\,\mathrm{Tr}\Big[F_{\alpha}^{\langle\overline{q}\mathcal{A}q\rangle}(y,u,x)\Gamma_{2}S_{0}(y,u)\gamma^{\alpha}S_{0}(u,x)\Gamma_{1}\Big]\nonumber\\
&+\big[(\Gamma_{1},a,x)\leftrightarrow(\Gamma_{2},b,y)\big]\,,\label{eq:2pt_quark-gluon_condensate_propagation_2}
\end{align}
where we need to integrate over the coordinate $u$ of the four-potential of the gluon field. This integration is easily performed by introducing momentum $k$, changing the integration for Fourier transform and setting $k=0$ at the end of the calculations, i.e. schematically
\begin{equation}
\int\mathrm{d}^{4}u=\lim_{k\rightarrow 0}\int\mathrm{d}^{4}u\,e^{-ik\cdot u}\,.\label{eq:integration}
\end{equation}

Performing the integration of \eqref{eq:2pt_quark-gluon_condensate_propagation_2} over the coordinate $u$ according to \eqref{eq:integration}, we get
\begin{align}
\Pi_{\mathcal{O}_{1}^{a}\mathcal{O}_{2}^{b}}^{\langle\overline{q}\mathcal{A}q\rangle\rightarrow\langle\overline{q}\sigma\cdot Gq\rangle}(x,y)=&-\frac{g_{s}\langle\overline{q}\sigma\hspace{-1pt}\cdot\hspace{-1pt}Gq\rangle}{2^{6}\cdot 3^{2}}\delta^{ab}\mathrm{Tr}\Big[\sigma_{\alpha\nu}\Gamma_{2}\gamma^{\mu}\gamma^{\nu}\gamma^{\rho}\Gamma_{1}\Big]\times\label{eq:2pt_quark-gluon_condensate_propagation_2_integrated}\\
&\hspace{-60pt}\times\bigg(\frac{1}{8\pi^{2}}g_{\alpha\rho}\big[F_{2}(x-y)\big]_{\mu}-i x_{\alpha}\big[F_{8}(x-y)\big]_{\mu\rho}+\big[F_{9}(x-y)\big]_{\alpha\mu\rho}\bigg)\nonumber\\
&+\big[(\Gamma_{1},a,x)\leftrightarrow(\Gamma_{2},b,y)\big]\,.\nonumber
\end{align}
Then, one can substitute for the specific Dirac matrices and finds out that the quark-gluon condensate contribution to the $\langle AP\rangle$ correlator, given by the propagation of the nonlocal quark-gluon condensate, is the opposite to \eqref{eq:AP_quark-gluon_result_1}:
\begin{equation}
\big[\Pi_{AP}^{\langle\overline{q}\mathcal{A}q\rangle\rightarrow\langle\overline{q}\sigma\cdot Gq\rangle}(x,y)\big]_{\mu}^{ab}=-\big[\Pi_{AP}^{\langle\overline{q}q\rangle\rightarrow\langle\overline{q}\sigma\cdot Gq\rangle}(x,y)\big]_{\mu}^{ab}\,.\label{eq:AP_quark-gluon_result_2}
\end{equation}

\begin{figure}[htb]
  \centering
    \includegraphics[scale=0.55]{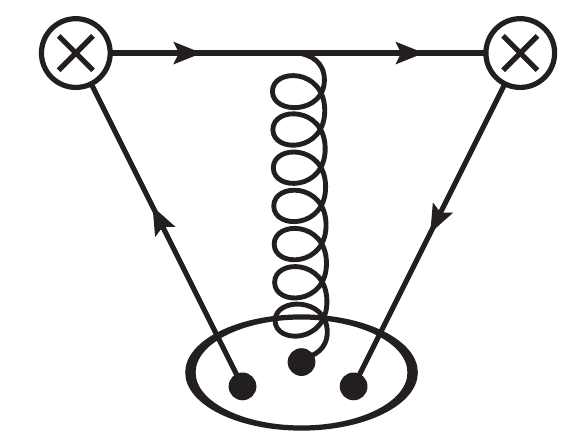}
    \caption{Feynman diagram of the contribution of the quark-gluon condensate to the two-point Green functions due to the effective propagation of the non-local quark-gluon condensate.}
    \label{fig:fig_2pt_quark-gluon_condensate_2}
\end{figure}

\subsection*{Results}
As can be seen from \eqref{eq:AP_quark-gluon_result_2}, both results \eqref{eq:2pt_quark-gluon_condensate_propagation_1} and \eqref{eq:2pt_quark-gluon_condensate_propagation_2_integrated}, obtained from the propagation of the nonlocal condensates, cancel each other,
\begin{equation}
\Pi_{\mathcal{O}_{1}^{a}\mathcal{O}_{2}^{b}}^{\langle\overline{q}q\rangle\rightarrow\langle\overline{q}\sigma\cdot Gq\rangle}(x,y)+\Pi_{\mathcal{O}_{1}^{a}\mathcal{O}_{2}^{b}}^{\langle\overline{q}\mathcal{A}q\rangle\rightarrow\langle\overline{q}\sigma\cdot Gq\rangle}(x,y)=0\,,
\end{equation}
i.e.
\begin{align}
\Pi_{AP}^{\langle\overline{q}\sigma\cdot Gq\rangle}(p^{2})=0\,,\label{eq:AP_quark-gluon_condensate_contribution}
\end{align}
as expected.

To conclude, there is no contribution from the quark-gluon condensate to any two-point Green function in the chiral limit.

\subsection{Four-quark Condensate}\label{ssec:four-quark_condensate}
Since the evaluation of the contribution of the four-quark condensate has been discussed thoroughly in the main text, we present here only the results. They are as follows:

\subsection*{Direct Contribution}
\begin{figure}[htb]
    \centering
    \begin{subfigure}[t]{0.23\textwidth}
        \hspace{8pt}\includegraphics[width=0.9\textwidth]{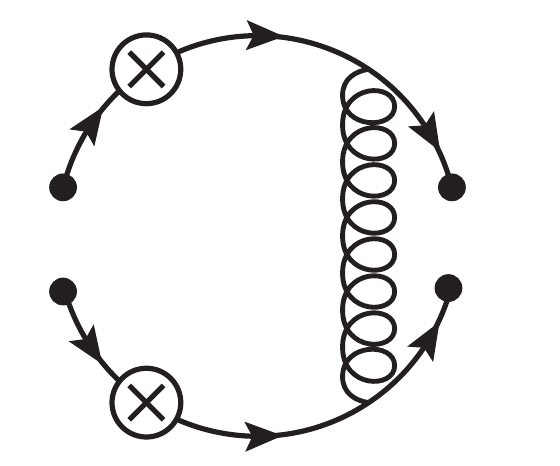}
        \caption{}
        \label{fig:fig_2pt_four-quark_condensate_1_a}
    \end{subfigure}
    \begin{subfigure}[t]{0.23\textwidth}
        \hspace{-3pt}\includegraphics[width=1\textwidth]{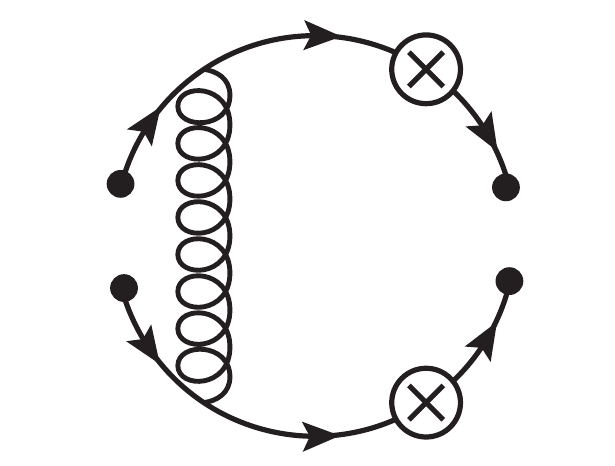}
        \caption{}
        \label{fig:fig_2pt_four-quark_condensate_1_b}
    \end{subfigure}   
    \begin{subfigure}[t]{0.23\textwidth}
        \hspace{-3pt}\includegraphics[width=1\textwidth]{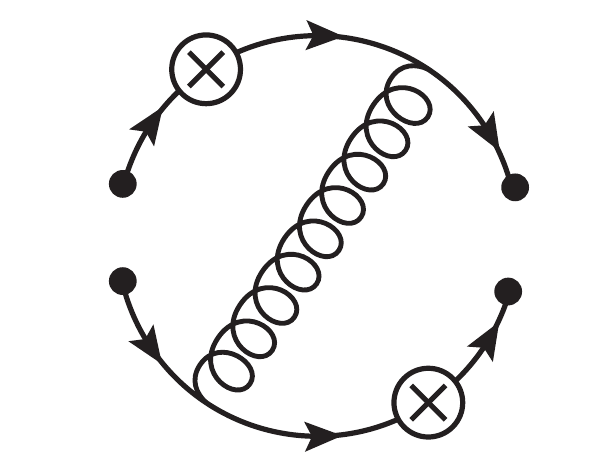}
        \caption{}
        \label{fig:fig_2pt_four-quark_condensate_1_c}
    \end{subfigure}
    \begin{subfigure}[t]{0.23\textwidth}
        \hspace{-3pt}\includegraphics[width=0.9\textwidth]{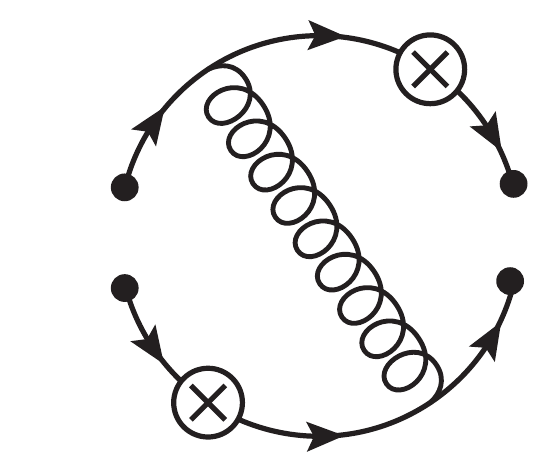}
        \caption{}
        \label{fig:fig_2pt_four-quark_condensate_1_d}
    \end{subfigure}
    \caption{Feynman diagrams of the direct contribution of the four-quark condensate to the two-point Green functions. The averaging over the quark fields is done according to \eqref{eq:4quark-averaging}.}
    \label{fig:fig_2pt_four-quark_condensate_1}
\end{figure}

Perturbative contribution to the four-quark condensate is of simple form. The relevant diagrams are shown at Fig.~\ref{fig:fig_2pt_four-quark_condensate_1} and their contributions read
\begin{align}
\big[\Pi_{\mathcal{O}_{1}^{a}\mathcal{O}_{2}^{b}}^{\mathrm{pert.}\rightarrow\langle\overline{q}q\rangle^{2}}(p)\big]_{(a)}=&-\frac{i\pi\alpha_{s}\langle\overline{q}q\rangle^{2}}{162p^{2}}\delta^{ab}\,\mathrm{Tr}\big[\gamma_{\alpha}S_{0}(p)\Gamma_{1}\gamma^{\alpha}S_{0}(-p)\Gamma_{2}\big]\,,\\
\big[\Pi_{\mathcal{O}_{1}^{a}\mathcal{O}_{2}^{b}}^{\mathrm{pert.}\rightarrow\langle\overline{q}q\rangle^{2}}(p)\big]_{(b)}=&-\frac{i\pi\alpha_{s}\langle\overline{q}q\rangle^{2}}{162p^{2}}\delta^{ab}\,\mathrm{Tr}\big[\Gamma_{1}S_{0}(-p)\gamma_{\alpha}\Gamma_{2}S_{0}(p)\gamma^{\alpha}\big]\,,\\
\big[\Pi_{\mathcal{O}_{1}^{a}\mathcal{O}_{2}^{b}}^{\mathrm{pert.}\rightarrow\langle\overline{q}q\rangle^{2}}(p)\big]_{(c)}=&-\frac{2i\pi\alpha_{s}\langle\overline{q}q\rangle^{2}}{81p^{2}}\delta^{ab}\,\mathrm{Tr}\big[S_{0}(p)\Gamma_{1}\Gamma_{2}S_{0}(p)\big]\,,\\
\big[\Pi_{\mathcal{O}_{1}^{a}\mathcal{O}_{2}^{b}}^{\mathrm{pert.}\rightarrow\langle\overline{q}q\rangle^{2}}(p)\big]_{(d)}=&-\frac{2i\pi\alpha_{s}\langle\overline{q}q\rangle^{2}}{81p^{2}}\delta^{ab}\,\mathrm{Tr}\big[\Gamma_{1}S_{0}(-p)S_{0}(-p)\Gamma_{2}\big]\,.
\end{align}

Summing up all the contributions above, and substituting for the individual $\Gamma$-matrices, leads to the following results:
\begin{alignat}{2}
\big[\Pi_{VV}^{\mathrm{pert.}\rightarrow\langle\overline{q}q\rangle^{2}}(p)\big]_{\mu\nu}^{ab}&=-\big[\Pi_{AA}^{\mathrm{pert.}\rightarrow\langle\overline{q}q\rangle^{2}}(p)\big]_{\mu\nu}^{ab}&&=\frac{16i\pi\alpha_{s}\langle\overline{q}q\rangle^{2}}{81p^{6}}(p^{2}g^{\mu\nu}-p^{\mu}p^{\nu})\delta^{ab}\,,\label{eq:2pt_four-quark_condensate_perturbative_result_VV_AA}\\
\big[\Pi_{SS}^{\mathrm{pert.}\rightarrow\langle\overline{q}q\rangle^{2}}(p)\big]^{ab}&=-\big[\Pi_{PP}^{\mathrm{pert.}\rightarrow\langle\overline{q}q\rangle^{2}}(p)\big]^{ab}&&=\frac{8i\pi\alpha_{s}\langle\overline{q}q\rangle^{2}}{27p^{4}}\delta^{ab}\,.\label{eq:2pt_four-quark_condensate_perturbative_result_SS_PP}
\end{alignat}
We alert the reader to notice the opposite sign between the contributions of the respective pairs of correlators.

\subsection*{Propagation of Nonlocal Quark Condensate}
\begin{figure}[htb]
    \centering
    \begin{subfigure}[t]{0.27\textwidth}
        \hspace{7pt}\includegraphics[width=0.85\textwidth]{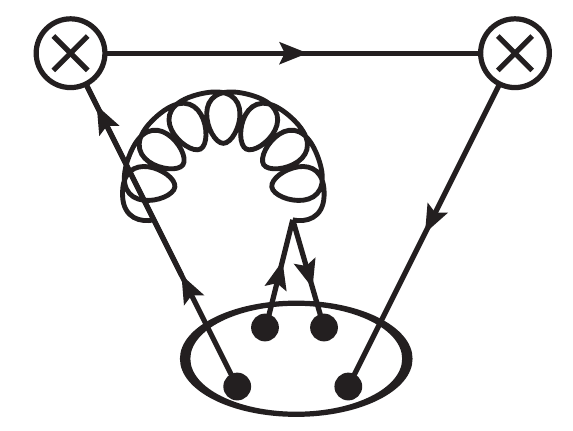}
        \caption{}
        \label{fig:fig_2pt_four-quark_condensate_2_a}
    \end{subfigure}
    \begin{subfigure}[t]{0.27\textwidth}
        \hspace{7pt}\includegraphics[width=0.85\textwidth]{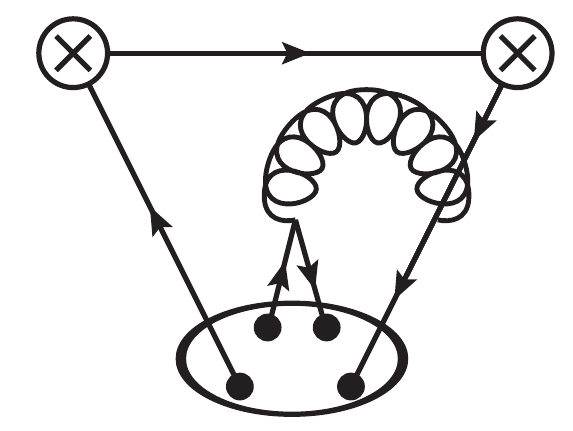}
        \caption{}
        \label{fig:fig_2pt_four-quark_condensate_2_b}
    \end{subfigure}
    \caption{Feynman diagrams of the contribution of the four-quark condensate to the two-point Green functions due to the effective propagation of the nonlocal quark condensate.}
    \label{fig:fig_2pt_four-quark_condensate_2}
\end{figure}

The contributing topologies are shown on the Fig.~\ref{fig:fig_2pt_four-quark_condensate_2}. Due to the same reasons as we have explained in the beginning of the previous section, we also here understand both graphs to be the same, and, obviously, contributing equally.

The contribution of such propagation is of simple form:
\begin{align}
\Pi_{\mathcal{O}_{1}^{a}\mathcal{O}_{2}^{b}}^{\langle\overline{q}q\rangle\rightarrow\langle\overline{q}q\rangle^{2}}(x,y)=&\quad\,\frac{i\pi\alpha_{s}\langle\overline{q}q\rangle^{2}}{2^{4}\cdot 3^{6}}\delta^{ab}\mathrm{Tr}\Big[G^{\langle\overline{q}q\rangle}(y,x)\Gamma_{2}S_{0}(y,x)\Gamma_{1}\Big]\\
&+\big[(\Gamma_{1},a,x)\leftrightarrow(\Gamma_{2},b,y)\big]\,,\nonumber
\end{align}
with $G^{\langle\overline{q}q\rangle}(x,y)$ given by \eqref{eq:quark-condensate-propagation-function}. However, after inserting the Dirac matrices and evaluating the trace, we get somewhat lengthy expressions for individual contributions. After assuring ourselves that these results, together with the results \eqref{eq:4q-condensate-propagace-2-vysledek-1}-\eqref{eq:4q-condensate-propagace-2-vysledek-2} of effective propagation of the quark-gluon condensate, are translation-invariant, it allows us to set $y=0$, which gives us
\begin{alignat}{2}
\big[\Pi_{VV}^{\langle\overline{q}q\rangle\rightarrow\langle\overline{q}q\rangle^{2}}(x)\big]_{\mu\nu}^{ab}&=\big[\Pi_{AA}^{\langle\overline{q}q\rangle\rightarrow\langle\overline{q}q\rangle^{2}}(x)\big]_{\mu\nu}^{ab}=&&-\frac{\alpha_{s}\langle\overline{q}q\rangle^{2}}{2916\pi}\delta^{ab}\bigg(g^{\mu\nu}-\frac{2x^{\mu}x^{\nu}}{x^{2}}\bigg)\,,\\
\big[\Pi_{SS}^{\langle\overline{q}q\rangle\rightarrow\langle\overline{q}q\rangle^{2}}(x)\big]^{ab}&=\big[\Pi_{PP}^{\langle\overline{q}q\rangle\rightarrow\langle\overline{q}q\rangle^{2}}(x)\big]^{ab}=&&\quad\,\frac{\alpha_{s}\langle\overline{q}q\rangle^{2}}{2916\pi}\delta^{ab}\,,
\end{alignat}
and which finally leads to the results in the momentum representation:
\begin{alignat}{2}
\big[\Pi_{VV}^{\langle\overline{q}q\rangle\rightarrow\langle\overline{q}q\rangle^{2}}(p)\big]_{\mu\nu}^{ab}&=\big[\Pi_{AA}^{\langle\overline{q}q\rangle\rightarrow\langle\overline{q}q\rangle^{2}}(p)\big]_{\mu\nu}^{ab}=&&-\frac{4i\pi\alpha_{s}\langle\overline{q}q\rangle^{2}}{729p^{6}}(p^{2}g^{\mu\nu}-4p^{\mu}p^{\nu})\delta^{ab}\,,\label{eq:2pt_four-quark_condensate_propagation_1_result_VV_AA}\\
\big[\Pi_{SS}^{\langle\overline{q}q\rangle\rightarrow\langle\overline{q}q\rangle^{2}}(p)\big]^{ab}&=\big[\Pi_{PP}^{\langle\overline{q}q\rangle\rightarrow\langle\overline{q}q\rangle^{2}}(p)\big]^{ab}=&&\quad\,0\,.\label{eq:2pt_four-quark_condensate_propagation_1_result_SS_PP}
\end{alignat}

\subsection*{Propagation of Nonlocal Quark-gluon Condensate}
\begin{figure}[htb]
  \centering
    \includegraphics[scale=0.6]{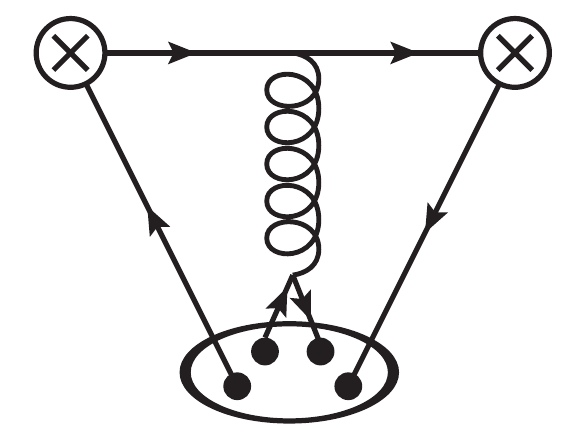}
    \caption{Feynman diagram of the contribution of the four-quark condensate to the two-point Green functions due to the effective propagation of the nonlocal quark-gluon condensate.}
    \label{fig:fig_2pt_four-quark_condensate_3}
\end{figure}

The contribution of the graphs at Fig.~\ref{fig:fig_2pt_four-quark_condensate_3} reads
\begin{align}
\Pi_{\mathcal{O}_{1}^{a}\mathcal{O}_{2}^{b}}^{\langle\overline{q}\mathcal{A}q\rangle\rightarrow\langle\overline{q}q\rangle^{2}}&(x,y)=-\frac{i\pi\alpha_{s}\langle\overline{q}q\rangle^{2}}{2^{2}\cdot 3^{5}}\delta^{ab}\times\\
&\hspace{-50pt}\times\int\mathrm{d}^{4}u\,\mathrm{Tr}\Big[G_{\alpha}^{\langle\overline{q}\mathcal{A}q\rangle}(y,u,x)\Gamma_{2}S_{0}(y,u)\gamma^{\alpha}S_{0}(u,x)\Gamma_{1}\Big]+\big[(\Gamma_{1},a,x)\leftrightarrow(\Gamma_{2},b,y)\big]\,,\nonumber
\end{align}
where $G_{\alpha}^{\langle\overline{q}\mathcal{A}q\rangle}(x,u,y)$ is given by \eqref{eq:4q_propagace_qAq-function_contracted}. Performing the integration over the gluon field is quite lengthy in this case, and the result reads
\begin{align}
\Pi_{\mathcal{O}_{1}^{a}\mathcal{O}_{2}^{b}}^{\langle\overline{q}\mathcal{A}q\rangle\rightarrow\langle\overline{q}q\rangle^{2}}&(x,y)=-\frac{\pi\alpha_{s}\langle\overline{q}q\rangle^{2}}{2^{2}\cdot 3^{5}}\delta^{ab}\times\\
&\times\bigg[-i\,\mathrm{Tr}\big[t_{\alpha\nu}^{(2)}(y,x)\Gamma_{2}\gamma^{\mu}\gamma^{\nu}\gamma^{\rho}\Gamma_{1}\big]\times\nonumber\\
&\hspace{20pt}\times\bigg(\frac{1}{8\pi^{2}}g_{\alpha\rho}\big[F_{2}(x-y)\big]_{\mu}-i x_{\alpha}\big[F_{8}(x-y)\big]_{\mu\rho}+\big[F_{9}(x-y)\big]_{\alpha\mu\rho}\bigg)\nonumber\\
&\hspace{20pt}-\frac{2}{3}\mathrm{Tr}\big[t_{\beta\alpha\nu}^{(1)}\Gamma_{2}\gamma^{\mu}\gamma^{\nu}\gamma^{\rho}\Gamma_{1}\big]\times\nonumber\\
&\hspace{20pt}\times\bigg(-\frac{i}{8\pi^{2}}(x_{\alpha}g_{\beta\rho}+x_{\beta}g_{\alpha\rho})\big[F_{2}(x-y)\big]_{\mu}-x_{\alpha}x_{\beta}\big[F_{8}(x-y)\big]_{\mu\rho}\nonumber\\
&\hspace{20pt}+\big[F_{12}(x-y)\big]_{\alpha\beta\mu\rho}-i x_{\alpha}\big[F_{9}(x-y)\big]_{\beta\mu\rho}-i x_{\beta}\big[F_{9}(x-y)\big]_{\alpha\mu\rho}\nonumber\\
&\hspace{20pt}-g_{\beta\rho}\big[F_{11}(x-y)\big]_{\alpha\mu}-g_{\alpha\rho}\big[F_{11}(x-y)\big]_{\beta\mu}-g_{\alpha\beta}\big[F_{11}(x-y)\big]_{\mu\rho}\bigg)\bigg]\nonumber\\
&+\big[(\Gamma_{1},a,x)\leftrightarrow(\Gamma_{2},b,y)\big]\,,\nonumber
\end{align}
where we have denoted the respective tensors as
\begin{align}
t_{\alpha\beta\mu}^{(1)}&=i(g^{\alpha\beta}\gamma^{\mu}-g^{\alpha\mu}\gamma^{\beta})\,,\\
t_{\alpha\beta}^{(2)}(x,y)&=\frac{i}{2}\big[(x+y)^{\alpha}\gamma^{\beta}-(x+y)^{\beta}\gamma^{\alpha}\big]-\frac{1}{2}\varepsilon^{(x-y)\alpha\beta\mu}\gamma_{\mu}\gamma_{5}\,.
\end{align}

However, after substituing for the specific Dirac matrices, the previous results simplifies a lot and we obtain
\begin{alignat}{3}
&\big[\Pi_{VV}^{\langle\overline{q}\mathcal{A}q\rangle\rightarrow\langle\overline{q}q\rangle^{2}}(x)\big]_{\mu\nu}^{ab}&&=\big[\Pi_{AA}^{\langle\overline{q}\mathcal{A}q\rangle\rightarrow\langle\overline{q}q\rangle^{2}}(x)\big]_{\mu\nu}^{ab}=&&\label{eq:4q-condensate-propagace-2-vysledek-1}\\
& && &&\hspace{-81pt}=-\frac{\alpha_{s}\langle\overline{q}q\rangle^{2}}{2916\pi}\delta^{ab}\bigg[\Big(6\log(-x^{2})-5\Big)g_{\mu\nu}-\frac{2x_{\mu}x_{\nu}}{x^{2}}\bigg]\,,\nonumber\\
&\big[\Pi_{SS}^{\langle\overline{q}\mathcal{A}q\rangle\rightarrow\langle\overline{q}q\rangle^{2}}(x)\big]^{ab}&&=\big[\Pi_{PP}^{\langle\overline{q}\mathcal{A}q\rangle\rightarrow\langle\overline{q}q\rangle^{2}}(x)\big]^{ab}=&&\frac{\alpha_{s}\langle\overline{q}q\rangle^{2}}{2916\pi}\delta^{ab}\Big(7+12\log(-x^{2})\Big)\,.\label{eq:4q-condensate-propagace-2-vysledek-2}
\end{alignat}

Performing the Fourier transform is now easy and the result in the momentum representation reads
\begin{alignat}{3}
&\big[\Pi_{VV}^{\langle\overline{q}\mathcal{A}q\rangle\rightarrow\langle\overline{q}q\rangle^{2}}(p)\big]_{\mu\nu}^{ab}&&=\big[\Pi_{AA}^{\langle\overline{q}\mathcal{A}q\rangle\rightarrow\langle\overline{q}q\rangle^{2}}(p)\big]_{\mu\nu}^{ab}=-&&\frac{4i\pi\alpha_{s}\langle\overline{q}q\rangle^{2}}{729p^{6}}(7p^{2}g^{\mu\nu}-4p^{\mu}p^{\nu})\delta^{ab}\,,\label{eq:2pt_four-quark_condensate_propagation_2_result_VV_AA}\\
&\big[\Pi_{SS}^{\langle\overline{q}\mathcal{A}q\rangle\rightarrow\langle\overline{q}q\rangle^{2}}(p)\big]^{ab}&&=\big[\Pi_{PP}^{\langle\overline{q}\mathcal{A}q\rangle\rightarrow\langle\overline{q}q\rangle^{2}}(p)\big]^{ab}=&&\frac{16i\pi\alpha_{s}\langle\overline{q}q\rangle^{2}}{243p^{4}}\delta^{ab}\,.\label{eq:2pt_four-quark_condensate_propagation_2_result_SS_PP}
\end{alignat}

\subsection*{Results}
Taking all the individual results \eqref{eq:2pt_four-quark_condensate_perturbative_result_VV_AA}-\eqref{eq:2pt_four-quark_condensate_perturbative_result_SS_PP}, \eqref{eq:2pt_four-quark_condensate_propagation_1_result_VV_AA}-\eqref{eq:2pt_four-quark_condensate_propagation_1_result_SS_PP} and \eqref{eq:2pt_four-quark_condensate_propagation_2_result_VV_AA}-\eqref{eq:2pt_four-quark_condensate_propagation_2_result_SS_PP} together, we arrive at the total contribution of the four-quark condensate to the individual two-point Green functions:
\begin{align}
\Pi_{VV}^{\langle\overline{q}q\rangle^{2}}(p^{2})=&\quad\,\frac{112}{729}\frac{i\pi\alpha_{s}\langle\overline{q}q\rangle^{2}}{p^{6}}\,,\label{eq:2pt_four-quark_condensate_VV}\\
\Pi_{AA}^{\langle\overline{q}q\rangle^{2}}(p^{2})=&-\frac{176}{729}\frac{i\pi\alpha_{s}\langle\overline{q}q\rangle^{2}}{p^{6}}\,,\label{eq:2pt_four-quark_condensate_AA}\\
\Pi_{SS}^{\langle\overline{q}q\rangle^{2}}(p^{2})=&\quad\,\frac{88}{243}\frac{i\pi\alpha_{s}\langle\overline{q}q\rangle^{2}}{p^{4}}\,,\label{eq:2pt_four-quark_condensate_SS}\\
\Pi_{PP}^{\langle\overline{q}q\rangle^{2}}(p^{2})=&-\frac{56}{243}\frac{i\pi\alpha_{s}\langle\overline{q}q\rangle^{2}}{p^{4}}\,.\label{eq:2pt_four-quark_condensate_PP}
\end{align}

%%%%%%%%%%%%%%%%%%%%%%%%%%%%%%%%%%%%%%%%%%%%%%%%%%%%%%%%%%%%%%%%%%%%%%%%%%%%%%%%%%%%%%%%%%%%%%%%%%%%%%%%%
%%%%%%%%%%%%%%%%%%%%%%%%%%%%%%%%%%%%%%%%%%%%%%%%%%%%%%%%%%%%%%%%%%%%%%%%%%%%%%%%%%%%%%%%%%%%%%%%%%%%%%%%%
%%% Section: On Decompositions of the VVA and AAA Green Function
%%%%%%%%%%%%%%%%%%%%%%%%%%%%%%%%%%%%%%%%%%%%%%%%%%%%%%%%%%%%%%%%%%%%%%%%%%%%%%%%%%%%%%%%%%%%%%%%%%%%%%%%%
%%%%%%%%%%%%%%%%%%%%%%%%%%%%%%%%%%%%%%%%%%%%%%%%%%%%%%%%%%%%%%%%%%%%%%%%%%%%%%%%%%%%%%%%%%%%%%%%%%%%%%%%%

\section{On Decompositions of the \texorpdfstring{\boldmath$\langle VVA\rangle$}{} and \texorpdfstring{\boldmath$\langle AAA\rangle$}{} Green Functions}\label{sec:VVA_AAA_decompositions}
In this appendix, we show the derivation of the decompositions of the transversal Lorentz parts of the $\langle VVA\rangle$ and $\langle AAA\rangle$ correlators \eqref{eq:vva-definition_transversal}-\eqref{eq:aaa-definition_transversal}.

In order to obtain a decomposition of any correlator, one has to take into account two requirements: the Bose symmetries and the Ward identities of such correlator. We now show explicit forms of these requirements for the two Green functions in question.

\subsection*{Bose Symmetry}
The $\langle VVA\rangle$ Green function, as defined in \eqref{eq:vva-definition}, has only one Bose symmetry, which is invariance with respect to the interchange $(\mu,a,p)\leftrightarrow(\nu,b,q)$. Due to the symmetry of the flavor part, the Bose symmetry dictates the following requirement on the Lorentz part:
\begin{equation}
\big[\Pi_{VVA}(p,q;r)\big]_{\mu\nu\rho}-\big[\Pi_{VVA}(q,p;r)\big]_{\nu\mu\rho}=0\,.\label{eq:VVA_bose_relation}
\end{equation}

On the other hand, the $\langle AAA\rangle$ Green function, as defined in \eqref{eq:aaa-definition}, is a bit more complicated since there are three relations due to the Bose symmetry:
\begin{align}
\big[\Pi_{AAA}(p,q;r)\big]_{\mu\nu\rho}-\big[\Pi_{AAA}(q,p;r)\big]_{\nu\mu\rho}&=0\,,\label{eq:AAA_bose_relation_1}\\
\big[\Pi_{AAA}(p,q;r)\big]_{\mu\nu\rho}-\big[\Pi_{AAA}(r,q;p)\big]_{\rho\nu\mu}&=0\,,\label{eq:AAA_bose_relation_2}\\
\big[\Pi_{AAA}(p,q;r)\big]_{\mu\nu\rho}-\big[\Pi_{AAA}(p,r;q)\big]_{\mu\rho\nu}&=0\,.\label{eq:AAA_bose_relation_3}
\end{align}
The first one is equivalent to \eqref{eq:VVA_bose_relation}, while \eqref{eq:AAA_bose_relation_2} and \eqref{eq:AAA_bose_relation_3} stand for the interchanges of $(\mu,a,p)\leftrightarrow(\rho,c,r)$ and $(\nu,b,q)\leftrightarrow(\rho,c,r)$, respectively.

As one can easily see that the longitudinal parts \eqref{eq:vva-definition_longitudinal}-\eqref{eq:aaa-definition_longitudinal} of these correlators satisfy the Bose symmetries instantly. Then, when searching for the complete decompositions \eqref{eq:vva-definition_longitudinal_and_transversal}-\eqref{eq:aaa-definition_longitudinal_and_transversal}, it is sufficient to demand the fulfillment of the Bose symmetries only for the transversal parts.

\subsection*{Ward Identities}
The full Ward identities of the $\langle VVA\rangle$ and $\langle AAA\rangle$ correlators are \eqref{eq:VVA-Ward_kopie_1}-\eqref{eq:VVA-Ward_kopie_3} and \eqref{eq:AAA-Ward_kopie_1}-\eqref{eq:AAA-Ward_kopie_3}, respectively. Naturally, the nonzero right-hand sides of these Ward identities are given by the longitudinal parts \eqref{eq:vva-definition_longitudinal}-\eqref{eq:aaa-definition_longitudinal},
\begin{align}
\lbrace p^{\mu},q^{\nu},r^{\rho}\rbrace\big[\Pi_{VVA}^{(L)}(p,q;r)\big]_{\mu\nu\rho}&=\bigg\lbrace 0,0,-\frac{i N_{c}}{8\pi^{2}}\varepsilon^{\mu\nu(p)(q)}\bigg\rbrace\,,\label{eq:VVA_ward_v2a}\\
\lbrace p^{\mu},q^{\nu},r^{\rho}\rbrace\big[\Pi_{AAA}^{(L)}(p,q;r)\big]_{\mu\nu\rho}&=\bigg\lbrace -\frac{i N_{c}}{24\pi^{2}}\varepsilon^{\nu\rho(p)(q)},\frac{i N_{c}}{24\pi^{2}}\varepsilon^{\mu\rho(p)(q)},-\frac{i N_{c}}{24\pi^{2}}\varepsilon^{\mu\nu(p)(q)}\bigg\rbrace\,,\label{eq:AAA_ward_v2a}
\end{align}
while the right-hand sides of the Ward identities for the transversal parts must vanish by definition,
\begin{align}
\lbrace p^{\mu},q^{\nu},r^{\rho}\rbrace\big[\Pi_{VVA}^{(T)}(p,q;r)\big]_{\mu\nu\rho}&=\lbrace 0,0,0\rbrace\,,\label{eq:VVA_ward_v2b}\\
\lbrace p^{\mu},q^{\nu},r^{\rho}\rbrace\big[\Pi_{AAA}^{(T)}(p,q;r)\big]_{\mu\nu\rho}&=\lbrace 0,0,0\rbrace\,.\label{eq:AAA_ward_v2b}
\end{align}

\subsection*{Tensor Base}
The decomposition of these correlators must be built out of momenta contracted with the Levi-Civita tensor in such way that the structures carry three Lorentz indices. Taking the conservation of momenta into account, we can reduce the number of tensors to 8: 
\begin{eqnarray}
&&p^{\mu}\varepsilon^{\nu\rho(p)(q)},\; p^{\nu}\varepsilon^{\mu\rho(p)(q)},\; q^{\mu}\varepsilon^{\nu\rho(p)(q)},\; q^{\nu}\varepsilon^{\mu\rho(p)(q)},\nonumber\\
&&p^{\rho}\varepsilon^{\mu\nu(p)(q)},\; q^{\rho}\varepsilon^{\mu\nu(p)(q)},\; \varepsilon^{\mu\nu\rho(p)},\; \varepsilon^{\mu\nu\rho(q)}.\label{eq:table_VVA_AAA}
\end{eqnarray}

\subsection{\texorpdfstring{$\langle VVA\rangle$}{} Green Function}
Our task now is to find the transversal part of the $\langle VVA\rangle$ Green function such that it satisfies the Bose symmetry \eqref{eq:VVA_bose_relation} and the Ward identities in the form \eqref{eq:VVA_ward_v2b}. We start with the requirement of the Bose symmetry.\footnote{The authors are grateful to Marc Knecht for sharing the manuscript of \cite{Knecht:2020xyr} before its publishing, where the derivation of the $\langle VVA\rangle$ decomposition can be also found. Our derivation is found on the same principles and follows the same procedure. However, in comparison with Ref.~\cite{Knecht:2020xyr}, we present an approach with slightly detailed explanations of individual steps. Also, our notation and normalization differs.}

\subsection*{Bose Symmetry}
Following introductory remarks in the paragraphs above, we can write down the transversal part of the $\langle VVA\rangle$ Green function in the form of a sum of the eight tensors shown in  (\ref{eq:table_VVA_AAA}), with respective formfactors that are functions of squares of momenta. However, to make the fulfillment of the Bose symmetry more apparent, we will take suitable combinations of these tensors into account.
Therefore, let us define the relevant tensors,
\begin{align}
t_{\mu\nu\rho}^{(1)}(p,q;r)&=p^{\mu}\varepsilon^{\nu\rho(p)(q)}+q^{\nu}\varepsilon^{\mu\rho(p)(q)}\,,\label{eq:VVA_tensors_14}\\
t_{\mu\nu\rho}^{(2)}(p,q;r)&=p^{\mu}\varepsilon^{\nu\rho(p)(q)}-q^{\nu}\varepsilon^{\mu\rho(p)(q)}\,,\label{eq:VVA_tensors_14b}\\
t_{\mu\nu\rho}^{(3)}(p,q;r)&=p^{\nu}\varepsilon^{\mu\rho(p)(q)}+q^{\mu}\varepsilon^{\nu\rho(p)(q)}\,,\label{eq:VVA_tensors_23}\\
t_{\mu\nu\rho}^{(4)}(p,q;r)&=p^{\nu}\varepsilon^{\mu\rho(p)(q)}-q^{\mu}\varepsilon^{\nu\rho(p)(q)}\,,\label{eq:VVA_tensors_23b}\\
t_{\mu\nu\rho}^{(5)}(p,q;r)&=p^{\rho}\varepsilon^{\mu\nu(p)(q)}+q^{\rho}\varepsilon^{\mu\nu(p)(q)}\,,\label{eq:VVA_tensors_56}\\
t_{\mu\nu\rho}^{(6)}(p,q;r)&=p^{\rho}\varepsilon^{\mu\nu(p)(q)}-q^{\rho}\varepsilon^{\mu\nu(p)(q)}\,,\label{eq:VVA_tensors_56b}\\
t_{\mu\nu\rho}^{(7)}(p,q;r)&=\varepsilon^{\mu\nu\rho(p)}+\varepsilon^{\mu\nu\rho(q)}\,,\label{eq:VVA_tensors_78}\\
t_{\mu\nu\rho}^{(8)}(p,q;r)&=\varepsilon^{\mu\nu\rho(p)}-\varepsilon^{\mu\nu\rho(q)}\,,\label{eq:VVA_tensors_78b}
\end{align}
and write down the transversal part of \eqref{eq:vva-definition_longitudinal_and_transversal} as
\begin{equation}
\big[\Pi_{VVA}^{(T)}(p,q;r)\big]_{\mu\nu\rho}=\sum_{i=1}^{8}a_{i}(p^{2},q^{2};r^{2})t_{\mu\nu\rho}^{(i)}(p,q;r)\,.\label{eq:vva-definition_transversal_1}
\end{equation}

The structure \eqref{eq:vva-definition_transversal_1}, consisted of eight terms, can be further simplified upon taking the following Schouten identities into account:
\begin{align}
p^{\mu}\varepsilon^{\nu\rho(p)(q)}-p^{\nu}\varepsilon^{\mu\rho(p)(q)}+p^{\rho}\varepsilon^{\mu\nu(p)(q)}-p^{2}\varepsilon^{\mu\nu\rho(q)}+(p\hspace{-1pt}\cdot\hspace{-1pt}q)\varepsilon^{\mu\nu\rho(p)}&=0\,,\label{eq:schouten_1}\\
q^{\mu}\varepsilon^{\nu\rho(p)(q)}-q^{\nu}\varepsilon^{\mu\rho(p)(q)}+q^{\rho}\varepsilon^{\mu\nu(p)(q)}+q^{2}\varepsilon^{\mu\nu\rho(p)}-(p\hspace{-1pt}\cdot\hspace{-1pt}q)\varepsilon^{\mu\nu\rho(q)}&=0\,.\label{eq:schouten_2}
\end{align}

In fact, we can rewrite two of the eight tensors \eqref{eq:VVA_tensors_14}-\eqref{eq:VVA_tensors_78} in terms of the remaining six. We choose to eliminate the following ones:
\begin{align}
t_{\mu\nu\rho}^{(2)}(p,q;r)=&\quad\,t_{\mu\nu\rho}^{(4)}(p,q;r)-t_{\mu\nu\rho}^{(5)}(p,q;r)+\frac{1}{2}(p^{2}-q^{2})t_{\mu\nu\rho}^{(7)}(p,q;r)\label{eq:VVA_tensor_14minus}\\
&-\frac{1}{2}(p+q)^{2}t_{\mu\nu\rho}^{(8)}(p,q;r)\,,\nonumber\\
t_{\mu\nu\rho}^{(3)}(p,q;r)=&\quad\,t_{\mu\nu\rho}^{(1)}(p,q;r)+t_{\mu\nu\rho}^{(6)}(p,q;r)-\frac{1}{2}(p-q)^{2}t_{\mu\nu\rho}^{(7)}(p,q;r)\label{eq:VVA_tensor_23plus}\\
&+\frac{1}{2}(p^{2}-q^{2})t_{\mu\nu\rho}^{(8)}(p,q;r)\,.\nonumber
\end{align}

Then, upon substituting \eqref{eq:VVA_tensor_14minus}-\eqref{eq:VVA_tensor_23plus} back into \eqref{eq:vva-definition_transversal_1}, we are left with the transversal part of the $\langle VVA\rangle$ correlator given by six terms, i.e.
\begin{align}
\big[\Pi_{VVA}^{(T)}(p,q;r)\big]_{\mu\nu\rho}=&\quad\,a_{9}(p^{2},q^{2};r^{2})t_{\mu\nu\rho}^{(1)}(p,q;r)+a_{10}(p^{2},q^{2};r^{2})t_{\mu\nu\rho}^{(4)}(p,q;r)\label{eq:vva-definition_transversal_2}\\
&+a_{11}(p^{2},q^{2};r^{2})t_{\mu\nu\rho}^{(6)}(p,q;r)+a_{12}(p^{2},q^{2};r^{2})t_{\mu\nu\rho}^{(5)}(p,q;r)\nonumber\\
&+a_{13}(p^{2},q^{2};r^{2})t_{\mu\nu\rho}^{(8)}(p,q;r)+a_{14}(p^{2},q^{2};r^{2})t_{\mu\nu\rho}^{(7)}(p,q;r)\,,\nonumber
\end{align}
where the new formfactors are linear combinations of the previous ones.

An advantage in choosing the transversal part of the $\langle VVA\rangle$ correlator in the form of \eqref{eq:vva-definition_transversal_1} is that the tensor structure $t^{(1)}$, $t^{(3)}$, $t^{(6)}$ are antisymmetric and $t^{(2)}$, $t^{(4)}$, $t^{(5)}$ symmetric upon exchanging $p,\mu\leftrightarrow q,\nu$. In other words, the Bose symmetry requires these conditions to be satisfied:
\begin{align}
a_{9}(p^{2},q^{2};r^{2})+a_{9}(q^{2},p^{2};r^{2})&=0\,,\label{eq:VVA_transversal_part_formfactor_1}\\
a_{10}(p^{2},q^{2};r^{2})-a_{10}(q^{2},p^{2};r^{2})&=0\,,\label{eq:VVA_transversal_part_formfactor_2}\\
a_{11}(p^{2},q^{2};r^{2})+a_{11}(q^{2},p^{2};r^{2})&=0\,,\label{eq:VVA_transversal_part_formfactor_3}\\
a_{12}(p^{2},q^{2};r^{2})-a_{12}(q^{2},p^{2};r^{2})&=0\,,\label{eq:VVA_transversal_part_formfactor_4}\\
a_{13}(p^{2},q^{2};r^{2})-a_{13}(q^{2},p^{2};r^{2})&=0\,,\label{eq:VVA_transversal_part_formfactor_5}\\
a_{14}(p^{2},q^{2};r^{2})+a_{14}(q^{2},p^{2};r^{2})&=0\,.\label{eq:VVA_transversal_part_formfactor_6}
\end{align}

\subsection*{Ward Identities}
Now, having the transversal structure \eqref{eq:vva-definition_transversal_2} of the $\langle VVA\rangle$ Green function with the formfactors satisfying the conditions \eqref{eq:VVA_transversal_part_formfactor_1}-\eqref{eq:VVA_transversal_part_formfactor_6}, one can try to further simplify the structure with the requirement of the fulfillment of the respective Ward identities. Indeed, the vanishing vector Ward identities lead to the relations
\begin{align}
p^{2}a_{9}(p^{2},q^{2};r^{2})-(p\hspace{-1pt}\cdot\hspace{-1pt}q)a_{10}(p^{2},q^{2};r^{2})-a_{13}(p^{2},q^{2};r^{2})+a_{14}(p^{2},q^{2};r^{2})&=0\,,\label{eq:VVA_vector_WI_1}\\
q^{2}a_{9}(p^{2},q^{2};r^{2})+(p\hspace{-1pt}\cdot\hspace{-1pt}q)a_{10}(p^{2},q^{2};r^{2})+a_{13}(p^{2},q^{2};r^{2})+a_{14}(p^{2},q^{2};r^{2})&=0\,,\label{eq:VVA_vector_WI_2}
\end{align}
which can be solved for the $a_{5}$ and $a_{6}$ formfactors as follows:
\begin{align}
a_{13}(p^{2},q^{2};r^{2})=&\quad\,\frac{1}{2}(p^{2}-q^{2})a_{9}(p^{2},q^{2};r^{2})-(p\hspace{-1pt}\cdot\hspace{-1pt}q)a_{10}(p^{2},q^{2};r^{2})\,,\label{eq:VVA_vector_WI_solution_1}\\
a_{14}(p^{2},q^{2};r^{2})=&-\frac{1}{2}(p^{2}+q^{2})a_{9}(p^{2},q^{2};r^{2})\,.\label{eq:VVA_vector_WI_solution_2}
\end{align}
Inserting the solutions \eqref{eq:VVA_vector_WI_solution_1}-\eqref{eq:VVA_vector_WI_solution_2} back into \eqref{eq:vva-definition_transversal_2} gives us the transversal structure made of four terms,
\begin{equation}
\big[\Pi_{VVA}^{(T)}(p,q;r)\big]_{\mu\nu\rho}=\sum_{i=9}^{12}a_{i}(p^{2},q^{2};r^{2})t^{(i)}_{\mu\nu\rho}(p,q;r)\,,\label{eq:vva-definition_transversal_3}
\end{equation}
with the tensors being
\begin{align}
t_{\mu\nu\rho}^{(9)}(p,q;r)&=p^{\mu}\varepsilon^{\nu\rho(p)(q)}+q^{\nu}\varepsilon^{\mu\rho(p)(q)}-p^{2}\varepsilon^{\mu\nu\rho(q)}-q^{2}\varepsilon^{\mu\nu\rho(p)}\,,\label{eq:vva-definition_transversal_3_tensor_1}\\
t_{\mu\nu\rho}^{(10)}(p,q;r)&=p^{\nu}\varepsilon^{\mu\rho(p)(q)}-q^{\mu}\varepsilon^{\nu\rho(p)(q)}-(p\hspace{-1pt}\cdot\hspace{-1pt}q)(\varepsilon^{\mu\nu\rho(p)}-\varepsilon^{\mu\nu\rho(q)})\,,\label{eq:vva-definition_transversal_3_tensor_2}\\
t_{\mu\nu\rho}^{(11)}(p,q;r)&=(p-q)^{\rho}\varepsilon^{\mu\nu(p)(q)}\,,\label{eq:vva-definition_transversal_3_tensor_3}\\
t_{\mu\nu\rho}^{(12)}(p,q;r)&=(p+q)^{\rho}\varepsilon^{\mu\nu(p)(q)}\,.\label{eq:vva-definition_transversal_3_tensor_4}
\end{align}
Finally, the axial Ward identity gives us the condition
\begin{equation}
a_{12}(p^{2},q^{2};r^{2})=\frac{p^{2}-q^{2}}{r^{2}}\big[a_{9}(p^{2},q^{2};r^{2})-a_{11}(p^{2},q^{2};r^{2})\big]-\frac{2(p\hspace{-1pt}\cdot\hspace{-1pt}q)}{r^{2}}a_{10}(p^{2},q^{2};r^{2})\,,\label{eq:VVA_axial_WI_solution}
\end{equation}
which not only reduces the number of formfactors, but also introduces kinematical zeros.

\subsection*{Result}
Substituting the solution \eqref{eq:VVA_axial_WI_solution} into \eqref{eq:vva-definition_transversal_3} leads, after some algebraical manipulations, to the final result for the transversal part of the $\langle VVA\rangle$ Green function in the form
\begin{align}
\big[\Pi_{VVA}^{(T)}&(p,q;r)\big]_{\mu\nu\rho}=\\
=&\quad\,a_{9}(p^{2},q^{2};r^{2})\big[t_{\mu\nu\rho}^{(9)}(p,q;r)+t_{\mu\nu\rho}^{(11)}(p,q;r)\big]\nonumber\\
&+a_{10}(p^{2},q^{2};r^{2})\bigg(t_{\mu\nu\rho}^{(10)}(p,q;r)-\frac{2(p\hspace{-1pt}\cdot\hspace{-1pt}q)}{r^{2}}t_{\mu\nu\rho}^{(12)}(p,q;r)\bigg)\nonumber\\
&+\big[a_{11}(p^{2},q^{2};r^{2})-a_{9}(p^{2},q^{2};r^{2})\big]\bigg(t_{\mu\nu\rho}^{(11)}(p,q;r)-\frac{p^{2}-q^{2}}{r^{2}}t_{\mu\nu\rho}^{(12)}(p,q;r)\bigg)\,.\nonumber
\end{align}

Finally, to obtain the transversal part of the $\langle VVA\rangle$ decomposition \eqref{eq:vva-definition_transversal}, it is now only necessary to denote
\begin{align}
\mathcal{F}_{VVA}(p^{2},q^{2};r^{2})&\equiv a_{10}(p^{2},q^{2};r^{2})\,,\\
\mathcal{G}_{VVA}(p^{2},q^{2};r^{2})&\equiv a_{11}(p^{2},q^{2};r^{2})-a_{9}(p^{2},q^{2};r^{2})\,,\\
\mathcal{H}_{VVA}(p^{2},q^{2};r^{2})&\equiv a_{9}(p^{2},q^{2};r^{2})\,,
\end{align}
and
\begin{align}
\mathcal{T}_{\mu\nu\rho}^{(1)}(p,q;r)&\equiv t_{\mu\nu\rho}^{(10)}(p,q;r)-\frac{2(p\hspace{-1pt}\cdot\hspace{-1pt}q)}{r^{2}}t_{\mu\nu\rho}^{(12)}(p,q;r)\,,\\
\mathcal{T}_{\mu\nu\rho}^{(2)}(p,q;r)&\equiv t_{\mu\nu\rho}^{(11)}(p,q;r)-\frac{p^{2}-q^{2}}{r^{2}}t_{\mu\nu\rho}^{(12)}(p,q;r)\,,\\
\mathcal{T}_{\mu\nu\rho}^{(3)}(p,q;r)&\equiv t_{\mu\nu\rho}^{(9)}(p,q;r)+t_{\mu\nu\rho}^{(11)}(p,q;r)\,.
\end{align}

After some manipulations, we indeed arrive at
\begin{align}
\mathcal{T}_{\mu\nu\rho}^{(1)}(p,q;r)&=p^{\nu}\varepsilon^{\mu\rho(p)(q)}-q^{\mu}\varepsilon^{\nu\rho(p)(q)}-\frac{p^{2}+q^{2}-r^{2}}{r^{2}}\bigg(\varepsilon^{\mu\nu(p)(q)}r^{\rho}-\frac{r^{2}}{2}\varepsilon^{\mu\nu\rho(p-q)}\bigg)\,,\\
\mathcal{T}_{\mu\nu\rho}^{(2)}(p,q;r)&=\varepsilon^{\mu\nu(p)(q)}(p-q)^{\rho}+\frac{p^{2}-q^{2}}{r^{2}}\varepsilon^{\mu\nu(p)(q)}r^{\rho}\,,\\
\mathcal{T}_{\mu\nu\rho}^{(3)}(p,q;r)&=p^{\nu}\varepsilon^{\mu\rho(p)(q)}+q^{\mu}\varepsilon^{\nu\rho(p)(q)}-\frac{p^{2}+q^{2}-r^{2}}{2}\varepsilon^{\mu\nu\rho(r)}\,,
\end{align}
by which we have finally derived the tensors \eqref{eq:VVA-transversal-tensor-1}-\eqref{eq:VVA-transversal-tensor-3}, with the following symmetry properties:
\begin{align}
\mathcal{T}_{\mu\nu\rho}^{(1)}(p,q;r)=&\quad\,\mathcal{T}_{\nu\mu\rho}^{(1)}(q,p;r)\,,\label{eq:VVA_tensor_1_symetrie}\\
\mathcal{T}_{\mu\nu\rho}^{(2)}(p,q;r)=&-\mathcal{T}_{\nu\mu\rho}^{(2)}(q,p;r)\,,\label{eq:VVA_tensor_2_symetrie}\\
\mathcal{T}_{\mu\nu\rho}^{(3)}(p,q;r)=&-\mathcal{T}_{\nu\mu\rho}^{(3)}(q,p;r)\,.\label{eq:VVA_tensor_3_symetrie}
\end{align}

\subsection{\texorpdfstring{$\langle AAA\rangle$}{} Green Function}
Similarly to the previous section, we will now construct the transversal part of the decomposition of the $\langle AAA\rangle$ Green function \eqref{eq:aaa-definition_longitudinal_and_transversal} such that the Bose symmetries \eqref{eq:AAA_bose_relation_1}-\eqref{eq:AAA_bose_relation_3} and the Ward identities \eqref{eq:AAA_ward_v2b} are satisfied. However, unlike in the case above, we will proceed with the construction in such a way that we will start with the requirement of the vanishing Ward identities.

\subsection*{Ward Identities}
As a first step let us determine the independent tensor structure. Using the Schouten identities \eqref{eq:schouten_1}-\eqref{eq:schouten_2}, we can get rid off two tensors out of the eight structures exactly as in the previous case of the $\langle VVA\rangle$. The general solution of the transversality conditions for the transverse part of the $\langle AAA\rangle$ correlator can be written as a linear combination of following three tensors
\begin{align}
t_{\mu\nu\rho}^{(13)}(p,q;r)=&-\frac{p^{\mu}\varepsilon^{\nu\rho(r)(s)}}{2p^{2}}+\frac{q^{\nu}\varepsilon^{\mu\rho(r)(s)}}{2q^{2}}+\frac{r^{\rho}\varepsilon^{\mu\nu(r)(s)}}{r^{2}}-\varepsilon^{\mu\nu\rho(s)}\,,\label{eq:AAA_tensor_structure_part_1_v2}\\
t_{\mu\nu\rho}^{(14)}(p,q;r)=&-\frac{p^{\mu}\varepsilon^{\nu\rho(r)(s)}}{2p^{2}}-\frac{q^{\nu}\varepsilon^{\mu\rho(r)(s)}}{2q^{2}}-\varepsilon^{\mu\nu\rho(r)}\,,\label{eq:AAA_tensor_structure_part_2_v2}\\
t_{\mu\nu\rho}^{(15)}(p,q;r)=&\quad\,\frac{1}{2}\bigg(\frac{r\hspace{-1pt}\cdot\hspace{-1pt}s}{r^{2}}r^{\rho}-s^{\rho}\bigg)\varepsilon^{\mu\nu(r)(s)}\,,\label{eq:AAA_tensor_structure_part_3_v2}
\end{align}
where we have introduced the momentum $s$
\begin{equation}
s\equiv p-q\,.
\end{equation}

\subsection*{Bose Symmetry}
Having the transversal tensors introduced above, let us make them subject to the simultaneous interchanges of the momenta and Lorentz indices in order to find out their behavior with respect to the Bose symmetries \eqref{eq:AAA_bose_relation_1}-\eqref{eq:AAA_bose_relation_3}. After some algebraic manipulations, we find out the following transformation properties. For $(p,\mu)\leftrightarrow (q,\nu)$ we have
\begin{align}
t_{\nu\mu\rho}^{(13)}(q,p;r)=&\quad\,t_{\mu\nu\rho}^{(13)}(p,q;r)\,,\label{eq:AAA_tensor_1_bose_1}\\
t_{\nu\mu\rho}^{(14)}(q,p;r)=&-t_{\mu\nu\rho}^{(14)}(p,q;r)\,,\label{eq:AAA_tensor_2_bose_1}\\
t_{\nu\mu\rho}^{(15)}(q,p;r)=&-t_{\mu\nu\rho}^{(15)}(p,q;r)\,,\label{eq:AAA_tensor_3_bose_1}
\end{align}
for $(p,\mu)\leftrightarrow (r,\rho)$ we find
\begin{align}
t_{\rho\nu\mu}^{(13)}(r,q;p)=&-\frac{1}{2}t_{\mu\nu\rho}^{(13)}(p,q;r)-\frac{3}{2}t_{\mu\nu\rho}^{(14)}(p,q;r)\,,\label{eq:AAA_tensor_1_bose_2}\\
t_{\rho\nu\mu}^{(14)}(r,q;p)=&-\frac{1}{2}t_{\mu\nu\rho}^{(13)}(p,q;r)+\frac{1}{2}t_{\mu\nu\rho}^{(14)}(p,q;r)\,,\label{eq:AAA_tensor_2_bose_2}\\
t_{\rho\nu\mu}^{(15)}(r,q;p)=&-t_{\mu\nu\rho}^{(15)}(p,q;r)-\frac{1}{2}(p^{2}-q^{2}-r^{2})t_{\mu\nu\rho}^{(13)}(p,q;r)\label{eq:AAA_tensor_3_bose_2}\\
&-\frac{1}{2}(p^{2}+3q^{2}-r^{2})t_{\mu\nu\rho}^{(14)}(p,q;r)\nonumber
\end{align}
and finally for $(q,\nu)\leftrightarrow (r,\rho)$ we obtain
\begin{align}
t_{\mu\rho\nu}^{(13)}(p,r;q)=&-\frac{1}{2}t_{\mu\nu\rho}^{(13)}(p,q;r)+\frac{3}{2}t_{\mu\nu\rho}^{(14)}(p,q;r)\,,\label{eq:AAA_tensor_1_bose_3}\\
t_{\mu\rho\nu}^{(14)}(p,r;q)=&\quad\,\frac{1}{2}t_{\mu\nu\rho}^{(13)}(p,q;r)+\frac{1}{2}t_{\mu\nu\rho}^{(14)}(p,q;r)\,,\label{eq:AAA_tensor_2_bose_3}\\
t_{\mu\rho\nu}^{(15)}(p,r;q)=&-t_{\mu\nu\rho}^{(15)}(p,q;r)-\frac{1}{2}(p^{2}-q^{2}+r^{2})t_{\mu\nu\rho}^{(13)}(p,q;r)\label{eq:AAA_tensor_3_bose_3}\\
&-\frac{1}{2}(3p^{2}+q^{2}-r^{2})t_{\mu\nu\rho}^{(14)}(p,q;r)\,.\nonumber
\end{align}

Based on the behavior \eqref{eq:AAA_tensor_1_bose_1}-\eqref{eq:AAA_tensor_3_bose_3} of the tensor structures \eqref{eq:AAA_tensor_structure_part_1_v2}-\eqref{eq:AAA_tensor_structure_part_3_v2}, we construct the following combinations of such tensors:
\begin{align}
\mathcal{T}_{\mu\nu\rho}^{(4)}(p,q;r)&\equiv(p^{2}+q^{2}-r^{2})t_{\mu\nu\rho}^{(14)}(p,q;r)+t_{\mu\nu\rho}^{(15)}(p,q;r)\,,\label{eq:AAA_tensor_Q}\\
\mathcal{T}_{\mu\nu\rho}^{(5)}(p,q;r)&\equiv(p^{2}-q^{2})t_{\mu\nu\rho}^{(13)}(p,q;r)-(p^{2}+q^{2}-2r^{2})t_{\mu\nu\rho}^{(14)}(p,q;r)\,,\label{eq:AAA_tensor_Q1}\\
\mathcal{T}_{\mu\nu\rho}^{(6)}(p,q;r)&\equiv\frac{1}{3}(p^{2}+q^{2}-2r^{2})t_{\mu\nu\rho}^{(13)}(p,q;r)+(p^{2}-q^{2})t_{\mu\nu\rho}^{(14)}(p,q;r)\,.\label{eq:AAA_tensor_Q2}
\end{align}
These are not only fully transversal but have also convenient symmetry properties. In detail, the tensors \eqref{eq:AAA_tensor_Q}-\eqref{eq:AAA_tensor_Q1} are antisymmetric upon the Bose symmetries, while the tensor \eqref{eq:AAA_tensor_Q2} is symmetric. Symbolically,
\begin{alignat}{4}
&\mathcal{T}_{\mu\nu\rho}^{(4)}(p,q;r)=-&&\mathcal{T}_{\nu\mu\rho}^{(4)}(q,p;r)=-&&\mathcal{T}_{\rho\nu\mu}^{(4)}(r,q;p)=-&&\mathcal{T}_{\mu\rho\nu}^{(4)}(p,r;q)\,,\label{eq:AAA_tensor_Q_bose}\\
&\mathcal{T}_{\mu\nu\rho}^{(5)}(p,q;r)=-&&\mathcal{T}_{\nu\mu\rho}^{(5)}(q,p;r)=-&&\mathcal{T}_{\rho\nu\mu}^{(5)}(r,q;p)=-&&\mathcal{T}_{\mu\rho\nu}^{(5)}(p,r;q)\,,\label{eq:AAA_tensor_Q1_bose}\\
&\mathcal{T}_{\mu\nu\rho}^{(6)}(p,q;r)=&&\mathcal{T}_{\nu\mu\rho}^{(6)}(q,p;r)=&&\mathcal{T}_{\rho\nu\mu}^{(6)}(r,q;p)=&&\mathcal{T}_{\mu\rho\nu}^{(6)}(p,r;q)\,.\label{eq:AAA_tensor_Q2_bose}
\end{alignat}

\subsection*{Result}
Then we simply construct the transversal part of the $\langle AAA\rangle$ Green function as
\begin{align}
\big[\Pi_{AAA}^{(T)}(p,q;r)\big]_{\mu\nu\rho}=&\quad\,\mathcal{F}_{AAA}(p^{2},q^{2};r^{2})\mathcal{T}_{\mu\nu\rho}^{(4)}(p,q;r)+\mathcal{G}_{AAA}(p^{2},q^{2};r^{2})\mathcal{T}_{\mu\nu\rho}^{(5)}(p,q;r)\label{eq:AAA_0}\\
&+\mathcal{H}_{AAA}(p^{2},q^{2};r^{2})\mathcal{T}_{\mu\nu\rho}^{(6)}(p,q;r)\,.\nonumber
\end{align}
Due to the behavior \eqref{eq:AAA_tensor_Q_bose}-\eqref{eq:AAA_tensor_Q2_bose} of the tensors \eqref{eq:AAA_tensor_Q}-\eqref{eq:AAA_tensor_Q2} under the Bose symmetries, the first two formfactors are antisymmetric under the Bose symmetries, while the third one is symmetric. See \eqref{eq:AAAformfactor1}-\eqref{eq:AAAformfactor3} for details.

%%%%%%%%%%%%%%%%%%%%%%%%%%%%%%%%%%%%%%%%%%%%%%%%%%%%%%%%%%%%%%%%%%%%%%%%%%%%%%%%%%%%%%%%%%%%%%%%%%%%%%%%%
%%%%%%%%%%%%%%%%%%%%%%%%%%%%%%%%%%%%%%%%%%%%%%%%%%%%%%%%%%%%%%%%%%%%%%%%%%%%%%%%%%%%%%%%%%%%%%%%%%%%%%%%%
%%% Section: On Decompositions of the AAV and VVV Green Functions
%%%%%%%%%%%%%%%%%%%%%%%%%%%%%%%%%%%%%%%%%%%%%%%%%%%%%%%%%%%%%%%%%%%%%%%%%%%%%%%%%%%%%%%%%%%%%%%%%%%%%%%%%
%%%%%%%%%%%%%%%%%%%%%%%%%%%%%%%%%%%%%%%%%%%%%%%%%%%%%%%%%%%%%%%%%%%%%%%%%%%%%%%%%%%%%%%%%%%%%%%%%%%%%%%%%

\section{On Decompositions of the \texorpdfstring{\boldmath$\langle AAV\rangle$}{} and \texorpdfstring{\boldmath$\langle VVV\rangle$}{} Green Functions}\label{sec:AAV_VVV_decompositions}
In this appendix, we get back to the $\langle AAV\rangle$ and $\langle VVV\rangle$ Green functions and present the derivation of their decompositions \eqref{eq:aav-definition}-\eqref{eq:vvv-definition_2}. 

\subsection*{Bose Symmetry}
Similarly as in the previous section, we start with the Bose symmetries of the Green functions in question.
The $\langle AAV\rangle$ Green function, as defined in \eqref{eq:aav-definition}, is invariant with respect to the interchange $(\mu,a,p)\leftrightarrow(\nu,b,q)$. Due to the antisymmetry of the flavor part, the Bose symmetry dictates:
\begin{equation}
\big[\Pi_{AAV}(p,q;r)\big]_{\mu\nu\rho}+\big[\Pi_{AAV}(q,p;r)\big]_{\nu\mu\rho}=0\,.\label{eq:AAV_bose_relation}
\end{equation}

Similarly for the $\langle VVV\rangle$ Green function, as defined in \eqref{eq:vvv-definition}, we have
\begin{align}
\big[\Pi_{VVV}(p,q;r)\big]_{\mu\nu\rho}+\big[\Pi_{VVV}(q,p;r)\big]_{\nu\mu\rho}&=0\,,\label{eq:VVV_bose_relation_1}\\
\big[\Pi_{VVV}(p,q;r)\big]_{\mu\nu\rho}+\big[\Pi_{VVV}(r,q;p)\big]_{\rho\nu\mu}&=0\,,\label{eq:VVV_bose_relation_2}\\
\big[\Pi_{VVV}(p,q;r)\big]_{\mu\nu\rho}+\big[\Pi_{VVV}(p,r;q)\big]_{\mu\rho\nu}&=0\,.\label{eq:VVV_bose_relation_3}
\end{align}

\subsection*{Ward Identities}
The Ward identities of the $\langle AAV\rangle$ and $\langle VVV\rangle$ Green functions have been presented in the main text, see \eqref{eq:AAV-Ward_kopie_1}-\eqref{eq:AAV-Ward_kopie_3} and \eqref{eq:VVV-Ward_kopie_1}-\eqref{eq:VVV-Ward_kopie_3}, respectively. After discarding the flavor part, we have for the Lorentz part of the $\langle AAV\rangle$ correlator the following Ward identities.
\begin{align}
p^{\mu}\big[\Pi_{AAV}&(p,q;r)\big]_{\mu\nu\rho}=\label{eq:AAV_ward_1_v2}\\
&=-\Pi_{AA}(q^{2})\big[q^{2}g_{\nu\rho}+(p+r)_{\nu}q_{\rho}\big]+\Pi_{VV}(r^{2})\big[r^{2}g_{\nu\rho}+r_{\nu}(p+q)_{\rho}\big]\,,\nonumber\\
q^{\nu}\big[\Pi_{AAV}&(p,q;r)\big]_{\mu\nu\rho}=\label{eq:AAV_ward_2_v2}\\
&=-\Pi_{VV}(r^{2})\big[r^{2}g_{\mu\rho}+r_{\mu}(p_{\rho}+q_{\rho})\big]+\Pi_{AA}(p^{2})\big[p^{2}g_{\mu\rho}+(q_{\mu}+r_{\mu})p_{\rho}\big]\,,\nonumber\\
r^{\rho}\big[\Pi_{AAV}&(p,q;r)\big]_{\mu\nu\rho}=\label{eq:AAV_ward_3_v2}\\
&=-\Pi_{AA}(p^{2})\big[p^{2}g_{\mu\nu}+(q_{\mu}+r_{\mu})p_{\nu}\big]+\Pi_{AA}(q^{2})\big[q^{2}g_{\mu\nu}+q_{\mu}(p_{\nu}+r_{\nu})\big]\,,\nonumber
\end{align}
written in this form anticipating the choice of the tensor basis below.
Similarly, for the Lorentz part of the $\langle VVV\rangle$:
\begin{align}
p^{\mu}\big[\Pi_{VVV}&(p,q;r)\big]_{\mu\nu\rho}=\label{eq:VVV_ward_1_v2}\\
&=-\Pi_{VV}(q^{2})\big[q^{2}g_{\nu\rho}+(p_{\nu}+r_{\nu})q_{\rho}\big]+\Pi_{VV}(r^{2})\big[r^{2}g_{\nu\rho}+r_{\nu}(p_{\rho}+q_{\rho})\big]\,,\nonumber\\
q^{\nu}\big[\Pi_{VVV}&(p,q;r)\big]_{\mu\nu\rho}=\label{eq:VVV_ward_2_v2}\\
&=-\Pi_{VV}(r^{2})\big[r^{2}g_{\mu\rho}+r_{\mu}(p_{\rho}+q_{\rho})\big]+\Pi_{VV}(p^{2})\big[p^{2}g_{\mu\rho}+(q_{\mu}+r_{\mu})p_{\rho}\big]\,,\nonumber\\
r^{\rho}\big[\Pi_{VVV}&(p,q;r)\big]_{\mu\nu\rho}=\label{eq:VVV_ward_3_v2}\\
&=-\Pi_{VV}(p^{2})\big[p^{2}g_{\mu\nu}+(q_{\mu}+r_{\mu})p_{\nu}\big]+\Pi_{VV}(q^{2})\big[q^{2}g_{\mu\nu}+q_{\mu}(p_{\nu}+r_{\nu})\big]\,.\nonumber
\end{align}

\subsection{\texorpdfstring{$\langle AAV\rangle$}{} Green Function}
Upon assuming the momentum conservation and using the trick of rewriting $p_{\mu}$, $q_{\nu}$ and $r_{\rho}$ as $-(q+r)_{\mu}$, $-(p+r)_{\nu}$ and $-(p+q)_{\rho}$, respectively, the Lorentz part of the $\langle AAV\rangle$ Green function \eqref{eq:aav-definition} can be written down as a linear combination of 14 tensors, i.e.
\begin{align}
\big[\Pi_{AAV}&(p,q;r)\big]_{\mu\nu\rho}=\label{eq:aav-definition_v2}\\
=&\quad\,c_{1}(p^{2},q^{2},r^{2})\,q_{\mu}p_{\nu}p_{\rho}+c_{2}(p^{2},q^{2},r^{2})\,q_{\mu}p_{\nu}q_{\rho}+c_{3}(p^{2},q^{2},r^{2})\,q_{\mu}r_{\nu}p_{\rho}\nonumber\\
&+c_{4}(p^{2},q^{2},r^{2})\,q_{\mu}r_{\nu}q_{\rho}+c_{5}(p^{2},q^{2},r^{2})\,r_{\mu}p_{\nu}p_{\rho}+c_{6}(p^{2},q^{2},r^{2})\,r_{\mu}p_{\nu}q_{\rho}\nonumber\\
&+c_{7}(p^{2},q^{2},r^{2})\,r_{\mu}r_{\nu}p_{\rho}+c_{8}(p^{2},q^{2},r^{2})\,r_{\mu}r_{\nu}q_{\rho}+c_{9}(p^{2},q^{2},r^{2})\,p_{\nu}g_{\mu\rho}\nonumber\\
&+c_{10}(p^{2},q^{2},r^{2})\,p_{\rho}g_{\mu\nu}+c_{11}(p^{2},q^{2},r^{2})\,q_{\mu}g_{\nu\rho}+c_{12}(p^{2},q^{2},r^{2})\,q_{\rho}g_{\mu\nu}\nonumber\\
&+c_{13}(p^{2},q^{2},r^{2})\,r_{\mu}g_{\nu\rho}+c_{14}(p^{2},q^{2},r^{2})\,r_{\nu}g_{\mu\rho}\,,\nonumber
\end{align}
to which we now apply the constraints that follow from the fundamental properties of the correlator, i.e. the Bose symmetry and Ward identities. We start with the former.

\subsection*{Bose Symmetry}
The respective Bose symmetry \eqref{eq:AAV_bose_relation} allows us to reduce the number of the formfactors from 14 to 7, since it leads to seven conditions for the formfactors:
\begin{align}
c_{2}(p^{2},q^{2},r^{2})&=-c_{1}(q^{2},p^{2},r^{2})\,,\\
c_{5}(p^{2},q^{2},r^{2})&=-c_{4}(q^{2},p^{2},r^{2})\,,\\
c_{6}(p^{2},q^{2},r^{2})&=-c_{3}(q^{2},p^{2},r^{2})\,,\\
c_{8}(p^{2},q^{2},r^{2})&=-c_{7}(q^{2},p^{2},r^{2})\,,\\
c_{11}(p^{2},q^{2},r^{2})&=-c_{9}(q^{2},p^{2},r^{2})\,,\\
c_{12}(p^{2},q^{2},r^{2})&=-c_{10}(q^{2},p^{2},r^{2})\,,\\
c_{14}(p^{2},q^{2},r^{2})&=-c_{13}(q^{2},p^{2},r^{2})\,.
\end{align}

We are thus allowed to rewrite the Lorentz structure \eqref{eq:aav-definition_v2} in a more compact form
\begin{align}
\big[\Pi_{AAV}&(p,q;r)\big]_{\mu\nu\rho}=\label{eq:aav_structure_bose}\\
=&\quad\,c_{1}(p^{2},q^{2},r^{2})q_{\mu}p_{\nu}p_{\rho}-c_{1}(q^{2},p^{2},r^{2})q_{\mu}p_{\nu}q_{\rho}+c_{3}(p^{2},q^{2},r^{2})q_{\mu}r_{\nu}p_{\rho}\nonumber\\
&-c_{3}(q^{2},p^{2},r^{2})r_{\mu}p_{\nu}q_{\rho}+c_{4}(p^{2},q^{2},r^{2})q_{\mu}r_{\nu}q_{\rho}-c_{4}(q^{2},p^{2},r^{2})r_{\mu}p_{\nu}p_{\rho}\nonumber\\
&+c_{7}(p^{2},q^{2},r^{2})r_{\mu}r_{\nu}p_{\rho}-c_{7}(q^{2},p^{2},r^{2})r_{\mu}r_{\nu}q_{\rho}+c_{9}(p^{2},q^{2},r^{2})p_{\nu}g_{\mu\rho}\nonumber\\
&-c_{9}(q^{2},p^{2},r^{2})q_{\mu}g_{\nu\rho}+c_{10}(p^{2},q^{2},r^{2})p_{\rho}g_{\mu\nu}-c_{10}(q^{2},p^{2},r^{2})q_{\rho}g_{\mu\nu}\nonumber\\
&+c_{13}(p^{2},q^{2},r^{2})r_{\mu}g_{\nu\rho}-c_{13}(q^{2},p^{2},r^{2})r_{\nu}g_{\mu\rho}\,,\nonumber
\end{align}
which is now Bose symmetrical. Now we need to make this structure satisfy the Ward identities.

\subsection*{Ward Identities}
Contracting the Lorentz structure \eqref{eq:aav_structure_bose} with the respective momenta leads to
\begin{align}
p^{\mu}\big[\Pi_{AAV}&(p,q;r)\big]_{\mu\nu\rho}=\label{eq:AAV_ward_1_v2_explicit}\\
=&\quad\,\big[(p\hspace{-1pt}\cdot\hspace{-1pt}q)c_{1}(p^{2},q^{2},r^{2})-(p\hspace{-1pt}\cdot\hspace{-1pt}r)c_{4}(q^{2},p^{2},r^{2})+c_{9}(p^{2},q^{2},r^{2})+c_{10}(p^{2},q^{2},r^{2})\big]p_{\nu}p_{\rho}\nonumber\\
&+\big[(p\hspace{-1pt}\cdot\hspace{-1pt}q)c_{3}(p^{2},q^{2},r^{2})+(p\hspace{-1pt}\cdot\hspace{-1pt}r)c_{7}(p^{2},q^{2},r^{2})-c_{13}(q^{2},p^{2},r^{2})\big]r_{\nu}p_{\rho}\nonumber\\
&-\big[(p\hspace{-1pt}\cdot\hspace{-1pt}q)c_{1}(q^{2},p^{2},r^{2})+(p\hspace{-1pt}\cdot\hspace{-1pt}r)c_{3}(q^{2},p^{2},r^{2})+c_{10}(q^{2},p^{2},r^{2})\big]p_{\nu}q_{\rho}\nonumber\\
&+\big[(p\hspace{-1pt}\cdot\hspace{-1pt}q)c_{4}(p^{2},q^{2},r^{2})-(p\hspace{-1pt}\cdot\hspace{-1pt}r)c_{7}(q^{2},p^{2},r^{2})\big]r_{\nu}q_{\rho}\nonumber\\
&-\big[(p\hspace{-1pt}\cdot\hspace{-1pt}q)c_{9}(q^{2},p^{2},r^{2})-(p\hspace{-1pt}\cdot\hspace{-1pt}r)c_{13}(p^{2},q^{2},r^{2})\big]g_{\nu\rho}\,,\nonumber\\
q^{\nu}\big[\Pi_{AAV}&(p,q;r)\big]_{\mu\nu\rho}=\label{eq:AAV_ward_2_v2_explicit}\\
=&-\big[(p\hspace{-1pt}\cdot\hspace{-1pt}q)c_{1}(q^{2},p^{2},r^{2})-(q\hspace{-1pt}\cdot\hspace{-1pt}r)c_{4}(p^{2},q^{2},r^{2})+c_{9}(q^{2},p^{2},r^{2})+c_{10}(q^{2},p^{2},r^{2})\big]q_{\mu}q_{\rho}\nonumber\\
&+\big[(p\hspace{-1pt}\cdot\hspace{-1pt}q)c_{1}(p^{2},q^{2},r^{2})+(q\hspace{-1pt}\cdot\hspace{-1pt}r)c_{3}(p^{2},q^{2},r^{2})+c_{10}(p^{2},q^{2},r^{2})\big]q_{\mu}p_{\rho}\nonumber\\
&-\big[(p\hspace{-1pt}\cdot\hspace{-1pt}q)c_{3}(q^{2},p^{2},r^{2})+(q\hspace{-1pt}\cdot\hspace{-1pt}r)c_{7}(q^{2},p^{2},r^{2})-c_{13}(p^{2},q^{2},r^{2})\big]r_{\mu}q_{\rho}\nonumber\\
&-\big[(p\hspace{-1pt}\cdot\hspace{-1pt}q)c_{4}(q^{2},p^{2},r^{2})-(q\hspace{-1pt}\cdot\hspace{-1pt}r)c_{7}(p^{2},q^{2},r^{2})\big]r_{\mu}p_{\rho}\nonumber\\
&+\big[(p\hspace{-1pt}\cdot\hspace{-1pt}q)c_{9}(p^{2},q^{2},r^{2})-(q\hspace{-1pt}\cdot\hspace{-1pt}r)c_{13}(q^{2},p^{2},r^{2})\big]g_{\mu\rho}\,,\nonumber\\
r^{\rho}\big[\Pi_{AAV}&(p,q;r)\big]_{\mu\nu\rho}=\label{eq:AAV_ward_3_v2_explicit}\\
=&\quad\,\big[(p\hspace{-1pt}\cdot\hspace{-1pt}r)c_{7}(p^{2},q^{2},r^{2})-(q\hspace{-1pt}\cdot\hspace{-1pt}r)c_{7}(q^{2},p^{2},r^{2})+c_{13}(p^{2},q^{2},r^{2})-c_{13}(q^{2},p^{2},r^{2})\big]r_{\mu}r_{\nu}\nonumber\\
&-\big[(q\hspace{-1pt}\cdot\hspace{-1pt}r)c_{3}(q^{2},p^{2},r^{2})+(p\hspace{-1pt}\cdot\hspace{-1pt}r)c_{4}(q^{2},p^{2},r^{2})-c_{9}(p^{2},q^{2},r^{2})\big]r_{\mu}p_{\nu}\nonumber\\
&+\big[(p\hspace{-1pt}\cdot\hspace{-1pt}r)c_{3}(p^{2},q^{2},r^{2})+(q\hspace{-1pt}\cdot\hspace{-1pt}r)c_{4}(p^{2},q^{2},r^{2})-c_{9}(q^{2},p^{2},r^{2})\big]q_{\mu}r_{\nu}\nonumber\\
&+\big[(p\hspace{-1pt}\cdot\hspace{-1pt}r)c_{1}(p^{2},q^{2},r^{2})-(q\hspace{-1pt}\cdot\hspace{-1pt}r)c_{1}(q^{2},p^{2},r^{2})\big]q_{\mu}p_{\nu}\nonumber\\
&+\big[(p\hspace{-1pt}\cdot\hspace{-1pt}r)c_{10}(p^{2},q^{2},r^{2})-(q\hspace{-1pt}\cdot\hspace{-1pt}r)c_{10}(q^{2},p^{2},r^{2})\big]g_{\mu\nu}\,,\nonumber
\end{align}
which gives us a system of three equations, to be compared with \eqref{eq:AAV_ward_1_v2}-\eqref{eq:AAV_ward_3_v2}.

Instead of solving the system all at once, it is easier to start with comparison of the two different forms of the first Ward identity. Then, upon comparing the scalar functions with the same tensors in \eqref{eq:AAV_ward_1_v2} and \eqref{eq:AAV_ward_1_v2_explicit}, we arrive at the following system of equations:
\begin{align}
(p\hspace{-1pt}\cdot\hspace{-1pt}q)c_{1}(p^{2},q^{2},r^{2})-(p\hspace{-1pt}\cdot\hspace{-1pt}r)c_{4}(q^{2},p^{2},r^{2})+c_{9}(p^{2},q^{2},r^{2})+c_{10}(p^{2},q^{2},r^{2})&=0\,,\\
(p\hspace{-1pt}\cdot\hspace{-1pt}q)c_{1}(q^{2},p^{2},r^{2})+(p\hspace{-1pt}\cdot\hspace{-1pt}r)c_{3}(q^{2},p^{2},r^{2})+c_{10}(q^{2},p^{2},r^{2})-\Pi_{AA}(q^{2})&=0\,,\\
(p\hspace{-1pt}\cdot\hspace{-1pt}q)c_{3}(p^{2},q^{2},r^{2})+(p\hspace{-1pt}\cdot\hspace{-1pt}r)c_{7}(p^{2},q^{2},r^{2})-c_{13}(q^{2},p^{2},r^{2})-\Pi_{VV}(r^{2})&=0\,,\\
(p\hspace{-1pt}\cdot\hspace{-1pt}q)c_{9}(q^{2},p^{2},r^{2})-(p\hspace{-1pt}\cdot\hspace{-1pt}r)c_{13}(p^{2},q^{2},r^{2})-q^{2}\Pi_{AA}(q^{2})+r^{2}\Pi_{VV}(r^{2})&=0\,,\\
(p\hspace{-1pt}\cdot\hspace{-1pt}q)c_{4}(p^{2},q^{2},r^{2})-(p\hspace{-1pt}\cdot\hspace{-1pt}r)c_{7}(q^{2},p^{2},r^{2})+\Pi_{AA}(q^{2})-\Pi_{VV}(r^{2})&=0\,.
\end{align}

The solution of this system can be found in such form that we are allowed to express the formfactors $c_{9}$, $c_{10}$ and $c_{13}$ in terms of $c_{1}$, $c_{3}$, $c_{4}$ and $c_{7}$. Specifically, we have
\begin{align}
c_{9}(p^{2},q^{2},r^{2})=&-\Pi_{AA}(p^{2})+(q\hspace{-1pt}\cdot\hspace{-1pt}r)c_{3}(p^{2},q^{2},r^{2})+(p\hspace{-1pt}\cdot\hspace{-1pt}r)c_{4}(q^{2},p^{2},r^{2})\,,\label{eq:AAV_ward_1_solution_1}\\
c_{10}(p^{2},q^{2},r^{2})=&\quad\,\Pi_{AA}(p^{2})-(p\hspace{-1pt}\cdot\hspace{-1pt}q)c_{1}(p^{2},q^{2},r^{2})-(q\hspace{-1pt}\cdot\hspace{-1pt}r)c_{3}(p^{2},q^{2},r^{2})\,,\label{eq:AAV_ward_1_solution_2}\\
c_{13}(p^{2},q^{2},r^{2})=&-\Pi_{VV}(r^{2})+(p\hspace{-1pt}\cdot\hspace{-1pt}q)c_{3}(q^{2},p^{2},r^{2})+(q\hspace{-1pt}\cdot\hspace{-1pt}r)c_{7}(q^{2},p^{2},r^{2})\,,\label{eq:AAV_ward_1_solution_3}
\end{align}
at the cost of having one additional condition for the formfactors $c_{4}$ and $c_{7}$,
\begin{equation}
(p\hspace{-1pt}\cdot\hspace{-1pt}q)c_{4}(p^{2},q^{2},r^{2})-(p\hspace{-1pt}\cdot\hspace{-1pt}r)c_{7}(q^{2},p^{2},r^{2})+\Pi_{AA}(q^{2})-\Pi_{VV}(r^{2})=0\,,\label{eq:AAV_ward_1_solution_4}
\end{equation}
which guarantees the fulfillment of the first Ward identity.

The solutions \eqref{eq:AAV_ward_1_solution_1}-\eqref{eq:AAV_ward_1_solution_3} can now be inserted into the second Ward identity \eqref{eq:AAV_ward_2_v2_explicit}, which is hereby solved identically, too, upon taking into account \eqref{eq:AAV_ward_1_solution_4}.

Finally, the previous solutions applied to the third Ward identity \eqref{eq:AAV_ward_3_v2_explicit} give us the requirement of the symmetry of the $c_{3}$ formfactor in the first two arguments, i.e.
\begin{equation}
c_{3}(p^{2},q^{2},r^{2})=c_{3}(q^{2},p^{2},r^{2})\,,\label{eq:AAV_ward_1_solution_5}
\end{equation}
and the condition
\begin{equation}
(p\hspace{-1pt}\cdot\hspace{-1pt}r)c_{1}(p^{2},q^{2},r^{2})-(q\hspace{-1pt}\cdot\hspace{-1pt}r)c_{1}(q^{2},p^{2},r^{2})+\Pi_{AA}(p^{2})-\Pi_{AA}(q^{2})=0\,.\label{eq:AAV_ward_1_solution_6}
\end{equation}

Now, the solutions \eqref{eq:AAV_ward_1_solution_1}-\eqref{eq:AAV_ward_1_solution_3} and \eqref{eq:AAV_ward_1_solution_5} could be inserted into the Lorentz structure \eqref{eq:aav_structure_bose}, which would give us the final result for the decomposition of the $\langle AAV\rangle$ correlator. However, keeping the additional constraints \eqref{eq:AAV_ward_1_solution_4} and \eqref{eq:AAV_ward_1_solution_6} would be unaesthetic. One may thus try to solve these contraints by carefully choosing the solution for the respective formfactors in such a way that these contraints are reduced into much easier relations, for example in the form of properties of the formfactors under the Bose symmetry.

Therefore, let us write down the general solution of the constraints \eqref{eq:AAV_ward_1_solution_4} and \eqref{eq:AAV_ward_1_solution_6} in the form
\begin{align}
c_{1}(p^{2},q^{2};r^{2})&=\frac{1}{(p\hspace{-1pt}\cdot\hspace{-1pt}r)}\Big[\Pi_{AA}(q^{2})+e_{1}(p^{2},q^{2};r^{2})\Big]\,,\label{eq:AAV_ward_1_solution_7}\\
c_{4}(p^{2},q^{2};r^{2})&=\frac{1}{(p\hspace{-1pt}\cdot\hspace{-1pt}q)}\Big[\Pi_{VV}(r^{2})+e_{4}(p^{2},q^{2};r^{2})\Big]\,,\label{eq:AAV_ward_1_solution_8}\\
c_{7}(p^{2},q^{2};r^{2})&=\frac{1}{(q\hspace{-1pt}\cdot\hspace{-1pt}r)}\Big[\Pi_{AA}(p^{2})+e_{7}(p^{2},q^{2};r^{2})\Big]\,,\label{eq:AAV_ward_1_solution_9}
\end{align}
where the newly introduced formfactors $b_i$ satisfy
\begin{align}
e_{1}(p^{2},q^{2};r^{2})&=e_{1}(q^{2},p^{2};r^{2})\,,\label{eq:AAV_ward_1_solution_10}\\
e_{7}(p^{2},q^{2};r^{2})&=e_{4}(q^{2},p^{2};r^{2})\,.\label{eq:AAV_ward_1_solution_11}
\end{align}
As one can see, the constraints \eqref{eq:AAV_ward_1_solution_4} and \eqref{eq:AAV_ward_1_solution_6} are thus reduced into simple relations \eqref{eq:AAV_ward_1_solution_10}-\eqref{eq:AAV_ward_1_solution_11}.

\subsection*{Result}
Applying the found solutions to the original Lorentz structure \eqref{eq:aav_structure_bose} allows us to write down its final form as
\begin{align}
\big[\Pi_{AAV}(p,q;r)\big]_{\mu\nu\rho}=&\quad\,e_{1}(p^{2},q^{2};r^{2})\tau_{\mu\nu\rho}^{(1)}(p,q;r)+c_{3}(p^{2},q^{2};r^{2})\tau_{\mu\nu\rho}^{(2)}(p,q;r)\label{eq:AAV_apendix_decomposition_result}\\
&+e_{4}(p^{2},q^{2};r^{2})\tau_{\nu\rho\mu}^{(1)}(q,r;p)-e_{4}(q^{2},p^{2};r^{2})\tau_{\mu\rho\nu}^{(1)}(p,r;q)\nonumber\\
&+\Pi_{AA}(p^{2})\tau_{\mu\nu\rho}^{(3)}(p,q;r)-\Pi_{AA}(q^{2})\tau_{\nu\mu\rho}^{(3)}(q,p;r)\nonumber\\
&-\Pi_{VV}(r^{2})\tau_{\rho\nu\mu}^{(3)}(r,q;p)\,,\nonumber
\end{align}
where we have used the following tensors:
\begin{align}
\tau_{\mu\nu\rho}^{(1)}(p,q;r)=&-\frac{\big[(p\hspace{-1pt}\cdot\hspace{-1pt}q)g_{\mu\nu}-q_{\mu}p_{\nu}\big]\big[(q\hspace{-1pt}\cdot\hspace{-1pt}r)p_{\rho}-(p\hspace{-1pt}\cdot\hspace{-1pt}r)q_{\rho}\big]}{(p\hspace{-1pt}\cdot\hspace{-1pt}r)(q\hspace{-1pt}\cdot\hspace{-1pt}r)}\,,\label{eq:AAV_apendix_tensor_1}\\
\tau_{\mu\nu\rho}^{(2)}(p,q;r)=&-q_{\mu}\big[(p\hspace{-1pt}\cdot\hspace{-1pt}r)g_{\nu\rho}-r_{\nu}p_{\rho}\big]+g_{\mu\nu}\big[(p\hspace{-1pt}\cdot\hspace{-1pt}r)q_{\rho}-(q\hspace{-1pt}\cdot\hspace{-1pt}r)p_{\rho}\big]\label{eq:AAV_apendix_tensor_2}\\
&+r_{\mu}\big[(p\hspace{-1pt}\cdot\hspace{-1pt}q)g_{\nu\rho}-p_{\nu}q_{\rho}\big]+g_{\mu\rho}\big[(q\hspace{-1pt}\cdot\hspace{-1pt}r)p_{\nu}-(p\hspace{-1pt}\cdot\hspace{-1pt}q)r_{\nu}\big]\,,\nonumber\\
\tau_{\mu\nu\rho}^{(3)}(p,q;r)=&-\frac{1}{q\hspace{-1pt}\cdot\hspace{-1pt}r}\Big(r_{\nu}\big[(p\hspace{-1pt}\cdot\hspace{-1pt}r)g_{\mu\rho}-r_{\mu}p_{\rho}\big]-q_{\rho}\big[(p\hspace{-1pt}\cdot\hspace{-1pt}q)g_{\mu\nu}-q_{\mu}p_{\nu}\big]\Big)\label{eq:AAV_apendix_tensor_3}\\
&-p_{\nu}g_{\mu\rho}+p_{\rho}g_{\mu\nu}\,.\nonumber
\end{align}

We remind the reader that the decomposition \eqref{eq:AAV_apendix_decomposition_result} possesses the symmetry constraints of its formfactors in the forms of \eqref{eq:AAV_ward_1_solution_5} and \eqref{eq:AAV_ward_1_solution_10}.

Finally, we note that to obtain the $\langle AAV\rangle$ decomposition \eqref{eq:aav-definition}, it is sufficient to redefine the tensors and formfactors as follows:
\begin{align}
\mathcal{T}_{\mu\nu\rho}^{(7)}(p,q;r)&=\tau_{\mu\nu\rho}^{(3)}(p,q;r)\,,\\
\mathcal{T}_{\mu\nu\rho}^{(8)}(p,q;r)&=\tau_{\mu\nu\rho}^{(1)}(p,q;r)\,,\\
\mathcal{T}_{\mu\nu\rho}^{(9)}(p,q;r)&=\tau_{\mu\nu\rho}^{(2)}(p,q;r)\,,
\end{align}
and
\begin{align}
\mathcal{F}_{AAV}(p^{2},q^{2};r^{2})&\equiv e_{1}(p^{2},q^{2};r^{2})\,,\\
\mathcal{G}_{AAV}(p^{2},q^{2};r^{2})&\equiv c_{3}(p^{2},q^{2};r^{2})\,,\\
\mathcal{H}_{AAV}(p^{2},q^{2};r^{2})&\equiv e_{4}(p^{2},q^{2};r^{2})\,.
\end{align}

\subsection{\texorpdfstring{$\langle VVV\rangle$}{} Green Function}
Since the first Bose symmetry \eqref{eq:VVV_bose_relation_1} is equivalent to the one of the $\langle AAV\rangle$ correlator, we can start with the result \eqref{eq:aav_structure_bose}, with the formfactors $c_{i}$'s replaced for $d_{i}$'s:
\begin{align}
\big[\Pi_{VVV}&(p,q;r)\big]_{\mu\nu\rho}=\label{eq:vvv_structure_bose_1}\\
=&\quad\,d_{1}(p^{2},q^{2},r^{2})q_{\mu}p_{\nu}p_{\rho}-d_{1}(q^{2},p^{2},r^{2})q_{\mu}p_{\nu}q_{\rho}+d_{3}(p^{2},q^{2},r^{2})q_{\mu}r_{\nu}p_{\rho}\nonumber\\
&-d_{3}(q^{2},p^{2},r^{2})r_{\mu}p_{\nu}q_{\rho}+d_{4}(p^{2},q^{2},r^{2})q_{\mu}r_{\nu}q_{\rho}-d_{4}(q^{2},p^{2},r^{2})r_{\mu}p_{\nu}p_{\rho}\nonumber\\
&+d_{7}(p^{2},q^{2},r^{2})r_{\mu}r_{\nu}p_{\rho}-d_{7}(q^{2},p^{2},r^{2})r_{\mu}r_{\nu}q_{\rho}+d_{9}(p^{2},q^{2},r^{2})p_{\nu}g_{\mu\rho}\nonumber\\
&-d_{9}(q^{2},p^{2},r^{2})q_{\mu}g_{\nu\rho}+d_{10}(p^{2},q^{2},r^{2})p_{\rho}g_{\mu\nu}-d_{10}(q^{2},p^{2},r^{2})q_{\rho}g_{\mu\nu}\nonumber\\
&+d_{13}(p^{2},q^{2},r^{2})r_{\mu}g_{\nu\rho}-d_{13}(q^{2},p^{2},r^{2})r_{\nu}g_{\mu\rho}\,,\nonumber
\end{align}
which has already the appropriate symmetry with respect to the first two arguments.

\subsection*{Bose Symmetry}
Using similar manipulations, as in the previous section, applied to the structure \eqref{eq:vvv_structure_bose_1} in order to satisfy the additional Bose symmetries \eqref{eq:VVV_bose_relation_2} and \eqref{eq:VVV_bose_relation_3}, allows us to rewrite the formfactors $d_{4}$, $d_{7}$ in terms of $d_{1}$ and $d_{10}$, $d_{13}$ in terms of $d_{9}$. On top of that, it also allows us to reduce the number of $d_{3}$ formfactors due to its symmetries. Specifically, the following properties of the respective formfactors can be found:
\begin{alignat}{3}
d_{1}(p^{2},q^{2};r^{2})&=&&d_{4}(r^{2},p^{2};q^{2})&&=d_{7}(q^{2},r^{2};p^{2})\,,\label{eq:VVV_bose_solution_1}\\
d_{9}(p^{2},q^{2};r^{2})&=-&&d_{10}(p^{2},r^{2};q^{2})&&=d_{13}(q^{2},r^{2};p^{2})\label{eq:VVV_bose_solution_2}\\
d_{3}(p^{2},q^{2};r^{2})&=&&d_{3}(r^{2},p^{2};q^{2})&&=d_{3}(q^{2},r^{2};p^{2})\,.\label{eq:VVV_bose_solution_3}
\end{alignat}

Inserting these relations into \eqref{eq:vvv_structure_bose_1} gives us the Lorentz part of the $\langle VVV\rangle$ correlator,
\begin{align}
\big[\Pi_{VVV}&(p,q;r)\big]_{\mu\nu\rho}=\label{eq:vvv_structure_bose_final}\\
=&\quad\,d_{1}(p^{2},q^{2},r^{2})q_{\mu}p_{\nu}p_{\rho}-d_{1}(p^{2},r^{2},q^{2})r_{\mu}p_{\nu}p_{\rho}+d_{1}(r^{2},p^{2},q^{2})r_{\mu}r_{\nu}p_{\rho}\nonumber\\
&-d_{1}(q^{2},p^{2},r^{2})q_{\mu}p_{\nu}q_{\rho}+d_{1}(q^{2},r^{2},p^{2})q_{\mu}r_{\nu}q_{\rho}-d_{1}(r^{2},q^{2},p^{2})r_{\mu}r_{\nu}q_{\rho}\nonumber\\
&+d_{3}(p^{2},q^{2},r^{2})q_{\mu}r_{\nu}p_{\rho}-d_{3}(q^{2},p^{2},r^{2})r_{\mu}p_{\nu}q_{\rho}-d_{9}(q^{2},p^{2},r^{2})q_{\mu}g_{\nu\rho}\nonumber\\
&+d_{9}(r^{2},p^{2},q^{2})r_{\mu}g_{\nu\rho}+d_{9}(p^{2},q^{2},r^{2})p_{\nu}g_{\mu\rho}-d_{9}(r^{2},q^{2},p^{2})r_{\nu}g_{\mu\rho}\nonumber\\
&-d_{9}(p^{2},r^{2},q^{2})p_{\rho}g_{\mu\nu}+d_{9}(q^{2},r^{2},p^{2})q_{\rho}g_{\mu\nu}\,,\nonumber
\end{align}
which satisfies all three Bose symmetries, upon taking the condition \eqref{eq:VVV_bose_solution_3} into account.

\subsection*{Ward Identities}
Now we perform the contractions of \eqref{eq:vvv_structure_bose_final} in order to make this structure satisfy the Ward identities. We have
\begin{align}
p^{\mu}\big[\Pi_{VVV}&(p,q;r)\big]_{\mu\nu\rho}=\label{eq:VVV_ward_1_v2_explicit}\\
=&\quad\,\big[(p\hspace{-1pt}\cdot\hspace{-1pt}q)d_{1}(p^{2},q^{2},r^{2})-(p\hspace{-1pt}\cdot\hspace{-1pt}r)d_{1}(p^{2},r^{2},q^{2})+d_{9}(p^{2},q^{2},r^{2})-d_{9}(p^{2},r^{2},q^{2})\big]p_{\nu}p_{\rho}\nonumber\\
&+\big[(p\hspace{-1pt}\cdot\hspace{-1pt}r)d_{1}(r^{2},p^{2},q^{2})+(p\hspace{-1pt}\cdot\hspace{-1pt}q)d_{3}(p^{2},q^{2},r^{2})-d_{9}(r^{2},q^{2},p^{2})\big]r_{\nu}p_{\rho}\nonumber\\
&-\big[(p\hspace{-1pt}\cdot\hspace{-1pt}q)d_{1}(q^{2},p^{2},r^{2})+(p\hspace{-1pt}\cdot\hspace{-1pt}r)d_{3}(q^{2},p^{2},r^{2})-d_{9}(q^{2},r^{2},p^{2})\big]p_{\nu}q_{\rho}\nonumber\\
&+\big[(p\hspace{-1pt}\cdot\hspace{-1pt}q)d_{1}(q^{2},r^{2},p^{2})-(p\hspace{-1pt}\cdot\hspace{-1pt}r)d_{1}(r^{2},q^{2},p^{2})\big]r_{\nu}q_{\rho}\nonumber\\
&-\big[(p\hspace{-1pt}\cdot\hspace{-1pt}q)d_{9}(q^{2},p^{2},r^{2})-(p\hspace{-1pt}\cdot\hspace{-1pt}r)d_{9}(r^{2},p^{2},q^{2})\big]g_{\nu\rho}\,,\nonumber\\
q^{\nu}\big[\Pi_{VVV}&(p,q;r)\big]_{\mu\nu\rho}=\label{eq:VVV_ward_2_v2_explicit}\\
=&-\big[(p\hspace{-1pt}\cdot\hspace{-1pt}q)d_{1}(q^{2},p^{2},r^{2})-(q\hspace{-1pt}\cdot\hspace{-1pt}r)d_{1}(q^{2},r^{2},p^{2})+d_{9}(q^{2},p^{2},r^{2})-d_{9}(q^{2},r^{2},p^{2})\big]q_{\mu}q_{\rho}\nonumber\\
&+\big[(p\hspace{-1pt}\cdot\hspace{-1pt}q)d_{1}(p^{2},q^{2},r^{2})+(q\hspace{-1pt}\cdot\hspace{-1pt}r)d_{3}(p^{2},q^{2},r^{2})-d_{9}(p^{2},r^{2},q^{2})\big]q_{\mu}p_{\rho}\nonumber\\
&-\big[(q\hspace{-1pt}\cdot\hspace{-1pt}r)d_{1}(r^{2},q^{2},p^{2})+(p\hspace{-1pt}\cdot\hspace{-1pt}q)d_{3}(q^{2},p^{2},r^{2})-d_{9}(r^{2},p^{2},q^{2})\big]r_{\mu}q_{\rho}\nonumber\\
&-\big[(p\hspace{-1pt}\cdot\hspace{-1pt}q)d_{1}(p^{2},r^{2},q^{2})-(q\hspace{-1pt}\cdot\hspace{-1pt}r)d_{1}(r^{2},p^{2},q^{2})\big]r_{\mu}p_{\rho}\nonumber\\
&+\big[(p\hspace{-1pt}\cdot\hspace{-1pt}q)d_{9}(p^{2},q^{2},r^{2})-(q\hspace{-1pt}\cdot\hspace{-1pt}r)d_{9}(r^{2},q^{2},p^{2})\big]g_{\mu\rho}\,,\nonumber\\
r^{\rho}\big[\Pi_{VVV}&(p,q;r)\big]_{\mu\nu\rho}=\label{eq:VVV_ward_3_v2_explicit}\\
=&\quad\,\big[(p\hspace{-1pt}\cdot\hspace{-1pt}r)d_{1}(r^{2},p^{2},q^{2})-(q\hspace{-1pt}\cdot\hspace{-1pt}r)d_{1}(r^{2},q^{2},p^{2})+d_{9}(r^{2},p^{2},q^{2})-d_{9}(r^{2},q^{2},p^{2})\big]r_{\mu}r_{\nu}\nonumber\\
&-\big[(p\hspace{-1pt}\cdot\hspace{-1pt}r)d_{1}(p^{2},r^{2},q^{2})+(q\hspace{-1pt}\cdot\hspace{-1pt}r)d_{3}(q^{2},p^{2},r^{2})-d_{9}(p^{2},q^{2},r^{2})\big]r_{\mu}p_{\nu}\nonumber\\
&+\big[(q\hspace{-1pt}\cdot\hspace{-1pt}r)d_{1}(q^{2},r^{2},p^{2})+(p\hspace{-1pt}\cdot\hspace{-1pt}r)d_{3}(p^{2},q^{2},r^{2})-d_{9}(q^{2},p^{2},r^{2})\big]q_{\mu}r_{\nu}\nonumber\\
&+\big[(p\hspace{-1pt}\cdot\hspace{-1pt}r)d_{1}(p^{2},q^{2},r^{2})-(q\hspace{-1pt}\cdot\hspace{-1pt}r)d_{1}(q^{2},p^{2},r^{2})\big]q_{\mu}p_{\nu}\nonumber\\
&-\big[(p\hspace{-1pt}\cdot\hspace{-1pt}r)d_{9}(p^{2},r^{2},q^{2})-(q\hspace{-1pt}\cdot\hspace{-1pt}r)d_{9}(q^{2},r^{2},p^{2})\big]g_{\mu\nu}\,.\nonumber
\end{align}

As in the previous case, we start with solving the first Ward identity, i.e. we compare \eqref{eq:VVV_ward_1_v2_explicit} with \eqref{eq:VVV_ward_1_v2}. This leads to the system of five equations:
\begin{align}
(p\hspace{-1pt}\cdot\hspace{-1pt}q)d_{1}(p^{2},q^{2};r^{2})-(p\hspace{-1pt}\cdot\hspace{-1pt}r)d_{1}(p^{2},r^{2};q^{2})+d_{9}(p^{2},q^{2};r^{2})-d_{9}(p^{2},r^{2};q^{2})&=0\,,\\
(p\hspace{-1pt}\cdot\hspace{-1pt}q)d_{9}(q^{2},p^{2};r^{2})-(p\hspace{-1pt}\cdot\hspace{-1pt}r)d_{9}(r^{2},p^{2};q^{2})-q^{2}\Pi_{VV}(r^{2})+r^{2}\Pi_{VV}(r^{2})&=0\,,\\
(p\hspace{-1pt}\cdot\hspace{-1pt}q)d_{1}(q^{2},p^{2};r^{2})+(p\hspace{-1pt}\cdot\hspace{-1pt}r)d_{3}(q^{2},p^{2};r^{2})-d_{9}(q^{2},r^{2};p^{2})-\Pi_{VV}(q^{2})&=0\,,\\
(p\hspace{-1pt}\cdot\hspace{-1pt}r)d_{1}(r^{2},p^{2};q^{2})+(p\hspace{-1pt}\cdot\hspace{-1pt}q)d_{3}(p^{2},q^{2};r^{2})-d_{9}(r^{2},q^{2};p^{2})-\Pi_{VV}(r^{2})&=0\,,\\
(p\hspace{-1pt}\cdot\hspace{-1pt}q)d_{1}(q^{2},r^{2};p^{2})-(p\hspace{-1pt}\cdot\hspace{-1pt}r)d_{1}(r^{2},q^{2};p^{2})+\Pi_{VV}(q^{2})-\Pi_{VV}(r^{2})&=0\,.
\end{align}

Solving this system leads to the expression for the formfactor $d_{9}$ in terms of $d_{1}$ and $d_{3}$,
\begin{equation}
d_{9}(p^{2},q^{2};r^{2})=(p\hspace{-1pt}\cdot\hspace{-1pt}r)d_{1}(p^{2},r^{2};q^{2})+(q\hspace{-1pt}\cdot\hspace{-1pt}r)d_{3}(p^{2},r^{2};q^{2})-\Pi_{VV}(p^{2})\,,\label{eq:VVV_ward_1_solution_1}
\end{equation}
and to the fact that $d_{3}$ is completely symmetrical. On the other hand, solving the system above gives us also the condition
\begin{equation}
(p\hspace{-1pt}\cdot\hspace{-1pt}r)d_{1}(p^{2},q^{2};r^{2})-(q\hspace{-1pt}\cdot\hspace{-1pt}r)d_{1}(q^{2},p^{2};r^{2})+\Pi_{VV}(p^{2})-\Pi_{VV}(q^{2})=0\,,\label{eq:VVV_ward_1_solution_2}
\end{equation}
which is somewhat equivalent to the condition \eqref{eq:AAV_ward_1_solution_6} in the case of the $\langle AAV\rangle$ correlator.

As it turns out, the solution \eqref{eq:VVV_ward_1_solution_1} and the requirement \eqref{eq:VVV_ward_1_solution_2}, together with the fully symmetric formfactor $d_{3}$, solves also the second and the third Ward identity.

A general solution of \eqref{eq:VVV_ward_1_solution_2} can be written as
\begin{equation}
d_{1}(p^{2},q^{2};r^{2})=\frac{1}{p\hspace{-1pt}\cdot\hspace{-1pt}r}\Big[\Pi_{VV}(q^{2})+f_{1}(p^{2},q^{2};r^{2})\Big]\,,
\end{equation}
where the new formfactor satisfies
\begin{equation}
f_{1}(p^{2},q^{2};r^{2})=f_{1}(q^{2},p^{2};r^{2})\,.
\end{equation}

\subsection*{Result}
Applying the solutions above into \eqref{eq:vvv_structure_bose_final} gives us the final form of the Lorentz part of the $\langle VVV\rangle$ correlator as follows:
\begin{align}
\big[\Pi_{VVV}(p,q;r)\big]_{\mu\nu\rho}=&\quad\,f_{1}(p^{2},q^{2};r^{2})\tau_{\mu\nu\rho}^{(1)}(p,q;r)+f_{1}(r^{2},p^{2};q^{2})\tau_{\rho\mu\nu}^{(1)}(r,p;q)\label{eq:VVV_apendix_decomposition_result}\\
&+f_{1}(q^{2},r^{2};p^{2})\tau_{\nu\rho\mu}^{(1)}(q,r;p)+d_{3}(p^{2},q^{2};r^{2})\tau_{\mu\nu\rho}^{(2)}(p,q;r)\nonumber\\
&+\Pi_{VV}(p^{2})\tau_{\mu\nu\rho}^{(3)}(p,q;r)-\Pi_{VV}(q^{2})\tau_{\nu\mu\rho}^{(3)}(q,p;r)\nonumber\\
&-\Pi_{VV}(r^{2})\tau_{\rho\nu\mu}^{(3)}(r,q;p)\,,\nonumber
\end{align}
where the individual tensors are given as \eqref{eq:AAV_apendix_tensor_1}-\eqref{eq:AAV_apendix_tensor_3}. As we have already mentioned, the formfactor $f_1$ is symmetric with respect to the first two arguments and $d_3$ is totally symmetric.

To obtain the $\langle VVV\rangle$ decomposition \eqref{eq:vvv-definition}, it is sufficient to redefine the tensors as follows:
\begin{align}
\mathcal{F}_{VVV}(p^{2},q^{2};r^{2})&\equiv f_{1}(p^{2},q^{2};r^{2})\,,\\
\mathcal{G}_{VVV}(p^{2},q^{2};r^{2})&\equiv d_{3}(p^{2},q^{2};r^{2})\,.
\end{align}

%%%%%%%%%%%%%%%%%%%%%%%%%%%%%%%%%%%%%%%%%%%%%%%%%%%%%%%%%%%%%%%%%%%%%%%%%%%%%%%%%%%%%%%%%%%%%%%%%%%%%%%%%
%%%%%%%%%%%%%%%%%%%%%%%%%%%%%%%%%%%%%%%%%%%%%%%%%%%%%%%%%%%%%%%%%%%%%%%%%%%%%%%%%%%%%%%%%%%%%%%%%%%%%%%%%
%%% Acknowledgments
%%%%%%%%%%%%%%%%%%%%%%%%%%%%%%%%%%%%%%%%%%%%%%%%%%%%%%%%%%%%%%%%%%%%%%%%%%%%%%%%%%%%%%%%%%%%%%%%%%%%%%%%%
%%%%%%%%%%%%%%%%%%%%%%%%%%%%%%%%%%%%%%%%%%%%%%%%%%%%%%%%%%%%%%%%%%%%%%%%%%%%%%%%%%%%%%%%%%%%%%%%%%%%%%%%%

\acknowledgments

The authors gratefully acknowledge useful discussions with Marc Knecht, especially during the stay of one of the authors (TK) at Centre de Physique Th\'{e}orique in Marseille. Further, TK appreciates correspondence with Hiren H. Patel, who kindly provided advices regarding \packageX. TK is also grateful to Tom G. Steele for pointing out an existence of the reference \cite{Elias:1987ac} and for an interesting correspondence. Our work was financially supported by The Czech Science Foundation (project GA\v{C}R no.~18-17224S).

%%%%%%%%%%%%%%%%%%%%%%%%%%%%%%%%%%%%%%%%%%%%%%%%%%%%%%%%%%%%%%%%%%%%%%%%%%%%%%%%%%%%%%%%%%%%%%%%%%%%%%%%%
%%%%%%%%%%%%%%%%%%%%%%%%%%%%%%%%%%%%%%%%%%%%%%%%%%%%%%%%%%%%%%%%%%%%%%%%%%%%%%%%%%%%%%%%%%%%%%%%%%%%%%%%%
%%% Bibliography
%%%%%%%%%%%%%%%%%%%%%%%%%%%%%%%%%%%%%%%%%%%%%%%%%%%%%%%%%%%%%%%%%%%%%%%%%%%%%%%%%%%%%%%%%%%%%%%%%%%%%%%%%
%%%%%%%%%%%%%%%%%%%%%%%%%%%%%%%%%%%%%%%%%%%%%%%%%%%%%%%%%%%%%%%%%%%%%%%%%%%%%%%%%%%%%%%%%%%%%%%%%%%%%%%%%

\end{document}